\documentclass[11pt]{article}
\pdfoutput=1  
\usepackage{%
  abstract,
  amsmath,
  amssymb,
  amsthm,
  array,
  color,
  comment,
  extarrows,
  latexsym,
  mathrsfs,
  mathtools,
  multirow,
  nccmath,
  rotating,
  setspace,
  slashed,
  textcomp,
  titlesec,
  upgreek,
}

\usepackage[sorting=none,backend=biber,sortcites,citestyle=numeric-comp]{biblatex}
\DeclareFieldFormat[misc]{title}{%
  {\mkbibquote{#1}}
}

\let\oldprintbibliography\printbibliography%
\renewcommand{\printbibliography}{\clearpage \oldprintbibliography}

\usepackage[pdfusetitle]{hyperref}
\hypersetup{
  linktocpage=true,
  breaklinks=true,
  pdfauthor={Author 1 and Author 2},
}

\makeatletter
\renewcommand{\sectionautorefname}{$\S$\@gobble}
\renewcommand{\subsectionautorefname}{$\S$\@gobble}
\renewcommand{\subsubsectionautorefname}{$\S$\@gobble}
\makeatother

\usepackage[multiple]{footmisc}
\usepackage[auth-lg,affil-it]{authblk}

\numberwithin{equation}{section}

\titleformat*{\section}{\bfseries\large}

\usepackage[a4paper,
textwidth=165mm,
tmargin=25mm,
bmargin=40mm,
]{geometry}

\allowdisplaybreaks[2]

\usepackage{%
  adjustbox, 
  graphicx,
  tikz,      
  tikz-cd,   
  subcaption
}

\usetikzlibrary{
  arrows,
  arrows.meta,
  automata,
  calc,
  decorations.pathmorphing,
  patterns.meta,
  positioning,
  shapes.misc
}

\newcommand{\dv}[2]{\frac{d #1}{d #2}}
\newcommand{\Dv}[2]{\frac{D #1}{D #2}}
\newcommand{\pdv}[2]{\frac{\partial{#1}}{\partial{#2}}}
\newcommand{\longto}{\longrightarrow}

\newcommand{\subtitle}[1]{\bigskip\noindent\textit{#1}\vspace*{0.5em}}

\makeatletter
\newcommand{\@saveenv}[2]{
  #2
  \protected@write\@mainaux{}{%
    \string\@savedenv{#1}{%
      \unexpanded{\unexpanded{#2}}%
    }%
  }%
}
\newcommand{\@savedenv}[2]{%
  \global\@namedef{SAVEDENV@#1}{#2}%
}
\newcommand{\@repeatenv}[1]{%
  \ifcsname SAVEDENV@#1\endcsname
    \@nameuse{SAVEDENV@#1}%
  \else
    \texttt{@saveenv Error}
    \GenericError{\\@saveenv Error}{%
      Labeled env not saved: `#1'%
    }{%
      See definition of \\@saveenv in the preamble.%
    }{%
      Usage: \\@saveenv{\#1}{\#2}, where \#1 is the label and \#2 is the math.%
    }%
    \notag%
  \fi
}
\newcommand{\saveboxed}[2]{%
  \@saveenv{#1}{\boxed{#2}}%
}
\newcommand{\repeatboxed}[1]{%
  \@repeatenv{#1}%
}
\newcommand{\savefig}[2]{%
  \@saveenv{#1}{#2}%
}
\newcommand{\repeatfig}[1]{%
  \@repeatenv{#1}%
}
\makeatother

\newcommand{\C}{\mathbb{C}}

\newcommand{\R}{\mathbb{R}}

\newcommand{\HH}{\mathbb{H}}
\newcommand{\OO}{\mathbb{O}}

\newcommand{\Tr}{\mathrm{Tr}}

\renewcommand{\Re}{\text{Re}}
\renewcommand{\Im}{\text{Im}}

\title{%
  \texorpdfstring{\vspace{1.5cm}}{}
  Topological 5d \texorpdfstring{$\mathcal{N} = 2$}{N = 2}  Gauge Theories: Mirror Symmetry and
  \texorpdfstring{\bigskip \\}{}
   Langlands Duality of \texorpdfstring{$A_\infty$}{A-infinity}-categories of Floer Homologies
  \texorpdfstring{\vspace{2.0cm}}{}
}%

\author{Arif Er\thanks{
    Email: \href{mailto:arif.er@u.nus.edu}{arif.er@u.nus.edu}
  }
}%

\author{Meng-Chwan Tan\thanks{
    Email: \href{mailto:mctan@nus.edu.sg}{mctan@nus.edu.sg}
  }
}%

\affil{%
  Department of Physics \\ \medskip%
  National University of Singapore \\ \medskip%
  2 Science Drive 3, Singapore 117551%
}%
\date{}

\begin{document}
\addtolength{\baselineskip}{1.5mm}


\maketitle
\pagenumbering{gobble} 

\begin{abstract}
  \normalsize%
  \singlespacing%
  \noindent%
  We explain why on certain five-manifolds, topological 5d $\mathcal{N} = 2$ gauge theory of Haydys-Witten twist with gauge group $G$, is dual to that of Geyer-Mülsch twist with gauge group $^LG$, where $G$ is a real, compact Lie group with Langlands dual $^LG$.
  In turn, via their 2d and 3d gauged A/B-twisted Landau-Ginzburg model interpretations, we can show that (i) a Fukaya-Seidel-type $A_\infty$-1-category of an HW$_4$-instanton Floer homology of three-manifolds and (ii) a Fueter-type $A_\infty$-2-category of an HW$_3$-instanton Floer homology of two-manifolds, are dual to (i) an Orlov-type $A_\infty$-1-category of a novel holomorphic $^LG_{\HH}$-flat Floer homology of three-manifolds and (ii) a Rozansky-Witten-type $A_\infty$-2-category of a novel holomorphic $^LG_{\OO}$-flat Floer homology of two-manifolds, respectively.
  We also derive their Atiyah-Floer-type correspondences to symplectic categories.
  Our work, which demonstrates a mirror symmetry and Langlands duality of (higher) $A_\infty$-categories of Floer homologies, therefore furnishes purely physical proofs and gauge-theoretic generalizations of the mathematical conjectures by Bousseau \cite{bousseau-2024-holom-floer} and Doan-Rezchikov \cite{doan-2022-holom-floer}, and more.
\end{abstract}

\clearpage
\pagenumbering{arabic} 
\tableofcontents


\section{Introduction and Motivation, Summary, and Acknowledgements}
\vspace{0.4cm}
\setlength{\parskip}{5pt}

\subtitle{Introduction and Motivation}

Inspired by our work in~\cite{ong-2023-vafa-witten-theor} where a novel Vafa-Witten (VW) Floer homology of three-manifolds was physically defined via a 4d topological gauge theory with sixteen supercharges that is VW theory~\cite{vafa-1994-stron-coupl}, a program was initiated in~\cite{er-2023-topol-n} to physically define other novel Floer homologies via other topological gauge theories with sixteen supercharges. Specifically, a 5d $\mathcal{N} = 2$ topological gauge theory that is Haydys-Witten (HW) theory was considered in~\cite{er-2023-topol-n}, where it was used to physically define a novel $\text{HW}_5$-instanton Floer homology of four-manifolds, as well as other novel Floer homologies of three and two-manifolds.

In fact, because HW theory is five and not four-dimensional, in~\cite{er-2023-topol-n}, we were furthermore able to physically realize a visionary mathematical construction~\cite{haydys-2015-fukay-seidel}  by Haydys of a \emph{gauge-theoretic} Fukaya-Seidel-type (FS-type) $A_{\infty}$-category of a novel Floer homology of three-manifolds. This was possible because HW theory could be recast as a 2d gauged A-twisted Landau-Ginzburg (LG) model whose BPS equation is fundamentally characterized by a holomorphic map.

Beyond these results, in~\cite{er-2023-topol-n}, we were also able to define novel symplectic Floer homologies of Hitchin moduli spaces, which, in turn, allowed us to derive novel Atiyah-Floer-type correspondences that relate gauge-theoretic Floer homologies to symplectic intersection Floer homologies of Hitchin moduli spaces. This was possible because HW theory could also be topologically reduced to a 3d A-twisted sigma model whose BPS equation is a Fueter map, which is a 3d version of a holomorphic map.

In the sequels \cite{er-2024-topol-gauge-theor, er-2024-topol-gauge} to \cite{er-2023-topol-n}, an 8d $\mathcal{N} = 1$ topological gauge theory that is Acharya-O’Loughlin-Spence (AOS) theory~\cite{acharya-1997-higher-dimen} was considered, where it was used to physically define novel Floer homologies of higher-dimensional manifolds. Similarly, we were able to physically realize (higher) $A_{\infty}$-categories of Floer homologies because AOS theory could be recast as a 4d, 3d, and 2d gauged A-twisted LG model whose BPS equations are fundamentally characterized by a Cauchy-Riemann-Fueter map, Fueter map, and holomorphic map, respectively, where a Cauchy-Riemann-Fueter map is a 4d version of a holomorphic map.

Beyond these results, in~\cite{er-2024-topol-gauge-theor, er-2024-topol-gauge}, we were again able to physically derive novel Atiyah-Floer-type correspondences between various gauge-theoretic Floer homologies and symplectic intersection Floer homologies of instanton moduli spaces, among other things. This was possible because AOS theory could also be topologically reduced to a 4d A-twisted sigma model whose BPS equation is a Cauchy-Riemann-Fueter map.

Clearly, 8d AOS and 5d HW theory can be regarded as ``A-twisted'' topological gauge theories. Indeed, the BPS equations that define them, which involve a self-dual field strength $F^+$  and  are given by $F^+ = 0$ and a generalization thereof, respectively, are of instanton-type, and instantons are mathematically established to be related to holomorphic maps~\cite{Atiyah1984Instantons} which characterize our aforementioned A-twisted LG and sigma models.

One might then ask if 8d and 5d  ``B-twisted'' topological gauge theories exist, whence applying the same computations to them would yield a B-version of our results in \cite{er-2024-topol-gauge-theor,er-2024-topol-gauge} and~\cite{er-2023-topol-n}. Such 8d and 5d ``B-twisted'' topological gauge theories ought to have BPS equations which are given by $F=0$ or some generalization thereof, i.e., they ought to be of flat-type -- from $F = F^+ + F^- = (DA)^+ + (DA)^-$, if instantons $(DA)^+ = 0$ are supposed to be related to holomorphic maps  $\bar \partial \Phi = 0$, then flat connections $(DA)^+ + (DA)^- = 0$  ought to be related to constant maps $d \Phi = 0$ which characterize B-twisted LG and sigma models.

Fortuitously, such a 5d ``B-twisted'' topological gauge theory exists, and it is an $\mathcal{N}=2$ theory that is Geyer-M\"ulsch (GM) theory~\cite{geyer-2003-higher-dimen} whose BPS equation is of flat-type given by a flat $G_\C$-connection!

In this paper, we will apply the same computations as before to GM theory, and obtain a B-version of our results in \cite{er-2023-topol-n} and \cite{er-2024-topol-gauge} via the resulting B-twisted LG and sigma models.

The computational techniques we employ are primarily those of standard Kaluza-Klein reduction; topological reduction as pioneered in~\cite{bershadsky-1995-topol-reduc}; recasting gauge theories as supersymmetric quantum mechanics as pioneered in~\cite{blau-1993-topol-gauge}; the physical realization of Floer homology groups via supersymmetric quantum mechanics in infinite-dimensional space as elucidated in~\cite{ong-2023-vafa-witten-theor}; and a physical realization of $A_{\infty}$-1-categories and $A_{\infty}$-2-categories via a string and membrane theory in infinite-dimensional space as developed in~\cite{er-2023-topol-n} and~\cite{er-2024-topol-gauge}.

In fact, because the computations, just like in HW theory, involve Hitchin moduli spaces which are Calabi-Yau with mirrors that are also Langlands dual~\cite{hausel-2021-enhan-mirror}, we have a B-mirror to and Langlands dual of our results in \cite{er-2023-topol-n} and \cite{er-2024-topol-gauge}. In other words, there is actually a mirror symmetry and Langlands duality between HW and GM theory!

Let us now give a brief plan and summary of the paper.

\subtitle{A Brief Plan and Summary of the Paper}


In \autoref{sec:hw}, we discuss general aspects of a topologically-twisted 5d $\mathcal{N} = 2$ gauge theory on $M_5 = M_4 \times \R$ resulting from the Haydys-Witten (HW) twist, where the gauge group $G$ is taken to be a real, simple, and compact Lie group.
We review the following important results from \cite{er-2023-topol-n} of HW theory on such a manifold.
When $M_4 = C \times \Sigma$, with $C$ a Riemann surface of genus $g(C) \geq 2$ and $\Sigma$ an arbitrary Riemann surface, we topologically reduce HW theory along $C$ using the Bershadsky-Johansen-Sadov-Vafa (BJSV) method \cite{bershadsky-1995-topol-reduc} to obtain a 3d $\mathcal{N} = 4$ A-model on $\Sigma \times \R$ with target the $G$-Hitchin moduli space $\mathcal{M}^G_{\text{H}}(C, \mathbf{K})$ on $C$ in complex structure $\mathbf{K}$.
When $M_4 = M_3 \times M_1$, with $M_3$ a closed three-manifold and $M_1$ an arbitrary one-manifold, we obtain, via a Heegaard split of $M_3$ at $C$, an equivalence between HW theory and the aforementioned 3d $\mathcal{N} = 4$ A-model on $I \times M_1 \times \R$ with target $\mathcal{M}^G_{\text{H}}(C, \mathbf{K})$.


In \autoref{sec:gm}, we discuss general aspects of another topologically-twisted 5d $\mathcal{N} = 2$ gauge theory on $M_5 = M_4 \times \R$ resulting from the Geyer-Mülsch (GM) twist, where the gauge group $G$ is also taken to be a real, simple, and compact Lie group.
When $M_4 = C \times \Sigma$, with $C$ and $\Sigma$ described as in \autoref{sec:hw}, we topologically reduce GM theory along $C$ using the BJSV method to obtain a 3d $\mathcal{N} = 4$ Rozansky-Witten (RW) B-model on $\Sigma \times \R$ with target $\mathcal{M}^G_{\text{H}}(C, \mathbf{J})$ in complex structure $\mathbf{J}$.
When $M_4 = M_3 \times M_1$, with $M_3$ and $M_1$ described as in \autoref{sec:hw}, we obtain, via a Heegaard split of $M_3$ at $C$, a \emph{novel} equivalence between GM theory and the aforementioned 3d $\mathcal{N} = 4$ B-model on $I \times M_1 \times \R$ with target $\mathcal{M}^G_{\text{H}}(C, \mathbf{J})$.



In \autoref{sec:m x r}, we study GM theory on $M_4 \times \R$, where $M_4$ can be a decomposable four-manifold.
First, when $M_4$ is closed, we recast GM theory as a 1d supersymmetric quantum mechanics (SQM) in the space $\mathfrak{B}_4$ of irreducible $(\overline{\mathcal{A}}, \overline{\mathcal{B}})$ fields on $M_4$ with action \eqref{eq:m4 x r:sqm:action}, where the $\overline{\mathcal{A}} \in \Omega^1(M_4, \text{ad}(G_{\C}))$ and $\overline{\mathcal{B}} \in \Omega^2(M_4, \text{ad}(G_{\C}))$ are $\text{ad}(G_{\C})$-valued gauge connections and two-forms, respectively.
This will in turn allow us to express the partition function as \eqref{eq:m4 x r:partition function}:
\begin{equation}
  \repeatboxed{eq:m4 x r:partition function}
\end{equation}
where $\text{HHF}^{\text{flat}}_{d_u}(M_4, G_{\C})$ is a \emph{novel} holomorphic $G_{\C}$-flat Floer homology class assigned to $M_4$, of degree $d_u$, defined by Floer differentials described by the gradient flow equations \eqref{eq:m4 x r:flow}:
\begin{equation}
  \repeatboxed{eq:m4 x r:flow}
\end{equation}
and holomorphic Morse functional \eqref{eq:m4 x r:morse function}:
\begin{equation}
  \repeatboxed{eq:m4 x r:morse function}
\end{equation}
The chains of the holomorphic $G_{\C}$-flat Floer complex are generated by fixed critical points of $V_4(\overline{\mathcal{A}}, \overline{\mathcal{B}})$, which correspond to \emph{time-invariant $G_{\C}$-BF configurations on $M_4$} given by time-independent solutions to the 4d equations \eqref{eq:m4 x r:critical points}:
\begin{equation}
  \repeatboxed{eq:m4 x r:critical points}
\end{equation}


Second, we let $M_4 = M_3 \times S^1$, where $M_3$ is a closed three-manifold, and perform a Kaluza-Klein (KK) dimensional reduction of GM theory on $M_4 \times \R$ by shrinking the $S^1$ circle to be infinitesimally small.
We obtain the corresponding 1d SQM in the space $\mathfrak{B}_3$ of irreducible $(\overline{\mathscr{A}}, \overline{\mathscr{B}})$ fields on $M_3$ with action \eqref{eq:m3 x r:action:sqm}, where the $\overline{\mathscr{A}} \in \Omega^1(M_3, \text{ad}(G_{\mathbb{H}}))$ and $\overline{\mathscr{B}} \in \Omega^1(M_3, \text{ad}(G_{\mathbb{H}}))$ are $\text{ad}(G_{\mathbb{H}})$-valued gauge connections and one-forms, respectively, that is equivalent to the resulting GM$_4$ theory on $M_3 \times \R$.
As before, this will allow us to express the partition function as \eqref{eq:m3 x r:partition function}:
\begin{equation}
  \repeatboxed{eq:m3 x r:partition function}
\end{equation}
where $\text{HHF}^{\text{flat}}_{d_v}(M_3, G_{\HH})$ is a \emph{novel} holomorphic $G_{\mathbb{H}}$-flat Floer homology class assigned to $M_3$, of degree $d_v$, defined by Floer differentials described by the gradient flow equations \eqref{eq:m3 x r:flow}:
\begin{equation}
  \repeatboxed{eq:m3 x r:flow}
\end{equation}
and holomorphic Morse functional \eqref{eq:m3 x r:morse functional}:
\begin{equation}
  \repeatboxed{eq:m3 x r:morse functional}
\end{equation}
The chains of the holomorphic $G_{\HH}$-flat Floer complex are generated by fixed critical points of $V_3(\overline{\mathscr{A}}, \overline{\mathscr{B}})$, which correspond to \emph{time-invariant $G_{\HH}$-BF configurations on $M_3$} given by time-independent solutions to the 3d equations \eqref{eq:m3 x r:critical points}:
\begin{equation}
  \repeatboxed{eq:m3 x r:critical points}
\end{equation}


Lastly, we further specialize to the case where $M_3 = M_2 \times S^1$, where $M_2$ is a closed two-manifold, and perform another KK dimensional reduction of GM$_4$ theory on $M_3 \times \R$ by shrinking the $S^1$ circle to be infinitesimally small.
We obtain the corresponding 1d SQM in the space $\mathfrak{B}_2$ of irreducible $(\overline{\mathsf{A}}, \overline{\mathsf{B}})$ fields on $M_2$ with action \eqref{eq:m2 x r:action:sqm}, where the $\overline{\mathsf{A}} \in \Omega^1(M_2, \text{ad}(G_{\OO}))$ and $\overline{\mathsf{B}} \in \Omega^0(M_2, \text{ad}(G_{\OO}))$ are $\text{ad}(G_{\OO})$-valued gauge connections and a scalar, respectively, that is equivalent to the resulting GM$_3$ theory on $M_2 \times \R$.
Again, this will also allow us to express the partition function as \eqref{eq:m2 x r:partition function}:
\begin{equation}
  \repeatboxed{eq:m2 x r:partition function}
\end{equation}
where each $\text{HHF}^{\text{flat}}_{d_w}(M_2, G_{\OO})$ is a \emph{novel} holomorphic $G_{\OO}$-flat Floer homology class assigned to $M_2$, of degree $d_w$, defined by Floer differentials described by the gradient flow equations \eqref{eq:m2 x r:flow}:
\begin{equation}
  \repeatboxed{eq:m2 x r:flow}
\end{equation}
and holomorphic Morse functional \eqref{eq:m2 x r:morse functional}:
\begin{equation}
  \repeatboxed{eq:m2 x r:morse functional}
\end{equation}
The chains of the holomorphic $G_{\OO}$-flat Floer complex are generated by fixed critical points of $V_2(\overline{\mathsf{A}}, \overline{\mathsf{B}})$, which correspond to \emph{time-invariant $G_{\OO}$-BF configurations on $M_2$} given by time-independent solutions to the 2d equations \eqref{eq:m2 x r:critical points}:
\begin{equation}
  \repeatboxed{eq:m2 x r:critical points}
\end{equation}


In \autoref{sec:m3 x r2}, we study GM theory on $M_3 \times \R^2$ and equivalently recast it as a 2d gauged B-twisted Landau-Ginzburg (LG) model on $\R^2$ with target $\mathfrak{B}_3$.
First, we further recast this 2d gauged B-twisted LG model as a 1d SQM theory in the path space $\mathfrak{P}(\R, \mathfrak{B}_3)$ of maps from $\R$ to $\mathfrak{B}_3$.
From the SQM and its critical points that can be interpreted as LG $\mathfrak{B}_3^{\theta}$-strings in the 2d gauged B-twisted LG model, we obtain \eqref{eq:m3 x r2:morphism as ext}:
\begin{equation}
  \label{summary:eq:m3 x r2:morphism as ext}
  \repeatboxed{eq:m3 x r2:morphism as ext}
\end{equation}
Here, $\text{Ext} ( \mathcal{T}^I_{V_3}, \mathcal{T}^J_{V_3} )$ is an Ext-group between singular fibers of the 2d LG superpotential $V_3$, whilst $\text{Hom} ( \mathfrak{C}^I_{\text{BF}_{\HH}}, \mathfrak{C}^J_{\text{BF}_{\HH}} )$ describes an LG $\mathfrak{B}_3^{\theta}$-string as a morphism between its endpoints $\mathfrak{C}^{*}_{\text{BF}_{\HH}}$ which correspond to ($\theta$-deformed) \emph{$G_{\HH}$-BF configurations on $M_3$}.
Furthermore, via this equivalent description of GM theory as a 2d gauged B-twisted LG model, we can interpret the normalized 5d GM partition function as a sum over tree-level scattering amplitudes of LG $\mathfrak{B}_3^{\theta}$-strings given by the composition map \eqref{eq:m3 x r2:mu-d maps}
\begin{equation}
  \label{summary:eq:m3 x r2:mu-d maps}
  \repeatboxed{eq:m3 x r2:mu-d maps}
\end{equation}
where $\text{Hom} ( \mathfrak{C}^*_{\text{BF}_{\HH}}, \mathfrak{C}^*_{\text{BF}_{\HH}} )_-$ and $\text{Hom} ( \mathfrak{C}^*_{\text{BF}_{\HH}}, \mathfrak{C}^*_{\text{BF}_{\HH}} )_+$ represent incoming and outgoing LG $\mathfrak{B}_3^{\theta}$-strings, respectively, as shown in \autoref{fig:m3 x r2:scattering amplitudes}.

Together, \eqref{summary:eq:m3 x r2:morphism as ext} and \eqref{summary:eq:m3 x r2:mu-d maps} underlie a \emph{novel} gauge-theoretic Orlov-type $A_{\infty}$-1-category whose 1-objects $\mathfrak{C}^{*}_{\text{BF}_{\HH}}$ correspond to ($\theta$-deformed) $G_{\HH}$-BF configurations on $M_3$.
As such configurations were shown in \autoref{sec:m3 x r} to generate $\text{HHF}^{\text{flat}}(M_3, G_{\HH})$, the Orlov-type $A_{\infty}$-1-category thus categorifies $\text{HHF}^{\text{flat}}(M_3, G_{\HH})$.


Next, applying the equivalence from \autoref{sec:gm:equivalence} between GM theory on $M_3 \times \R^2$ and the 3d RW model on $I \times \R^2$ with target $\mathcal{M}^G_{\text{H}, \theta}(C, \mathbf{J})$ via a Heegaard split $M_3 = M_3' \bigcup_C M_3''$, we obtain \eqref{eq:m3 x r2:correspondence amongst cats}:
\begin{equation}
  \label{summary:eq:m3 x r2:correspondence amongst cats}
  \repeatboxed{eq:m3 x r2:correspondence amongst cats}
\end{equation}
In particular, we have correspondences amongst the following categories:
(i) $\text{Orlov} \big( \text{HHF}^{\text{flat}}(M_3, G_{\HH}) \big)$, a gauge-theoretic Orlov-type $A_{\infty}$-1-category categorifying $\text{HHF}^{\text{flat}}(M_3, G_{\HH})$,
(ii) $\text{RW}_{(\sigma)} \big( \widehat{L}, \mathcal{M}^G_{\text{H}, \theta}(C, \mathbf{J}) \big)$, a Rozansky-Witten $A_{\infty}$-2-category of complex-Lagrangian branes $\widehat{L}$ in symplectic $\mathcal{M}^G_{\text{H}, \theta}(C, \mathbf{J})$,
(iii) $\text{Orlov}_{(\sigma)} \big( \mathscr{P}, \mathfrak{P}(\R, \mathcal{M}^G_{\text{H}, \theta}(C, \mathbf{J})) \big)$, an Orlov-type $A_{\infty}$-1-category of branes $\mathscr{P}$ in symplectic path space $\mathfrak{P}(\R, \mathcal{M}^G_{\text{H}, \theta}(C, \mathbf{J}))$,
and (iv) $\text{KRS}_{(\sigma)} \big( \widehat{L}, \mathcal{M}^G_{\text{H}, \theta}(C, \mathbf{J}) \big)$, a Kapustin-Rozansky-Saulina 2-category of complex-Lagrangian branes $\widehat{L}$ in symplectic $\mathcal{M}^G_{\text{H}, \theta}(C, \mathbf{J})$.
As the correspondences between the top and bottom lines of \eqref{summary:eq:m3 x r2:correspondence amongst cats} are correspondences between a gauge-theoretic category and symplectic categories via a Heegaard split of $M_3 = M_3' \bigcup_C M_3''$, they can be regarded as \emph{novel} Atiyah-Floer-type correspondences!

Note that Doan-Rezchikov (DR) conjectured in \cite[$\S$2.6]{doan-2022-holom-floer} a correspondence between a KRS 2-category of complex-Lagrangian branes in a hyperkähler $X$ and an Orlov-type $A_{\infty}$-1-category of branes in the path space $\mathfrak{P}(\R, X)$.
As $\mathcal{M}^G_{\text{H}, \theta}(C, \mathbf{J})$ is hyperkähler, the correspondence between (iii) and (iv) coincides with their conjecture.
Thus, we have furnished a purely physical proof of their mathematical conjecture!


In \autoref{sec:m2 x r3}, we study GM theory on $M_2 \times \R^3$ and equivalently recast it as
(i) a 3d gauged B-twisted LG model on $\R^3$ with target $\mathfrak{B}_2$, or
(ii) a 2d gauged B-twisted LG model on $\R^2$ with target the path space $\mathfrak{P}(\R, \mathfrak{B}_2)$ of maps from $\R$ to $\mathfrak{B}_2$,
which, in turn, can be recast as a 1d SQM theory in the double path space $\mathfrak{P}(\R^2, \mathfrak{B}_2)$ of maps from $\R^2$ to $\mathfrak{B}_2$.

From the SQM and its critical points that can be interpreted as LG $\mathfrak{P}^{\theta}(\R, \mathfrak{B}_2)$-strings in the 2d gauged B-twisted LG model, we obtain \eqref{eq:m2 x r3:2d-lg:hom as ext}:
\begin{equation}
  \label{summary:eq:m2 x r3:2d-lg:hom as ext}
  \repeatboxed{eq:m2 x r3:2d-lg:hom as ext}
\end{equation}
Here, $\text{Ext} \big( \mathfrak{T}^{IJ}_{W_2}, \mathfrak{T}^{KL}_{W_2} \big)_{\pm}$ is an Ext-group between singular fibers of the 2d LG superpotential $W_2$, whilst $\text{Hom} \big( \Gamma^{IJ}(\mathfrak{B}_2), \Gamma^{KL}(\mathfrak{B}_2) \big)_{\pm}$ describes an LG $\mathfrak{P}^{\theta}(\R, \mathfrak{B}_2)$-string as a 1-morphism  between its endpoints $\Gamma^{**}(\mathfrak{B}_2)$.
Furthermore, via the aforementioned equivalent description of GM theory as a 2d gauged B-twisted LG model, we can interpret the normalized 5d GM partition function as a sum over tree-level scattering amplitudes of LG $\mathfrak{P}^{\theta}(\R, \mathfrak{B}_2)$-strings given by the composition map \eqref{eq:m2 x r3:2d-lg:composition map}:
\begin{equation}
  \label{summary:eq:m2 x r3:2d-lg:composition map}
  \repeatboxed{eq:m2 x r3:2d-lg:composition map}
\end{equation}
where $\text{Hom} \big( \Gamma^{**}(\mathfrak{B}_2), \Gamma^{**}(\mathfrak{B}_2) \big)_-$ and $\text{Hom} \big( \Gamma^{**}(\mathfrak{B}_2), \Gamma^{**}(\mathfrak{B}_2) \big)_+$ represent incoming and outgoing LG $\mathfrak{P}^{\theta}(\R, \mathfrak{B}_2)$-strings, respectively.

The SQM and its critical points can also be interpreted as LG $\mathfrak{B}_2^{\theta}$-membranes in the 3d gauged B-twisted LG model, from which we obtain \eqref{eq:m2 x r3:3d-lg:homs as ext}:
\begin{equation}
  \label{summary:eq:m2 x r3:3d-lg:homs as ext}
  \repeatboxed{eq:m2 x r3:3d-lg:homs as ext}
\end{equation}
Here, $\text{Ext} \big( \mathcal{T}^{IJ}_{V_2}, \mathcal{T}^{KL}_{V_2} \big)_{\pm}$ is an Ext-group between strings whose endpoints are singular fibers of the 3d LG superpotential $V_2$.
At the same time, $\text{Hom} \big( \gamma^{IJ}(\tau, \mathfrak{B}_2), \gamma^{KL}(\tau, \mathfrak{B}_2) \big)_{\pm}$ describes an LG $\mathfrak{B}_2^{\theta}$-membrane as a 1-morphism between its edges $\gamma^{**}(\tau, \mathfrak{B}_2)$ corresponding to LG $\mathfrak{B}_2^{\theta}$-strings,
and $\text{Hom} \big( \text{Hom} \big( \mathfrak{C}^I_{\text{BF}_{\OO}}, \mathfrak{C}^J_{\text{BF}_{\OO}} \big), \big( \mathfrak{C}^K_{\text{BF}_{\OO}}, \mathfrak{C}^L_{\text{BF}_{\OO}} \big) \big)_{\pm}$ describes an LG $\mathfrak{B}_2^{\theta}$-membrane as a 2-morphism amongst its vertices $\mathfrak{C}^{*}_{\text{BF}_{\OO}}$ corresponding to ($\theta$-deformed) \emph{$G_{\OO}$-BF configurations on $M_2$}.
Furthermore, via the aforementioned equivalent description of GM theory as a 3d gauged B-twisted LG model, we can also interpret the normalized 5d GM partition function as a sum over tree-level scattering amplitudes of LG $\mathfrak{B}_2^{\theta}$-membranes given by the composition map \eqref{eq:m2 x r3:3d-lg:a-infinity structure}:
\begin{equation}
  \label{summary:eq:m2 x r3:3d-lg:a-infinity structure}
  \repeatboxed{eq:m2 x r3:3d-lg:a-infinity structure}
\end{equation}
where $\text{Hom} \big( \text{Hom} \big( \mathfrak{C}^{*}_{\text{BF}_{\OO}}, \mathfrak{C}^{*}_{\text{BF}_{\OO}} \big), \text{Hom} \big( \mathfrak{C}^{*}_{\text{BF}_{\OO}}, \mathfrak{C}^{*}_{\text{BF}_{\OO}} \big) \big)_-$ and $\text{Hom} \big( \text{Hom} \big( \mathfrak{C}^{*}_{\text{BF}_{\OO}}, \mathfrak{C}^{*}_{\text{BF}_{\OO}} \big), \text{Hom} \big( \mathfrak{C}^{*}_{\text{BF}_{\OO}}, \mathfrak{C}^{*}_{\text{BF}_{\OO}} \big) \big)_+$ represent incoming and outgoing LG $\mathfrak{B}_2^{\theta}$-membranes, respectively, as shown in \autoref{fig:m2 x r3:scattering membranes}.

Together, \eqref{summary:eq:m2 x r3:2d-lg:hom as ext} and \eqref{summary:eq:m2 x r3:2d-lg:composition map} underlie a \emph{novel} Orlov-type $A_{\infty}$-1-category of LG $\mathfrak{P}^{\theta}(\R, \mathfrak{B}_2)$-strings.
On the other hand, \eqref{summary:eq:m2 x r3:3d-lg:homs as ext} and \eqref{summary:eq:m2 x r3:3d-lg:a-infinity structure} underlie a \emph{novel} RW-type $A_{\infty}$-2-category whose 2-objects $\mathfrak{C}^{*}_{\text{BF}_{\OO}}$ correspond to ($\theta$-deformed) $G_{\OO}$-BF configurations on $M_2$.
As such configurations were shown in \autoref{sec:m2 x r} to generate $\text{HHF}^{\text{flat}}(M_2, G_{\OO})$, this RW-type $A_{\infty}$-2-category thus 2-categorifies $\text{HHF}^{\text{flat}}(M_2, G_{\OO})$!
Furthermore, as these two categories ought to correspond to each other, their correspondence can be regarded as a \emph{gauge-theoretic generalization} of DR's mathematical conjecture from \cite[$\S$2.6]{doan-2022-holom-floer}!


In \autoref{sec:dualities:theories}, we physically derive a mirror and Langlands duality between the relevant sigma models and gauge theories studied hitherto.
First, because of an enhanced Homological Mirror Symmetry (HMS) between 2d A and B-models in mirror Hitchin moduli spaces, we are able to obtain a \emph{novel} mirror duality between the 3d $\mathcal{N} = 4$ A-model with target $\mathcal{M}^G_{\text{H}}(C, \mathbf{K})$ and B-model with target $\mathcal{M}^{^{L}G}_{\text{H}}(C, \mathbf{J})$, on $\Sigma \times \R$, where $^{L}G$ is the Langlands dual of $G$, and $\mathcal{M}^{^{L}G}_{\text{H}}(C, \mathbf{J})$ is mirror to $\mathcal{M}^G_{\text{H}}(C, \mathbf{K})$.
In turn, since the 5d $\mathcal{N} = 2$ gauge theories on $M_3 \times M_1 \times \R$ (resp. $C \times \R^3$) are equivalent to the 3d $\mathcal{N} = 4$ sigma models on $I \times M_1 \times \R$ (resp. $\R^3$), it would mean that HW theory of $G$-type and GM theory of $^{L}G$-type, on $M_3 \times M_1 \times \R$ (resp. $C \times \R^3$), are Langlands dual.

In fact, we are able to provide an even more fundamental derivation of the Langlands duality between the 5d $\mathcal{N} = 2$ gauge theories on $M_3 \times M_1 \times \R$ (resp. $C \times \R^3$) via Kapustin-Witten's 4d $\mathcal{N} = 4$ S-duality \cite{kapustin-2006-elect-magnet}, from which the above mirror duality between the 3d $\mathcal{N} = 4$ A and B-models can be understood as a consequence thereof!

We summarize these dualities and their underlying physical derivations in \autoref{fig:summary:dualities:theories}, where $\widehat{C}$ is a Riemann surface of genus $g (\widehat C) \geq 2$ with variable size.
\begin{figure}[ht]
  \repeatfig{fig:dualities:theories}
  \label{fig:summary:dualities:theories}
\end{figure}


In \autoref{sec:dualities:floer}, we first elucidate the implications of the topological invariance of GM theory on the gauge-theoretic Floer homologies obtained in \autoref{sec:m x r}.
The result is given in \eqref{eq:dualities:floer:gm}:
\begin{equation}
  \repeatboxed{eq:dualities:floer:gm}
\end{equation}
where the $\widehat{S}^1$'s are circles of variable radii.

Second, we elucidate the implications of Tachikawa's 5d ``S-duality'' \cite{tachikawa-2011-s-dualit} of GM theory.
The result is given in \eqref{eq:dualities:s-duality:gm4}:
\begin{equation}
  \repeatboxed{eq:dualities:s-duality:gm4}
\end{equation}
and \eqref{eq:dualities:s-duality:gm3}:
\begin{equation}
  \repeatboxed{eq:dualities:s-duality:gm3}
\end{equation}
where
(i) $G$/$G_{\C}$ is a nonsimply-laced gauge group with loop group $LG$/$LG_{\C}$,
(ii) $(LG)^{\vee}$/$(LG_{\C})^{\vee}$ is Langlands dual to $LG$/$LG_{\C}$ (at the level of their loop algebras),
and (iii) $\text{H}^0_{\text{dol}}(\mathcal{M}^{LLG_{\C}}_{\text{BF}}(M_2))$ is a Dolbeault class of zero-forms in the $LLG_{\C}$-BF moduli space $\mathcal{M}^{LLG_{\C}}_{\text{BF}}(M_2)$ of $M_2$.

Lastly, by applying the Langlands duality between HW theory of $G$-type and GM theory of $^{L}G$-type, we obtain a web of Langlands dual relations amongst these results and the gauge-theoretic Floer homologies physically realized by HW theory in \cite{er-2023-topol-n}.
This is shown in~\autoref{summary:fig:dualities:floer}, where
(i) $\text{H}^0_{\text{dol}}(\mathcal{M}^{LLG_{\C}}_{\text{flat}}(M_2))$ therein is a de Rham class of zero-forms in the moduli space $\mathcal{M}^{LLG_{\C}}_{\text{flat}}(M_2)$ of flat $LLG_{\C}$-connections on $M_2$,
(ii) the radii of $\widehat{S}^1$ can be varied;
(iii) horizontal undashed lines indicate a relation that is due to a Langlands duality between HW and GM theory;
(iv) dashed lines indicate a relation that is due to an equivalence under dimensional reduction;
(v) squiggly lines indicate a relation that is due to a spectral equivalence of the underlying theory that realizes the (co)homology;
(vi) dotted lines indicate a relation that is due to an ``S-duality'' of the underlying 5d theory that realizes the Floer homology;
and (vii) finely dotted lines are used to specify the manifolds.
\begin{figure}
  \repeatfig{fig:dualities:floer}
  \label{summary:fig:dualities:floer}
\end{figure}


In \autoref{sec:dualities:cats}, we apply the Langlands duality between HW theory of $G$-type and GM theory of $^{L}G$-type to the $A_{\infty}$-categories obtained in this paper and in our previous works \cite{er-2023-topol-n, er-2024-topol-gauge}.
This will then allow us to furnish a Langlands dual \emph{gauge-theoretic generalization} of HMS for LG models \cite{orlov-2012-landau-ginzb}, as well as a physical proof and Langlands dual \emph{gauge-theoretic generalization} of Bousseau-Doan-Rezchikov's (B-DR) mathematical conjecture in \cite{bousseau-2024-holom-floer, doan-2022-holom-floer}!
We summarize the mirror symmetric and Langlands dual relations that we have obtained of these $A_{\infty}$-categories in \autoref{summary:fig:dualities:cats}.
\begin{figure}
  \repeatfig{fig:dualities:cats}
  \label{summary:fig:dualities:cats}
\end{figure}

\subtitle{Acknowledgements}

We would like to thank
S.~Gukov,
A.~Haydys,
N.~Hitchin,
D.~Joyce,
and R.P.~Thomas
for insightful discussions and technical assistance over the past three years with our preceding works \cite{ong-2023-vafa-witten-theor, er-2023-topol-n, er-2024-topol-gauge-theor, er-2024-topol-gauge} that have culminated in this paper.
M.-C.~Tan would also like to thank the organizers of ``ICBS 2025'' and ``Gauge Theory and String Geometry 2025''  for the opportunity to deliver a plenary talk on this paper.
Last but not least, we would also like to thank the ATMP referee for a careful reading of our paper, and for suggesting improvements to it.
Our work is supported in part by the MOE AcRF Tier 1 grant A-8003583-00-00.

\section{A 5d \texorpdfstring{$\mathcal{N} = 2$}{N = 2} Haydys-Witten-twisted Topological Gauge Theory}
\label{sec:hw}

In this section, we will study the 5d $\mathcal{N} = 2$ gauge theory on a five-manifold $M_5$ with the Haydys-Witten (HW) twist,\footnote{%
  We refer the reader to \cite{er-2023-topol-n, anderson-2013-five-dimen} for a detailed description of the HW twist.
  \label{ft:description of the HW twist}
} whose gauge group is a real, simple, compact Lie group $G$, and whose BPS equations that its path integral localizes onto are the HW equations \cite{haydys-2015-fukay-seidel, witten-2011-fiveb-knots}.
We will also consider several specializations of $M_5$ that allow us to derive a 3d $\mathcal{N} = 4$ A-model whose spectrum is effectively given by that of a 2d $\mathcal{N} = (4, 4)$ A-model.
Finally, via topological invariance, we will obtain an equivalence between the 5d $\mathcal{N} = 2$ HW-twisted gauge theory on $M_3 \times M_1 \times \R$ and a 3d A-model on $I \times M_1 \times \R$.

\subsection{HW Theory on \texorpdfstring{$M_5 = M_4 \times \R$}{M5 = M4 x R}}
\label{sec:hw:5d}

\subtitle{The Field Content}

The bosonic field content of HW theory on $M_4 \times \R$ consists of scalars $\upsilon, \bar{\upsilon}\in \Omega^0 (M_4 \times \R, \text{ad}(G))$, gauge connections $A_{\mu} \in \Omega^1 (M_4, \text{ad}(G)) \otimes \Omega^0(\R, \text{ad}(G))$ and $A_t \in \Omega^0(M_4, \text{ad}(G)) \otimes \Omega^1(\R, \text{ad}(G))$, and a self-dual two-form $B_{\mu\nu}\in \Omega^{2,+}(M_4, \text{ad}(G)) \otimes \Omega^0(\R, \text{ad}(G))$.
The fermionic field content consists of scalars $\eta, \tilde{\eta}\in \Omega^0 (M_4 \times \R, \text{ad}(G))$, one-forms $\psi_{\mu}, \tilde{\psi}_{\mu} \in \Omega^1(M_4, \text{ad}(G)) \otimes \Omega^0(\R, \text{ad}(G))$, and self-dual two-forms $\chi_{\mu\nu}, \tilde{\chi}_{\mu\nu}\in\Omega^{2, +}(M_4, \text{ad}(G)) \otimes \Omega^0(\R, \text{ad}(G))$.
Here, $\mu \in \{1, \dots, 4\}$ are the indices on $M_4$, $t$ is the coordinate along $\R$, and $\text{ad}(G)$ is the adjoint bundle of the underlying principal $G$-bundle.

As the supersymmetry generators transform in the same representation as the fermions above, the HW twist will result in two nilpotent \emph{scalar} supersymmetry generators $\mathcal{Q}$ and $\bar{\mathcal{Q}}$; we shall choose $\mathcal{Q}$ to define the twisted theory.
The supersymmetry transformations of the twisted fields generated by $\mathcal{Q}$ \cite{er-2023-topol-n} are
\begin{equation}
 \label{eq:hw susy variations}
  \begin{aligned}
    \delta_{\mathcal{Q}} \upsilon
    & = 0
      \, ,
    & \qquad
    \delta_{\mathcal{Q}} \bar{\upsilon}
    & = - i \tilde{\eta}
      \, ,
    \\
    \delta_{\mathcal{Q}} \eta
    & = - 2 i D_t \upsilon
      \, ,
    & \qquad
     \delta_{\mathcal{Q}} \tilde{\eta}
    & = - 2 [\upsilon, \bar{\upsilon}]
      \, ,
    \\
    \delta_{\mathcal{Q}} A_{\mu}
    & =  i \psi_{\mu}
      \, ,
    & \qquad
    \delta_{\mathcal{Q}} A_t
    & = \eta
      \, ,
    \\
    \delta_{\mathcal{Q}} \psi_{\mu}
    & = - 2 D_{\mu} \upsilon
      \, ,
    & \qquad
     \delta_{\mathcal{Q}} \tilde{\psi}_{\mu}
    & = i (F_{t\mu} + D^{\nu} B_{\nu\mu})
      \, ,
    \\
    \delta_{\mathcal{Q}} \chi_{\mu\nu}
    & = - i [B_{\mu\nu}, \upsilon]
      \, ,
    &\qquad
      \delta_{\mathcal{Q}} \tilde{\chi}_{\mu\nu}
    &= - \left(
        F^+_{\mu \nu}
        - \frac{1}{4} [B_{\mu \rho}, B^{\rho}_{\nu}]
        - \frac{1}{2} D_t B_{\mu \nu}
      \right)
      \, ,
    \\
    \delta_{\mathcal{Q}} B_{\mu\nu}
    &= 2 \chi_{\mu\nu}
      \, ,
  \end{aligned}
\end{equation}
where $\delta_{\mathcal{Q}}$ denotes a $\mathcal{Q}$-variation; $F^+_{\mu\nu} = \frac{1}{2} (F_{\mu\nu} + \frac{1}{2} \epsilon_{\mu\nu\rho\lambda} F^{\rho\lambda})$ is the self-dual component of $F_{\mu\nu}$; self-duality of the self-dual fields $\Phi \in \{B, \chi, \tilde{\chi}\}$ means that $\Phi_{\mu\nu} = \frac{1}{2} \epsilon_{\mu\nu\rho\lambda} \Phi^{\rho\lambda}$;
and $[B_{\mu\rho}, B_{\nu}^{\rho}] \equiv g^{\rho\varrho}[B_{\mu\rho}, B_{\nu\varrho}]$.

Notice that $\delta_{\mathcal{Q}}$ is nilpotent up to gauge transformations generated by $\upsilon$ when we add auxiliary fields to the theory off-shell.
As we wish to study the theory whereby the relevant moduli space is well-behaved (i.e., no reducible connections), we shall consider the case where $\upsilon$ has no zero-modes.

\subtitle{The BPS Equations}

The BPS equations of HW theory on $M_4 \times \R$ are obtained by setting to zero the $\mathcal{Q}$-variations of the fermions in \eqref{eq:hw susy variations}, i.e.,\footnote{%
  As we are only considering the case where $\upsilon$ has no zero-modes, we can take it to be zero in the variations of the fermions.
  \label{ft:hw:no nu zero-modes}
}
\begin{equation}
  \label{eq:hw:hw eqn}
  \begin{aligned}
    F_{t \mu}
    + D^{\nu} B_{\nu \mu}
    &= 0
    \, ,
    \\
    F^+_{\mu \nu}
    - \frac{1}{4} [B_{\mu \rho}, B^{\rho}_{\nu}]
    - \frac{1}{2} D_t B_{\mu \nu}
    &= 0
    \, .
  \end{aligned}
\end{equation}
These are the HW equations on $M_4 \times \R$.
Configurations in the space $\mathfrak{A}_5$ of all $(A, B)$ fields on $M_4 \times \R$ that satisfy \eqref{eq:hw:hw eqn},\footnote{%
  Since we will ultimately only consider gauge-inequivalent configurations, $\mathfrak{A}_5$ is more precisely the space of irreducible $(A, B)$ fields on $M_4 \times \R$ modulo gauge equivalence.
  Similar such spaces to appear in later sections should also be understood as spaces of fields modulo gauge equivalence.
  \label{ft:frak space modulo gauge equivalence}
} constitute a moduli space $\mathcal{M}_{\text{HW}}$ that the path integral of HW theory on $M_4 \times \R$ will localize onto.

\subtitle{The $\mathcal{Q}$-exact Action}

The $\mathcal{Q}$-exact topological action of HW theory on $M_4 \times \R$ is~\cite{er-2023-topol-n, anderson-2013-five-dimen}
\begin{equation}
  \label{eq:hw:action:q-exact}
  \begin{aligned}
    S_{\text{HW,}\mathcal{Q}}
    = \frac{1}{e^2}
    & \int_{M_4 \times \R} dt d^4x \, \Tr \bigg(
      \frac{1}{2} \left|
        F^+_{\mu\nu}
        - \frac{1}{4} [B_{\mu\rho}, B^{\rho}_{\nu}]
        - \frac{1}{2} D_t B_{\mu\nu}
      \right|^2
      + \frac{1}{2} \left| F_{t\mu} + D^{\nu} B_{\nu\mu} \right|^2
    \\
    & + 2 D_{\mu} \upsilon D^{\mu} \bar{\upsilon}
      + 2 D_t \upsilon D^t \bar{\upsilon}
      - 2 [\upsilon, \bar{\upsilon}]^2
      + \frac{1}{2} [B_{\mu\nu}, \upsilon] [B^{\mu\nu}, \bar{\upsilon}]
    \\
    & + \frac{1}{2} B^{\mu\nu} \{\tilde{\eta}, \chi_{\mu\nu}\}
      - \frac{1}{2} B^{\mu\nu} \{\eta, \tilde{\chi}_{\mu\nu}\}
      - B^{\mu\nu} \{\psi_{\mu}, \tilde{\psi}_{\nu}\}
      - B^{\mu\nu} \{\tilde{\chi}_{\mu\rho}, \chi^{\rho}_{\nu}\}
    \\
    & - i \tilde{\eta} D_{\mu} \psi^{\mu}
      - i \eta D_{\mu} \tilde{\psi}^{\mu}
      - 2 i \tilde{\psi}_{\mu} D_{\nu} \chi^{\mu\nu}
      - 2 i \psi_{\mu} D_{\nu} \tilde{\chi}^{\mu\nu}
      - \tilde{\eta} D_t \eta
      - \tilde{\psi}_{\mu} D_t \psi^{\mu}
      - \tilde{\chi}_{\mu\nu} D_t \chi^{\mu\nu}
    \\
    & - i \upsilon \{\tilde{\eta}, \tilde{\eta}\}
      - i \bar{\upsilon} \{\eta, \eta\}
      + i \upsilon \{\tilde{\psi}_{\mu}, \tilde{\psi}^{\mu}\}
      + i \bar{\upsilon} \{\psi_{\mu}, \psi^{\mu}\}
      - i \upsilon \{\tilde{\chi}_{\mu\nu}, \tilde{\chi}^{\mu\nu}\}
      - i \bar{\upsilon} \{\chi_{\mu\nu}, \chi^{\mu\nu}\}
      \bigg)
      \, .
  \end{aligned}
\end{equation}

Note that as we will ultimately only be interested in the $\mathcal{Q}$-cohomology spectrum of HW theory, we can (i) subtract from \eqref{eq:hw:action:q-exact} $\mathcal{Q}$-exact terms $\delta_{\mathcal{Q}}(\eta D^t \bar{\upsilon})$, $\delta_{\mathcal{Q}}(\bar{\eta} [ \upsilon, \bar{\upsilon}])$, and $\delta_{\mathcal{Q}}(\chi^{\mu \nu} [ B_{\mu \nu}, \bar{\upsilon}])$, and (ii) via the self-duality of $B$, topological terms involving $B$ and bilinear combinations of fermions.
This leaves us with
\begin{equation}
  \label{eq:hw:action}
  S_{\text{HW}}
  = \frac{1}{2e^2} \int_{M_4 \times \R} dt d^4 x \, \Tr \left(
    \left|
      F_{t \mu}
      + D^{\nu} B_{\nu \mu}
    \right|^2
    + \left|
      F^+_{\mu \nu}
      - \frac{1}{4} [B_{\mu \rho}, B^{\rho}_{\nu}]
      - \frac{1}{2} D_t B_{\mu \nu}
    \right|^2
    + \dots
  \right)
  \, ,
\end{equation}
where the ``$\dots$'' contain the terms involving $\upsilon, \bar{\upsilon}$ and fermions in the action.

\subtitle{A Balanced TQFT}

One important fact about the HW equations \eqref{eq:hw:hw eqn} (modulo gauge equivalence) is that on $M_4 \times S^1$, they are elliptic \cite{haydys-2015-fukay-seidel, witten-2011-fiveb-knots}.
In other words, the differential operator $\nabla_{\text{HW}}$ associated to the HW equations is an elliptic operator.\footnote{%
  The HW equations are nonlinear (differential) equations.
  However, the differential operator associated to a system of nonlinear (differential) equations is elliptic if the differential operator associated to its linearization, is.
  Therefore, the ellipticity of the HW equations is determined by the ellipticity of the differential operator $\nabla_{\text{HW}}$ associated to the linearization of the HW equations, which is a first-order linear differential operator in momentum space.
  Let us denote by $\sigma_{\text{HW}}(p)$ the principal symbol of $\nabla_{\text{HW}}$, and $\sigma^{\intercal}_{\text{HW}}(p)$ its transpose; these are matrix-valued linear functions of momentum $p$.
  For $\nabla_{\text{HW}}$ to be elliptic, $\sigma_{\text{HW}}(p)$ must be invertible for all nonzero real values of $p$.
  At any rate, we see in \eqref{eq:hw:action} that a mod-square of the HW equations \eqref{eq:hw:hw eqn}, after expansion, furnishes the terms in the action of (the bosonic sector of) a (twisted) 5d $\mathcal{N} = 2$ SYM.
  This means that the principal symbol $\sigma_{\text{5d-SYM}}(p)$ of the differential operator $\nabla_{\text{5d-SYM}}$ associated to the linearized second-order equations (modulo gauge equivalence) of 5d $\mathcal{N} = 2$ SYM, is equal to $\sigma^{\intercal}_{\text{HW}}(p) \sigma_{\text{HW}}(p)$.
  $\nabla_{\text{5d-SYM}}$ is certainly elliptic, i.e., invertible, as $\sigma_{\text{5d-SYM}}(p)$ is the identity matrix multiplied by the principal symbol $f(p) = p^2$ of the Laplace operator acting on scalars with momentum $p$ (which is indeed invertible for all nonzero real values of $p$).
  This invertibility of $\sigma^{\intercal}_{\text{HW}}(p) \sigma_{\text{HW}}(p)$ therefore implies the invertibility of $\sigma_{\text{HW}}(p)$, and thus the ellipticity of $\nabla_{\text{HW}}$ and the HW equations.
  \label{ft:why linearized hw eqns}
}

Since $\nabla_{\text{HW}}$ is an elliptic operator, its index can be used to calculate the difference in the number of fermion zero-modes of HW theory on $M_4 \times S^1$ -- it is 0.\footnote{%
  The index of an elliptic operator on an odd-dimensional compact manifold is 0 \cite[Prop. 9.2]{atiyah-1968-index-ellip-iii}.
  \label{ft:atiyah-singer on odd-dimensional manifold}
}
Such a TQFT whose difference in the number of fermion zero-modes is 0 is called a balanced TQFT.
As this difference is the same for HW theory on $M_4     \times \R$ and $M_4 \times S^1$, where $\R$ and $S^1$ are the directions of time,\footnote{%
  Note that fermion zero-modes in a TQFT have zero energy, i.e., they are static and therefore time-invariant. In particular, they have no $t$ dependence.
  Hence, they are insensitive to the geometry of the $t$-direction.
  So, if we were to replace $S^1$ with $\R$, the fermion zero-modes would not be affected -- specifically, their number, and thus difference, would not change.
  \label{ft:fermion zero modes}
} HW theory on $M_4 \times \R$ is also a balanced TQFT.
This will be relevant in later sections.

\subsection{A 3d \texorpdfstring{$\mathcal{N} = 4$}{N = 4} A-model and a 2d \texorpdfstring{$\mathcal{N} = (4, 4)$}{N = (4, 4)} A-model}
\label{sec:hw:3d}

Consider the specialization $M_4 = C \times \Sigma$, where  $C$ and $\Sigma$ are Riemann surfaces, and the genus of $C$ is $g(C) \geq 2$.
We will perform a topological reduction of HW theory on $M_5 = C \times \Sigma \times \R$ along $C$ using the Bershadsky-Johansen-Sadov-Vafa (BJSV) reduction method \cite{bershadsky-1995-topol-reduc}, as was done in \cite[$\S$6]{er-2023-topol-n}.
Let us give a brief description of this BJSV reduction.

\subtitle{A 3d Sigma Model}

The BJSV reduction process is effected by writing the metric on $M_5$ as a block diagonal with a vanishing scale parameter $\varepsilon$, i.e., $g_{M_5} = \varepsilon g_C \oplus g_{\Sigma \times \R}$, and taking the vanishing limit of $\varepsilon$.
In taking this vanishing limit, there will be terms in the action \eqref{eq:hw:action} that will diverge due to having negative powers of $\varepsilon$; they need to be set to zero to ensure the finiteness of the resulting 3d theory -- we find that these finiteness conditions are the $G$-Hitchin equations on $C$.
At the same time, there will be terms in the action \eqref{eq:hw:action} that will vanish due to having positive powers of $\varepsilon$; they do not play any role in the resulting 3d theory.

Finally, terms in the action \eqref{eq:hw:action} with no power of $\varepsilon$ will remain as the only terms that are relevant in the action of the resulting 3d theory; integrating out the auxiliary fields and performing the integration over $C$, we will arrive at the action of a 3d $\mathcal{N} = 4$ sigma model with target the $G$-Hitchin moduli space $\mathcal{M}^G_{\text{H}}(C, \mathbf{K})$ of $C$ in complex structure $\mathbf{K}$ \cite[$\S$6]{er-2023-topol-n},\footnote{%
  The complex structure of the target space in our sigma model can be determined as follows.
  The space of all $(A_C, B_C)$ fields on $C$, which are the $(A, B)$ fields restricted to $C$, is an infinite-dimensional affine space.
  Let us denote by $\mathfrak{d}$ the exterior derivative on this space; the basis of the $G$-Hitchin moduli space $\mathcal{M}^G_{\text{H}}(C)$ of $C$ is spanned by $(\mathfrak{d} A_C, \mathfrak{d} B_C)$.
  Note that $\mathcal{M}^G_{\text{H}}(C)$ is a hyperkähler manifold, i.e., it is equipped with three complex structures $(\mathbf{I}, \mathbf{J}, \mathbf{K})$ satisfying the quaternionic relations $\mathbf{I}^2 = \mathbf{J}^2 = \mathbf{K}^2 = -1$ and $\mathbf{I}\mathbf{J} = \mathbf{K}$, and three Kähler two-forms $(\omega^{(\mathbf{I})}, \omega^{(\mathbf{J})}, \omega^{(\mathbf{K})})$.
  From the pullback of the Kähler two-form of $\mathcal{M}^G_{\text{H}}(C)$ onto $\Sigma \times \R$ in the final expression of our sigma model's action, we find that this particular Kähler two-form is defined (up to a scaling factor) as $\omega = \int_C \mathfrak{d} A_C \wedge \mathfrak{d} B_C$, which, according to Kapustin-Witten's (KW) convention in \cite[eqn. (4.8)]{kapustin-2006-elect-magnet}, is the definition for $\omega^{(\mathbf{K})}$.
  We will use KW's convention throughout this paper.
  Therefore, the complex structure of $\mathcal{M}^G_{\text{H}}(C)$ in our sigma model is $\mathbf{K}$, so we shall henceforth denote it as $\mathcal{M}^G_{\text{H}}(C, \mathbf{K})$.
  \label{ft:3d sigma model in structure K}
} i.e.,
\begin{equation}
  \label{eq:hw:bjsv:sigma model:action}
  \begin{aligned}
    S_{\text{3d}}
    =
    & \frac{1}{e^2} \int_{\Sigma \times \R}
      dt |dw|^2 g_{\bar{\imath}j} \, \bigg(
      \partial_t X^i \partial_t X^{\bar{\jmath}}
      + \partial_t Y^i \partial_t Y^{\bar{\jmath}}
      - \lambda^i \nabla_t \bar{\eta}^{\bar{\jmath}}
      - \rho^i \nabla_t \bar{\zeta}^{\bar{\jmath}}
    \\
    & \qquad \qquad
      + \partial_w X^i \partial_{\bar{w}} X^{\bar{\jmath}}
      + \partial_w Y^i \partial_{\bar{w}} Y^{\bar{\jmath}}
      + 2i \left(
        \lambda^i \nabla_w \bar{\zeta}^{\bar{\jmath}}
        + \bar{\lambda}^{\bar{\imath}} \nabla_{\bar{w}} \zeta^j
        + \rho^i \nabla_{\bar{w}} \bar{\eta}^{\bar{\jmath}}
        + \bar{\rho}^{\bar{\imath}} \nabla_w \eta^j
      \right)
      \bigg)
    \\
      & + \frac{1}{e^2}
        \int_{\Sigma \times \R} \mathfrak{R}^{\bar{\imath}j}_{\bar{k}l} \left(
        \bar{\lambda}_{\bar{\imath}} \lambda_j
        - \bar{\rho}_{\bar{\imath}} \rho_j
      \right) \left(
        \bar{\eta}^{\bar{k}} \eta^l
        - \bar{\zeta}^{\bar{k}} \zeta^l
      \right)
      \, .
  \end{aligned}
\end{equation}
Here, (i) $w$ and $\bar{w}$ are complex coordinates of $\Sigma$;
(ii) $g_{\bar{\imath}j}$ is the metric on $\mathcal{M}^G_{\text{H}}(C, \mathbf{K})$;
and (iii) $\mathfrak{R}^{\bar{\imath}j}_{\bar{k}l}$ is the Riemann curvature tensor of $\mathcal{M}^G_{\text{H}}(C, \mathbf{K})$.
The bosonic worldsheet scalars $(X, Y)$ are maps $\Phi: \Sigma \times \R \rightarrow \mathcal{M}^G_{\text{H}}(C, \mathbf{K})$ corresponding to the $(A_C, B_C)$ fields on $C$.
The fermions are sections $(\eta, \zeta, \lambda, \rho) \in \Gamma(\Sigma \times \R, \Phi^{*}\mathcal{T})$ and $(\bar{\eta}, \bar{\zeta}, \bar{\lambda}, \bar{\rho}) \in \Gamma(\Sigma \times \R, \Phi^{*}\overline{\mathcal{T}})$ where $\mathcal{T} \coloneq T\mathcal{M}^G_{\text{H}}(C, \mathbf{K})$ is the tangent bundle of $\mathcal{M}^G_{\text{H}}(C, \mathbf{K})$ and $\overline{\mathcal{T}}$ is its complex conjugate.
Covariant derivatives of the fermions are defined as $\nabla_M \Psi^i = \partial_M \Psi^i + \varGamma^i_{jk} \Psi^j \partial_M(X^k + Y^k)$ and $\nabla_M \Psi^{\bar{\imath}} = \partial_M \Psi^{\bar{\imath}} + \varGamma^{\bar{\imath}}_{\bar{\jmath}\bar{k}} \Psi^{\bar{\jmath}} \partial_M(X^{\bar{k}} + Y^{\bar{k}})$, where $M \in \{t, w, \bar{w}\}$ and $\varGamma$ is the affine connection of $\mathcal{M}^G_{\text{H}}(C, \mathbf{K})$.

This 3d sigma model is invariant under the following supersymmetric transformations:
\begin{equation}
  \label{eq:hw:bjsv:sigma model:susy variations}
  \begin{aligned}
    \delta_{\mathcal{Q}} X^i
    &= \eta^i
      \, ,
    &\qquad
      \delta_{\mathcal{Q}} \eta^i
    &= 0
      \, ,
    &\qquad
      \delta_{\mathcal{Q}} \lambda^i
    &= - \partial_t X^i
      + \frac{1}{2} \partial_{\bar{w}} Y^i
      + i \varGamma^i_{jk} \lambda^j \left( \eta^k + \zeta^k \right)
      \, ,
    \\
    \delta_{\mathcal{Q}} X^{\bar{\imath}}
    &= \bar{\eta}^{\bar{\imath}}
      \, ,
    &\qquad
      \delta_{\mathcal{Q}} \bar{\eta}^{\bar{\imath}}
    &= 0
      \, ,
    &\qquad
      \delta_{\mathcal{Q}} \bar{\lambda}^{\bar{\imath}}
    &= - \partial_t X^{\bar{\imath}}
      + \frac{1}{2} \partial_w Y^{\bar{\imath}}
      + i \varGamma^{\bar{\imath}}_{\bar{\jmath}\bar{k}} \bar{\lambda}^{\bar{\jmath}} \left( \bar{\eta}^{\bar{k}} + \bar{\zeta}^{\bar{k}} \right)
      \, ,
    \\
    \delta_{\mathcal{Q}} Y^i
    &= \zeta^i
      \, ,
    &\qquad
      \delta_{\mathcal{Q}} \zeta^i
    &= 0
      \, ,
    &\qquad
      \delta_{\mathcal{Q}} \rho^i
    &= \partial_t Y^i
      + \frac{1}{2} \partial_w X^i
      + i \varGamma^i_{jk} \rho^j \left( \eta^k + \zeta^k \right)
      \, ,
    \\
    \delta_{\mathcal{Q}} Y^{\bar{\imath}}
    &= \bar{\zeta}^{\bar{\imath}}
      \, ,
    &\qquad
      \delta_{\mathcal{Q}} \bar{\zeta}^{\bar{\imath}}
    &= 0
      \, ,
    &\qquad
      \delta_{\mathcal{Q}} \bar{\rho}^{\bar{\imath}}
    &= \partial_t Y^{\bar{\imath}}
      + \frac{1}{2} \partial_{\bar{w}} X^{\bar{\imath}}
      + i \varGamma^{\bar{\imath}}_{\bar{\jmath}\bar{k}} \bar{\rho}^{\bar{\jmath}} \left( \bar{\eta}^{\bar{k}} + \bar{\zeta}^{\bar{k}} \right)
      \, .
  \end{aligned}
\end{equation}

In short, topological reduction of HW theory on $C \times \Sigma \times \R$ along $C$ results in a 3d $\mathcal{N} = 4$ topological sigma model on $\Sigma \times \R$ with target $\mathcal{M}^G_{\text{H}}(C, \mathbf{K})$, whose action is \eqref{eq:hw:bjsv:sigma model:action}, and is invariant under the supersymmetric transformations \eqref{eq:hw:bjsv:sigma model:susy variations}.

\subtitle{An Effective 2d A-model}

The 3d sigma model on $\Sigma \times \R$ with target $\mathcal{M}^G_{\text{H}}(C, \mathbf{K})$ can be equivalently recast as a 1d SQM on $\R$ with target the space of maps $\mathfrak{M}(\Sigma, \mathcal{M}^G_{\text{H}}(C, \mathbf{K}))$ from $\Sigma$ to $\mathcal{M}^G_{\text{H}}(C, \mathbf{K})$, which is a Riemannian manifold.\footnote{%
  In our 3d sigma model, $\mathfrak{M}(\Sigma, \mathcal{M}^G_{\text{H}}(C, \mathbf{K}))$ is more precisely the space of all $X, Y$ maps from $\Sigma$ to $\mathcal{M}^G_{\text{H}}(C, \mathbf{K})$; it is an infinite-dimensional affine space.
  Let us denote by $\delta$ the exterior derivative on $\mathfrak{M}(\Sigma, \mathcal{M}^G_{\text{H}}(C, \mathbf{K}))$.
  The flat Riemannian metric of $\mathfrak{M}(\Sigma, \mathcal{M}^G_{\text{H}}(C, \mathbf{K}))$ can be defined as $g_{\alpha\beta} = \int_{\Sigma} g_{i\bar{\jmath}} \left( \delta_{\alpha} X^i \wedge \star \delta_{\beta} X^{\bar{\jmath}} + \delta_{\alpha} Y^i \wedge \star \delta_{\beta} Y^{\bar{\jmath}} + ( \alpha \leftrightarrow \beta) \right)$, where $\wedge$ and $\star$ are the wedge product and Hodge star operator on $\Sigma$, respectively.
  \label{ft:riemann metric of maps from sigma to hitchin moduli space}
}
Generically, the $\mathcal{Q}$-cohomology of this equivalent 1d SQM is given by the de Rham cohomology of $\mathfrak{M}(\Sigma, \mathcal{M}^G_{\text{H}}(C, \mathbf{K}))$.
However, the actual $\mathcal{Q}$-cohomology that its partition function localizes onto corresponds to the de Rham cohomology of the space of \emph{holomorphic} maps $\mathfrak{M}^{\text{hol}}(\Sigma, \mathcal{M}^G_{\text{H}}(C, \mathbf{K}))$ from $\Sigma$ to $\mathcal{M}^G_{\text{H}}(C, \mathbf{K})$ \cite[$\S$6.3]{er-2023-topol-n}.
Since the 3d sigma model on $\Sigma \times \R$ and the 1d SQM on $\R$ are equivalent, this in turn means that the $\mathcal{Q}$-cohomology which the partition function of the 3d sigma model localizes onto can also be described by the de Rham cohomology of $\mathfrak{M}^{\text{hol}}(\Sigma, \mathcal{M}^G_{\text{H}}(C, \mathbf{K}))$.

Note that the only QFT whose $\mathcal{Q}$-cohomology is described by the de Rham cohomology of $\mathfrak{M}^{\text{hol}}(\Sigma, \mathcal{M}^G_{\text{H}}(C, \mathbf{K}))$ is a 2d A-model on $\Sigma$ with target $\mathcal{M}^G_{\text{H}}(C, \mathbf{K})$.
This therefore means that the $\mathcal{Q}$-cohomology of our 3d sigma model on $\Sigma \times \R$ is  \emph{effectively} the $\mathcal{Q}$-cohomology of a 2d A-model on $\Sigma$.

\subtitle{The 2d A-model from Another Perspective}

We can also understand this as follows.
The 3d sigma model is a \emph{topological} sigma model.
This means that its Hamiltonian is zero, i.e., its spectrum of states (given by its $\mathcal{Q}$-cohomology) are necessarily spanned by supersymmetric ground states that are time-invariant.
Therefore, to study the spectrum of the 3d sigma model, it suffices to consider a time-invariant hyperslice along $\R$, i.e., consider an instant in time of the 3d sigma model.
We do this by picking a specific value of $t$ along $\R$ and treat the fields as being constant in time, i.e., by inserting a delta-function $\delta(t - T_0)$ and setting $\partial_t \rightarrow 0$ in \eqref{eq:hw:bjsv:sigma model:action}.
The action on the hyperslice becomes
\begin{equation}
  \label{eq:hw:bjsv:sigma model:hyperslice:action}
  \begin{aligned}
    S_{\text{2d}}
    =
    & \frac{1}{e^2}
      \int_{\Sigma} |dw|^2 g_{\bar{\imath}j} \bigg(
      \partial_w X^i \partial_{\bar{w}} X^{\bar{\jmath}}
      + \partial_w Y^i \partial_{\bar{w}} Y^{\bar{\jmath}}
      + 2i \left(
        \lambda^i \nabla_w \bar{\zeta}^{\bar{\jmath}}
        + \bar{\lambda}^{\bar{\imath}} \nabla_{\bar{w}} \zeta^j
        + \rho^i \nabla_{\bar{w}} \bar{\eta}^{\bar{\jmath}}
        + \bar{\rho}^{\bar{\imath}} \nabla_w \eta^j
      \right)
      \bigg)
    \\
      & + \frac{1}{e^2}
        \int_{\Sigma} \mathfrak{R}^{\bar{\imath}j}_{\bar{k}l} \left(
        \bar{\lambda}_{\bar{\imath}} \lambda_j
        - \bar{\rho}_{\bar{\imath}} \rho_j
      \right) \left(
        \bar{\eta}^{\bar{k}} \eta^l
        - \bar{\zeta}^{\bar{k}} \zeta^l
      \right)
      \, ,
  \end{aligned}
\end{equation}
and is left invariant under the following supersymmetric transformations:
\begin{equation}
  \label{eq:hw:bjsv:sigma model:hyperslice:susy variations}
  \begin{aligned}
    \delta_{\mathcal{Q}} X^i
    &= \eta^i
      \, ,
    &\qquad
      \delta_{\mathcal{Q}} \eta^i
    &= 0
      \, ,
    &\qquad
      \delta_{\mathcal{Q}} \lambda^i
    &= \frac{1}{2} \partial_{\bar{w}} Y^i
      + i \varGamma^i_{jk} \lambda^j \zeta^k
      \, ,
    \\
    \delta_{\mathcal{Q}} X^{\bar{\imath}}
    &= \bar{\eta}^{\bar{\imath}}
      \, ,
    &\qquad
      \delta_{\mathcal{Q}} \bar{\eta}^{\bar{\imath}}
    &= 0
      \, ,
    &\qquad
      \delta_{\mathcal{Q}} \bar{\lambda}^{\bar{\imath}}
    &= \frac{1}{2} \partial_w Y^{\bar{\imath}}
      + i \varGamma^{\bar{\imath}}_{\bar{\jmath}\bar{k}} \bar{\lambda}^{\bar{\jmath}} \bar{\zeta}^{\bar{k}}
      \, ,
    \\
    \delta_{\mathcal{Q}} Y^i
    &= \zeta^i
      \, ,
    &\qquad
      \delta_{\mathcal{Q}} \zeta^i
    &= 0
      \, ,
    &\qquad
      \delta_{\mathcal{Q}} \rho^i
    &= \frac{1}{2} \partial_w X^i
      + i \varGamma^i_{jk} \rho^j \eta^k
      \, ,
    \\
    \delta_{\mathcal{Q}} Y^{\bar{\imath}}
    &= \bar{\zeta}^{\bar{\imath}}
      \, ,
    &\qquad
      \delta_{\mathcal{Q}} \bar{\zeta}^{\bar{\imath}}
    &= 0
      \, ,
    &\qquad
      \delta_{\mathcal{Q}} \bar{\rho}^{\bar{\imath}}
    &= \frac{1}{2} \partial_{\bar{w}} X^{\bar{\imath}}
      + i \varGamma^{\bar{\imath}}_{\bar{\jmath}\bar{k}} \bar{\rho}^{\bar{\jmath}} \bar{\eta}^{\bar{k}}
      \, .
  \end{aligned}
\end{equation}
Notice that we can regard the fermions
(i) $(\eta, \zeta) \in \Omega^0(\Sigma, \Phi^{*} \mathcal{T})$ and $(\bar{\eta}, \bar{\zeta}) \in \Omega^0(\Sigma, \Phi^{*} \overline{\mathcal{T}})$ as zero-forms on $\Sigma$;
(ii) $\bar{\lambda} \in \Omega^{(1, 0)}(\Sigma, \Phi^{*} \overline{\mathcal{T}})$ and $\rho \in \Omega^{(1, 0)}(\Sigma, \Phi^{*} \mathcal{T})$ as $(1, 0)$-forms on $\Sigma$;
and (iii) $\lambda \in \Omega^{(0, 1)}(\Sigma, \Phi^{*} \mathcal{T})$ and $\bar{\rho} \in \Omega^{(0, 1)}(\Sigma, \Phi^{*} \overline{\mathcal{T}})$ as $(0, 1)$-forms on $\Sigma$.
It is then clear that \eqref{eq:hw:bjsv:sigma model:hyperslice:action} is the action of a 2d $\mathcal{N} = (4, 4)$ A-model on $\Sigma$ with target $\mathcal{M}^G_{\text{H}}(C, \mathbf{K})$.

In short, the spectrum of the 3d $\mathcal{N} = 4$ topological sigma model on $\Sigma \times \R$ with target $\mathcal{M}^G_{\text{H}}(C, \mathbf{K})$ is indeed given by the spectrum of a 2d $\mathcal{N} = (4, 4)$ topological A-model on $\Sigma$ with target $\mathcal{M}^G_{\text{H}}(C, \mathbf{K})$.

\subtitle{A 3d A-model in $\mathcal{M}^G_{\text{H}}(C, \mathbf{K})$}

Hence, we can regard the 3d sigma model as an \emph{A-model}.
In other words, BJSV reduction of HW theory on $C \times \Sigma \times \R$ along $C$ results in a 3d $\mathcal{N} = 4$ topological A-model on $\Sigma \times \R$ with target $\mathcal{M}^G_{\text{H}}(C, \mathbf{K})$.

\subsection{An Equivalence Between HW Theory on \texorpdfstring{$M_3 \times M_1 \times \R$}{M3 x M1 x R} and the 3d A-model on \texorpdfstring{$I \times M_1 \times \R$}{I x M1 x R}}
\label{sec:hw:equivalence}

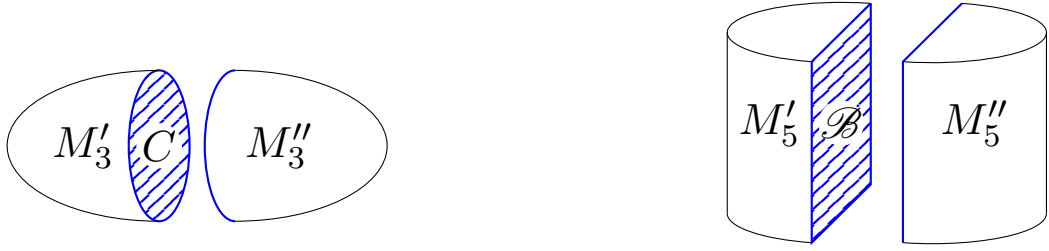
\begin{figure}
  \centering
  \begin{subfigure}{0.45\textwidth}
    \centering
    \begin{tikzpicture}[auto]
      \draw[blue, pattern={Lines[angle=45,distance=4pt]},pattern color=blue, thick] (0,0) ellipse (0.4 and 1);
      \draw (0,1) arc(90:270:2 and 1);
      \draw[fill=white,draw=none] (0,0) circle[radius=0.3] node [scale=1.5] (sigma-left) {$C$};
      \node [left of=sigma-left,scale=1.5] {$M_3'$};
      \draw[blue, thick] (1,1) arc(90:270:0.4 and 1);
      \draw (1,-1) arc(270:450:2 and 1);
      \node [right=0.5cm of sigma-left,scale=1.5] {$M_3''$};
      \node [below=1.3cm of sigma-left] {};
    \end{tikzpicture}
    \caption{$M_3$ as a connected sum of three-manifolds $M_3'$ and $M_3''$ along the Riemann surface $C$.}
    \label{fig:heegaard:m3}
  \end{subfigure}
  \hfill
  \begin{subfigure}{0.45\textwidth}
    \centering
    \begin{tikzpicture}[auto]
      \def \eliA {1.5} 
      \def \eliB {0.4} 
      \def \eliAngle {15} 
      \def \faceLength {1.2} 
      \def \sepLength {0.6} 
      \def \nodeScale {1.5} 
      \draw ({\eliA*sin(\eliAngle) - \sepLength}, {\eliB*cos(\eliAngle) + \faceLength})
      arc({90 - \eliAngle}:{270 - \eliAngle}:{\eliA} and {\eliB});
      \draw[blue, pattern={Lines[angle=45,distance=4pt]},pattern color=blue, thick]
      ({\eliA*sin(\eliAngle) - \sepLength}, {\eliB*cos(\eliAngle) + \faceLength})
      -- ({-\eliA*sin(\eliAngle) - \sepLength}, {-\eliB*cos(\eliAngle) + \faceLength})
      -- ({-\eliA*sin(\eliAngle) - \sepLength}, {-\eliB*cos(\eliAngle) - \faceLength})
      -- ({\eliA*sin(\eliAngle) - \sepLength}, {\eliB*cos(\eliAngle) - \faceLength})
      -- ({\eliA*sin(\eliAngle) - \sepLength}, {\eliB*cos(\eliAngle) + \faceLength});
      \draw[fill=white,draw=none] ({-\sepLength},0) circle[radius=0.3] node [scale=\nodeScale] (sigma-left) {$\mathscr{B}$};
      \draw ({-\eliA - \sepLength}, {\faceLength})
      -- ({-\eliA - \sepLength}, {-\faceLength});
      \draw ({-\eliA - \sepLength}, {-\faceLength})
      arc(180:{270 - \eliAngle}:{\eliA} and {\eliB});
      \node [scale=\nodeScale] at ({-\eliA*(1 + sin(\eliAngle))/2 - \sepLength}, 0) (M5-prime) {$M_5'$};
      \draw ({\eliA*sin(\eliAngle) + \sepLength}, {\eliB*cos(\eliAngle) + \faceLength})
      arc({360 + 90 - \eliAngle}:{270 - \eliAngle}:{\eliA} and {\eliB});
      \draw[blue, thick] ({\eliA*sin(\eliAngle) + \sepLength}, {\eliB*cos(\eliAngle) + \faceLength})
      -- ({-\eliA*sin(\eliAngle) + \sepLength}, {-\eliB*cos(\eliAngle) + \faceLength})
      -- ({-\eliA*sin(\eliAngle) + \sepLength}, {-\eliB*cos(\eliAngle) - \faceLength});
      \draw ({\eliA + \sepLength}, {\faceLength})
      -- ({\eliA + \sepLength}, {-\faceLength});
      \draw ({-\eliA*sin(\eliAngle) + \sepLength}, {-\eliB*cos(\eliAngle) - \faceLength}) arc({270 - \eliAngle}:360:{\eliA} and {\eliB});
      \node [scale=\nodeScale] at ({\eliA*(1 - sin(\eliAngle))/2 + \sepLength}, 0) (M5-primeprime) {$M_5''$};
      \node [below=1.5cm of M5-primeprime] {};
    \end{tikzpicture}
    \caption{$M_5$ splits into five-manifolds $M_5'$ and $M_5''$ along their common boundary $\mathscr{B}$.}
    \label{fig:heegaard:m5}
  \end{subfigure}
  \caption{Heegaard splits of $M_3$ and $M_5$.}
  \label{fig:heegaard}
\end{figure}

Consider now HW theory on $M_3 \times M_1 \times \R$, where $M_3$ is a closed three-manifold.
In \cite[$\S$7]{er-2023-topol-n}, we showed how this HW theory is equivalent to the 3d A-model on $I \times M_1 \times \R$ with target $\mathcal{M}^G_{\text{H}}(C, \mathbf{K})$ via a Heegaard split of $M_3$ along a Heegaard surface $C$, i.e., $M_3 = M_3' \bigcup_C M_3''$ (illustrated in \autoref{fig:heegaard:m3}).
Let us briefly describe the process.

First, a Heegaard split of HW theory on $M_3 \times M_1 \times \R$ along the Heegaard surface $C$ results in two copies of HW theory, one on $M_5' = M_3' \times M_1 \times \R$ and another on $M_5'' = M_3'' \times M_1 \times \R$, glued along their common boundary $\mathscr{B}$.
Topological invariance of HW theory means that we can perform a Weyl rescaling of the five-manifolds' metric, whence we can regard $M_5' = C \times I' \times M_1 \times \R$, $M_5'' = C \times I'' \times M_1 \times \R$, and thus $\mathscr{B} = C \times M_1 \times \R$.
This allows us to interpret HW theory on $M_3 \times M_1 \times \R$ as the gluing of two copies of HW theory along $\mathscr{B}$.
This is illustrated in \autoref{fig:heegaard:m5}.

\begin{figure}
  \centering
  \begin{tikzpicture}[%
    auto,%
    shorten >=-2pt,%
    shorten <=-2pt,%
    box/.style={rectangle, text centered, minimum height=10em,text width=8mm,draw,fill=gray!40},%
    shortwave/.style={*-*,thick,decorate,decoration={snake,amplitude=5pt,segment length=1.3cm}},%
    longwave/.style={*-*,thick,decorate,decoration={snake,amplitude=5pt,segment length=1.7cm}},%
    ]
    \node [box] (ori-l0) {$\mathscr{L}'_0$};
    \node [box, right=1.5cm of ori-l0] (ori-l1-prime) {$\mathscr{L}'_1$};
    \node [box, right=0cm of ori-l1-prime] (ori-l0-prime) {$\mathscr{L}''_0$};
    \node [box, right=1.5cm of ori-l0-prime] (ori-l1) {$\mathscr{L}''_1$};
    \path (ori-l1-prime) -- (ori-l0-prime) coordinate[midway] (R-aux);
    \node [below=5em of R-aux] (R) {$\mathscr{R}$};
    \draw [shortwave] (ori-l0) -- node[below=10pt] {$\R \times I'$} (ori-l1-prime);
    \draw [shortwave] (ori-l0-prime) -- node[below=10pt] {$\R \times I''$} (ori-l1);
    \node [box, right=3cm of ori-l1] (fin-l0) {$\mathscr{L}'_0$};
    \node [box, right=2cm of fin-l0] (fin-l1) {$\mathscr{L}''_1$};
    \draw [longwave] (fin-l0) -- node[below=10pt] {$\R \times I$} (fin-l1);
    \draw [-{Latex[length=3mm]}, thick, shorten >=10pt, shorten <=10pt] (ori-l1) -- (fin-l0);
  \end{tikzpicture}
  \caption{Union of 2d A-models along their common boundary $\mathscr{R}$.}
  \label{fig:a-models}
\end{figure}
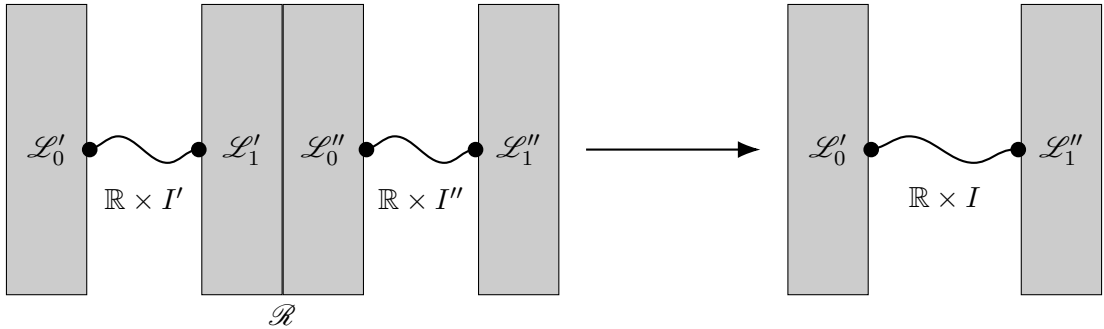

Next, we perform a topological reduction of the two copies of HW theory along $C$ to get 3d A-models with target $\mathcal{M}^G_{\text{H}}(C, \mathbf{K})$ on $I' \times M_1 \times \R$ and $I'' \times M_1 \times \R$.
These 3d A-models can be equivalently interpreted as 2d A-models on $I' \times \R$ and $I'' \times \R$ with target the space of maps $\mathfrak{M}(M_1, \mathcal{M}^G_{\text{H}}(C, \mathbf{K}))$ from $M_1$ to $\mathcal{M}^G_{\text{H}}(C, \mathbf{K})$, with branes $(\mathscr{L}'_0, \mathscr{L}'_1)$ and $(\mathscr{L}''_0, \mathscr{L}''_1)$, respectively.
The gluing of the two copies of HW theory along their common boundary $\mathscr{B}$ is realized as the union of the two 2d A-models along their common boundary $\mathscr{R} = \mathscr{L}'_1 = \mathscr{L}''_0$.
Here, the union can be interpreted as a concatenation of the intervals $I'$ and $I''$ of the two 2d A-models into a single interval $I = I' \oplus I''$ of a 2d A-model on $I \times \R$ with target $\mathfrak{M}(M_1, \mathcal{M}^G_{\text{H}}(C, \mathbf{K}))$.
This is illustrated in \autoref{fig:a-models}.
In turn, this final 2d A-model can be equivalently interpreted as a 3d A-model on $I \times M_1 \times \R$ with target $\mathcal{M}^G_{\text{H}}(C, \mathbf{K})$.

Lastly, since HW theory is invariant under the topological reduction along $C$, we have an equivalence between HW theory on $M_3 \times M_1 \times \R$ and the 3d A-model on $I \times M_1 \times \R$ with target $\mathcal{M}^G_{\text{H}}(C, \mathbf{K})$.

\section{A 5d \texorpdfstring{$\mathcal{N} = 2$}{N = 2} Geyer-M\"{u}lsch-twisted Topological Gauge Theory}
\label{sec:gm}

In this section, we will study the 5d $\mathcal{N} = 2$ gauge theory on a five-manifold $M_5 = M_4 \times \R$ with the Geyer-M\"{u}lsch (GM) twist,\footnote{%
  We refer the reader to \cite{geyer-2003-higher-dimen} for a detailed description of the GM twist.
  \label{ft:description of the GM twist}
} whose gauge group is a real, simple, compact Lie group $G$, and whose BPS equations that its path integral localizes onto are flat $G_\C$ connections on $M_5$ \cite{geyer-2003-higher-dimen}.
We will also consider several specializations of $M_5$ that allow us to derive a 3d $\mathcal{N} = 4$ B-model whose spectrum is effectively given by that of a 2d $\mathcal{N} = (4, 4)$ B-model.
Finally, via topological invariance, we will obtain an equivalence between the 5d $\mathcal{N} = 2$ GM-twisted gauge theory on $M_3 \times M_1 \times \R$ and a 3d B-model on $I \times M_1 \times \R$.

\subsection{GM Theory on \texorpdfstring{$M_5 = M_4 \times \R$}{M5 = M4 x R}}
\label{sec:gm:5d}

\subtitle{The Field Composition}

The bosonic field content of GM theory on $M_4 \times \R$ consists of an auxiliary scalar $H \in \Omega^0(M_5, \text{ad}(G))$, gauge connections $A_\mu \in \Omega^1(M_4, \text{ad}(G)) \otimes \Omega^0(\R, \text{ad}(G))$ and $A_t \in \Omega^0(M_4, \text{ad}(G)) \otimes \Omega^1(\R, \text{ad}(G))$, and one-forms $\phi_\mu \in \Omega^1(M_4, \text{ad}(G))  \otimes \Omega^0(\R, \text{ad}(G))$ and $\phi_t \in \Omega^0(M_4, \text{ad}(G)) \otimes \Omega^1(\R, \text{ad}(G))$.
The fermionic content of the theory consists of a scalar $\eta \in \Omega^0(M_4 \times \R, \text{ad}(G))$, one-forms $\psi_{\mu} \in \Omega^1(M_4, \text{ad}(G)) \otimes \Omega^0(\R, \text{ad}(G))$ and $\psi_t \in \Omega^0(M_4, \text{ad}(G)) \otimes \Omega^1(\R, \text{ad}(G))$, and two-forms $\chi_{\mu\nu} \in \Omega^2(M_4, \text{ad}(G_{\C})) \otimes \Omega^0(\R, \text{ad}(G_{\C}))$ and $\Lambda_{\mu} \in \Omega^1(M_4, \text{ad}(G_{\C})) \otimes \Omega^1(\R, \text{ad}(G_{\C}))$.
Here, $G_{\C}$ is the complexification of the real Lie group $G$.

As the supersymmetry generators transform in the same representation as the fermions above, the GM twist will result in a single nilpotent \emph{scalar} supersymmetry generator $\hat{\mathcal{Q}}$.
The supersymmetry transformations of the twisted fields generated by $\hat{\mathcal{Q}}$ \cite{geyer-2003-higher-dimen} are
\begin{equation}
  \label{eq:gm:susy variations}
  \begin{aligned}
    \delta_{\hat{Q}} A_t
    &= \psi_t
      \, ,
    &\qquad
      \delta_{\hat{Q}} \phi_t
    &= - i \psi_t
      \, ,
    \\
    \delta_{\hat{Q}} A_{\mu}
    &= \psi_{\mu}
      \, ,
    &\qquad
      \delta_{\hat{Q}} \phi_{\mu}
    &= - i \psi_{\mu}
      \, ,
    \\
    \delta_{\hat{Q}} \mathcal{A}_t
    &= 2 \psi_t
      \, ,
    &\qquad
      \delta_{\hat{Q}} \overline{\mathcal{A}}_t
    &= 0
      \, ,
    \\
    \delta_{\hat{Q}} \mathcal{A}_{\mu}
    &= 2 \psi_{\mu}
      \, ,
    &\qquad
      \delta_{\hat{Q}} \overline{\mathcal{A}}_{\mu}
    &= 0
      \, ,
    \\
    \delta_{\hat{Q}} \psi_t
    &= 0
      \, ,
    &\qquad
      \delta_{\hat{Q}} \psi_{\mu}
    &= 0
      \, ,
    \\
    \delta_{\hat{Q}} \Lambda_{\mu}
    &= - i \overline{\mathcal{F}}_{t\mu}
      \, ,
    &\qquad
      \delta_{\hat{Q}} \chi_{\mu\nu}
    &= - i \overline{\mathcal{F}}_{\mu\nu}
      \, ,
    \\
    \delta_{\hat{Q}} \eta
    &= H
      \, ,
    &\qquad
      \delta_{\hat{Q}} H
    &= 0
      \, ,
  \end{aligned}
\end{equation}
where $\delta_{\hat{\mathcal{Q}}}$ denotes a $\hat{\mathcal{Q}}$-variation;
(Ia) $\mathcal{A}_{\mu} \coloneq A_{\mu} + i \phi_{\mu} \in \Omega^1(M_4, \text{ad}(G_{\C})) \otimes \Omega^0(\R, \text{ad}(G_{\C}))$,
(Ib) $\mathcal{A}_t \coloneq A_t + i \phi_t \in \Omega^0(M_4, \text{ad}(G_{\C})) \otimes \Omega^1(\R, \text{ad}(G_{\C}))$,
(Ic) $\overline{\mathcal{A}}_{\mu} \coloneq A_{\mu} - i \phi_{\mu} \in \Omega^1(M_4, \text{ad}(G_{\C})) \otimes \Omega^0(\R, \text{ad}(G_{\C}))$,
and (Id) $ \overline{\mathcal{A}}_t \coloneq A_t - i \phi_t \in \Omega^0(M_4, \text{ad}(G_{\C})) \otimes \Omega^1(\R, \text{ad}(G_{\C}))$ are $G_{\C}$-connections;
and (II) $\overline{\mathcal{F}}_{t\mu} = \partial_t \overline{\mathcal{A}}_{\mu} - \partial_{\mu} \overline{\mathcal{A}}_t + [ \overline{\mathcal{A}}_t, \overline{\mathcal{A}}_{\mu} ]$ and $\overline{\mathcal{F}}_{\mu\nu} = \partial_{\mu} \overline{\mathcal{A}}_{\nu} - \partial_{\nu} \overline{\mathcal{A}}_{\mu} + [ \overline{\mathcal{A}}_{\mu}, \overline{\mathcal{A}}_{\nu} ]$ are field strengths of $\overline{\mathcal{A}}$.

Notice that unlike $\delta_{\mathcal{Q}}$ in \autoref{sec:hw:5d}, $\delta_{\hat{\mathcal{Q}}}$ is nilpotent.

\subtitle{The BPS Equations}

The BPS equations of GM theory on $M_4 \times \R$ are obtained by setting to zero the $\hat{\mathcal{Q}}$-variations of the fermions in \eqref{eq:gm:susy variations}, i.e.,
\begin{equation}
  \label{eq:gm:bps}
  \overline{\mathcal{F}}_{t\mu}
  = 0
    \, ,
  \qquad
  \overline{\mathcal{F}}_{\mu\nu}
  = 0
    \, .
\end{equation}
These are equations defining flat $G_{\C}$-connections on $M_4 \times \R$.
(We shall henceforth refer to similar such equations as $G_{\C}$-flat equations.)
Configurations in the space $\mathfrak{B}_5$ of all $\overline{\mathcal{A}}$ fields on $M_4 \times \R$ that satisfy \eqref{eq:gm:bps}, constitute a moduli space $\mathcal{M}_{\text{GM}}$ that the path integral of GM theory on $M_4 \times \R$ will localize onto.

\subtitle{The $\hat{\mathcal{Q}}$-exact Action}

The $\hat{\mathcal{Q}}$-exact topological action of GM theory on $M_4 \times \R$ is \cite{geyer-2003-higher-dimen}
\begin{equation}
  \label{eq:gm:m4 x r:action}
  \begin{aligned}
    S_{\text{GM}}
    = \frac{1}{e^2} \int_{M_4 \times \R} dt d^4x \, \Tr \bigg(
    & \frac{1}{2} \overline{\mathcal{F}}_{t\mu} \mathcal{F}^{t\mu}
      + \frac{1}{4} \overline{\mathcal{F}}_{\mu\nu} \mathcal{F}^{\mu\nu}
      - i \Lambda^{\mu} (\mathcal{D}_t \psi_{\mu} - \mathcal{D}_{\mu} \psi_t)
      - i \chi^{\mu\nu} \mathcal{D}_{\mu} \psi_{\nu}
    \\
    & - i \psi^t \overline{\mathcal{D}}_t \eta
      - i \psi^{\mu} \overline{\mathcal{D}}_{\mu} \eta
      - H D^t \phi_t
      - H D^{\mu} \phi_{\mu}
      - \frac{1}{2} H^2
    \\
    & - \frac{i}{8} \epsilon^{\mu\nu\rho\sigma} \left(
      \chi_{\mu\nu} \overline{\mathcal{D}}_t \chi_{\rho\sigma}
      + 2 \Lambda_{\mu} \overline{\mathcal{D}}_{\nu} \chi_{\rho\sigma}
      + 2 \chi_{\rho\sigma} \overline{\mathcal{D}}_{\nu} \Lambda_{\mu}
      \right)
      \bigg)
      \, .
  \end{aligned}
\end{equation}

Note that as we will ultimately only be interested in the $\hat{\mathcal{Q}}$-cohomology spectrum of GM theory, we can subtract $\hat{\mathcal{Q}}$-exact terms $\delta_{\hat{\mathcal{Q}}} (\eta D^t \phi_t)$, $\delta_{\hat{\mathcal{Q}}} (\eta D^{\mu} \phi_{\mu})$, and $\delta_{\hat{\mathcal{Q}}} (\eta H / 2)$.
This leaves us with
\begin{equation}
  \label{eq:gm:action}
  \begin{aligned}
    S_{\text{GM}}
    = & \frac{1}{e^2} \int_{M_4 \times \R} dt d^4 x \, \Tr \bigg(
        \frac{1}{2} \left|  \overline{\mathcal{F}}_{t\mu} \right|^2
        + \frac{1}{4} \left| \overline{\mathcal{F}}_{\mu\nu}  \right|^2
        + \dots
        \bigg)
        \, ,
  \end{aligned}
\end{equation}
where the ``$\dots$'' contain the fermion terms in the action.

\subtitle{A Balanced TQFT}

Recall the explanations in \autoref{sec:hw} which led to the ellipticity of the HW equations \eqref{eq:hw:hw eqn} (modulo gauge equivalence) and thus interpretation of HW theory on $M_4 \times \R$ as a balanced TQFT.
Applying it to the $G_{\C}$-flat equations \eqref{eq:gm:bps} (modulo gauge equivalence), we find that GM theory on $M_4 \times \R$, just like HW theory on $M_4 \times \R$, is also a balanced TQFT.
This will be relevant in later sections.

\subsection{A 3d \texorpdfstring{$\mathcal{N} = 4$}{N = 4} B-model and a 2d \texorpdfstring{$\mathcal{N} = (4, 4)$}{N = (4, 4)} B-model}
\label{sec:gm:3d}

Consider the specialization $M_4 = C \times \Sigma$, where $C$ and $\Sigma$ are Riemann surfaces, and the genus of $C$ is $g(C) \geq 2$.
We will perform a topological reduction of GM theory on $M_5 = C \times \Sigma \times \R$ along $C$ using the BJSV reduction method.

\subtitle{A 3d Sigma Model}

The BJSV reduction process \cite{bershadsky-1995-topol-reduc} is effected in the same manner as described in \autoref{sec:hw:3d}, where we write the metric on $M_5$ as a block diagonal with a vanishing scale parameter $\varepsilon$, i.e., $g_{M_5} = \varepsilon g_C \oplus g_{\Sigma \times \R}$.
The finiteness condition we obtain by setting to zero the divergent terms (in the vanishing limit of $\varepsilon \rightarrow 0$) in the action \eqref{eq:gm:action} is the $G_{\C}$-flat equation on $C$.

The terms in the action \eqref{eq:gm:action} with no power of $\varepsilon$ will remain as the only terms that are relevant in the action of the resulting 3d theory; integrating out the auxiliary fields and performing the integration over $C$, we will arrive at the action of a 3d $\mathcal{N} = 4$ sigma model with target $\mathcal{M}^{G_{\C}}_{\text{flat}}(C) \equiv \mathcal{M}^G_{\text{H}}(C, \mathbf{J})$,\footnote{%
  The moduli space $\mathcal{M}^{G_{\C}}_{\text{flat}}(C)$ of flat $G_{\C}$-connections on $C$ is equivalent to $\mathcal{M}^G_{\text{H}}(C)$ in a certain complex structure, which can be determined as follows.
  The space of $(\overline{\mathcal{A}}_C, \mathcal{A}_C)$ fields on $C$, which are the $G_{\C}$-connections $(\overline{\mathcal{A}}, \mathcal{A})$ restricted to $C$, is an infinite-dimensional affine space.
  Let us denote by $\hat{\mathfrak{d}}$ the exterior derivative on this space; the basis of $\mathcal{M}^G_{\text{H}}(C)$ is thus spanned by $(\hat{\mathfrak{d}} \overline{\mathcal{A}}_C, \hat{\mathfrak{d}} \mathcal{A}_C)$.
  Since $\mathcal{M}^G_{\text{H}}(C)$ is a hyperkähler manifold, i.e., a holomorphic symplectic manifold, it is equipped with three holomorphic symplectic two-forms $(\Upomega^{(\mathbf{I})}, \Upomega^{(\mathbf{J})}, \Upomega^{(\mathbf{K})})$.
  The holomorphic symplectic two-form that appears in the following action is defined (up to a scaling factor) as $\Upomega \coloneq \int_C \hat{\mathfrak{d}} \overline{\mathcal{A}}_C \wedge \hat{\mathfrak{d}} \overline{\mathcal{A}}_C$, which, according to KW's convention in \cite[eqn. (4.9)]{kapustin-2006-elect-magnet}, is the definition for $\Upomega^{(\mathbf{J})}$.
  Therefore, the complex structure of $\mathcal{M}^G_{\text{H}}(C)$ in our sigma model is $\mathbf{J}$, so we shall henceforth denote it as $\mathcal{M}^G_{\text{H}}(C, \mathbf{J})$.
} i.e.,
\begin{equation}
  \label{eq:gm:bjsv:sigma model:rw model}
  \begin{aligned}
    S_{\text{3d}}
    =
    & \frac{1}{e^2} \int_{\Sigma \times \R} dt d^2x \,
    g_{i\bar{\jmath}} \, \left(
      \frac{1}{2} \partial_M \overline{\mathcal{Z}}^i  \partial^M \mathcal{Z}^{\bar{\jmath}}
      -i \varrho_M^i \nabla^M \varsigma^{\bar{\jmath}}
    \right)
    \\
    & + \frac{i}{2e^2} \int_{\Sigma \times \R} \,
      \Upomega_{ij} \left(
        \varrho^i \wedge \nabla \varrho^j
        + \frac{1}{3} \mathfrak{R}^j_{kl\bar{m}} \varrho^i \wedge \varrho^k \wedge \varrho^l \wedge \varsigma^{\bar{m}}
      \right)
    \, .
  \end{aligned}
\end{equation}
Here, (i) $M$ is the index on $\Sigma \times \R$;
(ii) $g_{\bar{\imath}j}$ is the metric on $\mathcal{M}^G_{\text{H}}(C, \mathbf{J})$;
(iii) $\Upomega_{ij}$ is the holomorphic symplectic two-form of $\mathcal{M}^G_{\text{H}}(C, \mathbf{J})$,
and (iv) $\mathfrak{R}^j_{kl\bar{m}}$ is the Riemann curvature tensor of $\mathcal{M}^G_{\text{H}}(C, \mathbf{J})$.
The bosonic worldsheet scalars $(\overline{\mathcal{Z}}, \mathcal{Z})$ are maps $\Phi : \Sigma \times \R \rightarrow \mathcal{M}^G_{\text{H}}(C, \mathbf{J})$ corresponding to the $(\overline{\mathcal{A}}_C, \mathcal{A}_C)$ fields on $C$, respectively.
The fermions are $\varrho \in \Omega^1(\Sigma \times \R, \Phi^{*}\mathcal{T})$ and $\varsigma \in \Omega^0(\Sigma \times \R, \Phi^{*}\overline{\mathcal{T}})$, where $\mathcal{T} \coloneq T\mathcal{M}^G_{\text{H}}(C, \mathbf{J})$ is the tangent bundle of $\mathcal{M}^G_{\text{H}}(C, \mathbf{J})$ and $\overline{\mathcal{T}}$ is its complex conjugate.
Covariant derivatives of the fermions are defined as $\nabla_M \varrho^i_N = \partial_M \varrho^i_N + \varGamma^i_{jk} (\partial_M \overline{\mathcal{Z}}^j) \varrho^k_N$ and $\nabla_M \varsigma^{\bar{\imath}} = \partial_M \varsigma^{\bar{\imath}} + \varGamma^{\bar{\imath}}_{\bar{\jmath}\bar{k}} (\partial_M \mathcal{Z}^{\bar{\jmath}}) \varsigma^{\bar{k}}$, where $\varGamma$ is the affine connection of $\mathcal{M}^G_{\text{H}}(C, \mathbf{J})$.

This 3d sigma model is invariant under the following supersymmetric transformations:
\begin{equation}
  \label{eq:gm:bjsv:sigma model:susy variations}
  \begin{aligned}
    \delta_{\hat{\mathcal{Q}}} \overline{\mathcal{Z}}^i
    &= 0
      \, ,
    &\qquad
      \delta_{\hat{\mathcal{Q}}} \mathcal{Z}^{\bar{\imath}}
    &= \varsigma^{\bar{\imath}}
      \, ,
    \\
    \delta_{\hat{\mathcal{Q}}} \varrho^i
    &= d \phi^i
      \, ,
    &\qquad
      \delta_{\hat{\mathcal{Q}}} \varsigma^{\bar{\imath}}
    &= 0
      \, ,
  \end{aligned}
\end{equation}
where $d$ is the exterior derivative on $\Sigma \times \R$.
Together, \eqref{eq:gm:bjsv:sigma model:rw model} and \eqref{eq:gm:bjsv:sigma model:susy variations} describe the 3d Rozansky-Witten (RW) topological sigma model on $\Sigma \times \R$ \cite{rozansky-1996-hyper-kahler, kapustin-2009-three-dimen}.

In short, topological reduction of GM theory on $C \times \Sigma \times \R$ along $C$ results in a 3d $\mathcal{N} = 4$ RW topological sigma model on $\Sigma \times \R$ with target $\mathcal{M}^G_{\text{H}}(C, \mathbf{J})$, whose action \eqref{eq:gm:bjsv:sigma model:rw model} is invariant under the supersymmetric transformations \eqref{eq:gm:bjsv:sigma model:susy variations}.

\subtitle{An Effective 2d B-model}

The 3d RW model on $\Sigma \times \R$ with target $\mathcal{M}^G_{\text{H}}(C, \mathbf{J})$ can be equivalently recast as a 1d SQM on $\R$ with target the space of maps $\mathfrak{M}(\Sigma, \mathcal{M}^G_{\text{H}}(C, \mathbf{J}))$ from $\Sigma$ to $\mathcal{M}^G_{\text{H}}(C, \mathbf{J})$, which is a Kähler manifold.\footnote{%
  In our 3d RW model, $\mathfrak{M}(\Sigma, \mathcal{M}^G_{\text{H}}(C, \mathbf{J}))$ is more precisely the space of all $\overline{\mathcal{Z}}, \mathcal{Z}$ maps from $\Sigma$ to $\mathcal{M}^G_{\text{H}}(C, \mathbf{J})$, where $\overline{\mathcal{Z}}$ is complex conjugate to $\mathcal{Z}$ with respect to the complex structure of $\mathcal{M}^G_{\text{H}}(C, \mathbf{J})$.
  In other words, $\mathfrak{M}(\Sigma, \mathcal{M}^G_{\text{H}}(C, \mathbf{J}))$ is a complex infinite-dimensional affine space; let us denote by $\hat{\delta}$ the exterior derivative on this space.
  The flat Hermitian metric of $\mathfrak{M}(\Sigma, \mathcal{M}^G_{\text{H}}(C, \mathbf{J}))$ can be defined as $h_{\alpha\beta} = \int_{\Sigma} g_{i\bar{\jmath}} (\hat{\delta}_{\alpha} \overline{\mathcal{Z}}^i \wedge \star \hat{\delta}_{\beta} \mathcal{Z}^{\bar{\jmath}} + (\alpha \leftrightarrow \beta))$, and the Kähler two-form can be defined as $\omega_{\alpha\beta} = \int_{\Sigma} g_{i\bar{\jmath}} (\hat{\delta}_{\alpha} \overline{\mathcal{Z}}^i \wedge \star \hat{\delta}_{\beta} \mathcal{Z}^{\bar{\jmath}} - (\alpha \leftrightarrow \beta))$, where $\wedge$ and $\star$ are the wedge product and Hodge star operator on $\Sigma$, respectively.
  \label{ft:kahler metric of maps from sigma to GC-flat}
}
Generically, the $\hat{\mathcal{Q}}$-cohomology of this equivalent 1d SQM is given by the Dolbeault cohomology of $\mathfrak{M}(\Sigma, \mathcal{M}^G_{\text{H}}(C, \mathbf{J}))$.
However, the actual $\hat{\mathcal{Q}}$-cohomology that its partition function localizes onto corresponds to the Dolbeault cohomology of the space of \emph{constant} maps $\mathfrak{M}^{\text{const}}(\Sigma, \mathcal{M}^G_{\text{H}}(C, \mathbf{J}))$ from $\Sigma$ to $\mathcal{M}^G_{\text{H}}(C, \mathbf{J})$.
Since the 3d RW model on $\Sigma \times \R$ and the 1d SQM on $\R$ are equivalent, this in turn means that the $\hat{\mathcal{Q}}$-cohomology which the partition of the 3d sigma model localizes onto can also be described by the Dolbeault cohomology of $\mathfrak{M}^{\text{const}}(\Sigma, \mathcal{M}^G_{\text{H}}(C, \mathbf{J}))$.

Note that the only QFT whose $\hat{\mathcal{Q}}$-cohomology is described by the Dolbeault cohomology of $\mathfrak{M}^{\text{const}}(\Sigma, \mathcal{M}^G_{\text{H}}(C, \mathbf{J}))$ is a 2d B-model on $\Sigma$ with target $\mathcal{M}^G_{\text{H}}(C, \mathbf{J})$.
This therefore means that the $\hat{\mathcal{Q}}$-cohomology of our 3d RW model on $\Sigma \times \R$ is \emph{effectively} the $\hat{\mathcal{Q}}$-cohomology of a 2d B-model on $\Sigma$.

\subtitle{2d B-model from Another Perspective}

We also can understand this as follows.
The 3d RW model is a \emph{topological} sigma model.
This means that its Hamiltonian is zero, i.e., its spectrum of states (given by its $\hat{\mathcal{Q}}$-cohomology) are necessarily spanned by supersymmetric ground states that are time-invariant.
Therefore, just like the 3d A-model in \autoref{sec:hw:3d}, to study the spectrum of states of the 3d RW model, it suffices to consider a time-invariant hyperslice along $\R$ by inserting a delta-function $\delta(t - T_0)$ and setting $\partial_t \rightarrow 0$ in \eqref{eq:gm:bjsv:sigma model:rw model}.
The action on the hyperslice becomes
\begin{equation}
  \label{eq:gm:bjsv:sigma model:hyperslice:action}
  \begin{aligned}
    S_{\text{2d}}
    =
    & \frac{1}{e^2} \int_{\Sigma} d^2 x \, g_{\bar{\imath}j} \left(
      \frac{1}{2} \partial_\mu \overline{\mathcal{Z}}^i \partial^\mu \mathcal{Z}^{\bar{\jmath}}
      - i \tilde{\varrho}^i_\mu \nabla^\mu \varsigma^{\bar{\jmath}}
      \right)
      \\
    & + \frac{1}{e^2} \int_{\Sigma}
    g_{\bar{\imath}j} \left(
      \vartheta^{\bar{\imath}} \wedge \nabla \tilde{\varrho}^j
      + \frac{1}{3} \mathfrak{R}^j_{kl\bar{m}} \vartheta^{\bar{\imath}} \wedge \tilde{\varrho}^k \wedge \tilde{\varrho}^l \wedge \varsigma^{\bar{m}}
    \right)
    \, ,
  \end{aligned}
\end{equation}
where (i) $\mu$ is the index on the hyperslice $\Sigma$;
(ii) $\varsigma \in \Omega^0(\Sigma, \Phi^{*}\overline{\mathcal{T}})$;
(iii) $\vartheta^{\bar{\imath}} \coloneq g^{\bar{\imath}k} \Upomega_{lk} \varrho^l_t \in \Omega^0(\Sigma, \Phi^{*}\overline{\mathcal{T}})$ ($\varrho_t$ being the $t$-component of $\varrho$ in the 3d RW model);
and (iv) $\tilde{\varrho} \in \Omega^1(\Sigma, \Phi^{*}\mathcal{T})$ is the pullback of $\varrho$ in the 3d RW model onto $\Sigma$.
This is the action of a 2d topological B-model on $\Sigma$ with target $\mathcal{M}^G_{\text{H}}(C, \mathbf{J})$ that is invariant under the following supersymmetric transformations:
\begin{equation}
  \label{eq:gm:bjsv:sigma model:susy variations:2d b-model}
  \begin{gathered}
    \delta_{\hat{\mathcal{Q}}} \overline{\mathcal{Z}}^i
    = 0
      \, ,
    \qquad
      \delta_{\hat{\mathcal{Q}}} \mathcal{Z}^{\bar{\imath}}
    = \varsigma^{\bar{\imath}}
      \, ,
    \\
    \delta_{\hat{\mathcal{Q}}} \varrho^i
    = d \phi^i
      \, ,
    \qquad
      \delta_{\hat{\mathcal{Q}}} \varsigma^{\bar{\imath}}
    = 0
      \, ,
    \qquad
      \delta_{\hat{\mathcal{Q}}} \vartheta^{\bar{\imath}}
    = 0
      \, ,
  \end{gathered}
\end{equation}
with $d$ now being understood as the exterior derivative on $\Sigma$.

In short, the spectrum of the 3d $\mathcal{N} = 4$ RW model on $\Sigma \times \R$ with target $\mathcal{M}^G_{\text{H}}(C, \mathbf{J})$ is indeed given by the spectrum of a 2d $\mathcal{N} = (4, 4)$ topological B-model on $\Sigma$ with target $\mathcal{M}^G_{\text{H}}(C, \mathbf{J})$.

\subtitle{A 3d B-model in $\mathcal{M}^G_{\text{H}}(C, \mathbf{J})$}

Hence, we can regard the 3d RW model as a \emph{B-model}.
In other words, BJSV reduction of GM theory on $C \times \Sigma \times \R$ along $C$ results in a 3d $\mathcal{N} = 4$ B-model on $\Sigma \times \R$ with target $\mathcal{M}^G_{\text{H}}(C, \mathbf{J})$.

\subsection{An Equivalence Between GM Theory on \texorpdfstring{$M_3 \times M_1 \times \R$}{M3 x M1 x R} and the 3d B-model on \texorpdfstring{$I \times M_1 \times \R$}{I x M1 x R}}
\label{sec:gm:equivalence}

Consider now GM theory on $M_3 \times M_1 \times \R$, where $M_3$ is a closed three-manifold.
We can repeat the process described in \autoref{sec:hw:equivalence} of
(i) Heegaard splitting this GM theory along a Heegaard surface $C$ to obtain two copies of GM theory on $C \times I' \times M_1 \times \R$ and $C \times I'' \times M_1 \times \R$ glued along their common boundary $\mathscr{B} = C \times M_1 \times \R$,
and (ii) BSJV reducing along $C$ to get two copies of the 3d B-model with target $\mathcal{M}^G_{\text{H}}(C, \mathbf{J})$ on $I' \times M_1 \times \R$ and $I'' \times M_1 \times \R$ ``concatenated'' into a single 3d B-model on $I \times M_1 \times \R$, where $I = I' \oplus I''$.
Since GM theory is invariant under the BJSV reduction along $C$, this means that GM theory on $M_3 \times M_1 \times \R$ is equivalent to a 3d B-model on $I \times M_1 \times \R$ with target $\mathcal{M}^G_{\text{H}}(C, \mathbf{J})$.

Like in \autoref{sec:hw:equivalence}, we have an equivalence between a 5d topological gauge theory and a 3d topological sigma model.

\section{Novel Floer Homologies From GM Theory}
\label{sec:m x r}

In this section, we will study GM theory on $M_4 \times \R$, where $M_4$ is a closed and possibly decomposable four-manifold.
By recasting it as a 1d SQM, we will physically realize novel gauge-theoretic Floer homologies of four, three, and two-manifolds.

\subsection{A Holomorphic \texorpdfstring{$G_{\C}$}{G-C}-flat Floer Homology of Four-Manifolds}
\label{sec:m4 x r}

Let us consider GM theory on $M_5 = M_4 \times \R$, where the direction along $\R$ is to be understood as a temporal direction.

\subtitle{GM Theory as a 1d SQM}

The action of this theory is given by \eqref{eq:gm:action}.
In the \emph{real} temporal gauge, i.e., $A_t = 0$,\footnote{%
  We can appeal to this gauge because when recast as a 1d SQM, the real gauge field in ``time'' will eventually be non-dynamical and thus integrated out to furnish the terms that contribute to the Christoffel symbols in the kinetic terms of the fermions \cite{er-2023-topol-n}.
  \label{ft:m4 x r:reason for taking temporal gauge}
}
the action can be expressed as
\begin{equation}
  \label{eq:m4 x r:action:temporal gauge}
  \begin{aligned}
    & S_{\text{GM}, A_t = 0}
    \\
    & = \frac{1}{e^2} \int_{M_4 \times \R} dt d^4x \, \Tr \left(
    \frac{1}{2} \left|
      \dot{\overline{\mathcal{A}}^{\nu}} \mathcal{E}_{\nu\mu}
      - \frac{1}{2} \epsilon_{\mu\kappa\rho\sigma} \overline{\mathcal{D}}^{\kappa} \overline{\mathcal{B}}^{\rho\sigma}
      \right|^2
    + \frac{1}{4} \left|
      \frac{1}{2} \epsilon_{\rho\sigma\kappa\lambda} \dot{\overline{\mathcal{B}}}^{\mu\nu} \mathcal{E}_{\mu\pi} \mathcal{E}^{\pi\kappa} \mathcal{E}_{\nu}^{\; \lambda}
      + \overline{\mathcal{F}}_{\rho\sigma}
    \right|^2
    + \dots
    \right)
    \, ,
  \end{aligned}
\end{equation}
where
(i) $\dot{\overline{\mathcal{A}}}_{\mu} \coloneq \partial_t \overline{\mathcal{A}}_{\mu}$ and $\dot{\overline{\mathcal{B}}}_{\mu\nu} \coloneq \partial_t \overline{\mathcal{B}}_{\mu\nu}$;
(ii) $\mathcal{E}_{\mu\nu} \in \Omega^2(M_4) \otimes \Omega^0(\R)$ is a unit two-form on $M_4$ whose components are \emph{all} equal, i.e., $\mathcal{E}_{\mu\nu} \equiv 1$ for $\mu < \nu$;
(iii) $\frac{1}{2} (\Re(\overline{\mathcal{A}}_t) - \overline{\mathcal{A}}_t) \epsilon_{\mu\nu\varrho\varsigma} \mathcal{E}^{\varrho\varsigma} = \frac{i}{2} \phi_t \epsilon_{\mu\nu\varrho\varsigma} \mathcal{E}^{\varrho\varsigma} \eqcolon \overline{\mathcal{B}}_{\mu\nu} \in \Omega^2(M_4, \text{ad}(G_{\C})) \otimes \Omega^0(\R, \text{ad}(G_{\C}))$,\footnote{%
  \label{ft:m4 x r:field dependence}%
  Note that $\overline{\mathcal{B}}_{\mu\nu}$ actually depends on the coordinates on $M_4$, and is thus an unconstrained antisymmetric two-form on $M_4$.
  To understand this, note that $\phi_t = \phi_t(x^1, x^2, x^3, x^4, t)$ is actually coordinate-dependent, as with any physical field on $M_4 \times \R$; in turn, this means that $\overline{\mathcal{B}}_{\mu\nu} = \overline{\mathcal{B}}_{\mu\nu}(x^1, x^2, x^3, x^4,t)$ would also be coordinate-dependent and therefore unconstrained.
  In particular, $\overline{\mathcal{B}}_{\mu\nu}$ would be freely-varying, and as it varies freely, it would define an independent coordinate on the space of all $\overline{\mathcal{B}}_{\mu\nu}$ fields on $M_4$, where $\dot{\overline{\mathcal{B}}}_{\mu\nu}$ is nonzero in general.
}
and (iv) the above term $\epsilon_{\rho\sigma\kappa\lambda} \dot{\overline{\mathcal{B}}}^{\mu\nu} \mathcal{E}_{\mu\pi} \mathcal{E}^{\pi\kappa} \mathcal{E}_{\nu}^{\; \lambda}$ is trivially zero (even though $\dot{\overline{\mathcal{B}}}^{\mu\nu} \neq 0$) due to the identity $\epsilon_{\rho\sigma\kappa\lambda} \epsilon^{\mu\nu\varrho\varsigma} \mathcal{E}_{\varrho\varsigma} \mathcal{E}_{\mu\pi} \mathcal{E}^{\pi\kappa} \mathcal{E}_{\nu}^{\; \lambda} \equiv 0$, as required.

Note that the path integral of the GM theory action \eqref{eq:m4 x r:action:temporal gauge} localizes onto configurations that satisfy the vanishing of the expression within the squared terms above.
Hence, we can effectively write it as
\begin{equation}
  \label{eq:m4 x r:action:temporal gauge:inverse}
  \begin{aligned}
    & S_{\text{GM}, A_t = 0}
    \\
    & = \frac{1}{e^2} \int_{M_4 \times \R} dt d^4x \, \Tr \left(
    \frac{1}{2} \left|
      \dot{\overline{\mathcal{A}}}^{\nu'}
      - \frac{1}{2} (\mathcal{E}_{\nu'\mu})^{-1} \epsilon_{\mu\kappa\rho\sigma} \overline{\mathcal{D}}^{\kappa} \overline{\mathcal{B}}^{\rho\sigma}
      \right|^2
    + \frac{1}{4} \left|
    \dot{\overline{\mathcal{B}}}^{\mu'\nu'}
    + \frac{1}{2} (\mathcal{E}_{\mu'\pi} \mathcal{E}^{\pi\kappa} \mathcal{E}_{\nu'}^{\; \lambda})^{-1} \epsilon^{\rho\sigma\kappa\lambda} \overline{\mathcal{F}}_{\rho\sigma}
    \right|^2
    + \dots
    \right)
    \, ,
  \end{aligned}
\end{equation}
where $(\mathcal{E}_{\nu'\mu})^{-1}$ and $(\mathcal{E}_{\mu'\pi} \mathcal{E}^{\pi\kappa} \mathcal{E}_{\nu'}^{\; \lambda})^{-1}$ are inverses of $\mathcal{E}_{\nu'\mu}$ and $\mathcal{E}_{\mu'\pi} \mathcal{E}^{\pi\kappa} \mathcal{E}_{\nu'}^{\; \lambda}$, respectively.\footnote{%
  The inverses can be computed explicitly to be
  \begin{equation*}
    (\mathcal{E}_{\nu'\mu})^{-1}
    = \frac{1}{2} \epsilon^{\nu'\mu\varrho\varsigma} \mathcal{E}_{\varrho\varsigma}
    \quad
    \text{and}
    \quad
    (\mathcal{E}_{\mu'\pi} \mathcal{E}^{\pi\kappa} \mathcal{E}_{\nu'}^{\; \lambda})^{-1}
    = \frac{1}{8} g^{\nu'\varkappa} \epsilon_{\lambda\varkappa\varrho\varsigma} \epsilon_{\kappa\pi\rho\sigma} \epsilon^{\mu'\pi\alpha\beta} \mathcal{E}^{\rho\sigma} \mathcal{E}^{\varrho\varsigma} \mathcal{E}_{\alpha\beta}
    \, ,
  \end{equation*}
  such that
  \begin{equation*}
    \mathcal{E}_{\nu\mu} (\mathcal{E}_{\nu'\mu})^{-1}
    = \delta_{\nu}^{\; \nu'}
    \quad
    \text{and}
    \quad
    \mathcal{E}_{\mu\pi} \mathcal{E}^{\pi\kappa} \mathcal{E}_{\nu}^{\; \lambda} (\mathcal{E}_{\mu'\pi} \mathcal{E}^{\pi\kappa} \mathcal{E}_{\nu'}^{\; \lambda})^{-1}
    = \delta_{\mu}^{\; \mu'} \delta_{\nu}^{\; \nu'}
    \, .
  \end{equation*}
}

After suitable rescalings, we can recast \eqref{eq:m4 x r:action:temporal gauge:inverse} as a 1d SQM model, where its action will now read
\begin{equation}
  \label{eq:m4 x r:sqm:action}
  S_{\text{SQM}, \mathfrak{B}_4}
  = \frac{1}{e^2} \int_{\R} dt \left(
    \left|
      \dv{\overline{\mathcal{A}}^{\alpha}}{t}
      + g_{\mathfrak{B}_4}^{\alpha\beta} \pdv{V_4}{\overline{\mathcal{A}}^{\beta}}
    \right|^2
    + \left|
      \dv{\overline{\mathcal{B}}^{\alpha}}{t}
      + g_{\mathfrak{B}_4}^{\alpha\beta} \pdv{V_4}{\overline{\mathcal{B}}^{\beta}}
    \right|^2
    + \dots
  \right)
  \, .
\end{equation}
Here, (i) $(\overline{\mathcal{A}}^{\alpha}, \overline{\mathcal{B}}^{\alpha})$ and $(\alpha, \beta)$ are coordinates and indices, respectively, on the space $\mathfrak{B}_4$ of irreducible $(\overline{\mathcal{A}}_{\mu}, \overline{\mathcal{B}}_{\mu\nu})$ fields on $M_4$;
(ii) $g_{\mathfrak{B}_4}$ is a corresponding  metric on $\mathfrak{B}_4$;
and (iii) $V_4(\overline{\mathcal{A}}, \overline{\mathcal{B}})$ is the potential function.

\subtitle{The $\hat{\mathcal{Q}}$-cohomology of GM Theory}

In a TQFT, the Hamiltonian vanishes in the $\hat{\mathcal{Q}}$-cohomology of the theory.
This means that the non-vanishing operators that span the relevant $\hat{\mathcal{Q}}$-cohomology of states in GM theory are actually the supersymmetric ground states that are therefore time-invariant.
In particular, for GM theory on $M_4 \times \R$, its relevant spectrum is associated only with $M_4$.

With $M_5 = M_4 \times \R$, $M_4$ is the far boundary of the five-manifold, and one needs to specify ``boundary conditions'' on $M_4$ to compute the path integral.
This can be done by first defining a restriction of the fields to $M_4$, for which we shall define as $\Psi_{M_4}$, and then specifying boundary values for these restrictions.
Doing this is equivalent to inserting in the path integral, an operator functional $F_4(\Psi_{M_4})$ that is non-vanishing in the $\hat{\mathcal{Q}}$-cohomology (so that the path integral will continue to be topological).
This means that the corresponding partition function of GM theory can be computed as \cite[eqn. (4.12)]{witten-1988-topol-quant}\footnote{%
  GM theory is a balanced TQFT, whence just as in Vafa-Witten theory, one can define its partition function.
  \label{ft:m4 x r:partition function of gm theory}
}
\begin{equation}
  \label{eq:m4 x r:witten partition function}
  \langle 1 \rangle_{F_4(\Psi_{M_4})}
  = \int_{\mathcal{M}_{\text{GM}}} F_4(\Psi_{M_4}) \, e^{-S_{\text{GM}}}
  \, .
\end{equation}

Since we have seen that GM theory on $M_4 \times \R$ can be expressed as a 1d SQM in $\mathfrak{B}_4$, we can thus write the partition function as
\begin{equation}
  \label{eq:m4 x r:partition function:no homology}
  \mathcal{Z}_{\text{GM}, M_4 \times \R}(G)
  = \langle 1 \rangle_{F_4(\Psi_{M_4})}
  = \sum_u \mathfrak{F}^{G}_{\text{GM}}(\Psi_{M_4}^u)
  \, .
\end{equation}
Here, $\mathfrak{F}^{G}_{\text{GM}}(\Psi_{M_4}^u)$, in the $\hat{\mathcal{Q}}$-cohomology of GM theory, is the $u^{\text{th}}$ contribution to the partition function that depends on the expression of $F_4(\Psi_{M_4})$ in the fields on $M_4$, evaluated over the corresponding solutions to the $G_{\C}$-flat equation in \eqref{eq:gm:bps} restricted to $M_4$. The summation in `$u$' is over all presumably isolated and non-degenerate configurations on $M_4$ in $\mathfrak{B}_4$ that the equivalent SQM localizes onto.\footnote{%
  This presumption that the configurations are isolated and non-degenerate will be justified in \autoref{ft:m4 x r:isolation justification}.
  \label{ft:m4 x r:defer isolation justification}
}

Let us now ascertain what the $\mathfrak{F}^{G}_{\text{GM}}(\Psi_{M_4}^u)$'s correspond to.
To this end, we have to first determine the configurations that the SQM localizes onto.
These are configurations that minimize the SQM action \eqref{eq:m4 x r:sqm:action}, i.e., configurations that set the squared terms therein to zero.
They are therefore given by
\begin{equation}
  \label{eq:m4 x r:flow}
  \saveboxed{eq:m4 x r:flow}{
    \dv{\overline{\mathcal{A}}^{\alpha}}{t}
    = - g_{\mathfrak{B}_4}^{\alpha\beta} \pdv{V_4}{\overline{\mathcal{A}}^{\beta}}
    \qquad
    \dv{\overline{\mathcal{B}}^{\alpha}}{t}
    = - g_{\mathfrak{B}_4}^{\alpha\beta} \pdv{V_4}{\overline{\mathcal{B}}^{\beta}}
  }
\end{equation}
where the squaring argument \cite{blau-1993-topol-gauge} means that both the LHS and RHS of both equations are \emph{simultaneously} set to zero.
In other words, the configurations that the SQM localizes onto are fixed (i.e., time-invariant) critical points of the potential $V_4(\overline{\mathcal{A}}, \overline{\mathcal{B}})$ in $\mathfrak{B}_4$.

\subtitle{$G_{\C}$-BF Configurations on $M_4$ as Critical Points of the 1d SQM}

To determine the explicit form of $V_4$, note that the squared term in \eqref{eq:m4 x r:sqm:action} originates from the squared term in \eqref{eq:m4 x r:action:temporal gauge:inverse}.
Indeed, setting the expression within the squared terms in \eqref{eq:m4 x r:action:temporal gauge:inverse} to zero minimizes the underlying 5d action, and this is consistent with setting the expression within the squared terms in \eqref{eq:m4 x r:sqm:action} to zero to minimize the equivalent SQM action.
Therefore, we can deduce the explicit form of $V_4$ by comparing \eqref{eq:m4 x r:flow} with \eqref{eq:m4 x r:action:temporal gauge:inverse}.
Specifically, setting to zero the expression within the squared terms in \eqref{eq:m4 x r:action:temporal gauge:inverse} would give us
\begin{equation}
  \label{eq:m4 x r:gm flow}
  \dot{\overline{\mathcal{A}}}^{\nu'}
  = \frac{1}{2} (\mathcal{E}_{\nu'\mu})^{-1} \epsilon_{\mu\kappa\rho\sigma} \overline{\mathcal{D}}^{\kappa} \overline{\mathcal{B}}^{\rho\sigma}
  \, ,
  \qquad
  \dot{\overline{\mathcal{B}}}^{\mu'\nu'}
  = - \frac{1}{2} (\mathcal{E}_{\mu'\pi} \mathcal{E}^{\pi\kappa} \mathcal{E}_{\nu'}^{\; \lambda})^{-1} \epsilon^{\rho\sigma\kappa\lambda} \overline{\mathcal{F}}_{\rho\sigma}
  \, .
\end{equation}

Comparing \eqref{eq:m4 x r:gm flow} with \eqref{eq:m4 x r:flow}, we find that
\begin{equation}
  \label{eq:m4 x r:morse function}
  \saveboxed{eq:m4 x r:morse function}{
    V_4(\overline{\mathcal{A}}, \overline{\mathcal{B}})
    = \int_{M_4} \Tr \left(
      \overline{\mathcal{B}} \wedge \overline{\mathcal{F}}
    \right)
  }
\end{equation}
Thus, the summation in `$u$' in \eqref{eq:m4 x r:partition function:no homology} is over all isolated and non-degenerate critical points of \eqref{eq:m4 x r:morse function} in $\mathfrak{B}_4$ that are also fixed.\footnote{%
  As we will explain next, the aforementioned critical points correspond to $G_{\C}$-BF configurations on $M_4$.
  For them to be isolated, the actual dimension of their moduli space needs to be zero.
  This is indeed possible with an appropriate choice of $G$ and $M_4$ \cite[$\S$V]{cartas-fuentevilla-2011-dimen-modul}; we shall assume such a choice of $G$ and $M_4$ henceforth.
  As for their non-degeneracy, a suitable perturbation of $V_4(\overline{\mathcal{A}}, \overline{\mathcal{B}})$, which can be effected by introducing physically-trivial $\hat{\mathcal{Q}}$-exact terms to the action, would ensure this \cite[footnote 5]{er-2023-topol-n}.
  \label{ft:m4 x r:isolation justification}
}

Critical points of $V_4(\overline{\mathcal{A}}, \overline{\mathcal{B}})$ are configurations in $\mathfrak{B}_4$ that set the RHS's of \eqref{eq:m4 x r:flow} to zero, which in turn, correspond to configurations on $M_4$ that set the RHS's of \eqref{eq:m4 x r:gm flow} to zero.
Such configurations span the space of solutions to the 4d $G_{\C}$-BF equations on $M_4$, for which we shall henceforth refer to as $G_{\C}$-BF configurations.

In summary, the partition function \eqref{eq:m4 x r:partition function:no homology} is an algebraic sum of \emph{fixed} $G_{\C}$-BF configurations on $M_4$ in $\mathfrak{B}_4$.

\subtitle{The Holomorphic $G_{\C}$-flat Floer Homology}

Notice that \eqref{eq:m4 x r:flow} is a gradient flow equation, and that it governs the classical trajectory of the 1d SQM model from one time-invariant $G_{\C}$-BF configuration to another on $M_4$, in $\mathfrak{B}_4$.
Hence, just as in \cite{witten-1988-topol-quant, ong-2023-vafa-witten-theor, er-2023-topol-n, er-2024-topol-gauge-theor}, the equivalent 1d SQM model will physically realize a gauge-theoretic Floer homology.

Specifically, the \emph{time-invariant $G_{\C}$-BF configurations on $M_4$ in $\mathfrak{B}_4$}, i.e., time-independent solutions to the 4d equations
\begin{equation}
  \label{eq:m4 x r:critical points}
  \saveboxed{eq:m4 x r:critical points}{
    \overline{\mathcal{D}} \overline{\mathcal{B}}
    = 0
    \qquad
    \overline{\mathcal{F}}
    = 0
  }
\end{equation}
will generate the chains of a Floer complex with \emph{holomorphic Morse functional} $V_4(\overline{\mathcal{A}}, \overline{\mathcal{B}})$ in \eqref{eq:m4 x r:morse function}.
The $G_{\C}$-flat flow lines, described by time-varying solutions to the \emph{gradient flow equation} \eqref{eq:m4 x r:flow}, are the Floer differentials such that the number of outgoing flow lines at each time-invariant configuration obeying \eqref{eq:m4 x r:critical points} is the degree $d_u$ of the corresponding chain in the Floer complex.

In other words, we can also write \eqref{eq:m4 x r:partition function:no homology} as
\begin{equation}
  \label{eq:m4 x r:partition function}
  \saveboxed{eq:m4 x r:partition function}{
    \mathcal{Z}_{\text{GM}, M_4 \times \R}(G)
    = \sum_u \mathfrak{F}^{G}_{\text{GM}}(\Psi_{M_4}^u)
    = \sum_u \text{HHF}^{\text{flat}}_{d_u}(M_4, G_{\C})
    = \mathcal{Z}^{\text{Floer}}_{\text{flat}, M_4}(G_{\C})
  }
\end{equation}
where each $\mathfrak{F}^{G}_{\text{GM}}(\Psi_{M_4}^u)$ can be identified with a \emph{novel} gauge-theoretic \emph{holomorphic} Floer homology class $\text{HHF}^{\text{flat}}_{d_u}(M_4, G_{\C})$, that we shall henceforth name a holomorphic $G_{\C}$-flat Floer homology class, assigned to $M_4$ defined by \eqref{eq:m4 x r:flow}, \eqref{eq:m4 x r:morse function}, \eqref{eq:m4 x r:critical points}, and the description above.

\subsection{A Holomorphic \texorpdfstring{$G_{\HH}$}{G-H}-flat Floer Homology of Three-Manifolds}
\label{sec:m3 x r}

Next, let us specialize to $M_4 = M_3 \times S^1$, and perform a Kaluza-Klein (KK) dimensional reduction of GM theory by shrinking the $S^1$ circle to be infinitesimally small.

\subtitle{The Quaternionified BPS Equations of GM$_4$ Theory}

Let $x^4$ be the coordinate of the $S^1$ circle.
The conditions that minimize GM theory when we KK reduce along $S^1$ are effectively obtained by KK reduction of the conditions that minimize the action \eqref{eq:m4 x r:action:temporal gauge}, i.e., by the KK reduction of the BPS equation of GM theory on $M_3 \times S^1 \times \R$ in \emph{real} temporal gauge.
We wish to write the reduced BPS equations in a form that resembles \eqref{eq:m4 x r:gm flow}.

To this end, let us first temporarily restore the real temporal gauge field.
Performing a KK reduction of the $G_{\C}$-flat equations \eqref{eq:gm:bps} on $M_3 \times S^1 \times \R$ along the $S^1$ circle, we get
\begin{equation}
  \label{eq:m3 x r:kk reduction}
  \overline{\mathcal{F}}_{\mu\nu}
  = 0
  \, ,
  \qquad
  \overline{\mathcal{D}}_{\mu} \overline{\mathcal{A}}_4
  = 0
  \, ,
\end{equation}
where now $\{\mu, \nu\} \in \{t, 1, 2, 3\}$, and $\overline{\mathcal{A}}_4 \equiv A_4 - i \phi_4$, which can then be interpreted as a $G_{\C}$-scalar on $M_3 \times \R$.

Second, let us introduce a unit one-form $\mathcal{H} \in \Omega^1(M_3 \times \R)$ whose components are \emph{all} equal, i.e., $\mathcal{H}_{\mu} \equiv 1$.
This one-form allows us to define $\Omega^1(M_3 \times \R, \text{ad}(G_{\mathbb{C}})) \ni \mathcal{C} \coloneq \overline{\mathcal{A}}_4 \mathcal{H}$ whose components are also all equal.
This allows us to rewrite the BPS equations on $M_3 \times \R$ as
\begin{equation}
  \label{eq:m3 x r:kk reduction:with one-form}
  \overline{\mathcal{F}}_{\mu\nu}
  = 0
  \, ,
  \qquad
  \overline{\mathcal{D}}_{\mu} \mathcal{C}_{\nu}
  = 0
  \, ,
\end{equation}
since componentwise, it is \emph{equal} to \eqref{eq:m3 x r:kk reduction}.
As the BPS equations of a theory are the classical equations of motion of the theory, the fact that the interpretation of a scalar $\overline{\mathcal{A}}_4$ as a unit one-form $\mathcal{C}$ reproduces the same classical equations of motion means that this interpretation is also a physically equivalent one.

Third, let us complexify the difference of the LHS's of \eqref{eq:m3 x r:kk reduction:with one-form} as follows:\footnote{%
  Since all components of $\mathcal{C}$ are equal, the $\epsilon^{\mu\nu} [j \overline{\mathcal{C}}_{\mu}, j \overline{\mathcal{C}}_{\nu}] = \epsilon^{\mu\nu}[\mathcal{C}_{\mu} j, \mathcal{C}_{\nu} j]$ term in the following expression, where $\overline{\mathcal{C}}$ and $j$ are described after the equation, is trivially zero.
  \label{fn:m3 x r:addition of trivial zero}
}
\begin{equation}
  \label{eq:m3 x r:bps:complexify}
  \begin{aligned}
    & \frac{1}{2} \epsilon^{\mu\nu} \left(
      \overline{\mathcal{F}}_{\mu\nu}
      + [j \overline{\mathcal{C}}_{\mu}, j \overline{\mathcal{C}}_{\nu}]
      \right)
      - \epsilon^{\mu\nu} \overline{\mathcal{D}}_{\mu} (j \overline{\mathcal{C}}_{\nu})
    \\
    & =
      \frac{1}{2} \epsilon^{\mu\nu} \left(
      \partial_{\mu} (\overline{\mathcal{A}}_{\nu} - j \overline{\mathcal{C}}_{\nu})
      - \partial_{\nu} (\overline{\mathcal{A}}_{\mu} - j \overline{\mathcal{C}}_{\nu})
      + [\overline{\mathcal{A}}_{\mu} - j \overline{\mathcal{C}}_{\mu}, \overline{\mathcal{A}}_{\nu} - j \overline{\mathcal{C}}_{\nu}]_{\mathbb{H}}
      \right)
    \\
    & \eqcolon
      \frac{1}{2} \epsilon^{\mu\nu} \left(
      \partial_{\mu} \overline{\mathscr{A}}_{\nu}
      - \partial_{\nu} \overline{\mathscr{A}}_{\mu}
      + [\overline{\mathscr{A}}_{\mu}, \overline{\mathscr{A}}_{\nu}]_{\mathbb{H}}
      \right)
    \\
    & =
    \frac{1}{2} \epsilon^{\mu\nu} \overline{\mathscr{F}}_{\mu\nu}
      \, ,
  \end{aligned}
\end{equation}
where
(i) $\overline{\mathscr{A}} \coloneq \overline{\mathcal{A}} - j \overline{\mathcal{C}}$ is a complexification of $\overline{\mathcal{A}}$ and $\overline{\mathcal{C}} \coloneq (A_4 + i \phi_4) \mathcal{H}$;
(ii) $j$ satisfies the quaternionic relations with $i$, i.e., $ij = - ji \eqcolon k$ and $i^2 = j^2 = k^2 = -1$;
and
(iii) $\overline{\mathscr{F}}$ is the field strength of $\overline{\mathscr{A}}$.
and
(iv) the antisymmetric commutator bracket $[\cdot, \cdot]_{\mathbb{H}}$ with quaternionic entries is defined as
\begin{equation}
  \label{eq:m3 x r:bracket operator}
  \begin{aligned}
    [\overline{\mathcal{A}}_{\mu} - j \overline{\mathcal{C}}_{\mu}, \overline{\mathcal{A}}_{\nu} - j \overline{\mathcal{C}}_{\nu}]_{\mathbb{H}}
    &\coloneq \overline{\mathcal{A}}_{\mu} \overline{\mathcal{A}}_{\nu}
    - \overline{\mathcal{A}}_{\nu} \overline{\mathcal{A}}_{\mu}
    - \mathcal{C}_{\mu} \overline{\mathcal{C}}_{\nu}
    + \mathcal{C}_{\nu} \overline{\mathcal{C}}_{\mu}
    \\
    &\quad
    - j \left(
      \mathcal{A}_{\mu} \overline{\mathcal{C}}_{\nu}
      - \overline{\mathcal{C}}_{\nu} \overline{\mathcal{A}}_{\mu}
      + \mathcal{A}_{\nu} \overline{\mathcal{C}}_{\mu}
      - \overline{\mathcal{C}}_{\mu} \overline{\mathcal{A}}_{\nu}
    \right)
    \, .
  \end{aligned}
\end{equation}

Just as $\overline{\mathcal{A}}$ is understood as a $G_{\C}$-connection, i.e., a connection valued in the adjoint bundle $\text{ad}(G_{\C})$ of the underlying \emph{complexified} principal $G_{\C}$-bundle, $\overline{\mathscr{A}}$ can be understood as a $G_{\HH}$-connection, i.e., a connection valued in the adjoint bundle $\text{ad}(G_{\HH})$ of the underlying \emph{quaternionified} principal $G_{\HH}$-bundle.\footnote{%
  Fields valued in $\text{ad}(G_{\HH})$ are $\mathfrak{g}_{\mathbb{H}}$-valued, where $\mathfrak{g}_{\mathbb{H}}$ is the Lie algebra of the quaternionified Lie group $G_{\mathbb{H}}$.
  These quaternionified Lie algebras are equipped with an $\mathbb{R}$-bilinear, antisymmetric bracket $[\cdot, \cdot]_{\mathbb{H}}: \mathfrak{g}_{\mathbb{H}} \times \mathfrak{g}_{\mathbb{H}} \rightarrow \mathfrak{g}_{\mathbb{H}}$ described in \eqref{eq:m3 x r:bracket operator} using $\mathfrak{g}_{\mathbb{C}}$-valued fields, satisfying the Jacobi identity (see \autoref{app:jacobi:g-h}), and are first defined in \cite{tosiaki-2020-quart-compl} -- examples include $\mathfrak{gl}(n, \mathbb{H})$ and $\mathfrak{sp}(n)$.
}
Since the RHS's of \eqref{eq:m3 x r:kk reduction:with one-form} are all zero, \eqref{eq:m3 x r:bps:complexify} tells us that the BPS equation of the 4d theory resulting from the KK reduction of GM theory along an $S^1$ circle -- henceforth referred to as GM$_4$ theory -- is the $G_{\HH}$-flat equation on $M_3 \times \R$, i.e., $\overline{\mathscr{F}}_{\mu\nu} = 0$.

Finally, working in the \emph{real} temporal gauge, the $G_{\HH}$-flat equation on $M_3 \times \R$ can be expressed as
\begin{equation}
  \label{eq:m3 x r:gm4 flow}
  \dot{\overline{\mathscr{A}}}^n \mathscr{E}_{nm}
  = \epsilon_{mpq} \overline{\mathscr{D}}^p \overline{\mathscr{B}}^q
  \, ,
  \qquad
  \epsilon_{mpq} \dot{\overline{\mathscr{B}}}^n \mathscr{E}_{rs} \mathscr{E}^{sm} \mathscr{E}_n^{\; r}
  = - \overline{\mathscr{F}}_{pq}
  \, ,
\end{equation}
where (i) $\{m, n, p, q, r, s\} \in \{1, 2, 3\}$ are indices on $M_3$;
(ii) $\mathscr{E} \in \Omega^2(M_3) \otimes \Omega^0(\R)$ is a unit two-form on $M_3$ whose components are \emph{all} equal, i.e., $\mathscr{E}_{mn} \equiv 1$ for $m < n$;
(iii) $\frac{1}{2} (\Re(\overline{\mathscr{A}}_t) - \overline{\mathscr{A}}_t) \epsilon_{mpq} \mathscr{E}^{pq} = \frac{1}{2} (i \phi_t + j A_4 - k \phi_4) \epsilon_{mpq} \mathscr{E}^{pq}  \eqcolon \overline{\mathscr{B}}_m \in \Omega^1(M_3, \text{ad}(G_{\HH})) \otimes \Omega^0(\R, \text{ad}(G_{\HH}))$,
and (iv) the above term $\epsilon_{mpq} \dot{\overline{\mathscr{B}}}^n \mathscr{E}_{rs} \mathscr{E}^{sm} \mathscr{E}_n^{\; r}$ is trivially zero (even though $\dot{\overline{\mathscr{B}}}^n \neq 0$) due to the identity $\epsilon^{npq} \mathscr{E}_{pq} \mathscr{E}_{nm} \equiv 0$, as required.
Notice that \eqref{eq:m3 x r:gm4 flow} has the same form of equations as the expressions within the squared terms in \eqref{eq:m4 x r:action:temporal gauge}, but with the gauge group complexified from $G_{\C}$ to $G_{\HH}$.

In other words, GM theory on $M_3 \times S^1 \times \R$, upon a KK reduction along $S^1$ to GM$_4$ theory on $M_3 \times \R$, localizes onto $\text{ad}(G_{\HH})$-valued configurations that obey \eqref{eq:m3 x r:gm4 flow}.

\subtitle{GM$_4$ Theory as a 1d SQM in $\mathfrak{B}_3$}

The 4d action can thus be written (in the \emph{real} temporal gauge) as
\begin{equation}
  \label{eq:m3 x r:action}
  \begin{aligned}
    & S_{\text{GM$_4$}, A_t = 0}
    \\
    & = \frac{1}{e^2} \int_{M_3 \times \R} dt d^3x \Tr \left(
      \frac{1}{2} \left|
        \dot{\overline{\mathscr{A}}}^n \mathscr{E}_{nm}
        - \epsilon_{mpq} \overline{\mathscr{D}}^p \overline{\mathscr{B}}^q
      \right|^2
      + \frac{1}{4} \left|
        \epsilon_{mpq} \dot{\overline{\mathscr{B}}}^n \mathscr{E}_{rs} \mathscr{E}^{sm} \mathscr{E}_n^{\; r}
        + \overline{\mathscr{F}}_{pq}
      \right|^2
      + \dots
    \right)
    \, .
  \end{aligned}
\end{equation}

Via a similar calculation that got us from \eqref{eq:m4 x r:action:temporal gauge} to \eqref{eq:m4 x r:sqm:action}, we can recast \eqref{eq:m3 x r:action} as a 1d SQM model, where its action will now read
\begin{equation}
  \label{eq:m3 x r:action:sqm}
  S_{\text{SQM}, \mathfrak{B}_3}
  = \frac{1}{e^2} \int_{\R} dt \left(
    \left|
      \dv{\overline{\mathscr{A}}^{\alpha}}{t}
      + g_{\mathfrak{B}_3}^{\alpha\beta} \pdv{V_3}{\overline{\mathscr{A}}^{\beta}}
    \right|^2
    + \left|
      \dv{\overline{\mathscr{B}}^{\alpha}}{t}
      + g_{\mathfrak{B}_3}^{\alpha\beta} \pdv{V_3}{\overline{\mathscr{B}}^{\beta}}
    \right|^2
    + \dots
  \right)
  \, .
\end{equation}
Here, (i) $(\overline{\mathscr{A}}^{\alpha}, \overline{\mathscr{B}}^{\alpha})$ and $(\alpha, \beta)$ are coordinates and indices, respectively, on the space $\mathfrak{B}_3$ of irreducible $(\overline{\mathscr{A}}_m, \overline{\mathscr{B}}_m)$ fields on $M_3$;
(ii) $g_{\mathfrak{B}_3}$ is a corresponding  metric on $\mathfrak{B}_3$;
and (iii) $V_3(\overline{\mathscr{A}}, \overline{\mathscr{B}})$ is the potential function.

\subtitle{Localizing onto Time-invariant $G_{\HH}$-BF Configurations on $M_3$}

By the squaring argument \cite{blau-1993-topol-gauge} applied to \eqref{eq:m3 x r:action:sqm}, the configurations that the equivalent SQM localizes onto are those that set the LHS's and RHS's of \eqref{eq:m3 x r:gm4 flow} \emph{simultaneously} to zero.
Such configurations in $\mathfrak{B}_3$ correspond to time-invariant configurations that span the space of solutions to the 3d $G_{\HH}$-BF equations on $M_3$.
We shall, in the rest of this paper, refer to such configurations as $G_{\HH}$-BF configurations on $M_3$.

In summary, the equivalent SQM localizes onto time-invariant $G_{\HH}$-BF configurations on $M_3$ in $\mathfrak{B}_3$.

\subtitle{The Holomorphic $G_{\HH}$-flat Floer Homology}

Since the resulting 4d theory on $M_3 \times \R$ can be interpreted as a 1d SQM in $\mathfrak{B}_3$, its partition function can, like in \eqref{eq:m4 x r:partition function:no homology}, be written as
\begin{equation}
  \label{eq:m3 x r:partition function:no homology}
  \mathcal{Z}_{\text{GM}, M_3 \times \R}(G)
  = \langle 1 \rangle_{F_3(\Psi_{M_3})}
  = \sum_v \mathfrak{F}^{G}_{\text{GM$_4$}}(\Psi_{M_3}^v)
  \, ,
\end{equation}
where $\mathfrak{F}^{G}_{\text{GM$_4$}}(\Psi_{M_3}^v)$, in the $\hat{\mathcal{Q}}$-cohomology of GM$_4$ theory, is the $v^{\text{th}}$ contribution to the partition function that depends on the expression $F_3(\Psi_{M_3})$ in the bosonic fields on $M_3$, and the summation in `$v$' is over all isolated and non-degenerate $G_{\HH}$-BF configurations on $M_3$ in $\mathfrak{B}_3$ that the equivalent SQM localizes onto.\footnote{%
  This presumption that the configurations will be isolated and non-degenerate is justified because the ($\hat{\mathcal{Q}}$-cohomology of) GM theory is topological in all directions and therefore invariant when we shrink the $S^1$ circle.
  Thus, if $M_3$ (where $M_3 \times S^1 = M_4$) and $G$ are chosen such as to satisfy the conditions spelt out in \autoref{ft:m4 x r:isolation justification}, $\mathcal{Z}_{\text{GM}, M_3 \times \R}(G)$ will be a discrete and non-degenerate sum of contributions, just like $\mathcal{Z}_{\text{GM}, M_4 \times \R}(G)$.
  We shall henceforth assume such a choice of $M_3$ whence the presumption would hold.
  \label{ft:m3 x r:isolation and non-degeneracy of GH-flat}
}

Let us now ascertain what the $\mathfrak{F}^{G}_{\text{GM$_4$}}(\Psi_{M_3}^v)$'s correspond to.
Repeating here the analysis in \autoref{sec:m4 x r} with \eqref{eq:m3 x r:action:sqm} as the action for the equivalent SQM model, we find that we can also write \eqref{eq:m3 x r:partition function:no homology} as
\begin{equation}
  \label{eq:m3 x r:partition function}
  \saveboxed{eq:m3 x r:partition function}{
    \mathcal{Z}_{\text{GM}, M_3 \times \R}(G)
    = \sum_v \mathfrak{F}^{G}_{\text{GM$_4$}}(\Psi_{M_3}^v)
    = \sum_v \text{HHF}^{\text{flat}}_{d_v}(M_3, G_{\HH})
    = \mathcal{Z}^{\text{Floer}}_{\text{flat}, M_3}(G_{\HH})
  }
\end{equation}
where each $\mathfrak{F}^{G}_{\text{GM$_4$}}(\Psi_{M_3}^v)$ can be identified with a \emph{novel} gauge-theoretic \emph{holomorphic} Floer homology class $\text{HHF}^{\text{flat}}_{d_v}(M_3, G_{\HH})$, that we shall henceforth name a holomorphic $G_{\HH}$-flat Floer homology class, of degree $d_v$, assigned to $M_3$.

Specifically, the \emph{time-invariant $G_{\HH}$-BF configurations on $M_3$ in $\mathfrak{B}_3$} that obey the simultaneous vanishing of the LHS and RHS of the \emph{gradient flow equations}
\begin{equation}
  \label{eq:m3 x r:flow}
  \saveboxed{eq:m3 x r:flow}{
    \dv{\overline{\mathscr{A}}^{\alpha}}{t}
    = - g_{\mathfrak{B}_3}^{\alpha\beta} \pdv{V_3}{\overline{\mathscr{A}}^{\beta}}
    \qquad
    \dv{\overline{\mathscr{B}}^{\alpha}}{t}
    = - g_{\mathfrak{B}_3}^{\alpha\beta} \pdv{V_3}{\overline{\mathscr{B}}^{\beta}}
  }
\end{equation}
will generate the chains of the holomorphic $G_{\HH}$-flat Floer complex with \emph{holomorphic Morse functional}
\begin{equation}
  \label{eq:m3 x r:morse functional}
  \saveboxed{eq:m3 x r:morse functional}{
    V_3(\overline{\mathscr{A}}, \overline{\mathscr{B}})
    = \int_{M_3} \Tr \left(
      \overline{\mathscr{B}} \wedge \overline{\mathscr{F}}
    \right)
  }
\end{equation}
in $\mathfrak{B}_3$.
The $G_{\HH}$-flat flow lines, described by time-varying solutions to \eqref{eq:m3 x r:flow}, are the Floer differentials such that the degree $d_v$ of the corresponding chain in the holomorphic $G_{\HH}$-flat Floer complex is counted by the outgoing flow lines at each time-invariant $G_{\HH}$-BF configuration on $M_3$ in $\mathfrak{B}_3$.
Such a configuration corresponds to a time-independent solution to the 3d equations
\begin{equation}
  \label{eq:m3 x r:critical points}
  \saveboxed{eq:m3 x r:critical points}{
    \overline{\mathscr{D}} \overline{\mathscr{B}}
    = 0
    \qquad
    \overline{\mathscr{F}}
    = 0
  }
\end{equation}

\subsection{A Holomorphic \texorpdfstring{$G_{\OO}$}{G-O}-flat Floer Homology of Two-Manifolds}
\label{sec:m2 x r}

Let us now further specialize to $M_3 = M_2 \times S^1$, and perform a KK dimensional reduction of GM$_4$ theory by shrinking the $S^1$ circle to be infinitesimally small.

\subtitle{The Octonionified BPS Equations of GM$_3$ Theory}

The KK reduction of the $G_{\HH}$-flat BPS equation of GM$_4$ theory on $M_2 \times S^1 \times \R$, is algorithmically the same as the KK reduction of the $G_{\C}$-flat BPS equation of GM theory on $M_3 \times S^1 \times \R$ in \autoref{sec:m3 x r}.
Before we say more about reduction process, let us first relabel the quaternionic imaginary numbers $(i, j, k)$ as $(\mathbf{e}_{01}, \mathbf{e}_{02}, \mathbf{e}_{03})$, and introduce an imaginary number $\mathbf{e}_0$ that satisfies the octonionic relations with $(\mathbf{e}_{01} , \mathbf{e}_{02}, \mathbf{e}_{03})$, i.e.,
\begin{equation}
  \label{eq:m2 x r:octonionic relations}
  \begin{aligned}
    \mathbf{e}_{0i} \mathbf{e}_{0j}
    &= - \delta_{ij} + \epsilon_{ijk} \mathbf{e}_{0k}
      \, ,
    &\qquad
      \mathbf{e}_0 \mathbf{e}_i
    &= \mathbf{e}_{0i}
      = \frac{1}{2} \epsilon_{ijk} \mathbf{e}_j \mathbf{e}_k
      \, ,
    &\qquad
      \mathbf{e}_0 \mathbf{e}_{0i}
      = - \mathbf{e}_i
      \, ,
    \\
    \mathbf{e}_i \mathbf{e}_j
    &= - \delta_{ij} - \epsilon_{ijk} \mathbf{e}_{0k}
      \, ,
    &\qquad
      \mathbf{e}_i \mathbf{e}_{0j}
    &= \delta_{ij} \mathbf{e}_0 - \epsilon_{ijk} \mathbf{e}_k
      \, .
  \end{aligned}
\end{equation}
This will be useful shortly.

Let $x^3$ be the coordinate of the $S^1$ circle.
The reduction process produces a $G_{\HH}$-scalar $\overline{\mathscr{A}}_3 \equiv \overline{\mathcal{A}}_3 - \mathbf{e}_{02} \overline{\mathcal{C}}_3 \equiv A_3 - \mathbf{e}_{01} \phi_3 - \mathbf{e}_{02} A_4 + \mathbf{e}_{03} \phi_4$ that, when multiplied by a unit one-form $\mathscr{H}_m$, can be equivalently interpreted as a one-form $\overline{\mathscr{A}}_3 \mathscr{H}_m \eqcolon \mathscr{C}_m \in \Omega^1(M_2 \times \R, \text{ad}(G_{\HH}))$ whose components are \emph{all} equal, where $m \in \{t, 1, 2\}$.

The BPS equations of the 3d theory resulting from the KK reduction of GM$_4$ theory on $M_2 \times S^1 \times \R$  along the $S^1$ circle -- henceforth referred to as GM$_3$ theory -- can then be complexified \emph{à la} \eqref{eq:m3 x r:bps:complexify} to become the $G_{\OO}$-flat equation, $\overline{\mathsf{F}}_{mn} = 0$, on $M_2 \times \R$.
Here $\overline{\mathsf{F}}$ is the field strength of $\overline{\mathsf{A}} \coloneq \overline{\mathscr{A}} - \mathbf{e}_0 \overline{\mathscr{C}}$, a complexification of $\overline{\mathscr{A}}$ and $\overline{\mathscr{C}}$ (the quaternion conjugate of $\mathscr{C}$) which can be understood as a $G_{\OO}$-connection, i.e., a connection valued in the adjoint bundle $\text{ad}(G_{\OO})$ of the underlying \emph{octonionified} principal $G_{\OO}$-bundle.\footnote{%
  Fields valued in $\text{ad}(G_{\OO})$ are $\mathfrak{g}_{\OO}$-valued, where $\mathfrak{g}_{\OO}$ is the Lie algebra of the octonionified Lie group $G_{\OO}$.
  These octonionified Lie algebras are equipped with an $\mathbb{R}$-bilinear, antisymmetric bracket $[\cdot, \cdot]_{\mathbb{O}}: \mathfrak{g}_{\OO} \times \mathfrak{g}_{\OO} \rightarrow \mathfrak{g}_{\OO}$ that is described as
  \begin{equation*}
    \begin{aligned}
      [\overline{\mathsf{A}}_\mu, \overline{\mathsf{A}}_\nu]
      = [\overline{\mathscr{A}}_{\mu} - j \overline{\mathscr{C}}_{\mu}, \overline{\mathscr{A}}_{\nu} - j \overline{\mathscr{C}}_{\nu}]_{\OO}
      &\coloneq \overline{\mathscr{A}}_{\mu} \overline{\mathscr{A}}_{\nu}
      - \overline{\mathscr{A}}_{\nu} \overline{\mathscr{A}}_{\mu}
      - \mathscr{C}_{\mu} \overline{\mathscr{C}}_{\nu}
      + \mathscr{C}_{\nu} \overline{\mathscr{C}}_{\mu}
      \\
      &\quad
      - j \left(
        \mathscr{A}_{\mu} \overline{\mathscr{C}}_{\nu}
        - \overline{\mathscr{C}}_{\nu} \overline{\mathscr{A}}_{\mu}
        + \mathscr{A}_{\nu} \overline{\mathscr{C}}_{\mu}
        - \overline{\mathscr{C}}_{\mu} \overline{\mathscr{A}}_{\nu}
      \right)
    \end{aligned}
  \end{equation*}
  using $\mathfrak{g}_{\mathbb{H}}$-valued fields $\mathscr{A}, \overline{\mathscr{A}}, \mathscr{C}, \overline{\mathscr{C}}$, which satisfies the Jacobi identity (see \autoref{app:jacobi:g-o}).
}

Finally, working in the \emph{real} temporal gauge, the $G_{\OO}$-flat equation on $M_2 \times \R$ can be expressed as
\begin{equation}
  \label{eq:m2 x r:gm3 flow}
  \dot{\overline{\mathsf{A}}}^N \mathsf{E}_{NM}
  = \epsilon_{MN} \overline{\mathsf{D}}^N \overline{\mathsf{B}}
  \, ,
  \qquad
  \epsilon_{P[M} \dot{\overline{\mathsf{B}}} \mathsf{E}_{N]Q} \mathsf{E}^{QR} \mathsf{E}_R^{\;P}
  = - \overline{\mathsf{F}}_{MN}
  \, ,
\end{equation}
where (i) $\{M, N, P, Q\} \in \{1, 2\}$ are indices on $M_2$;
(ii) $\mathsf{E}_{MN} \in \Omega^2(M_2) \otimes \Omega^0(\R)$ is a unit two-form on $M_2$, i.e., $\mathsf{E}_{12} = 1 = - \mathsf{E}_{21}$;
(iii) $\frac{1}{2} (\Re(\overline{\mathsf{A}}_t) - \overline{\mathsf{A}}_t) \epsilon_{MN} \mathsf{E}^{MN} = \frac{1}{2} (\mathbf{e}_{01} \phi_t + \mathbf{e}_{02} A_4 - \mathbf{e}_{03} \phi_4 + \mathbf{e}_0 A_3 - \mathbf{e}_1 \phi_3 - \mathbf{e}_2 A_4 + \mathbf{e}_3 \phi_4) \epsilon_{MN} \mathsf{E}^{MN} \eqcolon \overline{\mathsf{B}} \in \Omega^0(M_2 \otimes \R, \text{ad}(G_{\OO}))$,
and (iv) the above term $\epsilon_{P[M} \dot{\overline{\mathsf{B}}} \mathsf{E}_{N]Q} \mathsf{E}^{QR} \mathsf{E}_R^{\;P}$ is trivially zero (even though $\dot{\overline{\mathsf{B}}} \neq 0$) due to the identity $\epsilon_{P[M} \mathsf{E}_{N]Q} \mathsf{E}^{QR} \mathsf{E}_R^{\;P} \equiv 0$, as required.
This set of equations has the same form as \eqref{eq:m3 x r:gm4 flow}, just with the gauge group complexified from $G_{\HH}$ to $G_{\OO}$.

In other words, GM$_4$ theory on $M_2 \times S^1 \times \R$, upon a KK reduction along $S^1$ to GM$_3$ theory on $M_2 \times \R$, localizes onto $\text{ad}(G_{\OO})$-valued configurations that obey \eqref{eq:m2 x r:gm3 flow}.

\subtitle{GM$_3$ Theory as a 1d SQM in $\mathfrak{B}_2$}

The 3d action can thus be written (in the \emph{real} temporal gauge) as
\begin{equation}
  \label{eq:m2 x r:action}
  \begin{aligned}
    & S_{\text{GM$_3$}, A_t = 0}
    \\
    & = \frac{1}{e^2} \int_{M_2 \times \R} dt d^2x \, \Tr \left(
      \frac{1}{2} \left|
      \dot{\overline{\mathsf{A}}}^N \mathsf{E}_{NM}
      - \epsilon_{MN} \overline{\mathsf{D}}^N \overline{\mathsf{B}}
      \right|^2
      + \frac{1}{4} \left|
      \epsilon_{P[M} \dot{\overline{\mathsf{B}}} \mathsf{E}_{N]Q} \mathsf{E}^{QR} \mathsf{E}_R^{\;P}
      + \overline{\mathsf{F}}_{MN}
      \right|^2
      + \dots
      \right)
      \, .
  \end{aligned}
\end{equation}

Via a similar calculation that got us from \eqref{eq:m4 x r:action:temporal gauge} to \eqref{eq:m4 x r:sqm:action}, we can recast \eqref{eq:m2 x r:action} as a 1d SQM model, where its action will now read
\begin{equation}
  \label{eq:m2 x r:action:sqm}
  S_{\text{SQM}, \mathfrak{B}_2}
  = \frac{1}{e^2} \int_{\R} dt \left(
    \left|
      \dv{\overline{\mathsf{A}}^{\alpha}}{t}
      + g_{\mathfrak{B}_2}^{\alpha\beta} \pdv{V_2}{\overline{\mathsf{A}}^{\beta}}
    \right|^2
    + \left|
      \dv{\overline{\mathsf{B}}^{\alpha}}{t}
      + g_{\mathfrak{B}_2}^{\alpha\beta} \pdv{V_2}{\overline{\mathsf{B}}^{\beta}}
    \right|^2
    + \dots
  \right)
  \, .
\end{equation}
Here, (i) $(\overline{\mathsf{A}}^{\alpha}, \overline{\mathsf{B}}^{\alpha})$ and $(\alpha, \beta)$ are coordinates and indices, respectively, on the space $\mathfrak{B}_2$ of irreducible $(\overline{\mathsf{A}}_M, \overline{\mathsf{B}})$ fields on $M_2$;
(ii) $g_{\mathfrak{B}_2}$ is a corresponding  metric on $\mathfrak{B}_2$;
and (iii) $V_2(\overline{\mathsf{A}}, \overline{\mathsf{B}})$ is the potential function.

\subtitle{Localizing onto Time-invariant $G_{\OO}$-BF Configurations on $M_2$}

By the squaring argument \cite{blau-1993-topol-gauge} applied to \eqref{eq:m2 x r:action:sqm}, the configurations that the equivalent SQM localizes onto are those that set the LHS's and RHS's of \eqref{eq:m2 x r:gm3 flow} \emph{simultaneously} to zero.
Such configurations in $\mathfrak{B}_2$ correspond to time-invariant configurations that span the space of solutions to the 2d $G_{\OO}$-BF equations on $M_2$.
We shall, in the rest of this paper, refer to such configurations as $G_{\OO}$-BF configurations on $M_2$.

In summary, the equivalent SQM localizes onto time-invariant $G_{\OO}$-BF configurations on $M_2$ in $\mathfrak{B}_2$.

\subtitle{The Holomorphic $G_{\OO}$-flat Floer Homology}

Since the resulting 3d theory on $M_2 \times \R$ can be interpreted as a 1d SQM in $\mathfrak{B}_3$, its partition function can, like in \eqref{eq:m4 x r:partition function:no homology}, be written as
\begin{equation}
  \label{eq:m2 x r:partition function:no homology}
  \mathcal{Z}_{\text{GM}, M_2 \times \R}(G)
  = \langle 1 \rangle_{F_2(\Psi_{M_2})}
  = \sum_w \mathfrak{F}^{G}_{\text{GM$_3$}}(\Psi_{M_2}^w)
  \, ,
\end{equation}
where $\mathfrak{F}^{G}_{\text{GM$_3$}}(\Psi_{M_2}^w)$, in the $\hat{\mathcal{Q}}$-cohomology of GM$_3$ theory, is the $w^{\text{th}}$ contribution to the partition function that depends on the expression $F_2(\Psi_{M_2})$ in the bosonic fields on $M_2$, and the summation in `$w$' is over all isolated and non-degenerate $G_{\OO}$-BF configurations on $M_2$ in $\mathfrak{B}_2$ that the equivalent SQM localizes onto.\footnote{%
  This presumption that the configurations will be isolated and non-degenerate is justified because the ($\hat{\mathcal{Q}}$-cohomology of) GM theory is topological in all directions and therefore invariant when we shrink another $S^1$ circle from GM$_4$ theory.
  Thus, if $M_2$ (where $M_2 \times T^2 = M_4$) and $G$ are chosen such as to satisfy the conditions spelt out in \autoref{ft:m4 x r:isolation justification}, $\mathcal{Z}_{\text{GM}, M_2 \times \R}(G)$ will be a discrete and non-degenerate sum of contributions, just like $\mathcal{Z}_{\text{GM}, M_4 \times \R}(G)$.
  We shall henceforth assume such a choice of $M_2$ whence the presumption would hold.
  \label{ft:m2 x r:isolation and non-degeneracy of GO-flat}
}

Let us now ascertain what the $\mathfrak{F}^G_{\text{GM$_3$}}(\Psi^w_{M_2})$'s correspond to.
Repeating here the analysis in \autoref{sec:m4 x r} with \eqref{eq:m2 x r:action:sqm} as the action for the equivalent SQM model, we find that we can also write \eqref{eq:m2 x r:partition function:no homology} as
\begin{equation}
  \label{eq:m2 x r:partition function}
  \saveboxed{eq:m2 x r:partition function}{
    \mathcal{Z}_{\text{GM}, M_2 \times \R}(G)
    = \sum_w \mathfrak{F}^{G}_{\text{GM$_3$}}(\Psi_{M_2}^w)
    = \sum_w \text{HHF}^{\text{flat}}_{d_w}(M_2, G_{\OO})
    = \mathcal{Z}^{\text{Floer}}_{\text{flat}, M_2}(G_{\OO})
  }
\end{equation}
where each $\mathfrak{F}^{G}_{\text{GM$_3$}}(\Psi_{M_2}^w)$ can be identified with a \emph{novel} gauge-theoretic \emph{holomorphic} Floer homology class $\text{HHF}^{\text{flat}}_{d_w}(M_2, G_{\OO})$, that we shall henceforth name a holomorphic $G_{\OO}$-flat Floer homology class, of degree $d_w$, assigned to $M_2$.

Specifically, the \emph{time-invariant $G_{\OO}$-BF configurations on $M_2$ in $\mathfrak{B}_2$} that obey the simultaneous vanishing of the LHS and RHS of the \emph{gradient flow equations}
\begin{equation}
  \label{eq:m2 x r:flow}
  \saveboxed{eq:m2 x r:flow}{
    \dv{\overline{\mathsf{A}}^{\alpha}}{t}
    = - g_{\mathfrak{B}_2}^{\alpha\beta} \pdv{V_2}{\overline{\mathsf{A}}^{\beta}}
    \qquad
    \dv{\overline{\mathsf{B}}^{\alpha}}{t}
    = - g_{\mathfrak{B}_2}^{\alpha\beta} \pdv{V_2}{\overline{\mathsf{B}}^{\beta}}
  }
\end{equation}
will generate the chains of the holomorphic $G_{\OO}$-flat Floer complex with \emph{holomorphic Morse functional}
\begin{equation}
  \label{eq:m2 x r:morse functional}
  \saveboxed{eq:m2 x r:morse functional}{
    V_2(\overline{\mathsf{A}}, \overline{\mathsf{B}})
    = \int_{M_2} \Tr \left(
      \overline{\mathsf{B}} \wedge \overline{\mathsf{F}}
    \right)
  }
\end{equation}
in $\mathfrak{B}_2$.
The $G_{\OO}$-flat flow lines, described by time-varying solutions to \eqref{eq:m2 x r:flow}, are the Floer differentials such that the degree $d_w$ of the corresponding chain in the holomorphic $G_{\OO}$-flat Floer complex is counted by the outgoing flow lines at each time-invariant $G_{\OO}$-BF configuration on $M_2$ in $\mathfrak{B}_2$.
Such a configuration corresponds to a time-independent solution to the 2d equations
\begin{equation}
  \label{eq:m2 x r:critical points}
  \saveboxed{eq:m2 x r:critical points}{
    \overline{\mathsf{D}} \overline{\mathsf{B}}
    = 0
    \qquad
    \overline{\mathsf{F}}
    = 0
  }
\end{equation}

\section{A Novel Orlov-type \texorpdfstring{$A_{\infty}$}{A-infinity}-1-category of Three-Manifolds}
\label{sec:m3 x r2}

In this section, we will study GM theory on $M_5 = M_3 \times \R^2$, where $M_3$ is a closed three-manifold.
We will recast it as a 2d gauged B-twisted Landau-Ginzburg (LG) model and a 1d LG SQM.
Following what we did in our previous work \cite[$\S$9]{er-2023-topol-n}, we will, via the 5d GM partition function and its equivalent 2d gauged B-twisted LG model, be able to physically realize a novel gauge-theoretic Orlov-type $A_{\infty}$-1-category that categorifies the holomorphic $G_{\HH}$-flat Floer homology of $M_3$ from \autoref{sec:m3 x r}.
Also, via a Heegaard split at $M_3$ along a Riemann surface $C$, we will be able to physically realize an RW $A_{\infty}$-2-category and a Kapustin-Rozansky-Saulina (KRS) 2-category of complex-Lagrangian branes in $\mathcal{M}^{G}_{\text{H}, \theta}(C, \mathbf{J})$.

\subsection{GM Theory on \texorpdfstring{$M_3 \times \R^2$}{M3 x R2} as a 2d Model on \texorpdfstring{$\R^2$}{R2} or 1d SQM}
\label{sec:m3 x r2:gm}

\subtitle{GM Theory on $M_3 \times \R^2$ as a 2d Gauged Sigma Model}

Let $t, \tau$ be the coordinates along $\R^2$.
The action of this theory is given by
\begin{equation}
  \label{eq:m3 x r2:action:full}
  \begin{aligned}
    S_{\text{GM}}
    = \frac{1}{e^2}\int_{M_3 \times \R^2} dt d\tau d^3 x \, \Tr \bigg(
    & \left| F_{t\tau} - i D_t \phi_{\tau} + i D_{\tau} \phi_t - [\phi_t, \phi_{\tau}] \vphantom{\overline{\mathcal{F}}} \right|^2
      + \left| \overline{\mathcal{F}}_{mn} \right|^2
    \\
    & + \left| D_t \overline{\mathcal{A}}_m - \partial_m A_t + i \overline{\mathcal{D}}_m \phi_t \right|^2
      + \left| D_{\tau} \overline{\mathcal{A}}_m - \partial_m A_{\tau} + i \overline{\mathcal{D}}_m \phi_{\tau} \right|^2
      + \dots
    \bigg)
      \, ,
  \end{aligned}
\end{equation}
where the conditions that minimize the action \eqref{eq:m3 x r2:action:full} are identified to be
\begin{equation}
  \label{eq:m3 x r2:bps}
  \begin{aligned}
    F_{t\tau} - i D_t \phi_{\tau} + i D_{\tau} \phi_t - [\phi_t, \phi_{\tau}]
    &= 0
      \, ,
    &\qquad
    D_t \overline{\mathcal{A}}_m - \partial_m A_t + i \overline{\mathcal{D}}_m \phi_t
    &= 0
      \, ,
    \\
    \overline{\mathcal{F}}_{mn}
    &= 0
      \, ,
    &\qquad
    D_{\tau} \overline{\mathcal{A}}_m - \partial_m A_{\tau} + i \overline{\mathcal{D}}_m \phi_{\tau}
    &= 0
      \, .
  \end{aligned}
\end{equation}

Notice that we can complexify \eqref{eq:m3 x r2:bps} in a similar manner as was done for the BPS equations in \autoref{sec:m3 x r}.
The complexified conditions are
\begin{equation}
  \label{eq:m3 x r2:bps:quaternionified}
  \begin{aligned}
    (D_{\tau} + j D_t) \widetilde{\mathscr{A}}^m \mathscr{E}_{mn}
    - D_{\tau} \widetilde{\mathscr{B}}^m \widetilde{\mathscr{H}}_{mn}
    + F_{t\tau} \mathscr{H}^m \mathscr{E}_{mn}
    - \partial^m(A_{\tau} + j A_t) \mathscr{E}_{mn}
    &= j \epsilon_{npq} \widetilde{\mathscr{D}}^p \widetilde{\mathscr{B}}^q
      \, ,
    \\
    D_{\tau} \widetilde{\mathscr{A}}^m \widetilde{\mathscr{H}}_{mn}
    + (D_{\tau} + j D_t) \widetilde{\mathscr{B}}^m \mathscr{E}_{pq} \mathscr{E}^q_{\; n} \mathscr{E}_m^{\;p}
    - \partial^m A_{\tau} \widetilde{\mathscr{H}}_{mn}
    & = - \frac{j}{2} \epsilon_{npq} \widetilde{\mathscr{F}}^{pq}
      \, ,
  \end{aligned}
\end{equation}
where
(i) $\widetilde{\mathscr{A}}_m \coloneq \overline{\mathcal{A}}_m - k \phi_{\tau} \mathscr{H}_m$,
(ii) $\widetilde{\mathscr{B}}_m \coloneq \frac{1}{2} (i \phi_t - k \phi_{\tau}) \epsilon_{mpq} \mathscr{E}^{pq}$,
and (iii) $\widetilde{\mathscr{H}}_{mn} = \epsilon_{mnp} \mathscr{H}^p$.\footnote{%
  Note that when compared to the fields in \autoref{sec:m3 x r}, $\widetilde{\mathscr{A}}_m = \overline{\mathscr{A}}_m + j A_\tau \mathscr{H}_m$ and $\widetilde{\mathscr{B}}_m = \overline{\mathscr{B}}_m - \frac{1}{2} \epsilon_{mpq} j A_{\tau} \mathscr{E}^{pq}$.
  We choose to use the $\widetilde{\mathscr{A}}$ and $\widetilde{\mathscr{B}}$ fields in this subsection because we want the real gauge fields of $\R^2$, i.e., $A_t$ and $A_{\tau}$, to be explicit in the equations.
  \label{ft:m3 x r2:comparing fields to those in floer hom}
}
Moreover, noting that we are physically free to rotate $\R^2$ about the origin, \eqref{eq:m3 x r2:bps:quaternionified} becomes
\begin{equation}
  \label{eq:m3 x r2:bps:quaternionified:rotated}
  \begin{aligned}
    (D_{\tau} + j D_t) \widetilde{\mathscr{A}}^m \mathscr{E}_{mn}
    - e^{j\theta} (\cos(\theta) D_{\tau} + \sin(\theta) D_t) \widetilde{\mathscr{B}}^m \mathscr{H}_{mn}
    + e^{j\theta} F_{t\tau} \mathscr{H}^m \mathscr{E}_{mn}
    + p_n(\theta)
    &= 0
      \, ,
    \\
    e^{j\theta} (\cos(\theta) D_{\tau} + \sin(\theta) D_t) \widetilde{\mathscr{A}}^m \widetilde{\mathscr{H}}_{mn}
    + (D_{\tau} + j D_t) \widetilde{\mathscr{B}}^m \mathscr{E}_{pq} \mathscr{E}^q_{\; n} \mathscr{E}_m^{\;p}
    + q_n(\theta)
    &= 0
      \, ,
  \end{aligned}
\end{equation}
and
\begin{equation}
  \label{eq:m3 x r2:bps:quaternionified:rotated:components}
  \begin{aligned}
    p_n(\theta)
    & = - \partial^m  (A_{\tau} + j A_t) \mathscr{E}_{mn}
      - j e^{j\theta} \epsilon_{npq} \widetilde{\mathscr{D}}^p \widetilde{\mathscr{B}}^q
      \, ,
    \\
    q_n(\theta)
    &= - e^{j\theta} \partial^m \Big( A_{\tau} \cos(\theta) + A_t \sin(\theta) \Big) \widetilde{\mathscr{H}}_{mn}
      + \frac{j e^{j\theta}}{2} \epsilon_{npq} \widetilde{\mathscr{F}}^{pq}
      \, .
  \end{aligned}
\end{equation}

Thus, we can also write \eqref{eq:m3 x r2:action:full} as
\begin{equation}
  \label{eq:m3 x r2:action:complexified}
  \begin{aligned}
    & S_{\text{GM}}
    \\
    & = \frac{1}{e^2} \int_{M_3 \times \R^2} dt d\tau d^3x \, \Tr \bigg(
      \left|
      e^{j\theta} (\cos(\theta) D_{\tau} + \sin(\theta) D_t) \widetilde{\mathscr{A}}^m \widetilde{\mathscr{H}}_{mn}
      + (D_{\tau} + j D_t) \widetilde{\mathscr{B}}^m \mathscr{E}_{pq} \mathscr{E}^q_{\; n} \mathscr{E}_m^{\;p}
      + q_n(\theta)
      \right|^2
    \\
    & \qquad
      + \left|
      (D_{\tau} + j D_t) \widetilde{\mathscr{A}}^m \mathscr{E}_{mn}
      - e^{j\theta} (\cos(\theta) D_{\tau} + \sin(\theta) D_t) \widetilde{\mathscr{B}}^m \mathscr{H}_{mn}
      + e^{j\theta} F_{t\tau} \mathscr{H}^m \mathscr{E}_{mn}
      + p_n(\theta)
      \right|^2
      + \dots
      \bigg)
      \, .
  \end{aligned}
\end{equation}

After suitable rescalings, we can recast \eqref{eq:m3 x r2:action:complexified} as a 2d model, where the action is\footnote{%
  To arrive at the following expression, we have employed Stokes' theorem and the fact that $M_3$ has no boundary to omit terms with $\partial_m A_{\{t, \tau\}}$ as they will vanish when integrated over $M_3$.
  \label{ft:m3 x r2:stokes' theorem to 2d model}
}
\begin{equation}
  \label{eq:m3 x r2:action:2d sigma}
  \begin{aligned}
    S_{\text{2d}, \mathfrak{B}_3}
    & = \frac{1}{e^2} \int_{\R^2} dt d\tau \, \bigg(
      \left|
      e^{j\theta} (\cos(\theta) D_{\tau} + \sin(\theta) D_t) \widetilde{\mathscr{A}}^a
      + (D_{\tau} + j D_t) \widetilde{\mathscr{B}}^a
      + q^a(\theta)
      \right|^2
    \\
    & \qquad
      + \left|
      (D_{\tau} + j D_t) \widetilde{\mathscr{A}}^a
      - e^{j\theta} (\cos(\theta) D_{\tau} + \sin(\theta) D_t) \widetilde{\mathscr{B}}^a
      + e^{j\theta} F_{t\tau}
      + p^a(\theta)
      \right|^2
      + \dots
      \bigg)
      \, .
  \end{aligned}
\end{equation}
Here, $(\widetilde{\mathscr{A}}^a, \widetilde{\mathscr{B}}^a)$  and $a$ are coordinates and indices on the space $\mathfrak{B}_3$ of irreducible $(\widetilde{\mathscr{A}}_m, \widetilde{\mathscr{B}}_m)$ fields on $M_3$, and
\begin{equation}
  \label{eq:m3 x r2:action:2d sigma:components}
  p^a
  = - j e^{j\theta} (\widetilde{\mathscr{D}} \widetilde{\mathscr{B}})^a
  \, ,
  \qquad
  q^a
  = j e^{j\theta} \widetilde{\mathscr{F}}^a
  \, ,
\end{equation}
with (i) $(\widetilde{\mathscr{D}} \widetilde{\mathscr{B}})^a$ and (ii) $\widetilde{\mathscr{F}}^a$ corresponding to (i) $\epsilon_{npq} \widetilde{\mathscr{D}}^p \widetilde{\mathscr{B}}^q$ and (ii) $\frac{1}{2} \epsilon_{npq}\widetilde{\mathscr{F}}^{pq}$, respectively, in the underlying 5d theory.
In other words, GM theory on $M_3 \times \R^2$ can be regarded as a 2d gauged sigma model along the $(t, \tau)$-directions with target $\mathfrak{B}_3$ and action \eqref{eq:m3 x r2:action:2d sigma}.

\subtitle{The 2d Model on $\R^2$ with Target $\mathfrak{B}_3$ as a 1d SQM}

Singling out $\tau$ as the direction in ``time'', one can, after suitable rescalings, recast \eqref{eq:m3 x r2:action:2d sigma} as the following equivalent SQM action\footnote{%
  In the resulting SQM, as $A_{\tau}$ has no field strength and is thus non-dynamical, it will be integrated out to furnish the Christoffel symbols for the fermions in the SQM \cite{er-2023-topol-n}, leaving us with an SQM without $A_{\tau}$.
  \label{ft:m3 x r2:integrate out gauge field to sqm}
}
\begin{equation}
  \label{eq:m3 x r2:action:1d sqm}
  S_{\text{SQM}, \mathfrak{P}(\R, \mathfrak{B}_3)}
  = \frac{1}{e^2} \int d\tau \left(
    \left| \dv{\varphi_+^{\alpha}}{\tau} + g^{\alpha\beta}_{\mathfrak{P}(\R, \mathfrak{B}_3)}\pdv{h_3}{\varphi_+^{\beta}} \right|^2
    + \left| \dv{\varphi_-^{\alpha}}{\tau} + g^{\alpha\beta}_{\mathfrak{P}(\R, \mathfrak{B}_3)}\pdv{h_3}{\varphi_-^{\beta}} \right|^2
    + \dots
  \right)
  \, ,
\end{equation}
where $\varphi_+^{\alpha} \coloneq e^{j\theta} \cos(\theta) \widetilde{\mathscr{A}}^{\alpha} + \widetilde{\mathscr{B}}^{\alpha}$ and $\varphi_-^{\alpha} \coloneq \widetilde{\mathscr{A}}^{\alpha} - e^{j\theta} \cos(\theta) \widetilde{\mathscr{B}}^{\alpha} - e^{j\theta} (A_t)^{\alpha}$.
Here, $(\widetilde{\mathscr{A}}^{\alpha}, \widetilde{\mathscr{B}}^{\alpha}, (A_t)^{\alpha})$ and $(\alpha, \beta)$ are coordinates and indices on the path space $\mathfrak{P}(\R, \mathfrak{B}_3)$ of maps from $\R$ to $\mathfrak{B}_3$ with $(A_t)^{\alpha}$ corresponding to $A_t$ in the underlying 2d model;
$g_{\mathfrak{P}(\R, \mathfrak{B}_3)}^{\alpha\beta}$ is the metric on $\mathfrak{P}(\R, \mathfrak{B}_3)$;
and $h_3(\varphi_+, \varphi_-)$ is the potential function.

In other words, GM theory on $M_3 \times \R^2$ can also be regarded as a 1d SQM along $\tau$ in $\mathfrak{P}(\R, \mathfrak{B}_3)$ whose action is \eqref{eq:m3 x r2:action:1d sqm}.

\subsection{Non-constant Paths, Strings, and \texorpdfstring{$G_{\HH}$-BF}{GH-BF} Configurations on \texorpdfstring{$M_3$}{M3}}
\label{sec:m3 x r2:gh-bf}

\subtitle{$\theta$-deformed, Non-constant Paths in the SQM}

The squaring argument \cite{blau-1993-topol-gauge} applied to \eqref{eq:m3 x r2:action:1d sqm} tells us that the equivalent SQM localizes onto configurations that set both the LHS and RHS of the expression within the squared term \emph{simultaneously} to zero, i.e., the SQM localizes onto $\tau$-invariant critical points of $h_3(\varphi_+, \varphi_-)$ that obey
\begin{equation}
  \label{eq:m3 x r2:non-constant path}
  \begin{aligned}
    \partial_t \left(
    e^{j\theta} \sin(\theta) \widetilde{\mathscr{A}}
    + j \widetilde{\mathscr{B}}
    \right)^{\alpha}
    &= - je^{j\theta} \widetilde{\mathscr{F}}^{\alpha}
      \, ,
    \\
    \partial_t \left(
    j \widetilde{\mathscr{A}}
    - e^{j\theta} \sin(\theta) \widetilde{\mathscr{B}}
    \right)^{\alpha}
    &= j e^{j\theta} (\widetilde{\mathscr{D}} \widetilde{\mathscr{B}})^{\alpha}
      \, .
  \end{aligned}
\end{equation}
These are \emph{$\tau$-invariant, $\theta$-deformed, non-constant} paths in $\mathfrak{P}(\R, \mathfrak{B}_3)$.

\subtitle{$\mathfrak{B}^{\theta}_3$-strings in the 2d Gauged Model}

By comparing \eqref{eq:m3 x r2:action:1d sqm} with \eqref{eq:m3 x r2:action:2d sigma}, we find that such $\tau$-invariant, $\theta$-deformed, non-constant paths in the SQM defined by \eqref{eq:m3 x r2:non-constant path}, will correspond, in the 2d gauged sigma model with target $\mathfrak{B}_3$, to configurations defined by
\begin{equation}
  \label{eq:m3 x r2:soliton eqn}
  \begin{aligned}
    \partial_t \left(
    e^{j\theta} \sin(\theta) \widetilde{\mathscr{A}}^a
    + j \widetilde{\mathscr{B}}^a
    \right)
    =
    & - \left[
      A_t,
      e^{j\theta} \sin(\theta) \widetilde{\mathscr{A}}^a
      + j \widetilde{\mathscr{B}}^a
      \right]_{\HH}
    \\
    & - \left[
      A_{\tau},
      e^{j\theta} \cos(\theta) \widetilde{\mathscr{A}}^a
      + \widetilde{\mathscr{B}}^a
      \right]_{\HH}
      - q^a(\theta)
      \, ,
    \\
    \partial_t \left(
    j \widetilde{\mathscr{A}}^a
    - e^{j\theta} \sin(\theta) \widetilde{\mathscr{B}}^a
    + e^{j\theta} A_{\tau}
    \right)
    =
    & - \left[
      A_t,
      j \widetilde{\mathscr{A}}^a
      - e^{j\theta} \sin(\theta) \widetilde{\mathscr{B}}^a
      + e^{j\theta} A_{\tau}
      \right]_{\HH}
    \\
    & - \left[
      A_{\tau},
      \widetilde{\mathscr{A}}^a
      - e^{j\theta} \cos(\theta) \widetilde{\mathscr{B}}^a
      \right]_{\HH}
      - p^a(\theta)
      \, .
  \end{aligned}
\end{equation}
These are $\tau$-invariant, $\theta$-deformed strings along the $t$-direction in the 2d gauged sigma model.
We shall henceforth refer to such strings as $\mathfrak{B}^{\theta}_3$-strings.

\subtitle{$\tau$-independent, $\theta$-deformed GM Configurations in GM Theory}

By comparing \eqref{eq:m3 x r2:action:2d sigma} with \eqref{eq:m3 x r2:action:complexified}, we find that the 2d configurations defined by \eqref{eq:m3 x r2:soliton eqn}, will correspond, in GM theory, to 5d configurations defined by
\begin{equation}
  \label{eq:m3 x r2:gm configs}
  \begin{aligned}
    & \partial_t \left(
      e^{j\theta} \sin(\theta) \widetilde{\mathscr{A}}^m \widetilde{\mathscr{H}}_{mn}
      + j \widetilde{\mathscr{B}}^m \mathscr{E}_{pq} \mathscr{E}^q_{\; n} \mathscr{E}_m^{\;p}
      \right)
    \\
    &= - \left[
      A_t,
      e^{j\theta} \sin(\theta) \widetilde{\mathscr{A}}^m \widetilde{\mathscr{H}}_{mn}
      + j \widetilde{\mathscr{B}}^m \mathscr{E}_{pq} \mathscr{E}^q_{\; n} \mathscr{E}_m^{\;p}
      \right]_{\HH}
    \\
    & \quad
      - [A_{\tau},
      e^{j\theta} \cos(\theta) \widetilde{\mathscr{A}}^m \widetilde{\mathscr{H}}_{mn}
      + \widetilde{\mathscr{B}}^m \mathscr{E}_{pq} \mathscr{E}^q_{\; n} \mathscr{E}_m^{\;p}]_{\HH}
      - q_n(\theta)
      \, ,
    \\
    \\
    & \partial_t \left(
    j \widetilde{\mathscr{A}}^m \mathscr{E}_{mn}
    - e^{j\theta} \sin(\theta) \widetilde{\mathscr{B}}^m \mathscr{H}_{mn}
    + e^{j\theta} A_{\tau} \mathscr{H}^m \mathscr{E}_{mn}
    \right)
    \\
    &= - \left[
      A_t,
      j \widetilde{\mathscr{A}}^m \mathscr{E}_{mn}
      - e^{j\theta} \sin(\theta) \widetilde{\mathscr{B}}^m \mathscr{H}_{mn}
      + e^{j\theta} A_{\tau} \mathscr{H}^m \mathscr{E}_{mn}
      \right]_{\HH}
    \\
    & \quad
      - \left[A_{\tau},
      \widetilde{\mathscr{A}}^m \mathscr{E}_{mn}
      - e^{j\theta} \cos(\theta) \widetilde{\mathscr{B}}^m \mathscr{H}_{mn}
      \right]_{\HH}
      - p_n(\theta)
      \, .
  \end{aligned}
\end{equation}
These are $\tau$-independent, $\theta$-deformed GM configurations on $M_3 \times \R^2$.

\subtitle{GM Configurations, $\mathfrak{B}^{\theta}_3$-strings, and Non-constant Paths}

In short, these $\tau$-independent, $\theta$-deformed GM configurations on $M_3 \times \R^2$ that are defined by \eqref{eq:m3 x r2:gm configs}, will correspond to the $\mathfrak{B}^{\theta}_3$-strings defined by \eqref{eq:m3 x r2:soliton eqn}, which, in turn, will correspond to the $\tau$-invariant, $\theta$-deformed, non-constant paths in $\mathfrak{P}(\R, \mathfrak{B}^{\theta}_3)$ defined by \eqref{eq:m3 x r2:non-constant path}.

\subtitle{$\mathfrak{B}^{\theta}_3$-string Endpoints Corresponding to $\theta$-deformed $G_{\mathbb{H}}$-BF Configurations on $M_3$}

Consider now the fixed endpoints of the $\mathfrak{B}^{\theta}_3$-strings at $t = \pm \infty$, where we also expect the finite-energy 2d gauge fields $A_t, A_{\tau}$ to decay to zero.
They are given by \eqref{eq:m3 x r2:soliton eqn} with $\partial_t \widetilde{\mathscr{A}}^a = 0 = \partial_t \widetilde{\mathscr{B}}^a$ and $A_t, A_{\tau} \rightarrow 0$, i.e., (via \eqref{eq:m3 x r2:action:2d sigma:components})
\begin{equation}
  \label{eq:m3 x r2:soliton:endpoints}
  e^{j\theta} (\widetilde{\mathscr{D}} \widetilde{\mathscr{B}})^a
  = 0
  \, ,
  \qquad
  e^{j\theta} \widetilde{\mathscr{F}}^a
  = 0
  \, .
\end{equation}
In turn, they will correspond, in GM theory, to $(t, \tau)$-independent, $\theta$-deformed configurations that obey
\begin{equation}
  \label{eq:m3 x r2:gh-bf configs}
  e^{j\theta} \epsilon_{npq} \widetilde{\mathscr{D}}^n \widetilde{\mathscr{B}}^q
  = 0
  \, ,
  \qquad
  e^{j\theta} \epsilon_{npq} \widetilde{\mathscr{F}}^{pq}
  = 0
  \, .
\end{equation}
Notice that \eqref{eq:m3 x r2:gh-bf configs} can also be obtained from \eqref{eq:m3 x r2:gm configs} and \eqref{eq:m3 x r2:bps:quaternionified:rotated:components} with $\partial_t \widetilde{\mathscr{A}}^m = 0 = \partial_t \widetilde{\mathscr{B}}_{mn}$ and $A_t, A_{\tau} \rightarrow 0$.

The equations in \eqref{eq:m3 x r2:gh-bf configs} are a $\theta$-deformed version $G_{\HH}$-BF equations on $M_3$.
At $\theta = 0, \pi$, they become the regular $G_{\HH}$-BF equations on $M_3$.
Configurations spanning the space of solutions to these equations shall, in the rest of this section, be referred to as $G_{\mathbb{H}}$-BF configurations on $M_3$.
We will also assume choices of $M_3$ and $G$ satisfying \autoref{ft:m3 x r:isolation and non-degeneracy of GH-flat} whereby such configurations are isolated and non-degenerate.\footnote{%
  At $\theta = 0$, the moduli space of such configurations is the moduli space of undeformed $G_{\HH}$-BF configurations on $M_3$.
  For such a choice of $M_3$ and $G$ (where $G$ is the gauge group that the underlying GM theory is originally defined with), this moduli space will be made of isolated and non-degenerate points.
  Therefore, at $\theta = 0$, the endpoints of the $\mathfrak{B}^{\theta}_3$-strings will be isolated and non-degenerate.
  As the physical theory is symmetric under a variation of $\theta$, this observation about the endpoints of the $\mathfrak{B}^{\theta}_3$-strings will continue to hold true for any value of $\theta$.
  Hence, the presumption that the moduli space of $\theta$-deformed $G_{\HH}$-BF configurations on $M_3$ will be made of isolated and non-degenerate points, is justified.
  \label{ft:m3 x r2:isolation and non-degeneracy of GH-flat}
}

In short, from the equivalent 1d SQM of GM theory on $M_3 \times \R^2$, the theory localizes onto $\tau$-invariant, $\theta$-deformed, non-constant paths in $\mathfrak{P}(\R, \mathfrak{B}_3)$, which, in turn, will correspond to $\mathfrak{B}^{\theta}_3$-strings in the 2d gauged sigma model whose endpoints correspond to $\theta$-deformed $G_{\HH}$-BF configurations on $M_3$.

\subsection{The 2d Model and an Open String Theory}
\label{sec:m3 x r2:2d}

\subtitle{Flow Lines of the SQM as BPS Worldsheets of the 2d Model}

The classical trajectories or flow lines of the equivalent SQM are governed by the gradient flow equations (defined by setting to zero the expression within the squared terms in \eqref{eq:m3 x r2:action:1d sqm}), i.e.,
\begin{equation}
  \label{eq:m3 x r2:flow}
  \dv{\varphi_+^{\alpha}}{\tau}
  = - g^{\alpha\beta}_{\mathfrak{P}(\R, \mathfrak{B}_3)} \pdv{h_3}{\varphi_+^{\beta}}
  \, ,
  \qquad
  \dv{\varphi_-^{\alpha}}{\tau}
  = - g^{\alpha\beta}_{\mathfrak{P}(\R, \mathfrak{B}_3)} \pdv{h_3}{\varphi_-^{\beta}}
  \, ,
\end{equation}
and they go from one $\tau$-invariant critical point of $h_3$ to another in $\mathfrak{P}(\R, \mathfrak{B}_3)$.
In the 2d gauged sigma model with target $\mathfrak{B}_3$, these flow lines will correspond to worldsheets that have, at $\tau = \pm \infty$, $\mathfrak{B}^{\theta}_3$-strings.\footnote{%
  The $\mathfrak{B}^{\theta}_3$-string can translate in the $\tau$-direction due to its ``center of mass'' motion, and because it is $\tau$-invariant, it is effectively degenerate.
  This reflects the fact that generically, each critical point of $h_3$ is degenerate and does not correspond to a point but a \emph{real line} in $\mathfrak{P}(\R, \mathfrak{B}_3)$.
  Nonetheless, one can perturb $h_3$ via the addition of physically-inconsequential $\hat{\mathcal{Q}}$-exact terms to the SQM action, and collapse the degeneracy such that the critical points really correspond to points in $\mathfrak{P}(\R, \mathfrak{B}_3)$.
  This is equivalent to factoring out the center of mass degree of freedom of the $\mathfrak{B}^{\theta}_3$-string, and fixing it at $\tau = \pm \infty$.
  \label{ft:m3 x r2:center of mass}
}
These strings shall be denoted as $\gamma_{\pm}(t, \theta, \mathfrak{B}_3)$, and are defined by \eqref{eq:m3 x r2:soliton eqn} with finite-energy \emph{real} gauge fields $A_t, A_{\tau} \rightarrow 0$, i.e.,
\begin{equation}
  \label{eq:m3 x r2:soliton eqn:no gauge}
  e^{j\theta} \sin(\theta) \dv{\widetilde{\mathscr{A}}^a}{t}
  + j \dv{\widetilde{\mathscr{B}}^a}{t}
  = - q^a(\theta)
  \, ,
  \qquad
  j \dv{\widetilde{\mathscr{A}}^a}{t}
  - e^{j\theta} \sin(\theta) \dv{\widetilde{\mathscr{B}}^a}{t}
  = - p^a(\theta)
  \, .
\end{equation}
Their endpoints $\gamma(\pm \infty, \theta, \mathfrak{B}_3)$ at $t = \pm \infty$ are defined by
\begin{equation}
  \label{eq:m3 x r2:soliton endpoints:no gauge}
  q^a(\theta)
  =
  0
  \, ,
  \qquad
  p^a(\theta)
  = 0
  \, ,
\end{equation}
which is simply \eqref{eq:m3 x r2:soliton eqn:no gauge} with $d_t \widetilde{\mathscr{A}}^a = 0 = d_t \widetilde{\mathscr{B}}^a$.

Note that the gradient flow equations which govern the flow lines are actually the BPS equations of the 1d SQM.
This means that the worldsheets which the flow lines will correspond to are governed by the BPS equations of the equivalent 2d gauged sigma model with target $\mathfrak{B}_3$.
They are defined by setting to zero the expression within the squared terms in \eqref{eq:m3 x r2:action:2d sigma}, i.e., (via \eqref{eq:m3 x r2:action:2d sigma:components})
\begin{equation}
  \label{eq:m3 x r2:worldsheet eqn}
  \begin{aligned}
    \Dv{}{\tau} \left(
    e^{j\theta} \cos(\theta) \widetilde{\mathscr{A}}^a
    + \widetilde{\mathscr{B}}^a
    \right)
    + \Dv{}{t} \left(
    e^{j\theta} \sin(\theta) \widetilde{\mathscr{A}}^a
    + j \widetilde{\mathscr{B}}^a
    \right)
    &= - j e^{j\theta} \widetilde{\mathscr{F}}^a
      \, ,
    \\
    \Dv{}{\tau} \left(
    \widetilde{\mathscr{A}}^a
    - e^{j\theta} \cos(\theta) \widetilde{\mathscr{B}}^a
    \right)
    + \Dv{}{t} \left(
    j \widetilde{\mathscr{A}}^a
    - e^{j\theta} \sin(\theta) \widetilde{\mathscr{B}}^a
    \right)
    + e^{j\theta} F_{t\tau}
    &= j e^{j\theta} (\widetilde{\mathscr{D}} \widetilde{\mathscr{B}})^a
      \, .
  \end{aligned}
\end{equation}
Such worldsheets which correspond to the classical trajectories of 2d gauged sigma models were coined as BPS worldsheets in \cite[$\S$9.3]{er-2023-topol-n}; we shall do the same here.

In short, the BPS worldsheets of the 2d gauged sigma model with target $\mathfrak{B}_3$ will be defined by \eqref{eq:m3 x r2:worldsheet eqn}.

\subtitle{BPS Worldsheets with Boundaries Corresponding to $\theta$-deformed $G_{\HH}$-BF Configurations on $M_3$}

The boundaries of the BPS worldsheets are traced out by the endpoints of the $\mathfrak{B}^{\theta}_3$-strings as they propagate in $\tau$.
As we have seen at the end of \autoref{sec:m3 x r2:gh-bf}, at $\theta = 0$, these endpoints correspond to $G_{\HH}$-BF configurations on $M_3$.
If there are `$v$' such configurations $\{\mathfrak{C}^1_{\text{BF}_{\HH}}(0), \mathfrak{C}^2_{\text{BF}_{\HH}}(0), \dots, \mathfrak{C}^v_{\text{BF}_{\HH}}(0)\}$, we can further specify the undeformed $\mathfrak{B}^0_3$-strings at $\tau = \pm \infty$ as $\gamma^{IJ}_{\pm}(t, 0, \mathfrak{B}_3)$, where $I, J \in \{1, \dots, v\}$ indicates that its left and right endpoints, given by $\gamma^I(- \infty, 0, \mathfrak{B}_3)$ and $\gamma^J(+ \infty, 0, \mathfrak{B}_3)$, would correspond to the configurations $\mathfrak{C}^I_{\text{BF}_{\HH}}(0)$ and $\mathfrak{C}^J_{\text{BF}_{\HH}}(0)$, respectively.
As the physical theory is symmetric under a variation of $\theta$, this would be true at any value of $\theta$.
In other words, we can also further specify any $\mathfrak{B}^{\theta}_3$-string at $\tau = \pm \infty$ as $\gamma^{IJ}_{\pm}(t, \theta, \mathfrak{B}_3)$, where its left and right endpoints, given by $\gamma^I(- \infty, \theta, \mathfrak{B}_3)$ and $\gamma^J(+ \infty, \theta, \mathfrak{B}_3)$, would correspond to the configurations $\mathfrak{C}^I_{\text{BF}_{\HH}}(\theta)$ and $\mathfrak{C}^J_{\text{BF}_{\HH}}(\theta)$, respectively, with the $\mathfrak{C}^{*}_{\text{BF}_{\HH}}(\theta)$'s being the $v$ number of $\theta$-deformed $G_{\HH}$-BF configurations on $M_3$.

\begin{figure}
  \centering
  \begin{tikzpicture}
    \coordinate (lt) at (0,4) {};
    \coordinate (rt) at (4,4) {}
    edge node[pos=0.5, above] {$\gamma^{IJ}_+(t, \theta, \mathfrak{B}_3)$}
    (lt) {};
    \coordinate (lb) at (0,0) {}
    edge node[pos=0.5, left] {$\mathfrak{C}^I_{\text{BF}_{\HH}}(\theta)$}
    (lt) {};
    \coordinate (rb) at (4,0) {}
    edge node[pos=0.5, below] {$\gamma^{IJ}_-(t, \theta, \mathfrak{B}_3)$}
    (lb) {}
    edge node[pos=0.5, right] {$\mathfrak{C}^J_{\text{BF}_{\HH}}(\theta)$}
    (rt) {};
    \draw (lb) -- (lt);
    \draw (rb) -- (rt);
    \coordinate (co) at (5,0);
    \coordinate (cx) at (5.5,0);
    \node at (cx) [right=2pt of cx] {$t$};
    \coordinate (cy) at (5,0.5);
    \node at (cy) [above=2pt of cy] {$\tau$};
    \draw[->] (co) -- (cx);
    \draw[->] (co) -- (cy);
  \end{tikzpicture}
  \caption[]{BPS worldsheet with strings $\gamma^{IJ}_\pm(t, \theta, \mathfrak{B}_3)$ and boundaries corresponding to $\mathfrak{C}^I_{\text{BF}_{\HH}}(\theta)$ and $\mathfrak{C}^J_{\text{BF}_{\HH}}(\theta)$.
  }
  \label{fig:m3 x r2:bps worldsheet}
\end{figure}
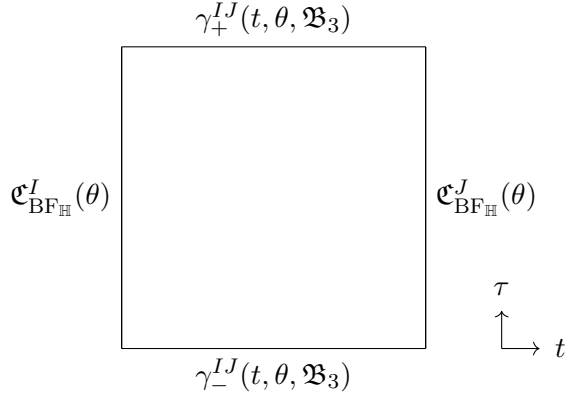

Since the $\mathfrak{C}^{*}_{\text{BF}_{\HH}}(\theta)$'s are $\tau$-independent and therefore, have the same values for all $\tau$, we have BPS worldsheets of the kind shown in \autoref{fig:m3 x r2:bps worldsheet}.

\subtitle{The 2d Model on $\R^2$ and an Open String Theory in $\mathfrak{B}_3$}

Hence, once can understand the 2d gauged sigma model on $\R^2$ with target $\mathfrak{B}_3$ to define an open string theory in $\mathfrak{B}_3$ with \emph{effective} worldsheet and boundaries shown in \autoref{fig:m3 x r2:bps worldsheet}, where $\tau$ and $t$ are the temporal and spatial directions, respectively.

\subsection{String Theory, the GM Partition Function, and an Orlov-type \texorpdfstring{$A_{\infty}$}{A-infinity}-1-category Categorifying the Holomorphic \texorpdfstring{$G_\HH$}{GH}-flat Floer Homology of \texorpdfstring{$M_3$}{M3}}
\label{sec:m3 x r2:orlov}

\subtitle{The 2d Model as a 2d Gauged B-twisted LG Model}

Notice that we can also express the action of the 2d gauged sigma model with target $\mathfrak{B}_3$ in \eqref{eq:m3 x r2:action:2d sigma} as
\begin{equation}
  \label{eq:m3 x r2:action:2d lg}
  \begin{aligned}
    &S_{\text{2d-LG}, \mathfrak{B}_3}
    \\
    & = \frac{1}{e^2} \int_{\R^2} dt d\tau \, \bigg(
      \left| \Dv{}{\tau} \left(
      e^{j\theta} \cos(\theta) \widetilde{\mathscr{A}}^a
      + \widetilde{\mathscr{B}}^a
      \right)
      + \Dv{}{t} \left(
      e^{j\theta} \sin(\theta) \widetilde{\mathscr{A}}^a
      + j \widetilde{\mathscr{B}}^a
      \right)
      + j e^{j\theta} \widetilde{\mathscr{F}}^a
      \right|^2
    \\
    & \qquad \qquad
      + \left| \Dv{}{\tau} \left(
      \widetilde{\mathscr{A}}^a
      - e^{j\theta} \cos(\theta) \widetilde{\mathscr{B}}^a
      \right)
      + \Dv{}{t} \left(
      j \widetilde{\mathscr{A}}^a
      - e^{j\theta} \sin(\theta) \widetilde{\mathscr{B}}^a
      \right)
      + e^{j\theta} F_{t\tau}
      - j e^{j\theta} (\widetilde{\mathscr{D}} \widetilde{\mathscr{B}})^a
      \right|^2
      + \dots
      \bigg)
    \\
    & = \frac{1}{e^2} \int_{\R^2} dt d\tau \, \bigg(
      \left| \Dv{}{\tau} \left(
      e^{j\theta} \cos(\theta) \widetilde{\mathscr{A}}^a
      + \widetilde{\mathscr{B}}^a
      \right)
      + \Dv{}{t} \left(
      e^{j\theta} \sin(\theta) \widetilde{\mathscr{A}}^a
      + j \widetilde{\mathscr{B}}^a
      \right)
      + j e^{j\theta} g^{ab}_{\mathfrak{B}_3} \pdv{V_3}{\widetilde{\mathscr{B}}^b}
      \right|^2
    \\
    & \qquad \qquad
      + \left| \Dv{}{\tau} \left(
      \widetilde{\mathscr{A}}^a
      - e^{j\theta} \cos(\theta) \widetilde{\mathscr{B}}^a
      \right)
      + \Dv{}{t} \left(
      j \widetilde{\mathscr{A}}^a
      - e^{j\theta} \sin(\theta) \widetilde{\mathscr{B}}^a
      \right)
      + e^{j\theta} F_{t\tau}
      + j e^{j\theta} g^{ab}_{\mathfrak{B}_3} \pdv{V_3}{\widetilde{\mathscr{A}}^b}
      \right|^2
      + \dots
      \bigg)
      \, ,
  \end{aligned}
\end{equation}
where $V_3(\widetilde{\mathscr{A}}, \widetilde{\mathscr{B}})$ is given in \eqref{eq:m3 x r:morse functional}.

Applying specific values of $\theta$ (in particular, applying $\theta = 0$, $\pi/2$, $\pi$, and $3\pi/2$) to the conditions within the squared terms of \eqref{eq:m3 x r2:action:2d lg} whilst noting that the physical theory ought to be symmetric under a variation of $\theta$, we find that \eqref{eq:m3 x r2:action:2d lg} can also be equivalently expressed as
\begin{equation}
  \label{eq:m3 x r2:action:2d lg:b-twist}
  S_{\text{2d-LG}, \mathfrak{B}_3}
  = \frac{1}{e^2} \int_{\R^2} \bigg(
  \left| D \widetilde{\mathscr{A}}^a \right|^2
  + \left| D \widetilde{\mathscr{B}}^a \right|^2
  + \left| j F - g^{ab}_{\mathfrak{B}_3} \pdv{V_3}{\widetilde{\mathscr{A}}^b} \right|^2
  + \left| g^{ab}_{\mathfrak{B}_3} \pdv{V_3}{\widetilde{\mathscr{B}}^b} \right|^2
  + \dots
  \bigg)
  \, ,
\end{equation}
where $D$ is the covariant derivative in the real 2d gauge fields, and $F$ is field strength of the real 2d gauge fields.
In other words, the 2d gauged sigma model with target $\mathfrak{B}_3$ can also be interpreted as a 2d gauged \emph{B-twisted} Landau-Ginzburg (LG) model in $\mathfrak{B}_3$ with holomorphic superpotential $V_3(\widetilde{\mathscr{A}}, \widetilde{\mathscr{B}})$.\footnote{%
  Note that the conditions that minimize the action \eqref{eq:m3 x r2:action:2d lg:b-twist} correspond to configurations that set to zero the expression within the squared terms  therein.
  These are \emph{gauge-theoretic generalizations} of the conditions that minimize the action of a 2d B-twisted LG model with holomorphic superpotential $V_3(\widetilde{\mathscr{A}}, \widetilde{\mathscr{B}})$, i.e., $d \widetilde{\mathscr{A}}^a = 0 = d \widetilde{\mathscr{B}}^a$ and $\partial V_3 / \partial \widetilde{\mathscr{A}}^a = 0 = \partial V_3 / \partial \widetilde{\mathscr{B}}^a$ \cite{vafa-1991-topol-landau}.
  \label{ft:m3 x r2:gauge-theoretic generalization of 2d B-LG}
}

By setting $d_{\tau} \widetilde{\mathscr{A}}^a = 0 = d_{\tau} \widetilde{\mathscr{B}}^a$ and $A_t, A_{\tau} \rightarrow 0$ in the expression within the squared terms in \eqref{eq:m3 x r2:action:2d lg}, we can read off the LG $\mathfrak{B}_3^{\theta}$-string equations corresponding to $\gamma^{IJ}_{\pm}(t, \theta, \mathfrak{B}_3)$ (that re-expresses \eqref{eq:m3 x r2:soliton eqn:no gauge}) as
\begin{equation}
  \label{eq:m3 x r2:lg:soliton eqn}
  e^{j\theta} \sin(\theta) \dv{\widetilde{\mathscr{A}}^a}{t}
  + j \dv{\widetilde{\mathscr{B}}^a}{t}
  = - j e^{j\theta} g^{ab}_{\mathfrak{B}_3} \pdv{V_3}{\widetilde{\mathscr{B}}^b}
  \, ,
  \quad
  j \dv{\widetilde{\mathscr{A}}^a}{t}
  - e^{j\theta} \sin(\theta) \dv{\widetilde{\mathscr{B}}^a}{t}
  = - j e^{j\theta} g^{ab}_{\mathfrak{B}_3} \pdv{V_3}{\widetilde{\mathscr{A}}^b}
  \, .
\end{equation}

By setting $d_t \widetilde{\mathscr{A}}^a = 0 = d_t \widetilde{\mathscr{B}}^a$ in \eqref{eq:m3 x r2:lg:soliton eqn}, we get the LG $\mathfrak{B}_3^{\theta}$-string endpoint equations corresponding to $\gamma^{IJ}_{\pm}(\pm \infty, \theta, \mathfrak{B}_3)$ (that re-expresses \eqref{eq:m3 x r2:soliton endpoints:no gauge}) as
\begin{equation}
  \label{eq:m3 x r2:lg:soliton endpoints}
  j e^{j\theta} g^{ab}_{\mathfrak{B}_3} \pdv{V_3}{\widetilde{\mathscr{B}}^b}
  = 0
  \, ,
  \qquad
  j e^{j\theta} g^{ab}_{\mathfrak{B}_3} \pdv{V_3}{\widetilde{\mathscr{A}}^b}
  = 0
  \, .
\end{equation}

Recall from the end of \autoref{sec:m3 x r2:gh-bf} that we are only considering $M_3$ and $G$ such that the endpoints $\gamma^{IJ}(\pm \infty, \theta, \mathfrak{B}_3)$ are isolated and non-degenerate.
Therefore, from their definition in \eqref{eq:m3 x r2:lg:soliton endpoints} as critical points of $V_3(\widetilde{\mathscr{A}}, \widetilde{\mathscr{B}})$, we conclude that $V_3(\widetilde{\mathscr{A}}, \widetilde{\mathscr{B}})$ is a (holomorphic) Morse function in $\mathfrak{B}_3$.

\subtitle{The Gauged LG Model as an LG SQM}

Last but not least, after suitable rescalings, we can recast \eqref{eq:m3 x r2:action:2d lg} as a 1d LG SQM (that re-expresses \eqref{eq:m3 x r2:action:1d sqm}), where its action will be given by\footnote{%
  Just as in \autoref{ft:m3 x r2:integrate out gauge field to sqm}, we have integrated out $A_{\tau}$ in the resulting SQM.
  \label{ft:m3 x r2:2d-lg:integrate out gauge field to sqm}
}
\begin{equation}
  \label{eq:m3 x r2:action:1d lg sqm}
  \begin{aligned}
    S_{\text{2d-LG SQM}, \mathfrak{P}(\R_t, \mathfrak{B}_3)}
    = \frac{1}{e^2} \int d\tau \left(
    \left| \dv{\varphi_+^{\alpha}}{\tau} + g^{\alpha\beta}_{\mathfrak{P}(\R, \mathfrak{B}_3)}\pdv{H_3}{\varphi_+^{\beta}} \right|^2
    + \left| \dv{\varphi_-^{\alpha}}{\tau} + g^{\alpha\beta}_{\mathfrak{P}(\R_t, \mathfrak{B}_3)}\pdv{H_3}{\varphi_-^{\beta}} \right|^2
    \right)^2
    \, ,
  \end{aligned}
\end{equation}
where $H_3(\varphi_+, \varphi_-)$ is the \emph{real-valued} potential in $\mathfrak{P}(\R, \mathfrak{B}_3)$, with $\varphi_+$ and $\varphi_-$ defined as in \autoref{sec:m3 x r2:gm}, and the subscript ``2d-LG SQM, $\mathfrak{P}(\R, \mathfrak{B}_3)$'' is to specify that it is a 1d SQM in $\mathfrak{P}(\R, \mathfrak{B}_3)$ obtained from the equivalent 2d LG model.
We will also refer to this \emph{1d} LG SQM as ``2d-LG SQM'' in the rest of this section.

The 2d-LG SQM will localize onto configurations that \emph{simultaneously} set to zero the LHS and RHS of the expression within the squared terms in \eqref{eq:m3 x r2:action:1d lg sqm}.
In other words, it will localize onto $\tau$-invariant critical points of $H_3(\varphi_+, \varphi_-)$ that will correspond to the LG $\mathfrak{B}_3^{\theta}$-strings defined by \eqref{eq:m3 x r2:lg:soliton eqn}.
For our choice of $M_3$ and $G$, just like their endpoints, the LG $\mathfrak{B}_3^{\theta}$-strings will be isolated and non-degenerate.
Thus, $H_3(\varphi_+, \varphi_-)$ can be regarded as a real-valued Morse functional in $\mathfrak{P}(\R, \mathfrak{B}_3)$.

\subtitle{String Theory from the 2d Gauged LG Model}

Just like the 2d gauged sigma model, the equivalent 2d gauged B-twisted LG model will define an open string theory in $\mathfrak{B}_3$ with effective worldsheets and boundaries shown in \autoref{fig:m3 x r2:bps worldsheet}, where $\tau$ and $t$ are the temporal and spatial directions, respectively.

The dynamics of this open string theory in $\mathfrak{B}_3$ will be governed by the BPS worldsheet equations determined by setting to zero the expression within the squared terms in \eqref{eq:m3 x r2:action:2d lg}, where $(\widetilde{\mathscr{A}}^a, \widetilde{\mathscr{B}}^a)$ are scalars on the worldsheet corresponding to the holomorphic coordinates of $\mathfrak{B}_3$.
At an arbitrary instant in time whence $d_{\tau} \widetilde{\mathscr{A}}^a = 0 = d_{\tau} \widetilde{\mathscr{B}}^a$ and $d_{\tau} A_t = 0$ therein, the dynamics of $(\widetilde{\mathscr{A}}^a, \widetilde{\mathscr{B}}^a)$ along $t$ will be governed by the string equations
\begin{equation}
  \label{eq:m3 x r2:soliton string}
  \begin{aligned}
    e^{j\theta} \sin(\theta) \dv{\widetilde{\mathscr{A}}^a}{t}
    + j \dv{\widetilde{\mathscr{B}}^a}{t}
    =
    & - [A_{\tau}, e^{j\theta} \cos(\theta) \widetilde{\mathscr{A}}^a + \widetilde{\mathscr{B}^a}]
      - [A_t, e^{j\theta} \sin(\theta) \widetilde{\mathscr{A}}^a + j \widetilde{\mathscr{B}}^a ]
    \\
    & - j e^{j\theta} g^{ab}_{\mathfrak{B}_3} \pdv{V_3}{\widetilde{\mathscr{B}}^b}
      \, ,
    \\
    j \dv{\widetilde{\mathscr{A}}^a}{t}
    - e^{j\theta} \sin(\theta) \dv{\widetilde{\mathscr{B}}^a}{t}
    + e^{j\theta} \dv{A_{\tau}}{t}
    =
    & - [A_{\tau}, \widetilde{\mathscr{A}}^a - e^{j\theta} \cos(\theta) \widetilde{\mathscr{B}}^a]
      - [A_t, j \widetilde{\mathscr{A}}^a - e^{j\theta} \sin(\theta) \widetilde{\mathscr{B}}^a + A_{\tau}]
    \\
    & - j e^{j\theta} g^{ab}_{\mathfrak{B}_3} \pdv{V_3}{\widetilde{\mathscr{A}}^b}
      \, .
  \end{aligned}
\end{equation}

\subtitle{Morphisms from $\mathfrak{C}^I_{G_{\HH}\text{-BF}}(\theta)$ to $\mathfrak{C}^I_{G_{\HH}\text{-BF}}(\theta)$ as Ext-groups}

It will now be useful to describe the LG $\mathfrak{B}_3^{\theta}$-string endpoint solutions satisfying \eqref{eq:m3 x r2:lg:soliton endpoints} as \emph{singular} fibers of $V_3$ over 0, which we shall henceforth denote as $V_{3, \text{sg}}^{-1}(0)$.\footnote{%
  The fiber space $V_3^{-1}$ over the $V_3$-plane is $\mathfrak{B}_3$, and the subset of singular fibers is a submanifold of $\mathfrak{B}_3$ containing the critical points of $V_3$.
  In our case where $V_3$ is given by \eqref{eq:m3 x r:morse functional}, its critical points will correspond to the value 0 in the $V_3$-plane, i.e., we will only have singular fibers over 0.
  \label{ft:m3 x r2:explain why singular fibers are over 0}
}
In particular, to an LG $\mathfrak{B}_3^{\theta}$-string $\gamma^{IJ}(t, \theta, \mathfrak{B}_3)$ (illustrated in \autoref{fig:m3 x r2:bps worldsheet}), its left endpoint $\gamma^I(- \infty, \theta, \mathfrak{B}_3) \eqcolon \mathcal{T}^I_{V_3}(\theta) \in V_{3, \text{sg}}^{-1}(0)$ can be described as a singular fiber which is also a zero-dimensional submanifold of $\mathfrak{B}_3$, and it corresponds to a $G_{\HH}$-BF configuration $\mathfrak{C}_{\text{BF}_{\HH}}^I(\theta)$ on $M_3$.
Likewise, its right endpoint $\gamma^J(+ \infty, \theta, \mathfrak{B}_3) \eqcolon \mathcal{T}^J_{V_3}(\theta) \in V_{3, \text{sg}}^{-1}(0)$ can also be described as a singular fiber which is a zero-dimensional submanifold of $\mathfrak{B}_3$, and it corresponds to a $G_{\HH}$-BF configuration $\mathfrak{C}_{\text{BF}_{\HH}}^J(\theta)$ on $M_3$.

Let us denote by $S^{IJ}_{\text{BF}_{\HH}}$ the collection of all such strings whose left and right endpoints correspond to $\mathfrak{C}^I_{\text{BF}_{\HH}}(\theta)$ and $\mathfrak{C}^J_{\text{BF}_{\HH}}(\theta)$.
Each LG $\mathfrak{B}_3^{\theta}$-string pair $\gamma^{IJ}_{\pm}(t, \theta, \mathfrak{B}_3)$ will therefore correspond to a pair of strings $p^{IJ}_{\text{BF}_{\HH}, \pm}(\theta) \in S^{IJ}_{\text{BF}_{\HH}}$.

As in \autoref{sec:m x r}, the 2d-LG SQM in $\mathfrak{P}(\R, \mathfrak{B}_3)$ with action \eqref{eq:m3 x r2:action:1d lg sqm} will physically realize a Floer homology that we shall name a $\mathfrak{B}_3$-2d-LG Floer homology.
The chains of the $\mathfrak{B}_3$-2d-LG Floer complex will be generated by LG $\mathfrak{B}_3^{\theta}$-strings which we can identify with $p^{**}_{\text{BF}_{\HH}, \pm}(\theta)$, and the $\mathfrak{B}_3$-2d-LG Floer differential will be realized by the flow lines governed by the gradient flow equations satisfied by $\tau$-varying configurations which set the expression within the squared terms in \eqref{eq:m3 x r2:action:1d lg sqm} to zero.
In particular, the SQM partition function of the 2d-LG SQM in $\mathfrak{P}(\R, \mathfrak{B}_3)$ will be given by\footnote{%
  The `$\theta$' label is omitted in the LHS of the following expression, as the physical theory is actually equivalent for all values of $\theta$.
  We will henceforth omit the `$\theta$' label in equations wherever relevant.
  \label{ft:m3 x r2:omission of theta}
}
\begin{equation}
  \label{eq:m3 x r2:lg:partition function}
  \mathcal{Z}_{\text{2d-LG SQM,}\mathfrak{P}(\R, \mathfrak{B}_3)}(G)
  = \sum_{I \neq J}^v
  \; \sum_{p^{IJ}_{\text{BF}_{\HH}, \pm}(\theta) \in S^{IJ}_{\text{BF}_{\HH}}}
  \text{HF}^G_{d_v} \left( p^{IJ}_{\text{BF}_{\HH}, \pm}(\theta) \right)
  \, ,
\end{equation}
where the contribution $\text{HF}^G_{d_k}(p^{IJ}_{\text{BF}_{\HH}, \pm}(\theta))$ can be identified with a homology class in a $\mathfrak{B}_3$-2d-LG Floer homology generated by LG $\mathfrak{B}_3^{\theta}$-strings whose endpoints correspond to singular fibers $V_{3, \text{sg}}^{-1}(0)$.
The degree of each chain is $d_v$, and is counted by the number of outgoing flow lines from the fixed critical points of $H_3(\widetilde{\mathscr{A}}, \widetilde{\mathscr{B}})$ in $\mathfrak{P}(\R, \mathfrak{B}_3)$ that can be identified as $p^{IJ}_{\text{BF}_{\HH}, \pm}(\theta)$.

Note that $p^{IJ}_{\text{BF}_{\HH}, \pm}(\theta)$, which corresponds to an LG $\mathfrak{B}_3^{\theta}$-string,  will be defined by (i) \eqref{eq:m3 x r2:lg:soliton eqn} with (ii) endpoints \eqref{eq:m3 x r2:lg:soliton endpoints}.
In other words, we can write
\begin{equation}
  \label{eq:m3 x r2:ext as hf}
  p^{IJ}_{\text{BF}_{\HH}, \pm}(\theta)
  \Longleftrightarrow
  \text{Ext} \left( \mathcal{T}^I_{V_3}(\theta), \mathcal{T}^J_{V_3}(\theta) \right)_{\pm}
  \, ,
\end{equation}
where $\text{Ext} (\mathcal{T}^I_{V_3}(\theta), \mathcal{T}^J_{V_3}(\theta))_{\pm}$ represents an LG $\mathfrak{B}_3^{\theta}$-string whose start and endpoints are described as singular fibers $\mathcal{T}^I_{V_3}(\theta)$ and $\mathcal{T}^J_{V_3}(\theta)$, respectively.

Equivalently, it is a $\mathfrak{B}_3^{\theta}$-string defined by (i) \eqref{eq:m3 x r2:soliton eqn} and $A_t, A_{\tau} \rightarrow 0$ with (ii) endpoints \eqref{eq:m3 x r2:soliton:endpoints}.
Note that a string can be regarded as a morphism between its endpoints.
Specifically, the $\gamma^{IJ}_{\pm}(t, \theta, \mathfrak{B}_3)$ $\mathfrak{B}_3^{\theta}$-string pair can be regarded as a pair of morphisms $\text{Hom}(\mathfrak{C}^I_{\text{BF}_{\HH}}(\theta), \mathfrak{C}^J_{\text{BF}_{\HH}}(\theta))_{\pm}$ from $\mathfrak{C}^I_{\text{BF}_{\HH}}(\theta)$ to $\mathfrak{C}^J_{\text{BF}_{\HH}}(\theta)$.
Thus, we have the following one-to-one identification
\begin{equation}
  \label{eq:m3 x r2:morphism as ext}
  \saveboxed{eq:m3 x r2:morphism as ext}{
    \text{Hom} \left( \mathfrak{C}^I_{\text{BF}_{\HH}}, \mathfrak{C}^J_{\text{BF}_{\HH}} \right)_{\pm}
    \Longleftrightarrow
    \text{Ext} \left( \mathcal{T}^I_{V_3}, \mathcal{T}^J_{V_3} \right)_{\pm}
  }
\end{equation}

Note that the RHS of \eqref{eq:m3 x r2:morphism as ext} can be regarded as morphisms in an Orlov-type triangulated category of singularities of $V_3$ \cite{orlov-2004-trian-categ}, i.e., we can write
\begin{equation}
  \label{eq:m3 x r2:morphism}
  \text{Hom} \left( \mathfrak{C}^I_{\text{BF}_{\HH}}, \mathfrak{C}^J_{\text{BF}_{\HH}} \right)
  \Longleftrightarrow
  \text{Hom}_{\mathbf{D}_{\text{sg}}(V_3^{-1}(0))} \left( \mathcal{T}^I_{V_3}(\theta), \mathcal{T}^J_{V_3}(\theta) \right)
  \, .
\end{equation}

\subtitle{The Normalized GM Partition Function, LG $\mathfrak{B}_3^{\theta}$-string Scattering, and Maps of an $A_{\infty}$-structure}

\begin{figure}
  \centering
  \begin{tikzpicture}[declare function={
      lenX(\legLength,\leftAngle,\rightAngle)
      = \legLength * cos((\leftAngle + \rightAngle)/2);
      lenY(\legLength,\leftAngle,\rightAngle)
      = \legLength * sin((\leftAngle + \rightAngle)/2);
      lenLX(\segAngle,\leftAngle,\rightAngle)
      = -2 * tan(\segAngle/2) * sin((\leftAngle + \rightAngle)/2);
      lenLY(\segAngle,\leftAngle,\rightAngle)
      = 2 * tan(\segAngle/2) * cos((\leftAngle + \rightAngle)/2);
    }]
    \def \NumSeg {8}                                
    \def \Rad {1}                                   
    \def \Leg {1.5}                                 
    \def \SegAngle {180/\NumSeg}                    
    \def \TpRtAngle {{(\NumSeg - 1)*\SegAngle/2}}   
    \def \TpLtAngle {{(\NumSeg + 1)*\SegAngle/2}}   
    \def \BaLtAngle {{(\NumSeg + 1)*\SegAngle}}     
    \def \BaRtAngle {{(\NumSeg + 2)*\SegAngle}}     
    \def \BbLtAngle {{(\NumSeg + 3)*\SegAngle}}     
    \def \BbRtAngle {{(\NumSeg + 4)*\SegAngle}}     
    \def \BcLtAngle {{(2 * \NumSeg - 2)*\SegAngle}} 
    \def \BcRtAngle {(2 * \NumSeg - 1)*\SegAngle}   
    \draw ([shift=({\BcRtAngle-360}:\Rad)]3,3) arc ({\BcRtAngle-360}:\TpRtAngle:\Rad);
    \draw ([shift=(\TpLtAngle:\Rad)]3,3) arc (\TpLtAngle:\BaLtAngle:\Rad);
    \draw ([shift=(\BaRtAngle:\Rad)]3,3) arc (\BaRtAngle:\BbLtAngle:\Rad);
    \draw[dashed] ([shift=(\BbRtAngle:\Rad)]3,3) arc (\BbRtAngle:\BcLtAngle:\Rad);
    \draw ([shift=(\TpRtAngle:\Rad)]3cm,3cm)
    -- node[right] {\footnotesize $\mathfrak{C}^{I_{n_k + 1}}_{\text{BF}_{\HH}}$}
    ++(
    {lenX(\Leg,\TpLtAngle,\TpRtAngle)},
    {lenY(\Leg,\TpLtAngle,\TpRtAngle)}
    )
    -- node[above] {$+$}
    ++(
    {lenLX(\SegAngle,\TpLtAngle,\TpRtAngle)},
    {lenLY(\SegAngle,\TpLtAngle,\TpRtAngle)}
    )
    -- node[left] {\footnotesize $\mathfrak{C}^{I_1}_{\text{BF}_{\HH}}$}
    ++(
    -{lenX(\Leg,\TpLtAngle,\TpRtAngle)},
    -{lenY(\Leg,\TpLtAngle,\TpRtAngle)}
    );
    \draw ([shift=(\BaLtAngle:\Rad)]3,3)
    -- node[above=3pt] {\footnotesize $\mathfrak{C}^{I_1}_{\text{BF}_{\HH}}$}
    ++(
    {lenX(\Leg,\BaLtAngle,\BaRtAngle)},
    {lenY(\Leg,\BaLtAngle,\BaRtAngle)}
    )
    -- node[left] {$-$}
    ++(
    {lenLX(\SegAngle,\BaLtAngle,\BaRtAngle)},
    {lenLY(\SegAngle,\BaLtAngle,\BaRtAngle)}
    )
    -- node[below right={-6pt and -4pt}] {\footnotesize $\mathfrak{C}^{I_2}_{\text{BF}_{\HH}}$}
    ++(
    -{lenX(\Leg,\BaLtAngle,\BaRtAngle)},
    -{lenY(\Leg,\BaLtAngle,\BaRtAngle)}
    );
    \draw ([shift=(\BbLtAngle:\Rad)]3,3)
    -- node[near end, left=-3pt] {\footnotesize $\mathfrak{C}^{I_2}_{\text{BF}_{\HH}}$}
    ++(
    {lenX(\Leg,\BbLtAngle,\BbRtAngle)},
    {lenY(\Leg,\BbLtAngle,\BbRtAngle)}
    )
    -- node[below] {$-$}
    ++(
    {lenLX(\SegAngle,\BbLtAngle,\BbRtAngle)},
    {lenLY(\SegAngle,\BbLtAngle,\BbRtAngle)}
    )
    -- node[near start, right=-2pt] {\footnotesize $\mathfrak{C}^{I_3}_{\text{BF}_{\HH}}$}
    ++(
    -{lenX(\Leg,\BbLtAngle,\BbRtAngle)},
    -{lenY(\Leg,\BbLtAngle,\BbRtAngle)}
    );
    \draw ([shift=(\BcLtAngle:\Rad)]3,3)
    -- node[near end, below] {\footnotesize $\mathfrak{C}^{I_{n_k}}_{\text{BF}_{\HH}}$}
    ++(
    {lenX(\Leg,\BcLtAngle,\BcRtAngle)},
    {lenY(\Leg,\BcLtAngle,\BcRtAngle)}
    )
    -- node[right] {$-$}
    ++(
    {lenLX(\SegAngle,\BcLtAngle,\BcRtAngle)},
    {lenLY(\SegAngle,\BcLtAngle,\BcRtAngle)}
    )
    -- node[near start, above right={-8pt and 0pt}] {\footnotesize $\mathfrak{C}^{I_{n_k + 1}}_{\text{BF}_{\HH}}$}
    ++(
    -{lenX(\Leg,\BcLtAngle,\BcRtAngle)},
    -{lenY(\Leg,\BcLtAngle,\BcRtAngle)}
    );
  \end{tikzpicture}
  \caption[]{Tree-level scattering BPS worldsheet of incoming ($-$) and outgoing ($+$) LG $\mathfrak{B}_3^{\theta}$-strings.}
  \label{fig:m3 x r2:scattering amplitudes}
\end{figure}
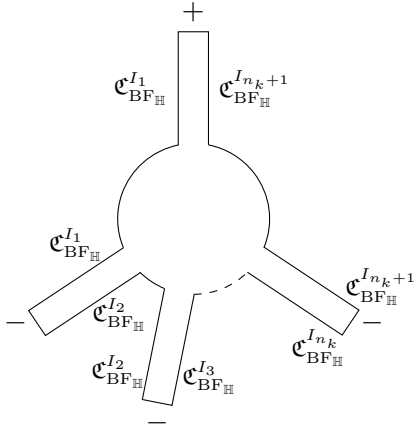

The spectrum of GM theory is given by the $\hat{\mathcal{Q}}$-cohomology of operators.
In particular, its normalized 5d partition function will be a sum over free-field correlation functions of these operators (see~\cite[footnote 37]{er-2023-topol-n}).
As our GM theory is semi-classical, these correlation functions will correspond to tree-level scattering only.
From the equivalent 2d-LG SQM and 2d gauged B-twisted LG model perspective, the $\hat{\mathcal{Q}}$-cohomology will be spanned by LG $\mathfrak{B}_3^{\theta}$-strings defined by \eqref{eq:m3 x r2:lg:soliton eqn}.
In turn, this means that the normalized GM partition function can also be regarded as a sum over tree-level scattering amplitudes of these LG $\mathfrak{B}_3^{\theta}$-strings.
The BPS worldsheet underlying such a tree-level scattering amplitude is shown in~\autoref{fig:m3 x r2:scattering amplitudes}.\footnote{%
  Here, we have exploited the topological and hence conformal invariance of the string theory to replace the outgoing LG $\mathfrak{B}_3^{\theta}$-strings with their vertex operators on the disc, then used their coordinate-independent operator products to reduce them to a single vertex operator, before finally translating it back as a single outgoing LG $\mathfrak{B}_3^{\theta}$-string.
  \label{ft:m3 x r2:reason for single outgoing string}
}

In other words, we can express the normalized GM partition function as
\begin{equation}
  \label{eq:m3 x r2:normalized partition function}
  \widetilde{\mathcal{Z}}_{\text{GM}, M_3 \times \R^2}(G)
  = \sum_{n_v} \mu^{n_v}_{\mathfrak{B}_3}
  \, ,
  \qquad
  n_v \in \{1, 2, \dots, v - 1\}
\end{equation}
where each
\begin{equation}
  \label{eq:m3 x r2:mu-d maps}
  \saveboxed{eq:m3 x r2:mu-d maps}{
    \mu^{n_v}_{\mathfrak{B}_3}:
    \bigotimes_{i = 1}^{n_v}
    \text{Hom}\left(
      \mathfrak{C}^{I_i}_{\text{BF}_{\HH}}, \mathfrak{C}^{I_{i + 1}}_{\text{BF}_{\HH}}
    \right)_-
    \longto
    \text{Hom}\left(
      \mathfrak{C}^{I_1}_{\text{BF}_{\HH}}, \mathfrak{C}^{I_{n_v+1}}_{\text{BF}_{\HH}}
    \right)_+
  }
\end{equation}
is a scattering amplitude of $n_v$ incoming LG $\mathfrak{B}_3^{\theta}$-strings $\text{Hom}(\mathfrak{C}^{*}_{\text{BF}_{\HH}}, \mathfrak{C}^{*}_{\text{BF}_{\HH}})_-$, and a single outgoing LG $\mathfrak{B}_3^{\theta}$-string $\text{Hom}(\mathfrak{C}^{I_1}_{\text{BF}_{\HH}}, \mathfrak{C}^{I_{n_v + 1}}_{\text{BF}_{\HH}})_+$, with left and right boundaries as labeled, whose underlying worldsheet is shown in \autoref{fig:m3 x r2:scattering amplitudes}.
That is, $\mu^{n_v}_{\mathfrak{B}_3}$ counts constant discs with $n_v + 1$ punctures at the boundary that are mapped to $\mathfrak{B}_3$ according to the BPS worldsheet equations (determined by setting to zero the expression within the squared terms in \eqref{eq:m3 x r2:action:2d lg}) that are given by \eqref{eq:m3 x r2:worldsheet eqn}.

In turn, this means that $\mu^{n_v}_{\mathfrak{B}_3}$ counts the moduli of solutions to \eqref{eq:m3 x r2:bps:quaternionified:rotated} (or equivalently \eqref{eq:m3 x r2:bps}) with $n_v + 1$ boundary conditions that can be described as follows.
First, note that we can regard $\R^2$ as the effective worldsheet in \autoref{fig:m3 x r2:scattering amplitudes} that we shall denote as $\Omega$, so $M_5$ can be interpreted as a trivial $M_3$ fibration over $\Omega$.
Then, at the $n_v + 1$ LG $\mathfrak{B}_3^{\theta}$-strings on $\Omega$ where $\tau = \pm \infty$, \eqref{eq:m3 x r2:bps:quaternionified:rotated} will become \eqref{eq:m3 x r2:gm configs} with $A_t, A_{\tau} \rightarrow 0$,
and over the LG $\mathfrak{B}_3^{\theta}$-string boundaries on $\Omega$ where $t = \pm \infty$, \eqref{eq:m3 x r2:bps:quaternionified:rotated} will become \eqref{eq:m3 x r2:gh-bf configs} which defines $\theta$-deformed $G_{\HH}$-BF configurations on $M_3$.

Note that the above description of the map  $\mu^{n_v}_{\mathfrak{B}_3}$ is similar to the definition of the map `$\mu^d$' in~\cite[$\S5$]{haydys-2015-fukay-seidel} that our previous work in~\cite{er-2023-topol-n} physically realizes. Indeed, there is a mirror symmetry and Langlands duality which relates them, as we shall elucidate in \autoref{sec:dualities:cats}.

At any rate, the collection of $\mu^{n_v}_{\mathfrak{B}_3}$ maps defined in \eqref{eq:m3 x r2:mu-d maps} can be regarded as composition maps defining an $A_{\infty}$-structure.

\subtitle{An Orlov-type $A_{\infty}$-1-category of Three-Manifolds}

Altogether, this means that the normalized partition function of GM theory on $M_3 \times \R^2$ as expressed in \eqref{eq:m3 x r2:normalized partition function}, manifests a \emph{novel} gauge-theoretic Orlov-type $A_{\infty}$-1-category whose (i) 1-objects are singular fibers $V_{3, \text{sg}}^{-1}(0)$ corresponding to $\theta$-deformed $G_{\HH}$-BF configurations on $M_3$, thereby categorifying the holomorphic $G_{\HH}$-flat Floer homology of $M_3$, (ii) 1-morphisms are $\text{Ext}$-groups corresponding to LG $\mathfrak{B}_3^{\theta}$-strings via the one-to-one identification \eqref{eq:m3 x r2:morphism as ext}, and (iii) $A_{\infty}$-structure is defined by the $\mu^{n_v}_{\mathfrak{B}_3}$ maps \eqref{eq:m3 x r2:mu-d maps}.

\subsection{An RW \texorpdfstring{$A_{\infty}$}{A-infinity}-2-category of Complex-Lagrangian Branes}
\label{sec:m3 x r2:rw}

Let us now consider the case where $M_3$ can be Heegaard split along a Heegaard surface $C$.
Via \autoref{sec:gm:equivalence} with $M_1 = \R$, we find that this GM theory on $M_3 \times \R^2$ is equivalent to the 3d B-model, i.e., the 3d RW model, on $I \times \R^2$ with target the $\theta$-generalized $\mathcal{M}^G_{\text{H}, \theta}(C, \mathbf{J})$, and with boundaries being complex-Lagrangian branes $\widehat{L}_{*}^{*}(\theta)$ of $\mathcal{M}^G_{\text{H}, \theta}(C, \mathbf{J})$.
Here $\theta$ is the angle of rotation about the origin of the $\R^2$-plane as per our formulation in this section thus far.

\subtitle{An Orlov $A_{\infty}$-1-category of Branes in Path Space}

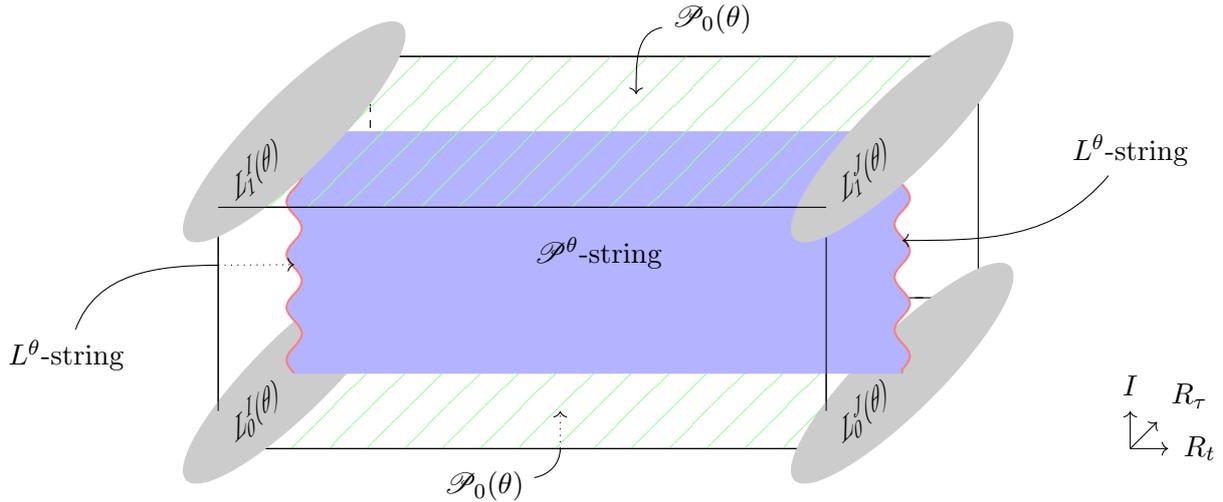
\begin{figure}
  \centering
  \begin{tikzpicture}[%
    auto,%
    relation/.style={scale=1, sloped, anchor=center, align=center, color=black},%
    vertRelation/.style={scale=1, anchor=center, align=center},%
    slanted/.style={rotate=\yAngle, xslant=0.8, scale=0.9},%
    coil/.style={thick,decorate,decoration={coil,aspect=0,segment length=\coilLength, amplitude=\coilAmp}},%
    ]
    \let\side\undefined 
    \newlength{\side}
    \setlength{\side}{2cm}
    \def\xyRatio{4}
    \def\yzRatio{1.6}
    \let\coordLength\undefined
    \newlength{\coordLength}
    \setlength{\coordLength}{0.5cm}
    \def\yAngle{45}
    \let\coilAmp\undefined
    \newlength{\coilAmp}
    \setlength{\coilAmp}{1mm}
    \let\coilLength\undefined
    \newlength{\coilLength}
    \setlength{\coilLength}{7mm}
    \coordinate (tll) at ($(- \xyRatio*\side/2, \yzRatio*\side)$); 
    \coordinate (trl) at ($(tll) + (\side, \side)$);
    \coordinate (tbl) at ($(tll)!0.5!(trl)$);
    \coordinate (brl) at ($(trl) + (0, - \yzRatio*\side)$);
    \coordinate (bll) at ($(tll) + (0, - \yzRatio*\side)$);
    \coordinate (bbl) at ($(bll)!0.5!(brl)$);
    \draw
    (bll) edge (tll)
    (tll) edge (trl)
    (trl) edge[dashed] (brl)
    (brl) edge[dashed] (bll)
    ;
    \coordinate (tlr) at ($(tll) + (\xyRatio*\side, 0)$);
    \coordinate (trr) at ($(tlr) + (\side, \side)$);
    \coordinate (tbr) at ($(tlr)!0.5!(trr)$);
    \coordinate (brr) at ($(trr) + (0, - \yzRatio*\side)$);
    \coordinate (blr) at ($(tlr) + (0, - \yzRatio*\side)$);
    \coordinate (bbr) at ($(blr)!0.5!(brr)$);
    \draw
    (blr) edge (tlr)
    (tlr) edge (trr)
    (trr) edge (brr)
    (brr) edge (blr)
    ;
    \coordinate (tlc) at ($(tll)!0.5!(tlr)$);
    \coordinate (trc) at ($(tlc) + (\side, \side)$);
    \coordinate (tbc) at ($(tlc)!0.5!(trc)$);
    \coordinate (blc) at ($(tlc) + (0, - \yzRatio*\side)$);
    \coordinate (brc) at ($(trc) + (0, - \yzRatio*\side)$);
    \coordinate (bbc) at ($(blc)!0.5!(brc)$);
    \draw
    (bll) edge (blr)
    (tll) edge (tlr)
    (brl) edge[dashed] (brr)
    (trl) edge (trr)
    ;
    \draw[pattern={Lines[angle=45,distance=10pt]},pattern color=green!40]
    (bll) -- (brl) -- (brr) -- (blr) -- (bll);
    \node (P0) at ($(blc) + (- \side/4, - \side/4)$) {$\mathscr{P}_0(\theta)$};
    \draw
    (P0) edge [out=10, in=270] ($(blc) + (\side/4, 0)$)
    ($(blc) + (\side/4, 0)$) edge [->, dotted] ($(bbc) + (- \side/4, - \side/4)$);
    \draw[fill=gray!40, gray!40, rotate around={45:(bbl)}] (bbl) ellipse (2 and 0.5);
    \node[slanted] at ($(bbl) + (- \side/4, - \side/4)$) {$L_0^I(\theta)$};
    \draw[fill=gray!40, gray!40, rotate around={45:(bbr)}] (bbr) ellipse (2 and 0.5);
    \node[slanted] at ($(bbr) + (- \side/4, - \side/4)$) {$L_0^J(\theta)$};
    \draw[fill=blue!30, blue!30]
    (bbl)
    decorate[coil] {-- (tbl)}
    -- (tbr)
    decorate[coil] {-- (bbr)}
    -- (bbl)
    ;
    \node at ($(tbc)!0.5!(bbc)$) {$\mathscr{P}^{\theta}$-string};
    \draw[pattern={Lines[angle=45,distance=10pt]},pattern color=green!40]
    (tll) -- (trl) -- (trr) -- (tlr) -- (tll);
    \node (P1) at ($(trc) + (\side/4, \side/4)$) {$\mathscr{P}_0(\theta)$};
    \draw [->]
    (P1) edge [out=190, in=90] ($(tbc) + (\side/4, \side/4)$);
    \draw (bbl) edge[coil, red!50] (tbl);
    \node (LI-string) at ($(tll)!0.3!(bll) + (- \side, - 0.5*\side)$) {$L^{\theta}$-string};
    \draw
    (LI-string) edge [out=70, in=180] ($(tll)!0.24!(bll)$)
    ($(tll)!0.24!(bll)$) edge [->, dotted] ($(tbl)!0.55!(bbl)$);
    \draw (tbr) edge[coil, red!50] (bbr);
    \node (LJ-string) at ($(trr)!0.7!(brr) + (\side, 0.5*\side)$) {$L^{\theta}$-string};
    \draw [->]
    (LJ-string) edge [out=230, in=0] ($(tbr)!0.45!(bbr)$);
    \draw[fill=gray!40, gray!40, rotate around={45:(tbl)}] (tbl) ellipse (2 and 0.5);
    \node[slanted] at ($(tbl) + (- \side/4, - \side/4)$) {$L_1^I(\theta)$};
    \draw[fill=gray!40, gray!40, rotate around={45:(tbr)}] (tbr) ellipse (2 and 0.5);
    \node[slanted] at ($(tbr) + (- \side/4, - \side/4)$) {$L_1^J(\theta)$};
    \draw
    (tll) edge (tlr)
    ($(blr) + (0, \side/4)$) edge ($(tlr) + (0, - \side/5)$)
    ($(bll) + (0, \side/4)$) edge ($(tll) + (0, - \side/5)$)
    ;
    \coordinate (cco) at ($(blr) + (2*\side, 0)$);
    \coordinate (ccx) at ($(cco) + (\coordLength, 0)$);
    \coordinate (ccy) at ($(cco) + cos(\yAngle)*(\coordLength, 0) + sin(\yAngle)*(0, \coordLength)$);
    \coordinate (ccz) at ($(cco) + (0, \coordLength)$);
    \node at ($(brl) + (- 2*\side, 0)$) {}; 
    \node at (ccx) [right=2pt of ccx] {$R_t$};
    \node at (ccy) [above right=2pt of ccy] {$R_{\tau}$};
    \node at (ccz) [above=2pt of ccz] {$I$};
    \draw[->] (cco) -- (ccx);
    \draw[->] (cco) -- (ccy);
    \draw[->] (cco) -- (ccz);
  \end{tikzpicture}
  \caption{%
    3d RW model on $I \times \R_t \times \R_{\tau}$ physically realizing
    (I) a derived category of coherent sheaves with {%
      (a) 1-objects being branes $\mathscr{P}_{\{0, 1\}}(\theta)$,
      and (b) 1-morphisms being $\mathscr{P}^{\theta}$-strings ending on the 1-objects;
    }%
    as well as (II) an RW 2-category with {%
      (i) 2-objects being complex-Lagrangian branes $\widehat{L}_{\{0, 1\}}^{*}(\theta)$,
      (ii) 2-morphisms being $\mathscr{P}^{\theta}$-strings with vertices being the 2-objects,
      (iii) 1-objects being $\widehat{L}^{\theta}$-strings,
      and (iv) 1-morphisms being $\mathscr{P}^{\theta}$-strings with edges being the 1-objects.
    }%
  }
  \label{fig:m3 x r2:rw cat}
\end{figure}

In the same way that GM theory on $M_3 \times \R^2$ is equivalent to a 2d model on $\R^2$ with target $\mathfrak{B}_3$, the 3d RW model on $I \times \R_t \times \R_{\tau}$ with target $\mathcal{M}^G_{\text{H}, \theta}(C, \mathbf{J})$ is also equivalent to a 2d B-model on $I \times R_{\tau}$ with target the affine path space $\mathfrak{P}(\R_t, \mathcal{M}^G_{\text{H}, \theta}(C, \mathbf{J}))$.
It is well known that such a model would physically realize the derived 1-category $\mathbf{D} \big( \text{coh} (\mathfrak{P} ( \R_t, \mathcal{M}^G_{\text{H}, \theta}(C, \mathbf{J}))) \big)$ of coherent sheaves on $\mathfrak{P} (\R_t, \mathcal{M}^G_{\text{H}, \theta}(C, \mathbf{J}))$.
In particular, the 1-objects are (complexes of these) coherent sheaves, corresponding to the branes $\mathscr{P}_0(\theta)$ and $\mathscr{P}_1(\theta)$ of $\mathfrak{P}(\R_t, \mathcal{M}^G_{\text{H}, \theta}(C, \mathbf{J}))$ at the start and end of the interval, respectively.
The 1-morphisms are Ext-groups amongst these 1-objects, corresponding to B-strings that start and end on $\mathscr{P}_0(\theta)$ and $\mathscr{P}_1(\theta)$, respectively; we shall call these $\mathscr{P}^{\theta}$-strings.
This is illustrated in \autoref{fig:m3 x r2:rw cat}.

Since GM theory on $M_3 \times \R^2$ and 3d RW theory on $I \times \R^2$ are equivalent via a Heegaard split of $M_3 = M_3' \bigcup_C M_3''$, we have a correspondence between $\mathbf{D}^b \big( \text{coh} (\mathfrak{P} ( \R_t, \mathcal{M}^G_{\text{H}, \theta}(C, \mathbf{J}))) \big)$ and the Orlov-type $A_{\infty}$-1-category from the previous subsection (derived from the aforementioned 2d model on $\R^2$).
Thus, we can compare the two, and in doing so,  we find that a $\mathscr{P}^{\theta}$-string would correspond to a $\mathfrak{B}_3^{\theta}$-string.
In other words, we have the one-to-one correspondence
\begin{equation}
  \label{eq:m3 x r2:Db hom = orlov hom}
  \text{Hom}^{(1)}_{\mathbf{D} \big( \text{coh} (\mathfrak{P} (\R_t, \mathcal{M}^G_{\text{H}, \theta}(C, \mathbf{J}))) \big)} \big( \mathscr{P}_0(\theta), \mathscr{P}_1(\theta) \big)
  \Longleftrightarrow
  \text{Hom} \left( \mathfrak{C}^I_{\text{BF}_{\HH}}(\theta), \mathfrak{C}^J_{\text{BF}_{\HH}}(\theta) \right)
  \, .
\end{equation}
That is, we have
(i) 1-objects which are complex-Lagrangian branes $\mathscr{P}_{*}(\theta)$ in $\mathfrak{P} (\R_t, \mathcal{M}^G_{\text{H}, \theta}(C, \mathbf{J}))$ that are related to $\{\mathfrak{C}^1_{\text{BF}_{\HH}}(\theta), \dots, \mathfrak{C}^v_{\text{BF}_{\HH}}(\theta)\}$ which correspond to $G_{\HH}$-BF configurations on $M_3$,
(ii) 1-morphisms which are $\mathscr{P}^{\theta}$-strings that are related to $\mathfrak{B}_3^{\theta}$-strings,
and (iii) an $A_{\infty}$-structure that is defined by applying \eqref{eq:m3 x r2:Db hom = orlov hom} to \eqref{eq:m3 x r2:mu-d maps}.

In short, we have an Orlov $A_{\infty}$-1-category of branes in path space.

\subtitle{An RW $A_{\infty}$-2-category}

Moreover, we can also interpret the $\mathscr{P}^{\theta}$-strings as open membranes states of the 3d B-model, which, in turn, can be interpreted as morphisms between open B-string states of two 2d B-models on $I \times \R_{\tau}$ with target $\mathcal{M}^G_{\text{H}, \theta}(C, \mathbf{J})$ and branes $\widehat{L}_0^{*}(\theta)$ and $\widehat{L}_1^{*}(\theta)$ at the start and end of the interval, respectively; we shall call these $\widehat{L}^{\theta}$-strings.
Since these $\widehat{L}^{\theta}$-strings can be regarded as 1-morphisms in $\mathbf{D}^b \big( \text{coh}(\mathcal{M}^G_{\text{H}, \theta}(C, \mathbf{J})) \big)$, the $\mathscr{P}^{\theta}$-strings can then be regarded as 1-morphisms of a 2-category whose 1-objects are the $\widehat{L}^{\theta}$-strings.

In other words, the 3d RW model on $I \times \R_t \times \R_{\tau}$ with target $\mathcal{M}^G_{\text{H}, \theta}(C, \mathbf{J})$ physically realizes a 2-category, that we shall name an RW 2-category, whose
(i) 2-objects are complex-Lagrangian branes $\widehat{L}_{*}^{*}(\theta)$,
(ii) 2-morphisms $\text{Hom}^{(2)}_{\text{RW}} \left( \text{Hom} \big( \widehat{L}_0^I(\theta), \widehat{L}_0^J(\theta) \big), \text{Hom} \big( \widehat{L}_1^I(\theta), \widehat{L}_1^J(\theta) \big) \right)$ are $\mathscr{P}^{\theta}$-strings as open membranes with vertices being the 2-objects,
(iii) 1-objects are $\widehat{L}^{\theta}$-strings that can be represented as morphisms $\text{Hom} \big( \widehat{L}_0^{*}(\theta), \widehat{L}_1^{*}(\theta) \big)$ between their endpoints,
and (iv) 1-morphisms $\text{Hom}^{(1)}_{\text{RW}} \big( \text{Hom} \big( \widehat{L}_0^I(\theta),$ $\widehat{L}_1^I(\theta) \big), \text{Hom} \big( \widehat{L}_0^J(\theta), \widehat{L}_1^J(\theta) \big) \big)$ are $\mathscr{P}^{\theta}$-strings as open membranes with edges being the 1-objects.
This is illustrated in \autoref{fig:m3 x r2:rw cat}.

Once again, we can compare RW 2-category with the Orlov-type $A_{\infty}$-1-category from the previous subsection; we find that a $\mathscr{P}^{\theta}$-string would correspond to a $\mathfrak{B}_3^{\theta}$-string, and a $\widehat{L}^{\theta}$-string would correspond to a singular fiber $V_{3, \text{sg}}^{-1}(0)$.
In other words, we have the one-to-one correspondence
\begin{equation}
  \label{eq:m3 x r2:rw hom = orlov hom}
  \text{Hom}^{(2)}_{\text{RW}} \left( \text{Hom} \big( \widehat{L}_0^I(\theta), \widehat{L}_1^I(\theta) \big), \text{Hom} \big( \widehat{L}_0^J(\theta), \widehat{L}_1^J(\theta) \big) \right)
  \Longleftrightarrow
  \text{Hom} \left( \mathfrak{C}^I_{\text{BF}_{\HH}}(\theta), \mathfrak{C}^J_{\text{BF}_{\HH}}(\theta) \right)
  \, .
\end{equation}
That is, we have (i) 1-objects which are $\widehat{L}^{\theta}$-strings,
(ii) 1-morphisms which are $\mathscr{P}^{\theta}$-strings,
and (iii) an $A_{\infty}$-structure that is defined by applying \eqref{eq:m3 x r2:rw hom = orlov hom} to \eqref{eq:m3 x r2:mu-d maps}.

In short, the RW 2-category is actually an $A_{\infty}$-2-category!

\subtitle{A KRS 2-category}

Note that a 2-category of complex-Lagrangian branes realized by the 3d RW model was first constructed by KRS in \cite{kapustin-2009-three-dimen}.
However, as the morphisms of this KRS 2-category are realized by surface and line defects, an $A_{\infty}$-structure cannot defined.
It is therefore different from our RW $A_{\infty}$-2-category.

\subtitle{Corresponding Categories}

In summary, we have correspondences amongst
(i) $\text{Orlov} \left( \text{HHF}^{\text{flat}}(M_3, G_{\HH}) \right)$, a gauge-theoretic Orlov-type $A_{\infty}$-1-category of $\text{HHF}^{\text{flat}}(M_3, G_{\HH})$,
(ii) $\text{RW}_{(\sigma)} \big( \widehat{L}, \mathcal{M}^G_{\text{H}, \theta}(C, \mathbf{J}) \big)$, an RW $A_{\infty}$-2-category of complex-Lagrangian branes $\widehat{L}(\theta) \subset \mathcal{M}^G_{\text{H}, \theta}(C, \mathbf{J})$,
(iii) $\text{KRS}_{(\sigma)} \big( \widehat{L}, \mathcal{M}^G_{\text{H}, \theta}(C, \mathbf{J}) \big)$, a KRS 2-category of complex-Lagrangian branes $\widehat{L}(\theta) \subset \mathcal{M}^G_{\text{H}, \theta}(C, \mathbf{J})$,
and (iv) $\text{Orlov}_{(\sigma)} \big( \mathscr{P}, \mathfrak{P}(\R, \mathcal{M}^G_{\text{H}, \theta}(C, \mathbf{J})) \big)$, an Orlov-type $A_{\infty}$-1-category of path space branes $\mathscr{P}(\theta) \subset \mathfrak{P}(\R, \mathcal{M}^G_{\text{H}, \theta}(C, \mathbf{J}))$:
\begin{equation}
  \label{eq:m3 x r2:correspondence amongst cats}
  \saveboxed{eq:m3 x r2:correspondence amongst cats}{%
    \begin{adjustbox}{
        max totalsize={\textwidth},%
      }
      \begin{tikzcd}[
        row sep=huge,%
        column sep=huge,%
        arrows=Leftrightarrow,%
        ampersand replacement=\&,%
        ]
        \&
        \text{Orlov} \left( \text{HHF}^{\text{flat}}(M_3, G_{\HH}) \right)
        \arrow[dl, "M_3 = M_3' \bigcup_C M_3''"{sloped}] 
        \arrow[d, "M_3 = M_3' \bigcup_C M_3''"] 
        \arrow[dr, "M_3 = M_3' \bigcup_C M_3''"{sloped}] 
        \&
        \\
        \text{KRS}_{(\sigma)} \left( \widehat{L}, \mathcal{M}^G_{\text{H}, \theta}(C, \mathbf{J}) \right)
        \arrow[r]
        \&
        \text{RW}_{(\sigma)} \left( \widehat{L}, \mathcal{M}^G_{\text{H}, \theta}(C, \mathbf{J}) \right)
        \arrow[r]
        \&
        \text{Orlov}_{(\sigma)} \left( \mathscr{P}, \mathfrak{P}(\R, \mathcal{M}^G_{\text{H}, \theta}(C, \mathbf{J})) \right)
      \end{tikzcd}%
    \end{adjustbox}
  }%
\end{equation}

\subtitle{A Physical Proof of Doan-Rezchikov's Mathematical Conjecture}

Note that a correspondence between a KRS 2-category of complex-Lagrangian branes in a hyperkähler manifold $X$ and an Orlov-type $A_{\infty}$-1-category of branes in the path space $\mathfrak{P}(\R, X)$, was conjectured by Doan-Rezchikov (DR) in \cite[$\S$2.6]{doan-2022-holom-floer}.
Since $\mathcal{M}^G_{\text{H}, \theta}(C, \mathbf{J})$ is hyperkähler, this correspondence is exactly what we have in the bottom line of \eqref{eq:m3 x r2:correspondence amongst cats} for $X = \mathcal{M}^G_{\text{H}, \theta}(C, \mathbf{J})$.

Thus, we have furnished a purely physical proof of their mathematical conjecture!

\subtitle{Atiyah-Floer-type Correspondences between $A_{\infty}$-categories}

Note that the Atiyah-Floer (AF) correspondence between a gauge-theoretic $G$-instanton Floer homology of $M_3$ and a symplectic intersection Floer homology of Lagrangian branes in $\mathcal{M}^G_{\text{flat}}(C)$, is actually a correspondence between (0-)categories defined via a Heegaard split of $M_3 = M_3' \bigcup_C M_3''$. 
Since \eqref{eq:m3 x r2:correspondence amongst cats} involves correspondences between a gauge-theoretic category and symplectic categories via a Heegaard split of $M_3 = M_3' \bigcup_C M_3''$, we can thus regard them as AF-type correspondences.

In other words, we have \emph{novel} AF-type correspondences between the gauge-theoretic Orlov-type $A_{\infty}$-1-category of $\text{HHF}^{\text{flat}}(M_3, G_{\HH})$ and
(i) the RW $A_{\infty}$-2-category of complex-Lagrangian branes in symplectic $\mathcal{M}^G_{\text{H}, \theta}(C, \mathbf{J})$,
(ii) the KRS 2-category of complex-Lagrangian branes in symplectic $\mathcal{M}^G_{\text{H}, \theta}(C, \mathbf{J})$,
and (iii) the Orlov-type $A_{\infty}$-1-category of branes in symplectic $\mathfrak{P}(\R, \mathcal{M}^G_{\text{H}, \theta}(C, \mathbf{J}))$!

\section{A Novel RW-type \texorpdfstring{$A_{\infty}$}{A-infinity}-2-category of Two-Manifolds}
\label{sec:m2 x r3}

In this section, we will study GM theory on $M_5 = M_2 \times \R^3$, where $M_2$ is a closed two-manifold.
We will recast it as a 3d gauged B-twisted LG model on $\R^3$, a 2d gauged B-twisted LG model on $\R^2$, and a 1d LG SQM.
Following what we did in \autoref{sec:m3 x r2}, we will, via the 5d GM partition function and its equivalent 2d gauged B-twisted LG model, be able to physically realize an Orlov-type $A_{\infty}$-1-category of strings whose endpoints correspond to $G_{\OO}$-BF configurations on $M_2$ that generate the holomorphic $G_{\OO}$-flat Floer homology of $M_2$ from \autoref{sec:m2 x r}.
Next, following what we did in our previous work \cite{er-2024-topol-gauge}, we will, via the 5d GM partition function and its equivalent 3d gauged B-twisted LG model, be able to also physically realize a novel gauge-theoretic RW-type $A_{\infty}$-2-category that 2-categorifies the aforementioned holomorphic $G_{\OO}$-flat Floer homology of $M_2$.

\subsection{GM Theory on \texorpdfstring{$M_2 \times \R^3$}{M2 x R3} as a 3d Model on \texorpdfstring{$\R^3$}{R3}, 2d Model on \texorpdfstring{$\R^2$}{R2}, or 1d SQM}
\label{sec:m2 x r3:gm}

Let $t, \tau, \xi$ be the coordinates along $\R^3$.
The action of this theory is given by
\begin{equation}
  \label{eq:m2 x r3:gm:action}
  S_{\text{GM}}
  = \frac{1}{e^2} \int_{M_2 \times \R^3} dt d\tau d\xi d^2x \, \Tr \bigg(
  L_1
  + L_2
  + \dots
  \bigg)
  \, ,
\end{equation}
where
\begin{equation}
  \label{eq:m2 x r3:gm:action:L1}
  \begin{aligned}
    L_1
    &= \left|
      F_{t\tau} - i D_t \phi_{\tau} + i D_{\tau} \phi_t - [\phi_t, \phi_{\tau}]
      \right|^2
      + \left|
      F_{t\xi} - i D_t \phi_{\xi} + i D_{\xi} \phi_t - [\phi_t, \phi_{\xi}]
      \right|^2
    \\
    & \quad
      + \left|
      F_{\tau\xi} - i D_{\tau} \phi_{\xi} + i D_{\xi} \phi_{\tau} - [\phi_{\tau}, \phi_{\xi}]
      \right|^2
      \, ,
  \end{aligned}
\end{equation}
and
\begin{equation}
  \label{eq:m2 x r3:gm:action:L2}
  \begin{aligned}
    L_2
    &= \left|
      D_t \overline{\mathcal{A}}_M - \partial_M A_t + i \overline{\mathcal{D}}_M \phi_t
      \right|^2
      + \left|
      D_{\tau} \overline{\mathcal{A}}_M - \partial_M A_{\tau} + i \overline{\mathcal{D}}_M \phi_{\tau}
      \right|^2
    \\
    & \quad
      + \left|
      D_{\xi} \overline{\mathcal{A}}_M - \partial_M A_{\xi} + i \overline{\mathcal{D}}_M \phi_{\xi}
      \right|^2
      + \left|
      \overline{\mathcal{F}}_{MN}
      \right|^2
      \, .
  \end{aligned}
\end{equation}

\subtitle{GM Theory as a 3d Model}

We now want to recast GM theory as a 3d model on $\R^3$.
To this end, first note that the conditions that minimize the action \eqref{eq:m2 x r3:gm:action} can be octonionified in a similar manner as was done for the BPS equations in \autoref{sec:m2 x r}.
Using the same notation for imaginary numbers as was used there, the octonionified conditions are
\begin{equation}
  \label{eq:m2 x r3:gm:octonionified equations}
  \begin{aligned}
    0 =
    & (D_{\xi} + \mathbf{e}_0 D_t + \hat{\mathbf{e}}_2 D_{\tau}) \widetilde{\mathsf{A}}^N \mathsf{E}_{NM}
      + (D_{\xi} + \hat{\mathbf{e}}_2 D_{\tau}) \widetilde{\mathsf{B}} \mathsf{H}^N \mathsf{E}_{NM}
      + (F_{t\xi} + \hat{\mathbf{e}}_2 F_{t\tau}) \mathsf{H}^N \mathsf{E}_{NM}
    \\
    & - \partial^N (A_{\xi} + \mathbf{e}_0 A_t + \hat{\mathbf{e}}_2 A_{\tau}) \mathsf{E}_{NM}
      - \mathbf{e}_0 \epsilon_{MN} \widetilde{\mathsf{D}}^N \widetilde{\mathsf{B}}
      \, ,
    \\
    0 =
    & (D_{\xi} + \mathbf{e}_0 D_t + \hat{\mathbf{e}}_2 D_{\tau}) \widetilde{\mathsf{B}} \mathsf{E}^M_{\;N} \mathsf{E}^N_{\;P} \mathsf{E}^P_{\;M}
      - (D_{\xi} + \hat{\mathbf{e}}_2 D_{\tau}) \widetilde{\mathsf{A}}^M \widetilde{\mathsf{H}}_M
      + \partial^M (A_{\xi} + \hat{\mathbf{e}}_2 A_{\tau}) \widetilde{\mathsf{H}}_M
    \\
    & + \frac{\mathbf{e}_0}{2} \epsilon_{MN} \widetilde{\mathsf{F}}^{MN}
      \, ,
  \end{aligned}
\end{equation}
where
(i) $\hat{\mathbf{e}}_j \coloneq \mathbf{e}_j + \mathbf{e}_{0j}$,
(ii) $\widetilde{\mathsf{A}}_M \coloneq \widetilde{\mathscr{A}_M} - (\mathbf{e}_1 + \mathbf{e}_3) \phi_{\tau} \mathsf{H}_M$,
(ii) $\widetilde{\mathsf{B}} \coloneq \frac{1}{2} (\mathbf{e}_{01} \phi_t - \mathbf{e}_0 \hat{\mathbf{e}}_3 \phi_{\tau} - \mathbf{e}_1 \phi_{\xi}) \epsilon_{MN} \mathsf{E}^{MN}$,
and (iii) $\widetilde{\mathsf{H}}_M = \epsilon_{MN} \mathsf{H}^N$.\footnote{%
  Note that when compared to the fields in \autoref{sec:m2 x r}, $\widetilde{\mathsf{A}}_M = \overline{\mathsf{A}}_M + (\mathbf{e}_0 A_{\xi} + \mathbf{e}_{02} A_{\tau}) \mathsf{H}_M$ and $\widetilde{\mathsf{B}} = \overline{\mathsf{B}} + \frac{1}{2} \epsilon^{MN} (\mathbf{e}_0 A_{\xi} + \mathbf{e}_{02} A_{\tau}) \mathsf{E}_{MN}$.
  These $\widetilde{\mathsf{A}}$ and $\widetilde{\mathsf{B}}$ fields are used in this section as we want to make explicit the real gauge fields of $\R^3$, i.e., $A_t$, $A_{\tau}$, and $A_{\xi}$, in the equations.
  \label{ft:m2 x r3:comparing fields to those in floer hom}
}

Next, note that we are physically free to rotate the $(\xi, t)$-subplane of $\R^3$ about the origin by an angle $\theta$, whence \eqref{eq:m2 x r3:gm:octonionified equations} becomes $X_M = 0$ and $Y = 0$, where
\begin{equation}
  \label{eq:m2 x r3:gm:octonionified equations:rotated}
  \begin{aligned}
    X_M
    =
    & (D_{\xi} + \mathbf{e}_0 D_t + e^{\mathbf{e}_0 \theta} \hat{\mathbf{e}}_2 D_{\tau}) \widetilde{\mathsf{A}}^N \mathsf{E}_{NM}
      + e^{\mathbf{e}_0 \theta} \left(  \cos(\theta) D_{\xi} + \sin(\theta) D_t + \hat{\mathbf{e}}_2 D_{\tau} \right) \widetilde{\mathsf{B}} \mathsf{H}^N \mathsf{E}_{NM}
    \\
    & + e^{\mathbf{e}_0 \theta} \left( F_{t\xi} + \hat{\mathbf{e}}_2 (\cos(\theta) F_{t\tau} - \sin(\theta) F_{\xi\tau}) \right) \mathsf{H}^N \mathsf{E}_{NM}
      + P_M(\theta)
      \, ,
    \\
    Y
    =
    & (D_{\xi} + \mathbf{e}_0 D_t + e^{\mathbf{e}_0 \theta} \hat{\mathbf{e}}_2 D_{\tau}) \widetilde{\mathsf{B}} \mathsf{E}^M_{\;N} \mathsf{E}^N_{\;P} \mathsf{E}^P_{\;M}
      - e^{\mathbf{e}_0 \theta} (\cos(\theta) D_{\xi} + \sin(\theta) D_t + \hat{\mathbf{e}}_2 D_{\tau}) \widetilde{\mathsf{A}}^M \widetilde{\mathsf{H}}_M
      + Q(\theta)
      \, ,
  \end{aligned}
\end{equation}
and
\begin{equation}
  \label{eq:m2 x r3:gm:octonionified equations:rotated:components}
  \begin{aligned}
    P_M(\theta)
    &= - \partial^N (A_{\xi} + \mathbf{e}_0 A_t + e^{\mathbf{e}_0 \theta} \hat{\mathbf{e}}_2 A_{\tau}) \mathsf{E}_{NM}
      - \mathbf{e}_0 e^{\mathbf{e}_0 \theta} \epsilon_{MN} \widetilde{\mathsf{D}}^N \widetilde{\mathsf{B}}
      \, ,
    \\
    Q(\theta)
    &= e^{\mathbf{e}_0 \theta} \partial^M (A_{\xi} \cos(\theta) + A_t \sin(\theta) + \hat{\mathbf{e}}_2 A_{\tau}) \widetilde{\mathsf{H}}_M
      +  \frac{\mathbf{e}_0 e^{\mathbf{e}_0 \theta}}{2} \epsilon_{MN} \widetilde{\mathsf{F}}^{MN}
      \, .
  \end{aligned}
\end{equation}
Thus, we can also write \eqref{eq:m2 x r3:gm:action} as
\begin{equation}
  \label{eq:m2 x r3:gm:action:rotated}
  S_{\text{GM}}
  = \frac{1}{e^2} \int_{M_2 \times \R^3} dt d\tau d\xi d^2x \, \Tr \left(
    \left| X_M \right|^2
    + \left| Y \right|^2
    + \dots
  \right)
  \, .
\end{equation}

Lastly, after suitable rescalings, we can recast \eqref{eq:m2 x r3:gm:action:rotated} as a 3d model, where the action is\footnote{%
  To arrive at the following expression, we have employed the fact that $M_2$ has no boundary to omit terms with $\partial_M A_{\{t, \tau, \xi\}}$ as they will vanish when integrated over $M_2$.
  \label{ft:m2 x r3:integrate out gauge fields to 3d model}
}
\begin{equation}
  \label{eq:m2 x r3:gm:3d model:action}
  S_{\text{3d}, \mathfrak{B}_2}
  = \frac{1}{e^2} \int_{\R^3} dt d\tau d\xi \, \left(
    \left| X^a \right|^2
    + \left| Y^a \right|^2
    + \dots
  \right)
  \, ,
\end{equation}
where
\begin{equation}
  \label{eq:m2 x r3:gm:3d model:bps}
  \begin{aligned}
    X^a
    =
    & (D_{\xi} + \mathbf{e}_0 D_t + e^{\mathbf{e}_0 \theta} \hat{\mathbf{e}}_2 D_{\tau}) \widetilde{\mathsf{A}}^a
      + e^{\mathbf{e}_0 \theta} \left(  \cos(\theta) D_{\xi} + \sin(\theta) D_t + \hat{\mathbf{e}}_2 D_{\tau} \right) \widetilde{\mathsf{B}}^a
    \\
    & + e^{\mathbf{e}_0 \theta} \left( F_{t\xi} + \hat{\mathbf{e}}_2 (\cos(\theta) F_{t\tau} - \sin(\theta) F_{\xi\tau}) \right)
      + P^a(\theta)
      \, ,
    \\
    Y^a
    =
    & (D_{\xi} + \mathbf{e}_0 D_t + e^{\mathbf{e}_0 \theta} \hat{\mathbf{e}}_2 D_{\tau}) \widetilde{\mathsf{B}}^a
      - e^{\mathbf{e}_0 \theta} (\cos(\theta) D_{\xi} + \sin(\theta) D_t + \hat{\mathbf{e}}_2 D_{\tau}) \widetilde{\mathsf{A}}^a
      + Q^a(\theta)
      \, ,
  \end{aligned}
\end{equation}
and
\begin{equation}
  \label{eq:m2 x r3:gm:3d model:bps:components}
  P^a(\theta)
  = - \mathbf{e}_0 e^{\mathbf{e}_0 \theta} (\widetilde{\mathsf{D}} \widetilde{\mathsf{B}})^a
  \, ,
  \qquad
  Q^a(\theta)
  = \mathbf{e}_0 e^{\mathbf{e}_0 \theta} \widetilde{\mathsf{F}}^a
  \, .
\end{equation}
Here, $(\widetilde{\mathsf{A}}^a, \widetilde{\mathsf{B}}^a)$ and $a$ are coordinates and indices on the space $\mathfrak{B}_2$ of irreducible $(\widetilde{\mathsf{A}}_M, \widetilde{\mathsf{B}})$ fields on $M_2$, with (i) $(\widetilde{\mathsf{D}} \widetilde{\mathsf{B}})^a$ and (ii) $\widetilde{\mathsf{F}}^a$ corresponding to (i) $\epsilon_{MN} \widetilde{\mathsf{D}}^N \widetilde{\mathsf{B}}$ and (ii) $\frac{1}{2} \epsilon_{MN} \widetilde{\mathsf{F}}^{MN}$, respectively, in the underlying 5d theory.

In other words, GM theory on $M_2 \times \R^3$ can be regarded as a 3d gauged sigma model along the $(t, \tau, \xi)$-directions with target $\mathfrak{B}_2$ and action \eqref{eq:m2 x r3:gm:3d model:action}.

\subtitle{GM Theory as a 2d Model}

Note that \eqref{eq:m2 x r3:gm:3d model:action}, after suitable rescalings, can be recast as the following equivalent 2d model action\footnote{%
  To arrive at the following expression, we have employed Stokes' theorem and the fact that the finite-energy 3d real gauge fields $A_{\{t, \tau, \xi\}}$ would vanish at $\tau \rightarrow \pm \infty$.
  \label{ft:m2 x r3:integrate out gauge fields to 2d model}
}
\begin{equation}
  \label{eq:m2 x r3:gm:2d model:action}
  S_{\text{2d}, \mathfrak{P}(\R_{\tau}, \mathfrak{B}_2)}
  = \frac{1}{e^2} \int_{\R^2} dt d\xi \left(
    \left| X^u \right|^2
    + \left| Y^u \right|^2
    + \dots
  \right)
  \, ,
\end{equation}
where
\begin{equation}
  \label{eq:m2 x r3:gm:2d model:bps}
  \begin{aligned}
    X^u
    =
    & (D_{\xi} + \mathbf{e}_0 D_t) \widetilde{\mathsf{A}}^u
      + e^{\mathbf{e}_0 \theta} \left(  \cos(\theta) D_{\xi} + \sin(\theta) D_t \right) \widetilde{\mathsf{B}}^u
    \\
    & + e^{\mathbf{e}_0 \theta} \left( F_{t\xi} + \hat{\mathbf{e}}_2 (\cos(\theta) D_t - \sin(\theta) D_{\xi}) (\mathring{A}_{\tau})^u \right)
      + p^u(\theta)
      + P^u(\theta)
      \, ,
    \\
    Y^u
    =
    & (D_{\xi} + \mathbf{e}_0 D_t) \widetilde{\mathsf{B}}^u
      - e^{\mathbf{e}_0 \theta} (\cos(\theta) D_{\xi} + \sin(\theta) D_t) \widetilde{\mathsf{A}}^u
      + q^u(\theta)
      + Q^u(\theta)
      \, .
  \end{aligned}
\end{equation}
Here, $(\widetilde{\mathsf{A}}^u, \widetilde{\mathsf{B}}^u, (\mathring{A}_{\tau})^u)$ and $u$ are coordinates and indices on the path space $\mathfrak{P}(\R_{\tau}, \mathfrak{B}_2)$ of maps from $\R_{\tau}$ to $\mathfrak{B}_2$, and
\begin{equation}
  \label{eq:m2 x r3:gm:2d model:bps:components}
  \begin{aligned}
    p^u(\theta)
    &= e^{\mathbf{e}_0 \theta} \hat{\mathbf{e}}_2 (\mathring{D}_{\tau} \widetilde{\mathsf{A}} + \mathring{D}_{\tau} \widetilde{\mathsf{B}})^u
      \, ,
    & \qquad
    q^u(\theta)
    &= e^{\mathbf{e}_0 \theta} \hat{\mathbf{e}}_2 (\mathring{D}_{\tau} \widetilde{\mathsf{B}} - \mathring{D}_{\tau} \widetilde{\mathsf{A}})^u
      \, ,
    \\
    P^u(\theta)
    & = - \mathbf{e}_0 e^{\mathbf{e}_0 \theta} (\widetilde{\mathsf{D}} \widetilde{\mathsf{B}})^u
      \, ,
    & \qquad
      Q^u(\theta)
    & = \mathbf{e}_0 e^{\mathbf{e}_0 \theta} \widetilde{\mathsf{F}}^u
      \, ,
  \end{aligned}
\end{equation}
with $(\mathring{A}_{\tau}, \mathring{D}_{\tau})$ in $\mathfrak{P}(\R_{\tau}, \mathfrak{B}_2)$ corresponding to $(A_{\tau}, D_{\tau})$ in the underlying 3d model.

In other words, GM theory on $M_2 \times \R^3$ can also be regarded as a 2d gauged sigma model along the $(t, \xi)$-directions with target $\mathfrak{P}(\R_{\tau}, \mathfrak{B}_2)$ and action \eqref{eq:m2 x r3:gm:2d model:action}.

\subtitle{GM Theory as a 1d SQM}

Singling out $\xi$ as the direction in ``time'', one can, after suitable rescalings, recast \eqref{eq:m2 x r3:gm:2d model:action} as the following equivalent SQM action\footnote{%
  The resulting SQM will not involve the non-dynamical gauge field $A_{\xi}$ for the same reason as explained in \autoref{ft:m3 x r2:integrate out gauge field to sqm} (where there, the non-dynamical gauge field was $A_{\tau}$).
  We have again applied Stokes' theorem and the fact that the fields corresponding to the finite-energy real gauge fields $A_{\{t, \tau\}}$ would vanish at $t \rightarrow \pm \infty$, to arrive at the following expression.
  \label{ft:m2 x r3:integrate out gauge field to sqm}
}
\begin{equation}
  \label{eq:m2 x r3:gm:sqm:action}
  \begin{aligned}
    & S_{\text{SQM}, \mathfrak{P}(\R_t, \mathfrak{P}(\R_{\tau}, \mathfrak{B}_2))}
    \\
    & = \frac{1}{e^2} \int d\xi \left(
      \left| \partial_{\xi} \varphi_+^{\alpha} + g^{\alpha\beta}_{\mathfrak{P}(\R_t, \mathfrak{P}(\R_{\tau}, \mathfrak{B}_2))} \pdv{h_2}{\varphi_+^{\beta}} \right|^2
      +  \left| \partial_{\xi} \varphi_-^{\alpha} + g^{\alpha\beta}_{\mathfrak{P}(\R_t, \mathfrak{P}(\R_{\tau}, \mathfrak{B}_2))} \pdv{h_2}{\varphi_-^{\beta}} \right|^2
      + \dots
      \right)
      \, ,
  \end{aligned}
\end{equation}
where $(\varphi_+^{\alpha}, \varphi_-^{\alpha})$ and $(\alpha, \beta)$ are coordinates and indices on the path space $\mathfrak{P}(\R_t, \mathfrak{P}(\R_{\tau}, \mathfrak{B}_2))$ of maps from $\R_t$ to $\mathfrak{P}(\R_{\tau}, \mathfrak{B}_2)$ with
(i) $\varphi_+^{\alpha} \coloneq \big( \widetilde{\mathsf{A}} + e^{\mathbf{e}_0 \theta} \big(  \widetilde{\mathsf{B}} \cos(\theta) - \mathring{A}_t - \mathring{A}_{\tau} \sin(\theta) \big) \big)^{\alpha}$
and (ii) $\varphi_-^{\alpha} \coloneq \big( \widetilde{\mathsf{B}} - e^{\mathbf{e}_0 \theta} \widetilde{\mathsf{A}} \cos(\theta) \big)^{\alpha}$
corresponding to
(i) $\widetilde{\mathsf{A}}^u + e^{\mathbf{e}_0 \theta} \big(  \widetilde{\mathsf{B}}^u \cos(\theta) - A_t - (\mathring{A}_{\tau})^u \sin(\theta) \big)$
and (ii) $\widetilde{\mathsf{B}}^u - e^{\mathbf{e}_0 \theta} \widetilde{\mathsf{A}}^a \cos(\theta)$
in the underlying 2d model,
and to (i) $\widetilde{\mathsf{A}}^a + e^{\mathbf{e}_0 \theta} \big(  \widetilde{\mathsf{B}}^a \cos(\theta) - A_t - A_{\tau} \sin(\theta) \big)$
and (ii) $\widetilde{\mathsf{B}}^a - e^{\mathbf{e}_0 \theta} \widetilde{\mathsf{A}}^a \cos(\theta)$
in the underlying 3d model;
$g_{\mathfrak{P}(\R_t, \mathfrak{P}(\R_{\tau}, \mathfrak{B}_2))}$ is the metric on $\mathfrak{P}(\R_t, \mathfrak{P}(\R_{\tau}, \mathfrak{B}_2))$;
and $h_2(\varphi_+, \varphi_-)$ is the SQM potential function.
Note that we can also interpret $\mathfrak{P}(\R_t, \mathfrak{P}(\R_{\tau}, \mathfrak{B}_2))$ as the double path space $\mathfrak{P}(\R^2, \mathfrak{B}_2)$ of maps from $\R^2$ to $\mathfrak{B}_2$.

In short, GM theory on $M_2 \times \R^3$ can also be regarded as a 1d SQM along the $\xi$-direction in $\mathfrak{P}(\R^2, \mathfrak{B}_2)$ whose action is \eqref{eq:m2 x r3:gm:sqm:action}.

\subsection{Non-constant Double Paths, Strings, Membranes, and \texorpdfstring{$G_{\OO}$}{GO}-BF Configurations on \texorpdfstring{$M_2$}{M2}}
\label{sec:m2 x r3:go-bf}

\subtitle{$\theta$-deformed, Non-constant Double Paths in the SQM}

Applying the squaring argument \cite{blau-1993-topol-gauge} to \eqref{eq:m2 x r3:gm:sqm:action}, we find that the path integral of the equivalent SQM will localize onto configurations that set the LHS and RHS of the expression within the squared terms therein \emph{simultaneously} to zero, i.e., the SQM path integral localizes onto \emph{$\xi$-invariant} critical points of $h_2(\varphi_+, \varphi_-)$ that obey
\begin{equation}
  \label{eq:m2 x r3:go-bf:double paths}
  \begin{aligned}
    & \left\lbrace \mathring{\partial}_t
        \left( \mathbf{e}_0 \widetilde{\mathsf{A}}
          + e^{\mathbf{e}_0 \theta} \widetilde{\mathsf{B}} \sin(\theta)
          + e^{\mathbf{e}_0 \theta} \hat{\mathbf{e}}_2 \mathring{A}_{\tau} \cos(\theta)
        \right)
      \right\rbrace^{\alpha}
      + e^{\mathbf{e}_0 \theta} \hat{\mathbf{e}}_2 \left\lbrace \mathring{\partial}_{\tau}
        \left( \widetilde{\mathsf{A}} + \widetilde{\mathsf{B}} \right)
      \right\rbrace^{\alpha}
    \\
    & = - \left(
        \left[ \mathring{A}_t,
          \mathbf{e}_0 \widetilde{\mathsf{A}}
          + e^{\mathbf{e}_0 \theta} \widetilde{\mathsf{B}} \sin(\theta)
          + e^{\mathbf{e}_0 \theta} \hat{\mathbf{e}}_2 \mathring{A}_{\tau} \cos(\theta)
        \right]
      \right)^{\alpha}
      - e^{\mathbf{e}_0 \theta} \hat{\mathbf{e}}_2 \left(
        \left[ \mathring{A}_{\tau},
          \widetilde{\mathsf{A}} + \widetilde{\mathsf{B}}
        \right]
      \right)^{\alpha}
      - P^{\alpha}(\theta)
      \, ,
    \\
    & \left\{ \mathring{\partial}_t
        \left( \mathbf{e}_0 \widetilde{\mathsf{B}} - e^{\mathbf{e}_0 \theta} \widetilde{\mathsf{A}} \sin(\theta) \right)
      \right\}^{\alpha}
      + e^{\mathbf{e}_0 \theta} \hat{\mathbf{e}}_2 \left\{
        \mathring{\partial}_{\tau} \left( \widetilde{\mathsf{B}} - \widetilde{\mathsf{A}} \right)
      \right\}^{\alpha}
    \\
    & = - \left(
        \left[ \mathring{A}_t,
          \mathbf{e}_0 \widetilde{\mathsf{B}} - e^{\mathbf{e}_0 \theta} \widetilde{\mathsf{A}} \sin(\theta)
        \right]
      \right)^{\alpha}
      - e^{\mathbf{e}_0 \theta} \hat{\mathbf{e}}_2 \left(
        \left[ \mathring{A}_{\tau},
          \widetilde{\mathsf{B}} - \widetilde{\mathsf{A}}
        \right]
      \right)^{\alpha}
      - Q^{\alpha}(\theta)
      \, .
  \end{aligned}
\end{equation}
These are $\xi$-invariant, $\theta$-deformed, non-constant double paths in $\mathfrak{P}(\R_t, \mathfrak{P}(\R_{\tau}, \mathfrak{B}_2))$.

\subtitle{$\mathfrak{P}^{\theta}(\R_{\tau}, \mathfrak{B}_2)$-strings in the 2d Gauged Model}

By comparing \eqref{eq:m2 x r3:gm:sqm:action} with \eqref{eq:m2 x r3:gm:2d model:action}, we find that such $\xi$-invariant, $\theta$-deformed, non-constant double paths in the SQM defined by \eqref{eq:m2 x r3:go-bf:double paths}, will correspond, in the 2d gauged sigma model with target $\mathfrak{P}(\R_{\tau}, \mathfrak{B}_2)$, to configurations defined by
\begin{equation}
  \label{eq:m2 x r3:go-bf:p-strings}
  \begin{aligned}
    \partial_t \left( \mathbf{e}_0 \widetilde{\mathsf{A}}
        + e^{\mathbf{e}_0 \theta} \widetilde{\mathsf{B}} \sin(\theta)
        + e^{\mathbf{e}_0 \theta} \hat{\mathbf{e}}_2 \mathring{A}_{\tau} \cos(\theta)
      \right)^u
    & = - \left[ A_t,
        \left(
          \mathbf{e}_0 \widetilde{\mathsf{A}}
          + e^{\mathbf{e}_0 \theta} \widetilde{\mathsf{B}} \sin(\theta)
          + e^{\mathbf{e}_0 \theta} \hat{\mathbf{e}}_2 \mathring{A}_{\tau} \cos(\theta)
        \right)^u
      \right]
    \\
    & \quad
      - \left[ A_{\xi},
        \left(
          \widetilde{\mathsf{A}}
          + e^{\mathbf{e}_0 \theta} \left(
            \widetilde{\mathsf{B}} \cos(\theta)
            - A_t
            - \mathring{A}_{\tau} \sin(\theta)
          \right)
        \right)^u
      \right]
    \\
    & \quad - p^u(\theta)
      - P^u(\theta)
      \, ,
    \\
    \partial_t \left( \mathbf{e}_0 \widetilde{\mathsf{B}} - e^{\mathbf{e}_0 \theta} \widetilde{\mathsf{A}} \sin(\theta) \right)^u
    & = - \left[ A_t,
        \left( \mathbf{e}_0 \widetilde{\mathsf{B}} - e^{\mathbf{e}_0 \theta} \widetilde{\mathsf{A}} \sin(\theta) \right)^u
      \right]
      - q^u(\theta)
      - Q^u(\theta)
      \, .
  \end{aligned}
\end{equation}
These are $\xi$-invariant, $\theta$-deformed strings along the $t$-direction of the 2d gauged sigma model with target $\mathfrak{P}(\R_{\tau}, \mathfrak{B}_2)$.
We shall henceforth refer to such strings as $\mathfrak{P}^{\theta}(\R_{\tau}, \mathfrak{B}_2)$-strings.

\subtitle{$\mathfrak{B}_2^{\theta}$-membranes in the 3d Gauged Model}

By further comparing \eqref{eq:m2 x r3:gm:2d model:action} with \eqref{eq:m2 x r3:gm:3d model:action}, we find that such $\mathfrak{P}^{\theta}(\R_{\tau}, \mathfrak{B}_2)$-strings in the 2d gauged sigma model defined by \eqref{eq:m2 x r3:go-bf:p-strings}, will correspond, in the 3d gauged sigma model with target $\mathfrak{B}_2$, to configurations defined by
\begin{equation}
  \label{eq:m2 x r3:go-bf:b-membranes}
  \begin{aligned}
    & \partial_t \left(
        \mathbf{e}_0 \widetilde{\mathsf{A}}^a
        + e^{\mathbf{e}_0 \theta} \left( \widetilde{\mathsf{B}}^a \sin(\theta) + A_{\xi} + \hat{\mathbf{e}}_2 A_{\tau} \cos(\theta) \right)
      \right)
      + e^{\mathbf{e}_0 \theta} \hat{\mathbf{e}}_2 \partial_{\tau} \left(
        \widetilde{\mathsf{A}}^a + \widetilde{\mathsf{B}}^a - \left( A_t \cos(\theta) - A_{\xi} \sin(\theta) \right)
      \right)
    \\
    &= - \left[A_t,
        \mathbf{e}_0 \widetilde{\mathsf{A}}^a
        + e^{\mathbf{e}_0 \theta} \left( \widetilde{\mathsf{B}}^a \sin(\theta) + A_{\xi} \right)
      \right]
      - e^{\mathbf{e}_0 \theta} \hat{\mathbf{e}}_2 \left[ A_{\tau},
        \widetilde{\mathsf{A}}^a + \widetilde{\mathsf{B}}^a - \left( A_t \cos(\theta) - A_{\xi} \sin(\theta) \right)
      \right]
    \\
    & \quad
      - \left[ A_{\xi},
        \widetilde{\mathsf{A}}^a
        + e^{\mathbf{e}_0 \theta} \widetilde{\mathsf{B}}^a \cos(\theta)
      \right]
      - P^a(\theta)
      \, ,
    \\
    & \partial_t \left(
        \mathbf{e}_0 \widetilde{\mathsf{B}}^a
        - e^{\mathbf{e}_0 \theta} \widetilde{\mathsf{A}}^a \sin(\theta)
      \right)
      + e^{\mathbf{e}_0 \theta} \hat{\mathbf{e}}_2 \partial_{\tau} \left(
        \widetilde{\mathsf{B}}^a
        - \widetilde{\mathsf{A}}^a
      \right)
    \\
    &= - \left[ A_t,
        \mathbf{e}_0 \widetilde{\mathsf{B}}^a
        - e^{\mathbf{e}_0 \theta} \widetilde{\mathsf{A}}^a \sin(\theta)
      \right]
      - e^{\mathbf{e}_0 \theta} \hat{\mathbf{e}}_2 \partial_{\tau} \left[ A_{\tau},
        \widetilde{\mathsf{B}}^a
        - \widetilde{\mathsf{A}}^a
      \right]
      - \left[ A_{\xi},
        \widetilde{\mathsf{B}}^a
        - e^{\mathbf{e}_0 \theta} \widetilde{\mathsf{A}}^a \cos(\theta)
      \right]
      - Q^a(\theta)
      \, .
  \end{aligned}
\end{equation}
These are $\xi$-invariant, $\theta$-deformed membranes along the $(t, \tau)$-directions in the 3d gauged sigma model with target $\mathfrak{B}_2$.
We shall henceforth refer to such membranes as $\mathfrak{B}_2^{\theta}$-membranes.

\subtitle{$\xi$-independent, $\theta$-deformed, GM Configurations in GM Theory}

By comparing \eqref{eq:m2 x r3:gm:3d model:action} with \eqref{eq:m2 x r3:gm:action:rotated}, we find that the 3d configurations defined by \eqref{eq:m2 x r3:go-bf:b-membranes} will correspond, in GM theory, to 5d configurations defined by
\begin{equation}
  \label{eq:m2 x r3:go-bf:gm configs:1}
  \begin{aligned}
    & \Big( \mathbf{e}_0 \partial_t + e^{\mathbf{e}_0 \theta} \hat{\mathbf{e}}_2 \partial_{\tau} \Big) \widetilde{\mathsf{A}}^N \mathsf{E}_{NM}
      + e^{\mathbf{e}_0 \theta} \Big( \sin(\theta) \partial_t + \hat{\mathbf{e}}_2 \partial_{\tau} \Big) \widetilde{\mathsf{B}} \mathsf{H}^N \mathsf{E}_{NM}
    \\
    & + e^{\mathbf{e}_0 \theta} \Big( \partial_t A_{\xi} + \hat{\mathbf{e}}_2 \big( \partial_t A_{\tau} \cos(\theta) - \partial_{\tau} A_t \cos(\theta) + \partial_{\tau} A_{\xi} \sin(\theta) \big) \Big) \mathsf{H}^N \mathsf{E}_{NM}
    \\
    &= - \left[ A_{\xi} + \mathbf{e}_0 A_t + e^{\mathbf{e}_0 \theta} \hat{\mathbf{e}}_2 A_{\tau}, \widetilde{\mathsf{A}}^N \mathsf{E}_{NM} \right]
      - e^{\mathbf{e}_0 \theta} \left[  A_{\xi} \cos(\theta) + A_t \sin(\theta) + \hat{\mathbf{e}}_2 A_{\tau}, \widetilde{\mathsf{B}} \mathsf{H}^N \mathsf{E}_{NM} \right]
    \\
    & \quad
      - e^{\mathbf{e}_0 \theta} \left( [A_t, A_{\xi}] + \hat{\mathbf{e}}_2 [A_t \cos(\theta) - A_{\xi} \sin(\theta), A_{\tau} ] \right) \mathsf{H}^N \mathsf{E}_{NM}
      - P_M(\theta)
      \, ,
  \end{aligned}
\end{equation}
and
\begin{equation}
  \label{eq:m2 x r3:go-bf:gm configs:2}
  \begin{aligned}
    & \Big( \mathbf{e}_0 \partial_t + e^{\mathbf{e}_0 \theta} \hat{\mathbf{e}}_2 \partial_{\tau} \Big) \widetilde{\mathsf{B}} \mathsf{E}^M_{\;N} \mathsf{E}^N_{\;P} \mathsf{E}^P_{\;M}
      - e^{\mathbf{e}_0 \theta} \Big( \sin(\theta) \partial_t + \hat{\mathbf{e}}_2 \partial_{\tau} \Big) \widetilde{\mathsf{A}}^M \widetilde{\mathsf{H}}_M
    \\
    & = - \left[ A_{\xi} + \mathbf{e}_0 A_t + e^{\mathbf{e}_0 \theta} \hat{\mathbf{e}}_2 A_{\tau},
        \widetilde{\mathsf{B}} \mathsf{E}^M_{\;N} \mathsf{E}^N_{\;P} \mathsf{E}^P_{\;M}
      \right]
      + e^{\mathbf{e}_0 \theta} \left[ A_{\xi} \cos(\theta) + A_t \sin(\theta) + \hat{\mathbf{e}}_2 A_{\tau},
        \widetilde{\mathsf{A}}^M \widetilde{\mathsf{H}}_M
      \right]
      - Q(\theta)
      \, .
  \end{aligned}
\end{equation}
These are $\xi$-independent, $\theta$-deformed GM configurations on $M_2 \times \R^3$.

\subtitle{GM Configurations, $\mathfrak{B}_2^{\theta}$-membranes, $\mathfrak{P}^{\theta}(\R_{\tau}, \mathfrak{B}_2)$-strings, and Non-constant Double Paths}

In short, these $\xi$-independent, $\theta$-deformed GM configurations on $M_2 \times \R^3$ that are defined by \eqref{eq:m2 x r3:go-bf:gm configs:1} and \eqref{eq:m2 x r3:go-bf:gm configs:2}, will correspond to the $\mathfrak{B}_2^{\theta}$-membranes defined by \eqref{eq:m2 x r3:go-bf:b-membranes}, which, in turn, will correspond to the $\mathfrak{P}^{\theta}(\R_{\tau}, \mathfrak{B}_2)$-strings defined by \eqref{eq:m2 x r3:go-bf:p-strings}, which, in turn, will correspond to the $\xi$-invariant, $\theta$-deformed, non-constant double paths in $\mathfrak{P}(\R^2, \mathfrak{B}_2)$ defined by \eqref{eq:m2 x r3:go-bf:double paths}.

\subtitle{$\mathfrak{P}^{\theta}(\R_{\tau}, \mathfrak{B}_2)$-string Endpoints Corresponding to Non-constant Paths}

Consider now the endpoints of the $\mathfrak{P}^{\theta}(\R_{\tau}, \mathfrak{B}_2)$-strings at $t = \pm \infty$, where we also expect the fields in the 2d gauged sigma model corresponding to the finite-energy 3d gauge fields $A_t$, $A_{\tau}$, and $A_{\xi}$ to decay to zero.
They are given by \eqref{eq:m2 x r3:go-bf:p-strings} with $\partial_t \widetilde{\mathsf{A}} = 0 = \partial_t \widetilde{\mathsf{B}}$ and $A_t, \mathring{A}_{\tau}, A_{\xi} \rightarrow 0$, where via \eqref{eq:m2 x r3:gm:2d model:bps:components}, we can write them as
\begin{equation}
  \label{eq:m2 x r3:go-bf:non-constant paths}
  e^{\mathbf{e}_0 \theta} \hat{\mathbf{e}}_2 (\mathring{\partial}_{\tau} \widetilde{\mathsf{A}} + \mathring{\partial}_{\tau} \widetilde{\mathsf{B}})^u
  = - P^u(\theta)
  \, ,
  \qquad
  e^{\mathbf{e}_0 \theta} \hat{\mathbf{e}}_2 (\mathring{\partial}_{\tau} \widetilde{\mathsf{B}} - \mathring{\partial}_{\tau} \widetilde{\mathsf{A}})^u
  = - Q^u(\theta)
  \, .
\end{equation}
These are $(\xi, t)$-invariant, $\theta$-deformed, non-constant paths in $\mathfrak{P}(\R_{\tau}, \mathfrak{B}_2)$.

\subtitle{$\mathfrak{B}_2^{\theta}$-membrane Edges Corresponding to $\mathfrak{B}_2^{\theta}$-strings in the 3d Gauged Model}

In turn, \eqref{eq:m2 x r3:go-bf:non-constant paths} will correspond, in the 3d gauged sigma model, to the fixed edges of the $\mathfrak{B}_2^{\theta}$-membranes at $t = \pm \infty$, i.e., $(\xi, t)$-invariant, $\theta$-deformed configurations that obey
\begin{equation}
  \label{eq:m2 x r3:go-bf:b-strings}
  e^{\mathbf{e}_0 \theta} \hat{\mathbf{e}}_2 \partial_{\tau} ( \widetilde{\mathsf{A}}^a + \widetilde{\mathsf{B}}^a )
  = - P^a(\theta)
  \, ,
  \qquad
  e^{\mathbf{e}_0 \theta} \hat{\mathbf{e}}_2 \partial_{\tau} ( \widetilde{\mathsf{B}}^a - \widetilde{\mathsf{A}}^a )
  = - Q^a(\theta)
  \, .
\end{equation}
These are $(\xi, t)$-invariant, $\theta$-deformed strings along the $\tau$-direction in the 3d gauged sigma model with target $\mathfrak{B}_2$.
Notice that \eqref{eq:m2 x r3:go-bf:b-strings} can also be obtained from \eqref{eq:m2 x r3:go-bf:b-membranes} with $\partial_t \widetilde{\mathsf{A}}^a = 0 = \partial_t \widetilde{\mathsf{B}}^a$ and $A_t, A_{\tau}, A_{\xi} \rightarrow 0$.
We shall henceforth refer to such strings as $\mathfrak{B}_2^{\theta}$-strings.

\subtitle{$\mathfrak{B}_2^{\theta}$-string Endpoints or $\mathfrak{B}_2^{\theta}$-membrane Vertices Corresponding to $\theta$-deformed $G_{\OO}$-BF Configurations on $M_2$}

Consider now
(i) the fixed endpoints of the $\mathfrak{B}_2^{\theta}$-strings at $\tau = \pm \infty$,
or equivalently (ii) the vertices of the $\mathfrak{B}_2^{\theta}$-membranes at $t, \tau = \pm \infty$.
They are given by (i) \eqref{eq:m2 x r3:go-bf:b-strings} with $\partial_{\tau} \widetilde{\mathsf{A}}^a = 0 = \partial_{\tau} \widetilde{\mathsf{B}}^{\alpha}$,
or equivalently (ii) \eqref{eq:m2 x r3:go-bf:b-membranes} with $\partial_{\{t, \tau\}} \widetilde{\mathsf{A}}^a = 0 = \partial_{\{t, \tau\}} \widetilde{\mathsf{B}}^{\alpha}$ and $A_t, A_{\tau}, A_{\xi} \rightarrow 0$,
i.e., (via \eqref{eq:m2 x r3:gm:3d model:bps:components})
\begin{equation}
  \label{eq:m2 x r3:go-bf:b-string endpoints}
   \mathbf{e}_0 e^{\mathbf{e}_0 \theta} (\widetilde{\mathsf{D}} \widetilde{\mathsf{B}})^a
   = 0
   \, ,
   \qquad
   \mathbf{e}_0 e^{\mathbf{e}_0 \theta} \widetilde{\mathsf{F}}^a
   = 0
   \, .
\end{equation}
In turn, they will correspond, in GM theory, to $(t, \tau, \xi)$-independent, $\theta$-deformed configurations that obey
\begin{equation}
  \label{eq:m2 x r3:go-bf:go-bf configs}
   \mathbf{e}_0 e^{\mathbf{e}_0 \theta} \epsilon_{MN} \widetilde{\mathsf{D}}^N \widetilde{\mathsf{B}}
   = 0
   \, ,
   \qquad
   \frac{\mathbf{e}_0 e^{\mathbf{e}_0 \theta}}{2} \epsilon_{MN} \widetilde{\mathsf{F}}^{MN}
   = 0
   \, .
\end{equation}
Notice that \eqref{eq:m2 x r3:go-bf:go-bf configs} can also be obtained from \eqref{eq:m2 x r3:go-bf:gm configs:1} and \eqref{eq:m2 x r3:go-bf:gm configs:2} with $\partial_{\{t, \tau\}} \widetilde{\mathsf{A}} = 0 = \partial_{\{t, \tau\}} \widetilde{\mathsf{B}}$ and $A_t, A_{\tau}, A_{\xi} \rightarrow 0$.

The equations in \eqref{eq:m2 x r3:go-bf:go-bf configs} are a $\theta$-deformed version of $G_{\OO}$-BF configurations on $M_2$.
At $\theta = \pm \pi/2$, they become regular $G_{\OO}$-BF configurations on $M_2$ which generate the holomorphic $G_{\OO}$-flat Floer homology of $M_2$ as seen in \autoref{sec:m2 x r}.
We shall also assume choices of $M_2$ and $G$ satisfying \autoref{ft:m2 x r:isolation and non-degeneracy of GO-flat} whereby such configurations are isolated and non-degenerate.\footnote{%
  We can apply a similar explanation as the one used in \autoref{ft:m3 x r2:isolation and non-degeneracy of GH-flat} to justify the presumption that the moduli space of $\theta$-deformed $G_{\OO}$-BF configurations on $M_2$ will be made of isolated and non-degenerate points.
  \label{ft:m2 x r3:go-bf:isolation and non-degeneracy of configs}
}

\subtitle{Non-constant Double Paths, $\mathfrak{P}^{\theta}(\R_{\tau}, \mathfrak{B}_2)$-strings, $\mathfrak{B}_2^{\theta}$-membranes, $\mathfrak{B}_2^{\theta}$-strings, and $\text{HHF}^{\text{flat}}(M_2, G_{\OO})$}

In short, from the equivalent 1d SQM of GM theory on $M_2 \times \R^3$, the theory localizes onto $\xi$-invariant, $\theta$-deformed, non-constant double paths in $\mathfrak{P}(\R^2, \mathfrak{B}_2)$, which, in turn, will correspond to $\mathfrak{P}^{\theta}(\R_{\tau}, \mathfrak{B}_2)$-strings in the 2d gauged sigma model whose endpoints are $(\xi, t)$-invariant, $\theta$-deformed, non-constant paths in $\mathfrak{P}(\R_{\tau}, \mathfrak{B}_2)$.
In the 3d gauged sigma model, these $\mathfrak{P}^{\theta}(\R_{\tau}, \mathfrak{B}_2)$-strings will correspond to $\mathfrak{B}_2^{\theta}$-membranes, whose edges are $\mathfrak{B}_2^{\theta}$-strings, and whose vertices will correspond to $\theta$-deformed $G_{\OO}$-BF configurations on $M_2$ which generate the holomorphic $G_{\OO}$-flat Floer homology of $M_2$.

\subsection{The 2d Model and Open Strings, the 3d Model and Open Membranes}
\label{sec:m2 x r3:membranes and strings}

By following the same analysis in \autoref{sec:m3 x r2:2d}, we find that the 2d gauged sigma model with target $\mathfrak{P}(\R_{\tau}, \mathfrak{B}_2)$ whose action is \eqref{eq:m2 x r3:gm:2d model:action}, will define an open \emph{string} theory in $\mathfrak{P}(\R_{\tau}, \mathfrak{B}_2)$.
Likewise, we find that the 3d gauged sigma model with target $\mathfrak{B}_2$ whose action is \eqref{eq:m2 x r3:gm:3d model:action}, will define an open \emph{membrane} theory in $\mathfrak{B}_2$.
We will now work out the details of the BPS worldsheets and their 3d BPS worldvolume analogues that would be necessary to define this open string and open membrane theory, respectively.

\subtitle{BPS Worldsheets of the 2d Model}

The BPS worldsheets of the 2d gauged sigma model with target $\mathfrak{P}(\R_{\tau}, \mathfrak{B}_2)$ correspond to its classical trajectories.
Specifically, these are defined by setting to zero the expression within the squared terms in \eqref{eq:m2 x r3:gm:2d model:action}, i.e.,
\begin{equation}
  \label{eq:m2 x r3:2d-3d:bps worldsheets}
  \begin{aligned}
    - p^u
    - P^u
    =
    & \Dv{}{\xi} \left(
      \widetilde{\mathsf{A}}^u
      + e^{\mathbf{e}_0 \theta} \widetilde{\mathsf{B}}^u \cos(\theta)
      - e^{\mathbf{e}_0 \theta} \hat{\mathbf{e}}_2 (\mathring{A}_{\tau})^u \sin(\theta)
    \right)
    \\
    & + \Dv{}{t} \left(
      \mathbf{e}_0 \widetilde{\mathsf{A}}^u
      + e^{\mathbf{e}_0 \theta} \widetilde{\mathsf{B}}^u \sin(\theta)
      + e^{\mathbf{e}_0 \theta} \hat{\mathbf{e}}_2 (\mathring{A}_{\tau})^u \cos(\theta)
    \right)
    + e^{\mathbf{e}_0 \theta} F_{t\xi}
      \, ,
    \\
    - q^u
    - Q^u
    =
    & \Dv{}{\xi} \left(
      \widetilde{\mathsf{B}}^u
      - e^{\mathbf{e}_0 \theta} \widetilde{\mathsf{A}}^u \cos(\theta)
    \right)
    + \Dv{}{t} \left(
      \mathbf{e}_0 \widetilde{\mathsf{B}}^u
      - e^{\mathbf{e}_0 \theta} \widetilde{\mathsf{A}}^u \sin(\theta)
    \right)
    \, .
  \end{aligned}
\end{equation}

\subtitle{BPS Worldsheets with Boundaries Labeled by Non-constant Paths in $\mathfrak{P}(\R_{\tau}, \mathfrak{B}_2)$}

The boundaries of the BPS worldsheets are traced out by the endpoints of the $\mathfrak{P}^{\theta}(\R_{\tau}, \mathfrak{B}_2)$-strings as they propagate in $\xi$; we shall, at $\xi = \pm \infty$, denote such $\mathfrak{P}^{\theta}(\R_{\tau}, \mathfrak{B}_2)$-strings as $\Sigma_{\pm}(t, \theta, \mathfrak{B}_2)$.
As we have seen at the end of \autoref{sec:m2 x r3:go-bf}, these endpoints correspond to $(\xi, t)$-invariant, $\theta$-deformed, non-constant paths in $\mathfrak{P}(\R_{\tau}, \mathfrak{B}_2)$ that we shall denote as $\Gamma(\theta, \mathfrak{B}_2)$.
In turn, they will correspond, in the 3d gauged sigma model with target $\mathfrak{B}_2$, to $\mathfrak{B}_2^{\theta}$-strings that we shall denote as $\gamma(\tau, \theta, \mathfrak{B}_2)$, whose endpoints will correspond to $\theta$-deformed $G_{\OO}$-BF configurations on $M_2$.

If there are $w \geq 4$ such configurations, $\{\mathfrak{C}^1_{\text{BF}_{\OO}}(\theta), \mathfrak{C}^2_{\text{BF}_{\OO}}(\theta), \dots, \mathfrak{C}^w_{\text{BF}_{\OO}}(\theta)\}$, just as in \autoref{sec:m3 x r2:2d}, we can further specify any $\mathfrak{B}_2^{\theta}$-string as $\gamma^{IJ}(\tau, \theta, \mathfrak{B}_2)$, where its start and end would correspond to $\mathfrak{C}^I_{\text{BF}_{\OO}}(\theta)$ and $\mathfrak{C}^J_{\text{BF}_{\OO}}(\theta)$, respectively.
Consequently, in the 2d gauged sigma model, we can further specify any $\mathfrak{P}^{\theta}(\R_{\tau}, \mathfrak{B}_2)$-string endpoint $\Gamma(\theta, \mathfrak{B}_2)$ as $\Gamma^{IJ}(\theta, \mathfrak{B}_2)$, where the latter will correspond to a $\gamma^{IJ}(\tau, \theta, \mathfrak{B}_2)$ $\mathfrak{B}_2^{\theta}$-string in the equivalent 3d gauged sigma model.

Since the endpoints of a $\mathfrak{P}^{\theta}(\R_{\tau}, \mathfrak{B}_2)$-string are now denoted as $\Gamma^{**}(\theta, \mathfrak{B}_2)$, we can also denote and specify the former at $\xi = \pm \infty$ as $\Sigma^{IJ, KL}_{\pm}(t, \theta, \mathfrak{B}_3)$,\footnote{%
  Just like the explanation in \autoref{ft:m3 x r2:center of mass}, the $\xi$-invariant $\mathfrak{P}^{\theta}(\R_{\tau}, \mathfrak{B}_2)$-strings can be fixed at $\xi = \pm \infty$ by adding physically-inconsequential $\hat{\mathcal{Q}}$-exact terms to the SQM action.
  \label{ft:m2 x r3:center of mass}
}
where its left and right endpoints would be $\Gamma^{IJ}(\theta, \mathfrak{B}_2)$ and $\Gamma^{KL}(\theta, \mathfrak{B}_2)$, respectively.

\begin{figure}
  \centering
  \begin{tikzpicture}
    \coordinate (lt) at (0,4) {};
    \coordinate (rt) at (4,4) {}
    edge node[pos=0.5, above] {$\Sigma^{IJ, KL}_+(t, \theta, \mathfrak{B}_2)$}
    (lt) {};
    \coordinate (lb) at (0,0) {}
    edge node[pos=0.5, left] {$\Gamma^{IJ}(\theta, \mathfrak{B}_2)$}
    (lt) {};
    \coordinate (rb) at (4,0) {}
    edge node[pos=0.5, below] {$\Sigma^{IJ, KL}_-(t, \theta, \mathfrak{B}_2)$}
    (lb) {}
    edge node[pos=0.5, right] {$\Gamma^{KL}(\theta, \mathfrak{B}_2)$}
    (rt) {};
    \draw (lb) -- (lt);
    \draw (rb) -- (rt);
    \coordinate (co) at (5,0);
    \coordinate (cx) at (5.5,0);
    \node at (cx) [right=2pt of cx] {$t$};
    \coordinate (cy) at (5,0.5);
    \node at (cy) [above=2pt of cy] {$\xi$};
    \draw[->] (co) -- (cx);
    \draw[->] (co) -- (cy);
  \end{tikzpicture}
  \caption{BPS worldsheet with $\mathfrak{P}^{\theta}(\R_\tau, \mathfrak{B}_2)$-strings $\Sigma^{IJ, KL}_\pm(t, \theta, \mathfrak{B}_2)$ and boundaries corresponding to $\Gamma^{IJ}(\theta, \mathfrak{B}_2)$ and $\Gamma^{KL}(\theta, \mathfrak{B}_2)$.
  }
  \label{fig:m2 x r3:bps worldsheet}
\end{figure}
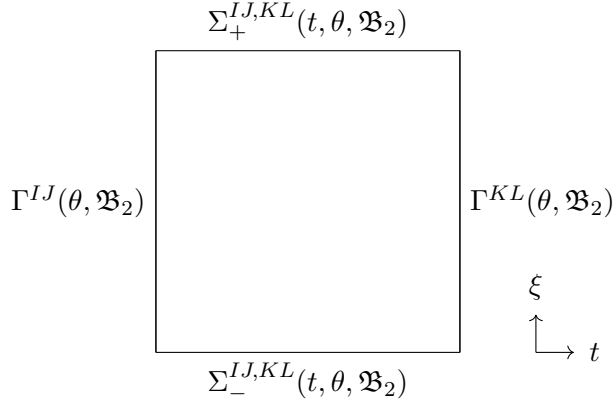

As the $\Gamma^{**}(\theta, \mathfrak{B}_2)$'s are $\xi$-invariant and therefore, have the same values for all $\xi$, we have BPS worldsheets of the kind shown in \autoref{fig:m2 x r3:bps worldsheet}.

\subtitle{The 2d Model on $\R^2$ and an Open String Theory in $\mathfrak{P}(\R_{\tau}, \mathfrak{B}_2)$}

Thus, like in \autoref{sec:m3 x r2:2d}, one can understand the 2d gauged sigma model with target $\mathfrak{P}(\R_{\tau}, \mathfrak{B}_2)$ to define an open string theory in $\mathfrak{P}(\R_{\tau}, \mathfrak{B}_2)$ with \emph{effective} worldsheet and boundaries shown in \autoref{fig:m2 x r3:bps worldsheet}, where $\xi$ and $t$ are the temporal and spatial directions, respectively.

\subtitle{BPS Worldvolumes of the 3d Model}

The 3d analogue of a BPS worldsheet is a BPS worldvolume.
In particular, BPS worldvolumes of the 3d gauged sigma model with target $\mathfrak{B}_2$ will correspond to its classical trajectories.
Specifically, these are defined by setting to zero the expression within the squared terms in \eqref{eq:m2 x r3:gm:3d model:action}, i.e.,
\begin{equation}
  \label{eq:m2 x r3:2d-3d:bps worldvolume}
  \begin{aligned}
    - P^a(\theta)
    =
    & \Dv{}{\xi} \left(
        \widetilde{\mathsf{A}}^a
        + e^{\mathbf{e}_0 \theta} \widetilde{\mathsf{B}}^a \cos(\theta)
      \right)
      + \Dv{}{t} \left(
        \mathbf{e}_0 \widetilde{\mathsf{A}}^a
        + e^{\mathbf{e}_0 \theta} \widetilde{\mathsf{B}}^a \sin(\theta)
      \right)
      + e^{\mathbf{e}_0 \theta} \hat{\mathbf{e}_2} \Dv{}{\tau} \left(
        \widetilde{\mathsf{A}}^a
        + \widetilde{\mathsf{B}}^a
      \right)
    \\
    & + e^{\mathbf{e}_0 \theta} F_{t\xi}
      + e^{\mathbf{e}_0 \theta} \hat{\mathbf{e}}_2 F_{t\tau} \cos(\theta)
      - e^{\mathbf{e}_0 \theta} \hat{\mathbf{e}}_2 F_{\xi\tau} \sin(\theta)
      \, ,
    \\
    - Q^a(\theta)
    =
    & \Dv{}{\xi} \left(
        \widetilde{\mathsf{B}}^a
        - e^{\mathbf{e}_0 \theta} \widetilde{\mathsf{A}}^a \cos(\theta)
      \right)
      + \Dv{}{t} \left(
        \mathbf{e}_0 \widetilde{\mathsf{B}}^a
        - e^{\mathbf{e}_0 \theta} \widetilde{\mathsf{A}}^a \sin(\theta)
      \right)
      + e^{\mathbf{e}_0 \theta} \hat{\mathbf{e}}_2 \Dv{}{\tau} \left(
        \widetilde{\mathsf{B}}^a
        - \widetilde{\mathsf{A}}^a
      \right)
      \, .
  \end{aligned}
\end{equation}

\subtitle{BPS Worldvolumes with Boundaries Labeled by $\mathfrak{B}_2^{\theta}$-strings, and Edges Labeled by $G_{\OO}$-BF Configurations on $M_2$}

The boundaries and edges of the BPS worldvolumes are traced out by the edges and vertices of the $\mathfrak{B}_2^{\theta}$-membranes, respectively, as they propagate in $\xi$.
As we have seen at the end of \autoref{sec:m2 x r3:go-bf}, these edges and vertices would correspond to $\mathfrak{B}_2^{\theta}$-strings and $\theta$-deformed $G_{\OO}$-BF configurations on $M_2$, respectively.

\begin{figure}
  \centering
  \begin{tikzpicture}[%
    auto,%
    every edge/.style={draw},%
    relation/.style={scale=1, sloped, anchor=center, align=center,%
      color=black},%
    vertRelation/.style={scale=1, anchor=center, align=center},%
    dot/.style={circle, fill, minimum size=2*\radius, node contents={}, inner sep=0pt, gray!40},%
    ]
    \let\radius\undefined
    \newlength{\radius}
    \setlength{\radius}{1mm}
    \node at (2, 2) {$\sigma^{IJ, KL}(\tau, t, \theta, \mathfrak{B}_2)$};
    \node (lt) at (0,4) [dot];
    \node (rt) at (4,4) [dot];
    \node (lb) at (0,0) [dot];
    \node (rb) at (4,0) [dot];
    \node at (lb) [below left]  {$\mathfrak{C}^I_{\text{BF}_{\OO}}(\theta)$};
    \node at (lt) [above left]  {$\mathfrak{C}^J_{\text{BF}_{\OO}}(\theta)$};
    \node at (rb) [below right]  {$\mathfrak{C}^K_{\text{BF}_{\OO}}(\theta)$};
    \node at (rt) [above right]  {$\mathfrak{C}^L_{\text{BF}_{\OO}}(\theta)$};
    \draw
    (lb)
    edge[red!40] node[pos=0.5, left, black] {$\gamma^{IJ}(\tau, \theta, \mathfrak{B}_2)$}
    (lt)
    (lt) -- (rt)
    (rt)
    edge[red!40] node[pos=0.5, right, black] {$\gamma^{KL}(\tau, \theta, \mathfrak{B}_2)$}
    (rb)
    (rb) -- (lb)
    ;
    \coordinate (co) at (7,0);
    \coordinate (cx) at (7.5,0);
    \node at (cx) [right=2pt of cx] {$t$};
    \coordinate (cy) at (7,0.5);
    \node at (cy) [above=2pt of cy] {$\tau$};
    \node at (-3,0) {};
    \draw[->] (co) -- (cx);
    \draw[->] (co) -- (cy);
  \end{tikzpicture}
  \caption{%
    $\mathfrak{B}_2^{\theta}$-membrane $\sigma^{IJ, KL}(\tau, t, \theta, \mathfrak{A}_2)$ with edges being $\mathfrak{B}_2^{\theta}$-strings $\gamma^{IJ}(\tau, \theta, \mathfrak{B}_2)$ and $\gamma^{KL}(\tau, \theta, \mathfrak{B}_2)$, and vertices corresponding to $\mathfrak{C}^I_{\text{BF}_{\OO}}(\theta)$, $\mathfrak{C}^J_{\text{BF}_{\OO}}(\theta)$, $\mathfrak{C}^K_{\text{BF}_{\OO}}(\theta)$, and $\mathfrak{C}^L_{\text{BF}_{\OO}}(\theta)$.
  }
  \label{fig:m2 x r3:frakB-sheet}
\end{figure}
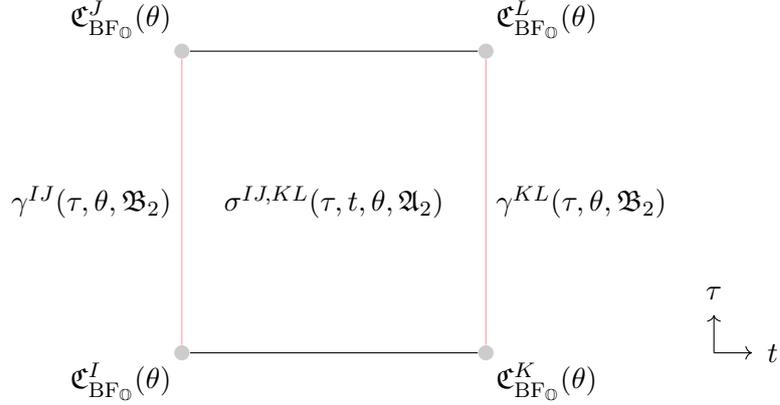

This means that we can denote and specify any $\mathfrak{B}_2^{\theta}$-membrane at $\xi = \pm \infty$ as $\sigma^{IJ, KL}_{\pm}(t, \tau, \theta, \mathfrak{B}_3)$,\footnote{%
  Just like the $\mathfrak{P}^{\theta}(\R_{\tau}, \mathfrak{B}_2)$-strings, the $\xi$-invariant $\mathfrak{B}_2^{\theta}$-membranes can be fixed at $\xi = \pm \infty$ by adding physically inconsequential $\hat{\mathcal{Q}}$-exact terms to the SQM action.
  \label{ft:m2 x r3:center of mass for b-membrane}
}
where
(i) its left and right edges correspond to the $\mathfrak{B}_2^{\theta}$-strings $\gamma^{IJ}(\tau, \theta, \mathfrak{B}_2)$ and $\gamma^{KL}(\tau, \theta, \mathfrak{B}_2)$, respectively,
and (ii) its four vertices would correspond to $\mathfrak{C}^I_{\text{BF}_{\OO}}(\theta)$, $\mathfrak{C}^J_{\text{BF}_{\OO}}(\theta)$, $\mathfrak{C}^K_{\text{BF}_{\OO}}(\theta)$, and $\mathfrak{C}^L_{\text{BF}_{\OO}}(\theta)$.
This is illustrated in \autoref{fig:m2 x r3:frakB-sheet}.

\begin{figure}
  \centering
  \begin{tikzpicture}[%
    auto,%
    relation/.style={scale=1, sloped, anchor=center, align=center, color=black},%
    vertRelation/.style={scale=1, anchor=center, align=center},%
    slanted/.style={rotate=\yAngle, xslant=0.8, scale=0.9},%
    coil/.style={thick,decorate,decoration={coil,aspect=0,segment length=\coilLength, amplitude=\coilAmp}},%
    ]
    \let\side\undefined 
    \newlength{\side}
    \setlength{\side}{2cm}
    \def\xyRatio{4}
    \def\yzRatio{1.6}
    \let\coordLength\undefined
    \newlength{\coordLength}
    \setlength{\coordLength}{0.5cm}
    \def\yAngle{45}
    \let\coilAmp\undefined
    \newlength{\coilAmp}
    \setlength{\coilAmp}{1mm}
    \let\coilLength\undefined
    \newlength{\coilLength}
    \setlength{\coilLength}{7mm}
    \coordinate (tll) at ($(- \xyRatio*\side/2, \yzRatio*\side)$); 
    \coordinate (trl) at ($(tll) + (\side, \side)$);
    \coordinate (tbl) at ($(tll)!0.5!(trl)$);
    \coordinate (brl) at ($(trl) + (0, - \yzRatio*\side)$);
    \coordinate (bll) at ($(tll) + (0, - \yzRatio*\side)$);
    \coordinate (bbl) at ($(bll)!0.5!(brl)$);
    \coordinate (tlr) at ($(tll) + (\xyRatio*\side, 0)$);
    \coordinate (trr) at ($(tlr) + (\side, \side)$);
    \coordinate (tbr) at ($(tlr)!0.5!(trr)$);
    \coordinate (brr) at ($(trr) + (0, - \yzRatio*\side)$);
    \coordinate (blr) at ($(tlr) + (0, - \yzRatio*\side)$);
    \coordinate (bbr) at ($(blr)!0.5!(brr)$);
    \coordinate (tlc) at ($(tll)!0.5!(tlr)$);
    \coordinate (trc) at ($(tlc) + (\side, \side)$);
    \coordinate (tbc) at ($(tlc)!0.5!(trc)$);
    \coordinate (blc) at ($(tlc) + (0, - \yzRatio*\side)$);
    \coordinate (brc) at ($(trc) + (0, - \yzRatio*\side)$);
    \coordinate (bbc) at ($(blc)!0.5!(brc)$);
    \coordinate (cco) at ($(blr) + (2*\side, 0)$);
    \coordinate (ccx) at ($(cco) + (\coordLength, 0)$);
    \coordinate (ccy) at ($(cco) + cos(\yAngle)*(\coordLength, 0) + sin(\yAngle)*(0, \coordLength)$);
    \coordinate (ccz) at ($(cco) + (0, \coordLength)$);
    \draw
    (cco) edge[->] (ccx)
    (cco) edge[->] (ccy)
    (cco) edge[->] (ccz)
    ;
    \node at ($(brl) + (- 2*\side, 0)$) {}; 
    \node at (ccx) [right=2pt of ccx] {$t$};
    \node at (ccy) [above right=2pt of ccy] {$\xi$};
    \node at (ccz) [above=2pt of ccz] {$\tau$};
    \draw
    (bll) edge (tll)
    (tll) edge (trl)
    (trl) edge[dashed] (brl)
    (brl) edge[dashed] (bll)
    (blr) edge (tlr)
    (tlr) edge (trr)
    (trr) edge (brr)
    (brr) edge (blr)
    (bll) edge (blr)
    (tll) edge (tlr)
    (brl) edge[dashed] (brr)
    (trl) edge (trr)
    ;
    \draw[pattern={Lines[angle={\yAngle+20},distance=4pt]},pattern color=red!40,draw=none]
    (tll) -- (trl) -- (brl) -- (bll) -- (tll);
    \node (gammaIJ) at ($(tll)!0.5!(bll) + (- \side, 0)$) {$\gamma^{IJ}(\tau, \theta, \mathfrak{B}_2)$};
    \draw
    (gammaIJ) edge [out=90, in=180] ($(tll)!0.15!(bll)$)
    ($(tll)!0.15!(bll)$) edge [dashed, ->] ($(tbl)!0.5!(bbl)$)
    ;
    \draw[pattern={Lines[angle=90,distance=4pt]},pattern color=green!40,draw=none]
    (trl) -- (trr) -- (brr) -- (brl) -- (trl);
    \node (sigma-plus) at ($(trl)!0.5!(trr) + (\side/2, \side/2)$) {$\sigma^{IJ, KL}_+(t, \tau, \theta, \mathfrak{B}_2)$};
    \draw
    (sigma-plus) edge ($(trl)!0.5!(trr)$)
    ($(trl)!0.5!(trr)$) edge[->, dashed] ($(trl)!0.5!(trr) + (-\side/2, -\side/2)$)
    ;
    \draw[pattern={Lines[angle={\yAngle+20},distance=4pt]},pattern color=red!40,draw=none]
    (tlr) -- (trr) -- (brr) -- (blr) -- (tlr);
    \node (gammaKL) at ($(trr)!0.5!(brr) + (\side, 0)$) {$\gamma^{KL}(\tau, \theta, \mathfrak{B}_2)$};
    \draw
    (gammaKL) edge [out=200, in=20, ->] ($(tbr)!0.5!(bbr)$);
    ;
    \draw[pattern={Lines[angle=0,distance=4pt]},pattern color=blue!40,draw=none]
    (tll) -- (tlr) -- (blr) -- (bll) -- (tll);
    \node (sigma-minus) at ($(bll)!0.5!(blr) + (-\side/2, -\side/2)$) {$\sigma^{IJ, KL}_-(t, \tau, \theta, \mathfrak{B}_2)$};
    \draw
    (sigma-minus) edge[->] ($(bll)!0.5!(blr) + (\side/2, \side/2)$)
    ;
    \draw[fill=gray!40, gray!40, rotate around={45:(bbl)}] (bbl) ellipse (2 and 0.5);
    \node[slanted] (CI) at (bbl) {$\mathfrak{C}^I_{\text{BF}_{\OO}}(\theta)$};
    \draw[fill=gray!40, gray!40, rotate around={45:(tbl)}] (tbl) ellipse (2 and 0.5);
    \node[slanted] (CJ) at (tbl) {$\mathfrak{C}^J_{\text{BF}_{\OO}}(\theta)$};
    \draw[fill=gray!40, gray!40, rotate around={45:(bbr)}] (bbr) ellipse (2 and 0.5);
    \node[slanted] (CK) at (bbr) {$\mathfrak{C}^K_{\text{BF}_{\OO}}(\theta)$};
    \draw[fill=gray!40, gray!40, rotate around={45:(tbr)}] (tbr) ellipse (2 and 0.5);
    \node[slanted] (CL) at (tbr) {$\mathfrak{C}^L_{\text{BF}_{\OO}}(\theta)$};
    \draw
    ($(tll) + (\side/4, 0)$) edge ($(tlr) + (- \side/4, 0)$)
    ($(blr) + (0, \side/4)$) edge ($(tlr) + (0, - \side/5)$)
    ($(bll) + (0, \side/4)$) edge ($(tll) + (0, - \side/5)$)
    ;
  \end{tikzpicture}
  \caption{%
    BPS worldvolume with $\mathfrak{B}_2^{\theta}$-membranes $\sigma^{IJ, KL}_\pm(t, \tau, \theta, \mathfrak{B}_2)$, boundaries labeled by $\mathfrak{B}_2^{\theta}$-strings $\gamma^{IJ}(\tau, \theta, \mathfrak{B}_2)$ and $\gamma^{KL}(\tau, \theta, \mathfrak{B}_2)$, and edges labeled by $\mathfrak{C}^I_{\text{BF}_{\OO}}(\theta)$, $\mathfrak{C}^J_{\text{BF}_{\OO}}(\theta)$, $\mathfrak{C}^K_{\text{BF}_{\OO}}(\theta)$, and $\mathfrak{C}^L_{\text{BF}_{\OO}}(\theta)$.
  }
  \label{fig:m2 x r3:bps worldvolume}
\end{figure}
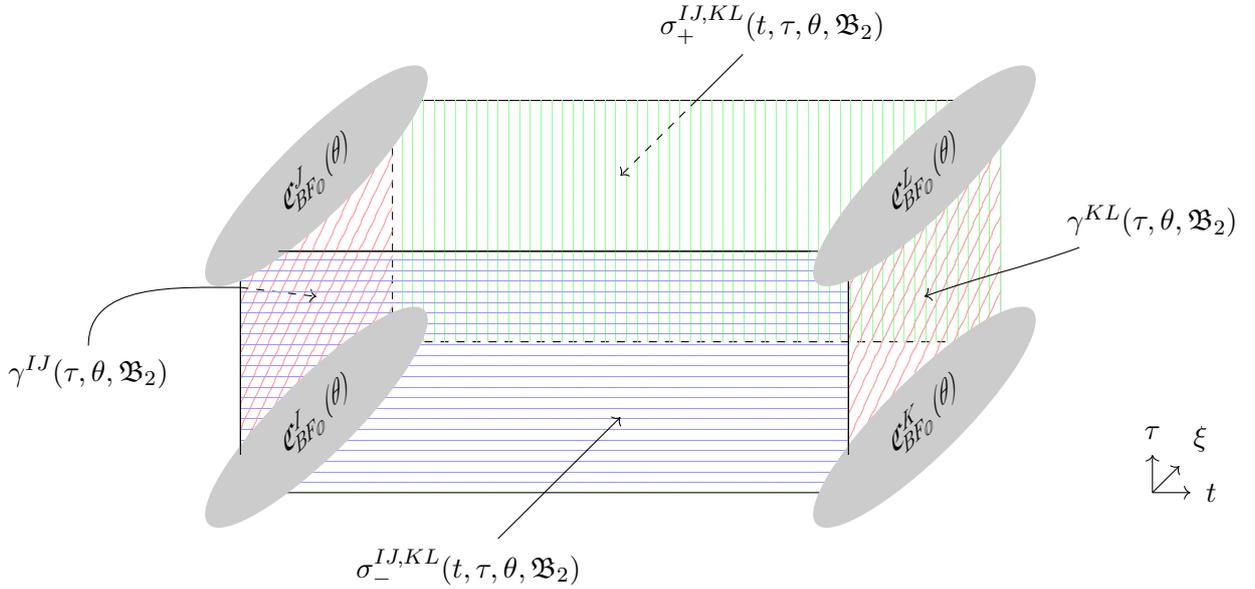

As the $\mathfrak{C}^{**}_{\text{BF}_{\OO}}(\theta)$'s and $\gamma^{**}(\tau, \theta, \mathfrak{B}_2)$'s are $\xi$-invariant and therefore, have the same values for all $\xi$, we have BPS worldvolumes of the kind shown in \autoref{fig:m2 x r3:bps worldvolume}

\subtitle{The 3d Model on $\R^3$ and an Open Membrane Theory in $\mathfrak{B}_2$}

Hence, one can understand the 3d gauged sigma model on $\R^3$ with target $\mathfrak{B}_2$ to define an open membrane theory in $\mathfrak{B}_2$, with \emph{effective} worldvolume, boundaries, and edges shown in \autoref{fig:m2 x r3:bps worldvolume}, where $\xi$ is the temporal direction, and $(t, \tau)$ are spatial directions, respectively.

\subtitle{SQM Flow Lines, BPS Worldsheets, and BPS Worldvolumes}

In short, the classical trajectories of GM theory on $M_2 \times \R^3$ are the SQM flow lines in the equivalent 1d SQM, which will correspond to BPS worldsheets in the 2d gauged sigma model with target $\mathfrak{P}(\R_{\tau}, \mathfrak{B}_2)$ of the kind shown in \autoref{fig:m2 x r3:bps worldsheet}, which, in turn, will correspond to BPS worldvolumes in the 3d gauged sigma model with target $\mathfrak{B}_2$ of the kind shown in \autoref{fig:m2 x r3:bps worldvolume}.

\subsection{String Theory, the GM Partition Function, and an Orlov-type \texorpdfstring{$A_{\infty}$}{A-infinity}-1-category of \texorpdfstring{$\mathfrak{B}_2^{\theta}$}{B2-theta}-strings}
\label{sec:m2 x r3:orlov}

\subtitle{The 2d Model as a 2d Gauged B-twisted LG Model}

Notice that we can also express \eqref{eq:m2 x r3:gm:2d model:action} as
\begin{equation}
  \label{eq:m2 x r3:2d-lg:action}
  \begin{aligned}
    S_{\text{2d-LG}, \mathfrak{P}(\R_{\tau}, \mathfrak{B}_2)}
    & = \frac{1}{e^2} \int_{\R^2} dt d\xi \left(
      \left| \mathcal{X}^u + p^u + P^u \right|^2
      + \left| \mathcal{Y}^u + q^u + Q^u \right|^2
      + \dots
      \right)
    \\
    & = \frac{1}{e^2} \int_{\R^2} dt d\xi \Bigg(
      \left| \mathcal{X}^u
      + \mathbf{e}_0 e^{\mathbf{e}_0 \theta} g_{\mathfrak{P}(\R_{\tau}, \mathfrak{B}_2)}^{uv} \left(
          \pdv{W_2}{\widetilde{\mathsf{A}}^v}
          + \pdv{W_2}{(\hat{\mathbf{e}}_2 (\mathring{A}_{\tau})^v)}
        \right)
      \right|^2
    \\
    & \qquad \qquad \qquad \quad
      + \left| \mathcal{Y}^u
        + \mathbf{e}_0 e^{\mathbf{e}_0 \theta} g_{\mathfrak{P}(\R_{\tau}, \mathfrak{B}_2)}^{uv} \left(
          \pdv{W_2}{\widetilde{\mathsf{B}}^v}
          + \pdv{W_2}{(\mathbf{e}_0 \hat{\mathbf{e}}_2 (\mathring{A}_{\tau})^v)}
        \right)
      \right|^2
      + \dots
      \Bigg)
      \, ,
  \end{aligned}
\end{equation}
where $W_2(\widetilde{\mathsf{A}}, \widetilde{\mathsf{B}}, \mathring{A}_{\tau})$ is a holomorphic superpotential, and
\begin{equation}
  \label{eq:m2 x r3:2d-lg:action:components}
  \begin{aligned}
    \mathcal{X}^u
    &= \Dv{}{\xi} \left(
        \widetilde{\mathsf{A}}^u
        + e^{\mathbf{e}_0 \theta} \widetilde{\mathsf{B}}^u \cos(\theta)
        - e^{\mathbf{e}_0 \theta} \hat{\mathbf{e}}_2 (\mathring{A}_{\tau})^u \sin(\theta)
      \right)
    \\
    & \quad
      + \Dv{}{t} \left(
        \mathbf{e}_0 \widetilde{\mathsf{A}}^u
        + e^{\mathbf{e}_0 \theta} \widetilde{\mathsf{B}}^u \sin(\theta)
        + e^{\mathbf{e}_0 \theta} \hat{\mathbf{e}}_2 (\mathring{A}_{\tau})^u \cos(\theta)
      \right)
      + e^{\mathbf{e}_0 \theta} F_{t\xi}
      \, ,
    \\
    \mathcal{Y}^u
    &= \Dv{}{\xi} \left(
        \widetilde{\mathsf{B}}^u
        - e^{\mathbf{e}_0 \theta} \widetilde{\mathsf{A}}^u \cos(\theta)
      \right)
      + \Dv{}{t} \left(
        \mathbf{e}_0 \widetilde{\mathsf{B}}^u
        - e^{\mathbf{e}_0 \theta} \widetilde{\mathsf{A}}^u \sin(\theta)
      \right)
      \, .
  \end{aligned}
\end{equation}

Just like in \autoref{sec:m3 x r2:orlov}, when we apply specific values of $\theta$ to the conditions obtained by setting to zero the expression within the squared terms of \eqref{eq:m2 x r3:2d-lg:action} whilst noting that the physical theory ought to be symmetric under a variation of $\theta$, we find that it can also be equivalently expressed as
\begin{equation}
  \label{eq:m2 x r3:2d-lg:action:b-twist}
  \begin{aligned}
    S_{\text{2d-LG}, \mathfrak{P}(\R_{\tau}, \mathfrak{B}_2)}
    = \frac{1}{e^2} \int_{\R^2} \Bigg(
    & \left| D \widetilde{\mathsf{A}} \right|^2
      + \left| D \widetilde{\mathsf{B}} \right|^2
      + \left| D \mathring{A}_{\tau} \right|^2
      + \left| \mathbf{e}_0 F
        - g_{\mathfrak{P}(\R_{\tau}, \mathfrak{B}_2)}^{uv} \left(
          \pdv{W_2}{\widetilde{\mathsf{A}}^v}
          + \pdv{W_2}{(\hat{\mathbf{e}}_2 (\mathring{A}_{\tau})^v)}
        \right)
      \right|^2
    \\
    & + \left| g_{\mathfrak{P}(\R_{\tau}, \mathfrak{B}_2)}^{uv} \left(
          \pdv{W_2}{\widetilde{\mathsf{B}}^v}
          + \pdv{W_2}{(\mathbf{e}_0 \hat{\mathbf{e}}_2 (\mathring{A}_{\tau})^v)}
        \right)
      \right|^2
      + \dots
      \Bigg)
      \, .
  \end{aligned}
\end{equation}
In other words, the 2d gauged sigma model with target $\mathfrak{P}(\R_{\tau}, \mathfrak{B}_2)$ can also be interpreted as a 2d gauged \emph{B-twisted} LG model in $\mathfrak{P}(\R_{\tau}, \mathfrak{B}_2)$ with holomorphic superpotential $W_2(\widetilde{\mathsf{A}}, \widetilde{\mathsf{B}}, \mathring{A}_{\tau})$.

By setting $d_{\xi} \widetilde{\mathsf{A}}^u = 0 = d_{\xi} \widetilde{\mathsf{B}}^u$ and $A_t, \mathring{A}_{\tau}, A_{\xi} \rightarrow 0$ in the expression within the squared terms in \eqref{eq:m2 x r3:2d-lg:action}, we can (via \eqref{eq:m2 x r3:2d-lg:action:components}) read off the LG $\mathfrak{P}^{\theta}(\R_{\tau}, \mathfrak{B}_2)$-string  (corresponding to $\Sigma^{IJ, KL}_{\pm}(t, \theta, \mathfrak{B}_2)$) equations (that re-express \eqref{eq:m2 x r3:go-bf:p-strings} with $A_t, \mathring{A}_{\tau}, A_{\xi} \rightarrow 0$) as
\begin{equation}
  \label{eq:m2 x r3:2d-lg:p-string}
  \begin{aligned}
    \dv{}{t} \left(
      \mathbf{e}_0 \widetilde{\mathsf{A}}^u
      + e^{\mathbf{e}_0 \theta} \widetilde{\mathsf{B}}^u \sin(\theta)
    \right)
    &= - \mathbf{e}_0 e^{\mathbf{e}_0 \theta} g_{\mathfrak{P}(\R_{\tau}, \mathfrak{B}_2)}^{uv} \left(
        \pdv{W_2}{\widetilde{\mathsf{A}}^v}
        + \pdv{W_2}{(\hat{\mathbf{e}}_2 (\mathring{A}_{\tau})^v)}
      \right)_{\mathring{A}_{\tau} = 0}
      \, ,
    \\
    \dv{}{t} \left(
      \mathbf{e}_0 \widetilde{\mathsf{B}}^u
      - e^{\mathbf{e}_0 \theta} \widetilde{\mathsf{A}}^u \sin(\theta)
    \right)
    &= - \mathbf{e}_0 e^{\mathbf{e}_0 \theta} g_{\mathfrak{P}(\R_{\tau}, \mathfrak{B}_2)}^{uv} \left(
        \pdv{W_2}{\widetilde{\mathsf{B}}^v}
        + \pdv{W_2}{(\mathbf{e}_0 \hat{\mathbf{e}}_2 (\mathring{A}_{\tau})^v)}
      \right)_{\mathring{A}_{\tau} = 0}
      \, ,
  \end{aligned}
\end{equation}
where the subscript ``$\mathring{A}_{\tau} = 0$'' means to set $\mathring{A}_{\tau}$ to zero in those terms.

By setting $d_t \widetilde{\mathsf{A}}^u = 0 = d_t \widetilde{\mathsf{B}}^u$ in \eqref{eq:m2 x r3:2d-lg:p-string}, we get the LG $\mathfrak{P}^{\theta}(\R_{\tau}, \mathfrak{B}_2)$-string endpoint (corresponding to $\Sigma^{IJ, KL}(\pm \infty, \theta, \mathfrak{B}_2)$) equations (that re-express \eqref{eq:m2 x r3:go-bf:non-constant paths}) as
\begin{equation}
  \label{eq:m2 x r3:2d-lg:p-string:endpoint}
  \begin{aligned}
    \mathbf{e}_0 e^{\mathbf{e}_0 \theta} g_{\mathfrak{P}(\R_{\tau}, \mathfrak{B}_2)}^{uv} \left(
      \pdv{W_2}{\widetilde{\mathsf{A}}^v}
      + \pdv{W_2}{(\hat{\mathbf{e}}_2 (\mathring{A}_{\tau})^v)}
    \right)_{\mathring{A}_{\tau} = 0}
    &= 0
      \, ,
    \\
    \mathbf{e}_0 e^{\mathbf{e}_0 \theta} g_{\mathfrak{P}(\R_{\tau}, \mathfrak{B}_2)}^{uv} \left(
      \pdv{W_2}{\widetilde{\mathsf{B}}^v}
      + \pdv{W_2}{(\mathbf{e}_0 \hat{\mathbf{e}}_2 (\mathring{A}_{\tau})^v)}
    \right)_{\mathring{A}_{\tau} = 0}
    &= 0
      \, .
  \end{aligned}
\end{equation}

Recall from the end of \autoref{sec:m2 x r3:go-bf} that we are only considering certain $M_2$ and $G$ such that the $\theta$-deformed $G_{\OO}$-BF configurations on $M_2$ are isolated and non-degenerate.
Next, recall also that such configurations will correspond to the endpoints of the $\mathfrak{B}_2^{\theta}$-strings; therefore, just like their endpoints, these $\mathfrak{B}_2^{\theta}$-strings would be isolated and non-degenerate.
As these $\mathfrak{B}_2^{\theta}$-strings will correspond here, to the endpoints of the LG $\mathfrak{P}^{\theta}(\R_{\tau}, \mathfrak{B}_2)$-strings, this means that the endpoints and thus the configurations satisfying \eqref{eq:m2 x r3:2d-lg:p-string:endpoint} will also be isolated and non-degenerate.
From \eqref{eq:m2 x r3:2d-lg:p-string:endpoint}, one can see that these configurations are critical points of $W_2(\widetilde{\mathsf{A}}, \widetilde{\mathsf{B}}, \mathring{A}_{\tau})$; that they are isolated and non-degenerate means that $W_2(\widetilde{\mathsf{A}}, \widetilde{\mathsf{B}}, \mathring{A}_{\tau})$ can be regarded as a (holomorphic) Morse function in $\mathfrak{P}(\R_{\tau}, \mathfrak{B}_2)$.

\subtitle{The 2d Gauged LG Model as an LG SQM}

With suitable rescalings, we can recast \eqref{eq:m2 x r3:2d-lg:action} as a 1d LG SQM (that re-expresses \eqref{eq:m2 x r3:gm:sqm:action}), where its action will be given by\footnote{%
  The same steps in \autoref{ft:m2 x r3:integrate out gauge field to sqm} are applied to arrive at the following expression.
  \label{ft:m2 x r3:2d-lg:sqm no gauge}
}
\begin{equation}
  \label{eq:m2 x r3:2d-lg:sqm:action}
  \begin{aligned}
    & S_{\text{2d-LG SQM}, \mathfrak{P}(\R_t, \mathfrak{P}(\R_{\tau}, \mathfrak{B}_2))}
    \\
    & = \frac{1}{e^2} \int d\xi \left(
      \left| \partial_{\xi} \varphi_+^{\alpha} + g^{\alpha\beta}_{\mathfrak{P}(\R_t, \mathfrak{P}(\R_{\tau}, \mathfrak{B}_2))} \pdv{H_2}{\varphi_+^{\beta}} \right|^2
      +  \left| \partial_{\xi} \varphi_-^{\alpha} + g^{\alpha\beta}_{\mathfrak{P}(\R_t, \mathfrak{P}(\R_{\tau}, \mathfrak{B}_2))} \pdv{H_2}{\varphi_-^{\beta}} \right|^2
      + \dots
      \right)
      \, ,
  \end{aligned}
\end{equation}
where $H_2(\varphi_+, \varphi_-)$ is the \emph{real-valued} potential in $\mathfrak{P}(\R_t, \mathfrak{P}(\R_{\tau}, \mathfrak{B}_2))$ with $\varphi_+$ and $\varphi_-$ defined as in \autoref{sec:m2 x r3:gm}, and the subscript ``2d-LG SQM, $\mathfrak{P}(\R_t, \mathfrak{P}(\R_{\tau}, \mathfrak{B}_2))$'' is to specify that it is a 1d SQM in $\mathfrak{P}(\R_t, \mathfrak{P}(\R_{\tau}, \mathfrak{B}_2))$ obtained from the equivalent 2d LG model.
We will also refer to this \emph{1d} LG SQM as ``2d-LG SQM'' in the rest of this subsection.

The 2d-LG SQM will localize onto configurations that \emph{simultaneously} set to zero the LHS and RHS of the expression within the squared terms in \eqref{eq:m2 x r3:2d-lg:sqm:action}.
In other words, it will localize onto $\xi$-invariant critical points of $H_2(\varphi_+, \varphi_-)$ that will correspond, when $A_t, \mathring{A}_{\tau}, A_{\xi} \rightarrow 0$, to the LG $\mathfrak{P}^{\theta}(\R_{\tau}, \mathfrak{B}_2)$-strings defined by \eqref{eq:m2 x r3:2d-lg:p-string}.
For our choice of $M_2$ and $G$,  just like their endpoints, the LG $\mathfrak{P}^{\theta}(\R_{\tau}, \mathfrak{B}_2)$-strings will be isolated and non-degenerate.
Thus, $H_2(\varphi_+, \varphi_-)$ can be regarded as a real-valued Morse functional in $\mathfrak{P}(\R_t, \mathfrak{P}(\R_{\tau}, \mathfrak{B}_2))$.

\subtitle{String Theory from the 2d B-twisted LG Model}

Just like in \autoref{sec:m3 x r2:orlov}, the 2d gauged B-twisted LG model in $\mathfrak{P}(\R_{\tau}, \mathfrak{B}_2)$ with action \eqref{eq:m2 x r3:2d-lg:action} can be interpreted as a open string theory in $\mathfrak{P}(\R_{\tau}, \mathfrak{B}_2)$.

The dynamics of this open string theory in $\mathfrak{P}(\R_{\tau}, \mathfrak{B}_2)$ will be governed by the BPS worldsheet equations determined by setting to zero the expression within the squared terms in \eqref{eq:m2 x r3:2d-lg:action}, where $(\widetilde{\mathsf{A}}^u, \widetilde{\mathsf{B}}^u, (\mathring{A}_{\tau})^u)$ are scalars on the worldsheet corresponding to the holomorphic coordinates of $\mathfrak{P}(\R_{\tau}, \mathfrak{B}_2)$.
At an arbitrary instant in time whence $d_{\xi} \widetilde{\mathsf{A}}^u = d_{\xi} \widetilde{\mathsf{B}}^u = 0 = d_{\xi} (\mathring{A}_{\tau})^u = d_{\xi} A_t$ therein, the dynamics of $(\widetilde{\mathsf{A}}^u, \widetilde{\mathsf{B}}^u, (\mathring{A}_{\tau})^u)$ and the 2d gauge fields $(A_t, A_{\xi})$ along $t$ will be governed by the string equations
\begin{equation}
  \label{eq:m2 x r3:2d-lg:bps string}
  \begin{aligned}
    & \dv{}{t} \left(
      \mathbf{e}_0 \widetilde{\mathsf{A}}^u
      + e^{\mathbf{e}_0 \theta} \widetilde{\mathsf{B}}^u \sin(\theta)
      + e^{\mathbf{e}_0 \theta} \hat{\mathbf{e}}_2 (\mathring{A}_{\tau})^u \cos(\theta)
      + e^{\mathbf{e}_0 \theta} A_{\xi}
    \right)
    \\
    & = - \left[ A_{\xi},
        \widetilde{\mathsf{A}}^u
        + e^{\mathbf{e}_0 \theta} \widetilde{\mathsf{B}}^u \cos(\theta)
        - e^{\mathbf{e}_0 \theta} \hat{\mathbf{e}}_2 (\mathring{A}_{\tau})^u \sin(\theta)
      \right]
      - \mathbf{e}_0 e^{\mathbf{e}_0 \theta} g_{\mathfrak{P}(\R_{\tau}, \mathfrak{B}_2)}^{uv} \left(
        \pdv{W_2}{\widetilde{\mathsf{A}}^v}
        + \pdv{W_2}{(\hat{\mathbf{e}}_2 (\mathring{A}_{\tau})^v)}
      \right)
    \\
    & \quad
      - \left[ A_t,
        \mathbf{e}_0 \widetilde{\mathsf{A}}^u
        + e^{\mathbf{e}_0 \theta} \widetilde{\mathsf{B}}^u \sin(\theta)
        + e^{\mathbf{e}_0 \theta} \hat{\mathbf{e}}_2 (\mathring{A}_{\tau})^u \cos(\theta)
        + e^{\mathbf{e}_0 \theta} A_{\xi}
      \right]
      \, ,
    \\
    & \dv{}{t} \left(
      \mathbf{e}_0 \widetilde{\mathsf{B}}^u
      - e^{\mathbf{e}_0 \theta} \widetilde{\mathsf{A}}^u \sin(\theta)
    \right)
    \\
    & =
      - \left[ A_{\xi},
        \widetilde{\mathsf{B}}^u
        - e^{\mathbf{e}_0 \theta} \widetilde{\mathsf{A}}^u \cos(\theta)
      \right]
      - \mathbf{e}_0 e^{\mathbf{e}_0 \theta} g_{\mathfrak{P}(\R_{\tau}, \mathfrak{B}_2)}^{uv} \left(
        \pdv{W_2}{\widetilde{\mathsf{B}}^v}
        + \pdv{W_2}{(\mathbf{e}_0 \hat{\mathbf{e}}_2 (\mathring{A}_{\tau})^v)}
      \right)
    \\
    & \quad
      - \left[ A_t,
        \mathbf{e}_0 \widetilde{\mathsf{B}}^u
        - e^{\mathbf{e}_0 \theta} \widetilde{\mathsf{A}}^u \sin(\theta)
      \right]
      \, .
  \end{aligned}
\end{equation}

\subtitle{Morphisms from $\Gamma^{IJ}(\theta, \mathfrak{B}_2)$ to $\Gamma^{KL}(\theta, \mathfrak{B}_2)$ as Ext-groups}

Repeating here the analysis in \autoref{sec:m3 x r2:orlov} with \eqref{eq:m2 x r3:2d-lg:sqm:action} as the action of the 2d-LG SQM, we find that we can interpret the LG $\mathfrak{P}^{\theta}(\R_{\tau}, \mathfrak{B}_2)$-string solution corresponding to $\Sigma^{IJ, KL}_{\pm}(t, \theta, \mathfrak{B}_2)$ as Ext-groups.
Specifically, a $\Sigma^{IJ, KL}_{\pm}(t, \theta, \mathfrak{B}_2)$-string pair, whose left and right endpoints correspond to $\Gamma^{IJ}(\theta, \mathfrak{B}_2)$ and $\Gamma^{KL}(\theta, \mathfrak{B}_2)$, respectively, can be identified as an LG $\mathfrak{P}^{\theta}(\R_{\tau}, \mathfrak{B}_2)$-string pair $\mathfrak{q}^{IJ, KL}_{W_2, \pm}(\theta) \in S^{IJ, KL}_{W_2}$, starting and ending on \emph{singular} fibers $\mathfrak{T}^{IJ}_{W_2}(\theta), \mathfrak{T}^{KL}_{W_2}(\theta) \in W_{2, \text{sg}}^{-1}(0)$, respectively, which are also described as zero-dimensional submanifolds of $\mathfrak{P}(\R_{\tau}, \mathfrak{B}_2)$.

This means that the 2d LG-SQM in $\mathfrak{P}(\R_t, \mathfrak{P}(\R_{\tau}, \mathfrak{B}_2))$ with action \eqref{eq:m2 x r3:2d-lg:sqm:action}, will physically realize a Floer homology that we shall name a $\mathfrak{P}(\R_{\tau}, \mathfrak{B}_2)$-2d-LG Floer homology.
The chains of the $\mathfrak{P}(\R_{\tau}, \mathfrak{B}_2)$-2d-LG Floer complex will be generated by LG $\mathfrak{P}^{\theta}(\R_{\tau}, \mathfrak{B}_2)$-strings which we can identify with $\mathfrak{q}^{**, **}_{W_2, \pm}(\theta)$, and the $\mathfrak{P}(\R_{\tau}, \mathfrak{B}_2)$-2d-LG Floer differential will be realized by the flow lines governed by the gradient flow equations satisfied by the $\xi$-varying configurations which set the expression within the squared terms in \eqref{eq:m2 x r3:2d-lg:sqm:action} to zero.
The partition function of the 2d-LG SQM in $\mathfrak{P}(\R_t, \mathfrak{P}(\R_{\tau}, \mathfrak{B}_2))$ will then be given by
\begin{equation}
  \label{eq:m2 x r3:2d-lg:lg partition function}
  \mathcal{Z}_{\text{2d-LG SQM}, \mathfrak{P}(\R_t, \mathfrak{P}(\R_{\tau}, \mathfrak{B}_2))}(G)
  = \sum_{I \neq J \neq K \neq L = 1}^w
  \;
  \sum_{\mathfrak{q}^{IJ, KL}_{W_2, \pm} \in S^{IJ, KL}_{W_2}}
  \text{HF}^G_{d_w} \left( \mathfrak{q}^{IJ, KL}_{W_2, \pm}(\theta) \right)
  \, .
\end{equation}
Here, the contribution $\text{HF}^G_{d_w} \big( \mathfrak{q}^{IJ, KL}_{W_2, \pm}(\theta) \big)$ can be identified with a homology class in a $\mathfrak{P}(\R_{\tau}, \mathfrak{B}_2)$-2d-LG Floer homology generated by LG $\mathfrak{P}^{\theta}(\R_{\tau}, \mathfrak{B}_2)$-strings whose endpoints correspond to $\theta$-deformed, non-constant paths in $\mathfrak{P}(\R_{\tau}, \mathfrak{B}_2)$.
The degree of each chain in the complex is $d_w$, and is counted by the number of outgoing flow lines from the fixed critical points of $H_2(\varphi_+, \varphi_-)$ in $\mathfrak{P}(\R_t, \mathfrak{P}(\R_{\tau}, \mathfrak{B}_2))$ which can also be identified as $\mathfrak{q}^{IJ, KL}_{W_2, \pm}(\theta)$.

Note that $\mathfrak{q}^{IJ, KL}_{W_2, \pm}(\theta)$, which corresponds to an LG $\mathfrak{P}^{\theta}(\R_{\tau}, \mathfrak{B}_2)$-string, will be defined by (i) \eqref{eq:m2 x r3:2d-lg:p-string} with (ii) endpoints defined by \eqref{eq:m2 x r3:2d-lg:p-string:endpoint}.
In other words, we can write
\begin{equation}
  \label{eq:m2 x r3:2d-lg:ext = hf}
  \mathfrak{q}^{IJ, KL}_{W_2, \pm}(\theta)
  \Longleftrightarrow
  \text{Ext} \left( \mathfrak{T}^{IJ}_{W_2}(\theta), \mathfrak{T}^{KL}_{W_2}(\theta) \right)_{\pm}
  \, ,
\end{equation}
where $\text{Ext} \left( \mathfrak{T}^{IJ}_{W_2}(\theta), \mathfrak{T}^{KL}_{W_2}(\theta) \right)_{\pm}$ represents an LG $\mathfrak{P}^{\theta}(\R_{\tau}, \mathfrak{B}_2)$-string whose start and end are described as singular fibers $\mathfrak{T}^{IJ}_{W_2}(\theta)$ and $\mathfrak{T}^{KL}_{W_2}(\theta)$, respectively.

Equivalently, it is a $\mathfrak{P}^{\theta}(\R_{\tau}, \mathfrak{B}_2)$-string defined by (i) \eqref{eq:m2 x r3:go-bf:p-strings} and $A_t, \mathring{A}_{\tau}, A_{\xi} \rightarrow 0$ with (ii) endpoints defined by \eqref{eq:m2 x r3:go-bf:non-constant paths} corresponding to $\mathfrak{B}_2^{\theta}$-strings.
This will lead us to the following one-to-one identification
\begin{equation}
  \label{eq:m2 x r3:2d-lg:hom as ext}
  \saveboxed{eq:m2 x r3:2d-lg:hom as ext}{%
    \text{Hom} \left( \Gamma^{IJ}(\mathfrak{B}_2), \Gamma^{KL}(\mathfrak{B}_2) \right)_{\pm}
    \Longleftrightarrow
    \text{Ext} \left( \mathfrak{T}^{IJ}_{W_2}, \mathfrak{T}^{KL}_{W_2} \right)_{\pm}
  }
\end{equation}

Just like in \autoref{sec:m3 x r2:orlov}, the RHS of \eqref{eq:m2 x r3:2d-lg:hom as ext} can be regarded as morphisms in an Orlov-type triangulated category of singularities of $W_2$.

\subtitle{The Normalized GM Partition Function, LG $\mathfrak{P}^{\theta}(\R_{\tau}, \mathfrak{B}_2)$-string Scattering, and Maps of an $A_{\infty}$-structure}

Since our GM theory is semi-classical, its normalized 5d partition function will be a sum over tree-level scattering amplitudes of the LG $\mathfrak{P}^{\theta}(\R_{\tau}, \mathfrak{B}_2)$-strings defined by \eqref{eq:m2 x r3:2d-lg:p-string}.
The BPS worldsheet underlying such a tree-level scattering is similar to \autoref{fig:m3 x r2:scattering amplitudes}, where instead of endpoints of each string being labeled by $\mathfrak{C}^{*}_{\text{BF}_{\HH}}(\theta)$, they will now be labeled $\Gamma^{**}(\theta, \mathfrak{B}_2)$.

In other words, we can, like in \eqref{eq:m3 x r2:normalized partition function}, express the normalized GM partition function as
\begin{equation}
  \label{eq:m2 x r3:2d-lg:normalized partition function}
  \widetilde{\mathcal{Z}}_{\text{GM}, M_2 \times \R^3}(G)
  = \sum_{\mathfrak{N}_w} \mu_{\mathfrak{P}(\mathbb{R}_{\tau}, \mathfrak{B}_2)}^{\mathfrak{N}_w}
  \, ,
  \qquad
  \mathfrak{N}_w = 1, 2, \dots, \left\lfloor \frac{w - 2}{2} \right\rfloor
  \, ,
\end{equation}
where each
\begin{equation}
  \label{eq:m2 x r3:2d-lg:composition map}
  \saveboxed{eq:m2 x r3:2d-lg:composition map}{%
    \mu_{\mathfrak{P}(\mathbb{R}_{\tau}, \mathfrak{B}_2)}^{\mathfrak{N}_w}: \bigotimes_{i = 1}^{\mathfrak{N}_w}
    \text{Hom} \left( \Gamma^{I_{2i - 1} I_{2i}}(\mathfrak{B}_2), \Gamma^{I_{2(i + 1) - 1} I_{2(i + 1)}}(\mathfrak{B}_2) \right)_-
    \rightarrow
    \text{Hom} \left( \Gamma^{I_1 I_2}(\mathfrak{B}_2), \Gamma^{I_{2\mathfrak{N}_w + 1} I_{2\mathfrak{N}_w + 2}}(\mathfrak{B}_2) \right)_+
  }
\end{equation}
is a scattering amplitude of $\mathfrak{N}_w$ incoming LG $\mathfrak{P}^{\theta}(\R_{\tau}, \mathfrak{B}_2)$-strings $\text{Hom} \left( \Gamma^{**}(\mathfrak{B}_2), \Gamma^{**}(\mathfrak{B}_2) \right)_-$, and a single outgoing LG $\mathfrak{P}^{\theta}(\R_{\tau}, \mathfrak{B}_2)$-string $\text{Hom} \left( \Gamma^{I_1 I_2}(\mathfrak{B}_2), \Gamma^{I_{2\mathfrak{N}_w + 1} I_{2\mathfrak{N}_w + 2}}(\mathfrak{B}_2) \right)_+$, with left and right boundaries as labeled, whose underlying worldsheet can be regarded as a constant disc with $\mathfrak{N}_w + 1$ vertex operators at the boundary.
In short, $\mu^{\mathfrak{N}_w}_{\mathfrak{B}_2}$ counts constant discs with $\mathfrak{N}_w + 1$ punctures at the boundary that are mapped to $\mathfrak{P}(\R_{\tau}, \mathfrak{B}_2)$ according to the BPS worldsheet equations (determined by setting to zero the expression within the squared terms in \eqref{eq:m2 x r3:2d-lg:action}) that are given by \eqref{eq:m2 x r3:2d-3d:bps worldsheets}.

At any rate, just as in \autoref{sec:m3 x r2:orlov}, the collection of $\mu^{\mathfrak{N}_w}_{\mathfrak{B}_2}$ in \eqref{eq:m2 x r3:2d-lg:composition map} can be regarded as composition maps defining an $A_{\infty}$-structure.

\subtitle{An Orlov-type $A_{\infty}$-1-category of $\mathfrak{B}_2^{\theta}$-strings}

Altogether, this means that the normalized partition function of GM theory on $M_2 \times \R^3$ as expressed in \eqref{eq:m2 x r3:2d-lg:normalized partition function}, manifests a \emph{novel} Orlov-type $A_{\infty}$-1-category whose
(i) 1-objects $\Gamma^{**}(\theta, \mathfrak{B}_2)$ are singular fibers $W_{2, \text{sg}}^{-1}(0)$ corresponding to $\mathfrak{B}_2^{\theta}$-strings,
(ii) 1-morphisms are Ext-groups corresponding to LG $\mathfrak{P}^{\theta}(\R_{\tau}, \mathfrak{B}_2)$-strings ending on the 1-objects via the one-to-one identification \eqref{eq:m2 x r3:2d-lg:hom as ext},
and (iii) $A_{\infty}$-structure is defined by the $\mu^{\mathfrak{N}_w}_{\mathfrak{B}_2}$ maps \eqref{eq:m2 x r3:2d-lg:composition map}.

\subsection{Membrane Theory, the GM Partition Function, and an RW-type \texorpdfstring{$A_{\infty}$}{A-infinity}-2-category Categorifying the Holomorphic \texorpdfstring{$G_{\OO}$}{G-OO}-flat Floer Homology of \texorpdfstring{$M_2$}{M2}}
\label{sec:m2 x r3:rw}

\subtitle{The 3d Model as a 3d Gauged B-twisted LG Model}

Note that we can also express \eqref{eq:m2 x r3:gm:3d model:action} as
\begin{equation}
  \label{eq:m2 x r3:3d-lg:action}
  \begin{aligned}
    S_{\text{3d-LG}, \mathfrak{B}_2}
    & = \frac{1}{e^2} \int_{\R^3} dt d\tau d\xi \, \left(
      \left| \mathcal{X}^a + P^a(\theta) \right|^2
      + \left| \mathcal{Y}^a + Q^a(\theta) \right|^2
      + \dots
      \right)
    \\
    & = \frac{1}{e^2} \int_{\R^3} dt d\tau d\xi \, \left(
      \left| \mathcal{X}^a + \mathbf{e}_0 e^{\mathbf{e}_0 \theta} g_{\mathfrak{B}_2}^{ab} \pdv{V_2}{\widetilde{\mathsf{A}}^b} \right|^2
      + \left| \mathcal{Y}^a + \mathbf{e}_0 e^{\mathbf{e}_0 \theta} g_{\mathfrak{B}_2}^{ab} \pdv{V_2}{\widetilde{\mathsf{B}}^b} \right|^2
      + \dots
      \right)
      \, ,
  \end{aligned}
\end{equation}
where $V_2(\widetilde{\mathsf{A}}, \widetilde{\mathsf{B}})$ is given by \eqref{eq:m2 x r:morse functional}, and
\begin{equation}
  \label{eq:m2 x r3:3d-lg:action:components}
  \begin{aligned}
    \mathcal{X}^a
    =
    & (D_{\xi} + \mathbf{e}_0 D_t + e^{\mathbf{e}_0 \theta} \hat{\mathbf{e}}_2 D_{\tau}) \widetilde{\mathsf{A}}^a
      + e^{\mathbf{e}_0 \theta} \left(  \cos(\theta) D_{\xi} + \sin(\theta) D_t + \hat{\mathbf{e}}_2 D_{\tau} \right) \widetilde{\mathsf{B}}^a
    \\
    & + e^{\mathbf{e}_0 \theta} F_{t\xi}
      + e^{\mathbf{e}_0 \theta} \hat{\mathbf{e}}_2 F_{t\tau} \cos(\theta)
      - e^{\mathbf{e}_0 \theta} \hat{\mathbf{e}}_2 F_{\xi\tau} \sin(\theta)
      \, ,
    \\
    \mathcal{Y}^a
    =
    & (D_{\xi} + \mathbf{e}_0 D_t + e^{\mathbf{e}_0 \theta} \hat{\mathbf{e}}_2 D_{\tau}) \widetilde{\mathsf{B}}^a
      - e^{\mathbf{e}_0 \theta} (\cos(\theta) D_{\xi} + \sin(\theta) D_t + \hat{\mathbf{e}}_2 D_{\tau}) \widetilde{\mathsf{A}}^a
      \, .
  \end{aligned}
\end{equation}

Applying specific values of $\theta$ to the conditions obtained by setting to zero the expression within the squared terms of \eqref{eq:m2 x r3:3d-lg:action} whilst noting that the physical theory ought to be symmetric under a variation of $\theta$, we find that it can also be equivalently expressed as
\begin{equation}
  \label{eq:m2 x r3:3d-lg:b-twist}
  S_{\text{3d-LG}, \mathfrak{B}_2}
  = \frac{1}{e^2} \int_{\R^3} \left(
    \left| D \widetilde{\mathsf{A}} \right|^2
    + \left| D \widetilde{\mathsf{B}} \right|^2
    + \left| \mathbf{e}_0 F - g_{\mathfrak{B}_2}^{ab} \pdv{V_2}{\widetilde{\mathsf{A}}^b} \right|^2
    + \left| g_{\mathfrak{B}_2}^{ab} \pdv{V_2}{\widetilde{\mathsf{B}}^b} \right|^2
    + \dots
  \right)
  \, ,
\end{equation}
where now, $D$ is the covariant derivative in the real 3d gauge fields, and $F$ is the field strength of the real 3d gauge fields.
As the conditions that minimize \eqref{eq:m2 x r3:3d-lg:b-twist} involve constant maps, we can interpret \eqref{eq:m2 x r3:3d-lg:b-twist} as the action of a ``B-twisted'' model.
In other words, the 3d gauged sigma model with target $\mathfrak{B}_2$ can also be interpreted as a 3d gauged \emph{B-twisted} LG model in $\mathfrak{B}_2$ with holomorphic superpotential $V_2(\widetilde{\mathsf{A}}, \widetilde{\mathsf{B}})$.

By setting $d_{\xi} \widetilde{\mathsf{A}}^a = 0 = d_{\xi} \widetilde{\mathsf{B}}^a$ and $A_t, A_{\tau}, A_{\xi} \rightarrow 0$ in the expression within the squared terms in \eqref{eq:m2 x r3:3d-lg:action}, we can read off (via \eqref{eq:m2 x r3:3d-lg:action:components}) the LG $\mathfrak{B}_2^{\theta}$-membrane  (corresponding to $\sigma^{IJ, KL}_{\pm}(t, \tau, \theta, \mathfrak{B}_2)$) equations (that re-express \eqref{eq:m2 x r3:go-bf:b-membranes} with $A_t, A_{\tau}, A_{\xi} \rightarrow 0$) as
\begin{equation}
  \label{eq:m2 x r3:3d-lg:b-membrane}
  \begin{aligned}
    \dv{}{t} \left(
    \mathbf{e}_0 \widetilde{\mathsf{A}}^a
    + e^{\mathbf{e}_0 \theta} \hat{\mathbf{e}}_2 \widetilde{\mathsf{B}}^a \sin(\theta)
    \right)
    + \dv{}{\tau} \left(
    e^{\mathbf{e}_0 \theta} \hat{\mathbf{e}}_2 \widetilde{\mathsf{A}}^a
    + e^{\mathbf{e}_0 \theta} \hat{\mathbf{e}}_2 \widetilde{\mathsf{B}}^a
    \right)
    &= - \mathbf{e}_0 e^{\mathbf{e}_0 \theta} g_{\mathfrak{B}_2}^{ab} \pdv{V_2}{\widetilde{\mathsf{A}}^b}
      \, ,
    \\
    \dv{}{t} \left(
    \mathbf{e}_0 \widetilde{\mathsf{B}}^a
    - e^{\mathbf{e}_0 \theta} \widetilde{\mathsf{A}}^a \sin(\theta)
    \right)
    + \dv{}{\tau} \left(
    e^{\mathbf{e}_0 \theta} \hat{\mathbf{e}}_2 \widetilde{\mathsf{B}}^a
    - e^{\mathbf{e}_0 \theta} \hat{\mathbf{e}}_2 \widetilde{\mathsf{A}}^a
    \right)
    &= - \mathbf{e}_0 e^{\mathbf{e}_0 \theta} g_{\mathfrak{B}_2}^{ab} \pdv{V_2}{\widetilde{\mathsf{B}}^b}
      \, .
  \end{aligned}
\end{equation}

By setting $d_t \widetilde{\mathsf{A}}^a = 0 = d_t \widetilde{\mathsf{B}}^a$ in \eqref{eq:m2 x r3:3d-lg:b-membrane}, we can read off the LG $\mathfrak{B}_2^{\theta}$-string (corresponding to $\gamma^{IJ}(\tau, \theta, \mathfrak{B}_2)$ and $\gamma^{KL}(\tau, \theta, \mathfrak{B}_2)$) equations, or equivalently, the LG $\mathfrak{B}_2^{\theta}$-membrane \emph{edge} (corresponding to $\sigma^{IJ, KL}(\pm \infty, \tau, \theta, \mathfrak{B}_2)$) equations (that re-express \eqref{eq:m2 x r3:go-bf:b-strings}) as
\begin{equation}
  \label{eq:m2 x r3:3d-lg:b-string}
  \begin{aligned}
    \dv{}{\tau} \left(
      e^{\mathbf{e}_0 \theta} \hat{\mathbf{e}}_2 \widetilde{\mathsf{A}}^a
      + e^{\mathbf{e}_0 \theta} \hat{\mathbf{e}}_2 \widetilde{\mathsf{B}}^a
    \right)
    &= - \mathbf{e}_0 e^{\mathbf{e}_0 \theta} g_{\mathfrak{B}_2}^{ab} \pdv{V_2}{\widetilde{\mathsf{A}}^b}
      \, ,
    \\
    \dv{}{\tau} \left(
      e^{\mathbf{e}_0 \theta} \hat{\mathbf{e}}_2 \widetilde{\mathsf{B}}^a
      - e^{\mathbf{e}_0 \theta} \hat{\mathbf{e}}_2 \widetilde{\mathsf{A}}^a
    \right)
    &= - \mathbf{e}_0 e^{\mathbf{e}_0 \theta} g_{\mathfrak{B}_2}^{ab} \pdv{V_2}{\widetilde{\mathsf{B}}^b}
      \, .
  \end{aligned}
\end{equation}

By setting $d_{\tau} \widetilde{\mathsf{A}}^a = 0 = d_{\tau} \widetilde{\mathsf{B}}^a$ in \eqref{eq:m2 x r3:3d-lg:b-string}, we can read off the LG $\mathfrak{B}_2^{\theta}$-string endpoint (corresponding to $\gamma^{**}(\pm \infty, \theta, \mathfrak{B}_2)$) equations, or equivalently, the LG $\mathfrak{B}_2^{\theta}$-membrane \emph{vertex} (corresponding to $\sigma^{IJ, KL}(\pm \infty, \pm \infty, \theta, \mathfrak{B}_2)$ and $\sigma^{IJ, KL}(\pm \infty, \mp \infty, \theta, \mathfrak{B}_2)$) equations (that re-express \eqref{eq:m2 x r3:go-bf:b-string endpoints}) as
\begin{equation}
  \label{eq:m2 x r3:3d-lg:b-string endpoints}
  \mathbf{e}_0 e^{\mathbf{e}_0 \theta} g_{\mathfrak{B}_2}^{ab} \pdv{V_2}{\widetilde{\mathsf{A}}^b}
  = 0
  \, ,
  \qquad
  \mathbf{e}_0 e^{\mathbf{e}_0 \theta} g_{\mathfrak{B}_2}^{ab} \pdv{V_2}{\widetilde{\mathsf{B}}^b}
  = 0
  \, .
\end{equation}

Recall from the end of \autoref{sec:m2 x r3:go-bf} that we are only considering certain $M_2$ and $G$ such that the vertices $\sigma^{IJ, KL}(\pm \infty, \pm \infty,$ $\theta, \mathfrak{B}_2)$ and $\sigma^{IJ, KL}(\pm \infty, \mp \infty, \theta, \mathfrak{B}_2)$ are isolated and non-degenerate.
Therefore, from their definition in \eqref{eq:m2 x r3:3d-lg:b-string endpoints} as critical points of $V_2(\widetilde{\mathsf{A}}, \widetilde{\mathsf{B}})$, it would mean that $V_2(\widetilde{\mathsf{A}}, \widetilde{\mathsf{B}})$ is a (holomorphic) Morse function in $\mathfrak{B}_2$.

\subtitle{The 3d Gauged B-twisted LG Model as an LG SQM}

Last but not least, after suitable rescalings, we can recast \eqref{eq:m2 x r3:3d-lg:action} as a 1d LG SQM (that re-expresses \eqref{eq:m2 x r3:gm:sqm:action}), where its action will be given by\footnote{%
  The same steps in \autoref{ft:m2 x r3:integrate out gauge field to sqm} are applied to arrive at the following expression.
  \label{ft:m2 x r3:3d-lg:sqm no gauge}
}
\begin{equation}
  \label{eq:m2 x r3:3d-lg:sqm:action}
  \begin{aligned}
    & S_{\text{3d-LG SQM}, \mathfrak{P}(\R^2, \mathfrak{B}_2)}
    \\
    & = \frac{1}{e^2} \int d\xi \left(
      \left| \partial_{\xi} \varphi_+^{\alpha} + g^{\alpha\beta}_{\mathfrak{P}(\R^2, \mathfrak{B}_2)} \pdv{\mathfrak{h}_2}{\varphi_+^{\beta}} \right|^2
      +  \left| \partial_{\xi} \varphi_-^{\alpha} + g^{\alpha\beta}_{\mathfrak{P}(\R^2, \mathfrak{B}_2)} \pdv{\mathfrak{h}_2}{\varphi_-^{\beta}} \right|^2
      + \dots
      \right)
      \, ,
  \end{aligned}
\end{equation}
where $\mathfrak{h}_2(\varphi_+, \varphi_-)$ is the \emph{real-valued} potential in $\mathfrak{P}(\R^2, \mathfrak{B}_2)$, with $\varphi_+$ and $\varphi_-$ defined as in \autoref{sec:m2 x r3:gm}, and the subscript ``3d-LG SQM, $\mathfrak{P}(\R^2, \mathfrak{B}_2)$'' is to specify that it is a 1d SQM with target $\mathfrak{P}(\R^2, \mathfrak{B}_2)$ obtained from the equivalent 3d LG model.
We will also refer to this \emph{1d} LG SQM as ``3d-LG SQM'' in the rest of this subsection.

The 3d-LG SQM will localize onto configurations that \emph{simultaneously} set to zero the LHS and RHS of the expression within the squared terms in \eqref{eq:m2 x r3:3d-lg:sqm:action}.
In other words, it will localize onto $\xi$-invariant critical points of $\mathfrak{h}_2(\varphi_+, \varphi_-)$ that will correspond, when $A_t, A_{\tau}, A_{\xi} \rightarrow 0$, to the LG $\mathfrak{B}_2^{\theta}$-membranes defined by \eqref{eq:m2 x r3:3d-lg:b-membrane}.
For our choice of $M_2$ and $G$, just like their vertices, the LG $\mathfrak{B}_2^{\theta}$-membranes  will be isolated and non-degenerate.
Thus, $\mathfrak{h}_2(\varphi_+, \varphi_-)$ can be regarded as a \emph{real-valued} Morse functional in $\mathfrak{P}(\R^2, \mathfrak{B}_2)$.

\subtitle{Membrane Theory from the 3d LG Model}

Just like in \autoref{sec:m2 x r3:membranes and strings}, the equivalent 3d gauged B-twisted LG model will define an open membrane theory in $\mathfrak{B}_2$ with effective worldvolumes and boundaries shown in \autoref{fig:m2 x r3:bps worldvolume}, where $\xi$ is the temporal direction and $(t, \tau)$ are the spatial directions.

The dynamics of this open membrane theory in $\mathfrak{B}_2$ will be governed by the BPS worldvolume equations determined by setting to zero the expression within the squared terms in \eqref{eq:m2 x r3:3d-lg:action}, where $(\widetilde{\mathsf{A}}^a, \widetilde{\mathsf{B}}^a)$ are scalars on the worldvolume corresponding to the holomorphic coordinates of $\mathfrak{B}_2$.
At an arbitrary instant in time whence $d_{\xi} \widetilde{\mathsf{A}}^a = d_{\xi} \widetilde{\mathsf{B}}^a = 0 = d_{\xi} A_{\{t, \tau\}}$ therein, the dynamics of $(\widetilde{\mathsf{A}}^a, \widetilde{\mathsf{B}}^a)$ and the 3d gauge fields along $(t, \tau)$ will be governed by the 2d membrane equations
\begin{equation}
  \label{eq:m2 x r3:3d-lg:bps membrane}
  \begin{aligned}
    & \dv{}{t} \left(
      \mathbf{e}_0 \widetilde{\mathsf{A}}^a
      + e^{\mathbf{e}_0 \theta} \widetilde{\mathsf{B}}^a \sin(\theta)
      + e^{\mathbf{e}_0 \theta} A_{\xi}
      + e^{\mathbf{e}_0 \theta} \hat{\mathbf{e}}_2 A_{\tau} \cos(\theta)
    \right)
    + e^{\mathbf{e}_0 \theta} \hat{\mathbf{e}_2} \dv{}{\tau} \left(
      \widetilde{\mathsf{A}}^a
      + \widetilde{\mathsf{B}}^a
      + A_{\xi} \sin(\theta)
      - A_t \cos(\theta)
    \right)
    \\
    &= - \left[ A_t,
      \mathbf{e}_0 \widetilde{\mathsf{A}}^a
      + e^{\mathbf{e}_0 \theta} \widetilde{\mathsf{B}}^a \sin(\theta)
      + e^{\mathbf{e}_0 \theta} A_{\xi}
    \right]
    + e^{\mathbf{e}_0 \theta} \hat{\mathbf{e}_2} \left[ A_{\tau},
      \widetilde{\mathsf{A}}^a
      + \widetilde{\mathsf{B}}^a
      + A_{\xi} \sin(\theta)
    \right]
    \\
    & \quad
      - \left[ A_{\xi},
        \widetilde{\mathsf{A}}^a
        + e^{\mathbf{e}_0 \theta} \widetilde{\mathsf{B}}^a \cos(\theta)
      \right]
      - e^{\mathbf{e}_0 \theta} \hat{\mathbf{e}}_2 [A_t, A_{\tau}] \cos(\theta)
      - \mathbf{e}_0 e^{\mathbf{e}_0 \theta} g_{\mathfrak{B}_2}^{ab} \pdv{V_2}{\widetilde{\mathsf{A}}^b}
      \, ,
    \\
    & \dv{}{t} \left(
        \mathbf{e}_0 \widetilde{\mathsf{B}}^a
        - e^{\mathbf{e}_0 \theta} \widetilde{\mathsf{A}}^a \sin(\theta)
      \right)
      + e^{\mathbf{e}_0 \theta} \hat{\mathbf{e}}_2 \dv{}{\tau} \left(
        \widetilde{\mathsf{B}}^a
        - \widetilde{\mathsf{A}}^a
      \right)
    \\
    &= - \left[ A_t,
        \mathbf{e}_0 \widetilde{\mathsf{B}}^a
        - e^{\mathbf{e}_0 \theta} \widetilde{\mathsf{A}}^a \sin(\theta)
      \right]
      - e^{\mathbf{e}_0 \theta} \hat{\mathbf{e}}_2 \left[ A_{\tau}
        \widetilde{\mathsf{B}}^a
        - \widetilde{\mathsf{A}}^a
      \right]
      - \left[ A_{\xi},
        \widetilde{\mathsf{B}}^a
        - e^{\mathbf{e}_0 \theta} \widetilde{\mathsf{A}}^a \cos(\theta)
      \right]
      - \mathbf{e}_0 e^{\mathbf{e}_0 \theta} g_{\mathfrak{B}_2}^{ab} \pdv{V_2}{\widetilde{\mathsf{B}}^b}
      \, .
  \end{aligned}
\end{equation}

\subtitle{Morphisms between LG $\mathfrak{B}_2^{\theta}$-strings as Ext-groups}

At this juncture, note that the endpoints of an LG $\mathfrak{B}_2^{\theta}$-string (defined by \eqref{eq:m2 x r3:3d-lg:b-string endpoints}) can be described as \emph{singular} fibers $V_{2, \text{sg}}^{-1}(0)$ that correspond to $G_{\OO}$-BF configurations on $M_2$.
In particular, to an LG $\mathfrak{B}_2^{\theta}$-string $\gamma^{IJ}(\tau, \theta, \mathfrak{B}_2)$ (illustrated as the left edge in \autoref{fig:m2 x r3:frakB-sheet}), its (i) bottom endpoint $\gamma^{IJ}(- \infty, \theta, \mathfrak{B}_2) \eqcolon \mathcal{T}^I_{V_2}(\theta) \in V_{2, \text{sg}}^{-1}(0)$ and (ii) top endpoint $\gamma^{IJ}(+ \infty, \theta, \mathfrak{B}_2) \eqcolon \mathcal{T}^J_{V_2}(\theta) \in V_{2, \text{sg}}^{-1}(0)$, are described as singular fibers which are also zero-dimensional submanifolds of $\mathfrak{B}_2$, and correspond to $\theta$-deformed $G_{\OO}$-BF configurations (i) $\mathfrak{C}^I_{\text{BF}_{\OO}}(\theta)$ and (ii) $\mathfrak{C}^J_{\text{BF}_{\OO}}(\theta)$ on $M_2$.
We shall denote by $\mathcal{T}^{IJ}_{V_2}(\theta)$ an LG $\mathfrak{B}_2^{\theta}$-string $\gamma^{IJ}(\tau, \theta, \mathfrak{B}_2)$ so as to indicate that its endpoints are singular fibers $V_{2, \text{sg}}^{-1}(0)$.

Let
(i) $S^{IJ}_{\text{BF}_{\OO}}$ and (ii) $S^{KL}_{\text{BF}_{\OO}}$
be the collection of all such strings whose bottom and top endpoints correspond to
(i) $\mathfrak{C}^I_{\text{BF}_{\OO}}(\theta)$ and $\mathfrak{C}^J_{\text{BF}_{\OO}}(\theta)$ and (ii) $\mathfrak{C}^K_{\text{BF}_{\OO}}(\theta)$ and $\mathfrak{C}^L_{\text{BF}_{\OO}}(\theta)$.
Each
(i) LG $\gamma^{IJ}(\tau, \theta, \mathfrak{B}_3)$-string and (ii) LG $\gamma^{KL}(\tau, \theta, \mathfrak{B}_3)$-string,
will then correspond to a string
(i) $p^{IJ}_{\text{BF}_{\OO}}(\theta) \in S^{IJ}_{\text{BF}_{\OO}}$ and (ii) $p^{KL}_{\text{BF}_{\OO}}(\theta) \in S^{KL}_{\text{BF}_{\OO}}$.
This means that an LG $\mathfrak{B}_2^{\theta}$-membrane pair $\sigma^{IJ, KL}_{\pm}(t, \tau, \theta, \mathfrak{B}_2)$ with left edge $\sigma^{IJ, KL}_{\pm}(- \infty, \tau, \theta, \mathfrak{B}_2) = \gamma^{IJ}(\tau, \theta, \mathfrak{B}_2)$ and right edge $\sigma^{IJ, KL}_{\pm}(+ \infty, \tau, \theta, \mathfrak{B}_2) = \gamma^{KL}(\tau, \theta, \mathfrak{B}_2)$, can be identified by a pair of strings $q^{IJ, KL}_{\text{BF}_{\OO}, \pm} \coloneq \{p^{IJ}_{\text{BF}_{\OO}}, p^{KL}_{\text{BF}_{\OO}}\}_{\pm} \in S^{IJ}_{\text{BF}_{\OO}} \otimes S^{KL}_{\text{BF}_{\OO}}$ that define their edges.

At any rate, the 3d-LG SQM in $\mathfrak{P}(\R^2, \mathfrak{B}_2)$ with action \eqref{eq:m2 x r3:3d-lg:sqm:action} will physically realize a Floer homology that we shall name a $\mathfrak{B}_2$-3d-LG Floer homology.
The chains of the $\mathfrak{B}_2$-3d-LG Floer complex are generated by LG $\mathfrak{B}_2^{\theta}$-membranes which, as detailed above, can be identified by $q^{**, **}_{\text{BF}_{\OO}, \pm}(\theta)$, and the $\mathfrak{B}_2$-3d-LG Floer differential will be realized by the flow lines governed by the gradient flow equations satisfied by the $\xi$-varying configurations which set the expression within the squared terms of \eqref{eq:m2 x r3:3d-lg:sqm:action} to zero.
The partition function of the 3d-LG SQM in $\mathfrak{P}(\R^2, \mathfrak{B}_2)$ will be given by
\begin{equation}
  \label{eq:m2 x r3:3d-lg:sqm:partition function}
  \mathcal{Z}_{\text{3d-LG SQM}, \mathfrak{P}(\R^2, \mathfrak{B}_2)}(G)
  = \sum_{I \neq J \neq K \neq L = 1}^w
  \;
  \sum_{q^{IJ, KL}_{\text{BF}_{\OO}, \pm} \in S^{IJ}_{\text{BF}_{\OO}} \otimes S^{KL}_{\text{BF}_{\OO}}}
  \text{HF}^G_{d_w} \left( q^{IJ, KL}_{\text{BF}_{\OO}, \pm} (\theta) \right)
  \, .
\end{equation}
Here, the contribution $\text{HF}^G_{d_w} \big( q^{IJ, KL}_{\text{BF}_{\OO}, \pm} (\theta) \big)$ can be identified with a homology class in a $\mathfrak{B}_2$-3d-LG Floer homology generated by LG $\mathfrak{B}_2^{\theta}$-membranes, whose edges correspond to LG $\mathfrak{B}_2^{\theta}$-strings, and whose vertices are described as singular fibers $V_{2, \text{sg}}^{-1}(0)$.
The degree of each chain is $d_w$, and is counted by the number of outgoing flow lines from the fixed critical points of $\mathfrak{h}_2(\varphi_+, \varphi_-)$ in $\mathfrak{P}(\R^2, \mathfrak{B}_2)$ which can also be identified by $q^{IJ, KL}_{\text{BF}_{\OO}, \pm} (\theta)$.

Note that an LG $\mathfrak{B}_2^{\theta}$-membrane in the 3d LG model defined by (i) \eqref{eq:m2 x r3:3d-lg:b-membrane} with (ii) edges being LG $\mathfrak{B}_2^{\theta}$-strings \eqref{eq:m2 x r3:3d-lg:b-string} and (iii) vertices \eqref{eq:m2 x r3:3d-lg:b-string endpoints}, will correspond to an LG $\mathfrak{P}^{\theta}(\R_{\tau}, \mathfrak{B}_2)$-string in the 2d LG model defined by \eqref{eq:m2 x r3:2d-lg:p-string} with (ii) endpoints \eqref{eq:m2 x r3:2d-lg:p-string:endpoint} being singular fibers $W_{2, \text{sg}}^{-1}(0)$, accordingly.
This means that just as how an LG $\mathfrak{P}^{\theta}(\R_{\tau}, \mathfrak{B}_2)$-string pair $\Sigma^{IJ, KL}_{\pm}(t, \theta, \mathfrak{B}_2)$ can be identified as an Ext-group between singular fibers $\mathfrak{T}^{IJ}_{W_2}(\theta)$ and $\mathfrak{T}^{KL}_{W_2}(\theta)$ describing the endpoints, an LG $\mathfrak{B}_2^{\theta}$-membrane pair $\sigma^{IJ, KL}_{\pm}(t, \tau, \theta, \mathfrak{B}_3)$ can be identified as an Ext-group between $\mathcal{T}^{IJ}_{V_2}(\theta)$ and $\mathcal{T}^{KL}_{V_2}(\theta)$ describing the edges.
In other words, we can write
\begin{equation}
  \label{eq:m2 x r3:3d-lg:ext = hf}
  q^{IJ, KL}_{\text{BF}_{\OO}, \pm} (\theta)
  \Longleftrightarrow
  \text{Ext} \left( \mathcal{T}^{IJ}_{V_2}(\theta), \mathcal{T}^{KL}_{V_2}(\theta) \right)_{\pm}
  \, ,
\end{equation}
where $\text{Ext} \left( \mathcal{T}^{IJ}_{V_2}(\theta), \mathcal{T}^{KL}_{V_2}(\theta) \right)_{\pm}$ is an Ext group representing a $\mathfrak{B}_2^{\theta}$-membrane, whose left and right edges correspond to $\mathfrak{B}_2^{\theta}$-strings described by $\mathcal{T}^{IJ}_{V_2}(\theta)$ and $\mathcal{T}^{KL}_{V_2}(\theta)$, respectively.

Equivalently, it is a $\mathfrak{B}_2^{\theta}$-membrane defined by (i) \eqref{eq:m2 x r3:go-bf:b-membranes} with (ii) edges \eqref{eq:m2 x r3:go-bf:b-strings} and (iii) vertices \eqref{eq:m2 x r3:go-bf:b-string endpoints}.
Note that a membrane can be regarded as a morphism between its edges, which, in turn, can be regarded as morphisms between their endpoints.
In other words, a $\sigma^{IJ, KL}(t, \tau, \theta, \mathfrak{B}_2)$-membrane can be regarded as a 1-morphism $\text{Hom} \big( \gamma^{IJ}(\tau, \theta, \mathfrak{B}_2), \gamma^{KL}(\tau, \theta, \mathfrak{B}_2) \big)$ from its left edge to its right edge, where the (A) $\gamma^{IJ}(\tau, \theta, \mathfrak{B}_2)$-string and (B) $\gamma^{KL}(\tau, \theta, \mathfrak{B}_2)$-string can themselves be regarded as 1-morphisms (A) $\text{Hom} \big( \mathfrak{C}^I_{\text{BF}_{\OO}}(\theta), \mathfrak{C}^J_{\text{BF}_{\OO}}(\theta) \big)$ and (B) $\text{Hom} \big( \mathfrak{C}^K_{\text{BF}_{\OO}}(\theta), \mathfrak{C}^L_{\text{BF}_{\OO}}(\theta) \big)$, from their bottom to top endpoints.
This is illustrated in \autoref{fig:m2 x r3:frakB-sheet}.
Thus, we have the following one-to-one identifications
\begin{equation}
  \label{eq:m2 x r3:3d-lg:homs as ext}
  \saveboxed{eq:m2 x r3:3d-lg:homs as ext}{
    \begin{gathered}
      \text{Ext} \left( \mathcal{T}^{IJ}_{V_2}, \mathcal{T}^{KL}_{V_2} \right)_{\pm}
      \\
      \Updownarrow
      \\
      \text{Hom} \left( \gamma^{IJ}(\tau, \mathfrak{B}_2), \gamma^{KL}(\tau, \mathfrak{B}_2) \right)_{\pm}
      \\
      \Updownarrow
      \\
      \text{Hom} \left(
        \text{Hom} \left( \mathfrak{C}^I_{\text{BF}_{\OO}}, \mathfrak{C}^J_{\text{BF}_{\OO}} \right),
        \text{Hom} \left( \mathfrak{C}^I_{\text{BF}_{\OO}}, \mathfrak{C}^J_{\text{BF}_{\OO}} \right)
      \right)_{\pm}
    \end{gathered}
  }
\end{equation}

\subtitle{The Normalized GM Partition Function, LG $\mathfrak{B}_2^{\theta}$-membrane Scattering, and Maps of an $A_{\infty}$-structure}

\begin{figure}
  \centering
  \begin{tikzpicture}[%
    auto,%
    every edge/.style={draw},%
    relation/.style={scale=1, sloped, anchor=center, align=center,%
      color=black},%
    vertRelation/.style={scale=1, anchor=center, align=center},%
    dot/.style={circle, fill, minimum size=2*\radius, node contents={},%
      inner sep=0pt},%
    ]
    \def \WsLength {0.8}  
    \def \WsDepth {0.2}   
    \draw (-2,0) arc (180:360:2 and 0.6);
    \draw[dashed] (2,0) arc (0:180:2 and 0.6);
    \draw[white, thick] (-2,0) arc (180:215:2 and 0.6);
    \draw[white, thick] (2,0) arc (360:325:2 and 0.6);
    \draw ({2*cos(10)}, {2*sin(10)}) arc (10:80:2);
    \coordinate (out-ws-bottom-front-left) at (-0.8, 1.5);
    \coordinate (out-ws-bottom-front-right) at (0, 1.5);  
    \coordinate (out-ws-bottom-back-left) at (-0.1, 1.7);
    \coordinate (out-ws-bottom-back-right) at (0.7, 1.9);
    \coordinate (out-ws-top-front-left) at (-0.8, 3.5);
    \coordinate (out-ws-top-front-right) at (0, 3.5);
    \coordinate (out-ws-top-back-left) at (-0.1, 3.7);
    \coordinate (out-ws-top-back-right) at (0.7, 3.7);
    \draw (out-ws-bottom-front-left)
    -- (out-ws-top-front-left)
    node[below left] {\footnotesize $\mathfrak{C}^{I_1}_{\text{BF}_{\OO}}$}
    -- (out-ws-top-back-left)
    -- (out-ws-top-back-right)
    -- (out-ws-top-front-right)
    -- (out-ws-top-front-left)
    ;
    \draw (out-ws-bottom-front-right)
    -- (out-ws-top-front-right)
    ;
    \draw (out-ws-bottom-back-right)
    -- (out-ws-top-back-right)
    node[below right] {\footnotesize $\mathfrak{C}^{I_{2 \mathfrak{N}_w + 2}}_{\text{BF}_{\OO}}$}
    ;
    \draw[dashed] (out-ws-bottom-back-left)
    -- (out-ws-top-back-left);
    \draw (-0.1, 3.8) node[above] {$+$}
    ;
    \draw[dashed] ({2*cos(170)}, {2*sin(170)}) arc (170:100:2);
    \draw ({2*cos(170)}, {2*sin(170)}) arc (170:115:2);
    \coordinate (in-ws1-bottom-front-left) at (-1.5, -0.8);
    \coordinate (in-ws1-bottom-front-right) at (-1.5, 0.0);
    \coordinate (in-ws1-bottom-back-left) at (-1.6, -0.1);
    \coordinate (in-ws1-bottom-back-right) at (-1.85, 0.7);
    \coordinate (in-ws1-top-front-left) at (-3.5, -0.8);
    \coordinate (in-ws1-top-front-right) at (-3.5, 0.0);
    \coordinate (in-ws1-top-back-left) at (-3.7, -0.1);
    \coordinate (in-ws1-top-back-right) at (-3.7, 0.7);
    \draw (in-ws1-bottom-front-right)
    -- (in-ws1-top-front-right)
    node[above right] {\footnotesize $\mathfrak{C}^{I_1}_{\text{BF}_{\OO}}$}
    ;
    \draw (in-ws1-top-front-left)
    node[below right] {\footnotesize $\mathfrak{C}^{I_3}_{\text{BF}_{\OO}}$}
    -- (in-ws1-bottom-front-left)
    ;
    \draw (in-ws1-bottom-back-right)
    -- (in-ws1-top-back-right)
    node[above right] {\footnotesize $\mathfrak{C}^{I_2}_{\text{BF}_{\OO}}$}
    -- (in-ws1-top-front-right)
    -- (in-ws1-top-front-left)
    -- (in-ws1-top-back-left)
    -- (in-ws1-top-back-right)
    ;
    \draw[dashed] (in-ws1-bottom-back-left)
    -- (in-ws1-top-back-left);
    ;
    \draw (-3.9, 0) node[left] {$-$}
    ;
    \draw[dashed] ({2*cos(192)}, {2*sin(192)}) arc (192:228:2)
    ;
    \draw ({2*cos(203)}, {2*sin(203)}) arc (203:228:2)
    ;
    \coordinate (in-ws2-bottom-front-left) at (-1.52, -1.26);
    \coordinate (in-ws2-bottom-front-right) at (-0.7, -1.6);
    \coordinate (in-ws2-bottom-back-left) at (-0.7, -1.2);
    \coordinate (in-ws2-bottom-back-right) at (0, -1.6);
    \coordinate (in-ws2-top-front-left) at (-2.5, -2.9);
    \coordinate (in-ws2-top-front-right) at (-1.8, -3.3);
    \coordinate (in-ws2-top-back-left) at (-1.8, -2.9);
    \coordinate (in-ws2-top-back-right) at (-1.1, -3.3);
    \draw (in-ws2-bottom-front-left)
    -- node[relation, above] {\footnotesize $\mathfrak{C}^{I_4}_{\text{BF}_{\OO}}$}
    (in-ws2-top-front-left)
    ;
    \draw (in-ws2-bottom-back-right)
    -- node[relation, below] {\footnotesize $\mathfrak{C}^{I_5}_{\text{BF}_{\OO}}$}
    (in-ws2-top-back-right)
    ;
    \draw (in-ws2-bottom-back-left)
    -- node[relation, above] {\footnotesize $\mathfrak{C}^{I_3}_{\text{BF}_{\OO}}$}
    (in-ws2-top-back-left)
    -- (in-ws2-top-front-left)
    -- (in-ws2-top-front-right)
    -- (in-ws2-top-back-right)
    -- (in-ws2-top-back-left)
    ;
    \draw[dashed] (in-ws2-bottom-front-right)
    -- (in-ws2-top-front-right)
    ;
    \draw (-2, -3.5) node[left] {$-$}
    ;
    \draw[dashed] ({2*cos(255)}, {2*sin(255)}) arc (255:350:2)
    ;
    \coordinate (in-ws3-bottom-front-right) at (1.5, -0.8);
    \coordinate (in-ws3-bottom-front-left) at (1.5, 0.0);
    \coordinate (in-ws3-bottom-back-right) at (1.6, -0.1);
    \coordinate (in-ws3-bottom-back-left) at (1.85, 0.7);
    \coordinate (in-ws3-top-front-right) at (3.5, -0.8);
    \coordinate (in-ws3-top-front-left) at (3.5, 0.0);
    \coordinate (in-ws3-top-back-right) at (3.3, -0.1);
    \coordinate (in-ws3-top-back-left) at (3.3, 0.7);
    \draw (in-ws3-bottom-front-left)
    -- (in-ws3-top-front-left)
    node[above left={-0.1cm} and {0.1cm}] {\footnotesize $\mathfrak{C}^{I_{2 \mathfrak{N}_w + 1}}_{\text{BF}_{\OO}}$}
    ;
    \draw (in-ws3-top-front-right)
    node[below left] {\footnotesize $\mathfrak{C}^{I_{2 \mathfrak{N}_w - 1}}_{\text{BF}_{\OO}}$}
    -- (in-ws3-bottom-front-right)
    ;
    \draw (in-ws3-bottom-back-left)
    -- (in-ws3-top-back-left)
    node[above left] {\footnotesize $\mathfrak{C}^{I_{2 \mathfrak{N}_w + 2}}_{\text{BF}_{\OO}}$}
    -- (in-ws3-top-front-left)
    -- (in-ws3-top-front-right)
    ;
    \draw[dashed] (in-ws3-top-back-left)
    -- (in-ws3-top-back-right)
    -- (in-ws3-top-front-right)
    ;
    \draw[dashed] (in-ws3-bottom-back-right)
    -- (in-ws3-top-back-right)
    ;
    \draw (4.3, 0) node[left] {$-$}
    ;
  \end{tikzpicture}
  \caption{%
    Tree-level scattering BPS worldvolume of incoming ($-$) and outgoing ($+$) LG $\mathfrak{B}_2^{\theta}$-membranes.
  }
  \label{fig:m2 x r3:scattering membranes}
\end{figure}
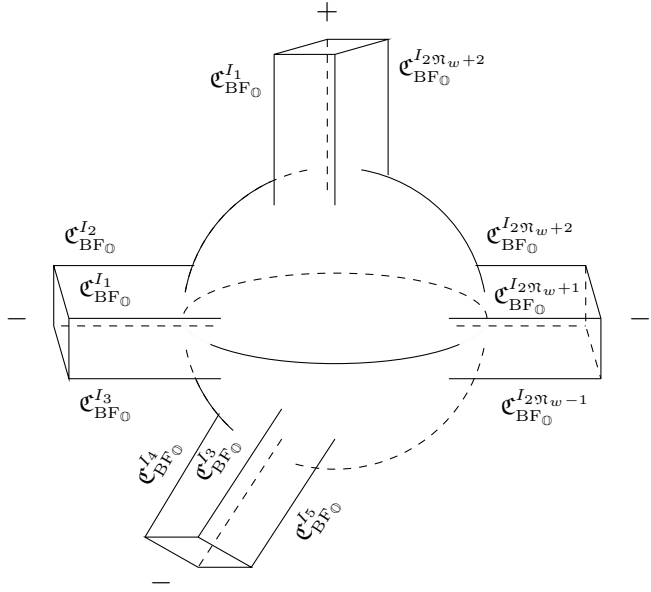

Just like in \autoref{sec:m2 x r3:orlov}, the normalized GM partition function will once again be a sum over the free-field correlation functions of operators that are in the $\hat{\mathcal{Q}}$-cohomology.
From the equivalent 3d-LG SQM and the 3d gauged B-twisted LG model perspective, the $\hat{\mathcal{Q}}$-cohomology will be spanned by LG $\mathfrak{B}_2^{\theta}$-membranes.
In turn, this means that the normalized GM partition function can also be regarded as a sum over tree-level scattering amplitudes of LG $\mathfrak{B}_2^{\theta}$-membranes as shown in \autoref{fig:m2 x r3:scattering membranes}.

In other words, we can express the normalized GM partition function as
\begin{equation}
  \label{eq:m2 x r3:3d-lg:normalized partition function}
  \widetilde{\mathcal{Z}}_{\text{GM}, M_2 \times \R^3}(G)
  = \sum_{\mathfrak{N}_w} \varPi^{\mathfrak{N}_w}_{\mathfrak{B}_2}
  \, ,
  \qquad
  \mathfrak{N}_w = 1, 2, \dots, \left\lfloor \frac{w - 2}{2} \right\rfloor
  \, ,
\end{equation}
where each
\begin{equation}
  \label{eq:m2 x r3:3d-lg:a-infinity structure}
  \saveboxed{eq:m2 x r3:3d-lg:a-infinity structure}{
    \begin{aligned}
      \varPi^{\mathfrak{N}_w}_{\mathfrak{B}_2}: \bigotimes_{i = 1}^{\mathfrak{N}_w}
      & \text{Hom} \left(
          \text{Hom} \left( \mathfrak{C}^{I_{2i - 1}}_{\text{BF}_{\OO}}, \mathfrak{C}^{I_{2i}}_{\text{BF}_{\OO}} \right),
          \text{Hom} \left( \mathfrak{C}^{I_{2(i + 1) - 1}}_{\text{BF}_{\OO}}, \mathfrak{C}^{I_{2(i + 1)}}_{\text{BF}_{\OO}} \right)
        \right)_-
      \\
      &\longto
        \text{Hom} \left(
          \text{Hom} \left( \mathfrak{C}^{I_1}_{\text{BF}_{\OO}}, \mathfrak{C}^{I_2}_{\text{BF}_{\OO}} \right),
          \text{Hom} \left( \mathfrak{C}^{I_{2 \mathfrak{N}_w + 1}}_{\text{BF}_{\OO}}, \mathfrak{C}^{I_{2 \mathfrak{N}_w + 2}}_{\text{BF}_{\OO}} \right)
        \right)_+
    \end{aligned}
  }
\end{equation}
is a scattering amplitude of $\mathfrak{N}_w$ incoming LG $\mathfrak{B}_2^{\theta}$-membranes $\text{Hom} \big( \text{Hom} \big( \mathfrak{C}^{*}_{\text{BF}_{\OO}}, \mathfrak{C}^{*}_{\text{BF}_{\OO}} \big), \text{Hom} \big( \mathfrak{C}^{*}_{\text{BF}_{\OO}}, $
\\ $\mathfrak{C}^{*}_{\text{BF}_{\OO}} \big) \big)_-$, and a single outgoing LG $\mathfrak{B}_2^{\theta}$-membrane $\text{Hom} \big( \text{Hom} \big( \mathfrak{C}^{I_1}_{\text{BF}_{\OO}}, \mathfrak{C}^{I_2}_{\text{BF}_{\OO}} \big), \text{Hom} \big( \mathfrak{C}^{I_{2\mathfrak{N}_w + 1}}_{\text{BF}_{\OO}}, \mathfrak{C}^{I_{2\mathfrak{N}_w + 2}}_{\text{BF}_{\OO}} \big) \big)_+$, with vertices as labeled, whose underlying worldvolume is as shown in \autoref{fig:m2 x r3:scattering membranes}. That is, $\varPi^{\mathfrak{N}_w}_{\mathfrak{B}_2}$ counts constant balls with $\mathfrak{N}_w + 1$ punctures at the boundary that are mapped to $\mathfrak{B}_2$ according to the BPS worldvolume equations (determined by setting to zero the expression within the squared terms in \eqref{eq:m2 x r3:3d-lg:action}) that are given by \eqref{eq:m2 x r3:2d-3d:bps worldvolume}.

In turn, this means that $\varPi^{\mathfrak{N}_w}_{\mathfrak{B}_2}$ counts the moduli space of solutions to \eqref{eq:m2 x r3:gm:octonionified equations:rotated} set to zero (or equivalently, the expression within the squared terms of \eqref{eq:m2 x r3:gm:action:L1} and \eqref{eq:m2 x r3:gm:action:L2} set to zero) with $(\mathfrak{N}_w + 1)$ boundary conditions that can be described as follows.
First, note that we can regard $\R^3$ as the effective worldvolume in \autoref{fig:m2 x r3:scattering membranes} that we shall denote as $\varXi$, so $M_5$ can be interpreted as a trivial fibration of $M_2$ over $\varXi$.
Next, at the $\mathfrak{N}_w + 1$ LG $\mathfrak{B}_2^{\theta}$-membranes on $\varXi$ where $\xi = \pm \infty$, \eqref{eq:m2 x r3:gm:octonionified equations:rotated} set to zero will become \eqref{eq:m2 x r3:go-bf:gm configs:1} and \eqref{eq:m2 x r3:go-bf:gm configs:2} with $A_t, A_{\tau}, A_{\xi} \rightarrow 0$.
Then, at the LG $\mathfrak{B}_2^{\theta}$-strings on $\varXi$ where $t = \pm \infty$, \eqref{eq:m2 x r3:gm:octonionified equations:rotated} set to zero will become \eqref{eq:m2 x r3:go-bf:gm configs:1} and \eqref{eq:m2 x r3:go-bf:gm configs:2} with $\partial_t \rightarrow 0$ and $A_t, A_{\tau}, A_{\xi} \rightarrow 0$.
Lastly, over the $\mathfrak{B}_2^{\theta}$-string boundaries on $\varXi$ where $\tau = \pm \infty$, \eqref{eq:m2 x r3:gm:octonionified equations:rotated} set to zero will become \eqref{eq:m2 x r3:go-bf:go-bf configs} which defines ($\theta$-deformed) $G_{\OO}$-BF configurations on $M_2$.

Note that the above description of the map $\varPi^{\mathfrak{N}_w}_{\mathfrak{B}_2}$ is similar to the description of the `$\varPi^{\mathfrak{N}_*}_{\mathfrak{A}_*}$' maps in our previous work \cite{er-2024-topol-gauge}.
Indeed, there is a mirror symmetry and Langlands duality which relates them, as we shall elucidate in \autoref{sec:dualities:cats}.

At any rate, notice that the $\varPi^{\mathfrak{N}_w}_{\mathfrak{B}_2}$ maps in \eqref{eq:m2 x r3:3d-lg:a-infinity structure} (like the $\mu^{\mathfrak{N}_w}_{\mathfrak{B}_2}$ maps in \eqref{eq:m2 x r3:2d-lg:composition map} of the 2d gauged B-twisted LG model) can also be regarded as composition maps defining an $A_{\infty}$-structure of a 2-category.

\subtitle{An RW-type $A_{\infty}$-2-category}

Altogether, this means that the normalized partition function of GM theory on $M_2 \times \R^3$ as expressed in \eqref{eq:m2 x r3:3d-lg:normalized partition function}, manifests a \emph{novel} gauge-theoretic RW-type $A_{\infty}$-2-category whose
(i) 2-objects are singular fibers $V_{2, \text{sg}}^{-1}(0)$ corresponding to $\theta$-deformed $G_{\OO}$-BF configurations on $M_2$, thereby 2-categorifying the holomorphic $G_{\OO}$-flat Floer homology of $M_2$,
(ii) 2-morphisms are Ext-groups corresponding to LG $\mathfrak{B}_2^{\theta}$-membranes, whose vertices are the 2-objects,
(iii) 1-objects are LG $\mathfrak{B}_2^{\theta}$-strings whose endpoints are the 2-objects,
(iv) 1-morphisms are Ext-groups corresponding to LG $\mathfrak{B}_2^{\theta}$-membranes, whose edges are the 1-objects,
and (v) $A_{\infty}$-structure is defined by the $\varPi^{\mathfrak{N}_m}_{\mathfrak{B}_2}$ maps \eqref{eq:m2 x r3:3d-lg:a-infinity structure}. These are all captured in \eqref{eq:m2 x r3:3d-lg:homs as ext}.

\subtitle{A Correspondence Between an Orlov-type $A_{\infty}$-1-category and an RW-type $A_{\infty}$-2-category}

Recall from \autoref{sec:m2 x r3:orlov} that the normalized partition function of GM theory on $M_2 \times \R^3$ also manifests the Orlov-type $A_{\infty}$-1-category of $\mathfrak{B}_2^{\theta}$-strings.
This means that we have a \emph{novel} correspondence between the RW-type $A_{\infty}$-2-category of the holomorphic $G_{\OO}$-flat Floer homology of $M_2$ and the Orlov-type $A_{\infty}$-1-category of $\mathfrak{B}_2^{\theta}$-strings!

\subtitle{A Gauge-theoretic Generalization of DR's Mathematical Conjecture}

In \autoref{sec:m3 x r2:rw}, we furnished a physical proof of DR's mathematical conjecture in \cite[$\S$6]{doan-2022-holom-floer} which (from the bottom line of \eqref{eq:m3 x r2:correspondence amongst cats}) can be interpreted as a correspondence between an RW $A_{\infty}$-2-category of branes in $X$ and an Orlov-type $A_{\infty}$-1-category of branes in $\mathfrak{P}(\R, X)$.
Therefore, the correspondence between the gauge-theoretic RW-type $A_{\infty}$-2-category of objects in $\mathfrak{B}_2$ and the gauge-theoretic Orlov-type $A_{\infty}$-1-category of objects in $\mathfrak{P}(\R, \mathfrak{B}_2)$,
would mean that we actually have a \emph{gauge-theoretic generalization} of DR's mathematical conjecture!

\section{A Mirror Duality of Sigma Models and Langlands Duality of Gauge Theories}
\label{sec:dualities:theories}

In this section, we will first explain how the 3d $\mathcal{N} = 4$ A and B-model in Hitchin moduli spaces can be mirror dual via enhanced Homological Mirror Symmetry (HMS), and why this will imply a Langlands duality between the 5d $\mathcal{N} = 2$ HW and GM theory.
We will then explain how 5d $\mathcal{N} = 2$ HW and GM theory can be Langlands dual via 4d $\mathcal{N} = 4$ S-duality, and why this will lead to the above mirror duality between the 3d $\mathcal{N} = 4$ A and B-models, thereby furnishing a more fundamental derivation of these dualities.

\subsection{A Mirror Duality Between the 3d \texorpdfstring{$\mathcal{N} = 4$}{N = 4} A-model and B-model in Hitchin Moduli Spaces via Enhanced HMS}
\label{sec:dualities:thoeries:3d}

The duality between a 2d $\mathcal{N} = (4, 4)$ A and B-model with mirror hyperkähler target spaces, and its relevance to the geometric Langlands program, was first studied by Kapustin-Witten (KW) in \cite{kapustin-2006-elect-magnet}, where recently, it has also been formulated rigorously as enhanced Homological Mirror Symmetry (HMS) by Hausel in \cite{hausel-2021-enhan-mirror}.\footnote{%
  Hausel's enhanced HMS involves the additional information of (i) how hyperkähler branes are matched under mirror symmetry, and (ii) the symmetry between Wilson and Hecke (or t'Hooft) operators.
  \label{ft:enhanced hms}
}
In particular, when the hyperkähler target spaces are Hitchin moduli spaces, KW showed that a 2d $\mathcal{N} = (4, 4)$ A-model with target $\mathcal{M}^G_{\text{H}}(C, \mathbf{K})$ is dual to a 2d $\mathcal{N} = (4, 4)$ B-model with a mirror target $\mathcal{M}^{^{L}G}_{\text{H}}(C, \mathbf{J})$, where $^{L}G$ is the Langlands dual of $G$.

Recall from
(i) \autoref{sec:hw:3d} and
(ii) \autoref{sec:gm:3d} that the spectrum of states of
(i) the 3d $\mathcal{N} = 4$ A-model with target $\mathcal{M}^G_{\text{H}}(C, \mathbf{K})$ on $\Sigma \times \R$,
and (ii) the 3d $\mathcal{N} = 4$ B-model with target $\mathcal{M}^G_{\text{H}}(C, \mathbf{J})$ on $\Sigma \times \R$, are given by the spectrum of states of
(i) a 2d $\mathcal{N} = (4, 4)$ A-model with target $\mathcal{M}^G_{\text{H}}(C, \mathbf{K})$ on $\Sigma$,
and (ii) a 2d $\mathcal{N} = (4, 4)$ B-model with target $\mathcal{M}^G_{\text{H}}(C, \mathbf{J})$ on $\Sigma$, respectively.

The enhanced HMS of the 2d $\mathcal{N} = (4, 4)$ A and B-models therefore implies that
the 3d $\mathcal{N} = 4$ A-model with target $\mathcal{M}^G_{\text{H}}(C, \mathbf{K})$ on $\Sigma \times \R$ is dual to the 3d $\mathcal{N} = 4$ B-model with target $\mathcal{M}^{^LG}_{\text{H}}(C, \mathbf{J})$ on $\Sigma \times \R$.

In other words, we have a mirror duality between the 3d $\mathcal{N} = 4$ A and B-model in mirror Hitchin moduli spaces.

\subsection{A Langlands Duality Between the 5d \texorpdfstring{$\mathcal{N} = 2$}{N = 2} HW and GM Theory on \texorpdfstring{$M_3 \times M_1 \times \R$}{M3 x M1 x R} via Enhanced HMS}
\label{sec:dualities:theories:5d}

Recall from (i) \autoref{sec:hw:equivalence} and (ii) \autoref{sec:gm:equivalence} that the
(i) 3d A-model with target $\mathcal{M}^G_{\text{H}}(C, \mathbf{K})$ on $I \times M_1 \times \R$ and
(ii) 3d B-model with target $\mathcal{M}^G_{\text{H}}(C, \mathbf{J})$ on $I \times M_1 \times \R$, are equivalent to
(i) HW theory of $G$ on $M_3 \times M_1 \times \R$ and
(ii) GM theory of $G$ on $M_3 \times M_1 \times \R$, respectively.

The duality between the 3d A and B-model in mirror Hitchin moduli spaces in \autoref{sec:dualities:thoeries:3d} therefore implies a duality between  HW theory of $G$ on $M_3 \times M_1 \times \R$ and GM theory of $^{L}G$ on $M_3 \times M_1 \times \R$.

In other words, we have a Langlands duality between the 5d $\mathcal{N} = 2$ HW and GM theory on $M_3 \times M_1 \times \R$ of $G$ and $^{L}G$.

\subsection{A Langlands Duality Between the 5d \texorpdfstring{$\mathcal{N} = 2$}{N = 2} HW and GM Theory on \texorpdfstring{$C \times \R^3$}{C x R3} via Enhanced HMS}
\label{sec:dualities:5d:r3}

Consider now (i) HW theory of $G$ and (ii) GM theory of $G$, on $C \times \R^3$.
Performing a BJSV reduction along the Riemann surface $C$ results in a
(i) 3d A-model with target $\mathcal{M}^G_{\text{H}}(C, \mathbf{K})$ on $\R^3$ whose action is \eqref{eq:hw:bjsv:sigma model:action} (where $\Sigma = \R^2$),
and (ii) 3d B-model with target $\mathcal{M}^G_{\text{H}}(C, \mathbf{J})$ on $\R^3$ whose action is \eqref{eq:gm:bjsv:sigma model:rw model} (where $\Sigma = \R^2$).

The duality between the 3d A and B-model in mirror Hitchin moduli spaces in \autoref{sec:dualities:thoeries:3d} therefore implies a duality between HW theory of $G$ on $C \times \R^3$ and GM theory of $^{L}G$ on $C \times \R^3$.

In other words, we also have a Langlands duality between the 5d $\mathcal{N} = 2$ HW and GM theory on $C \times \R^3$ of $G$ and $^{L}G$.

\subsection{A Langlands Duality Between the 5d \texorpdfstring{$\mathcal{N} = 2$}{N = 2} HW and GM Theory on \texorpdfstring{$M_3 \times M_1 \times \R$}{M3 x M1 x R} or \texorpdfstring{$C \times \R^3$}{C x R3} via 4d \texorpdfstring{$\mathcal{N} = 4$}{N = 4} S-duality}
\label{sec:dualities:theories:5d:s-dual}

We will now furnish a more fundamental derivation than that in \autoref{sec:dualities:theories:5d} and \autoref{sec:dualities:5d:r3} of the Langlands duality between the 5d $\mathcal{N} = 2$ gauge theories, from which the mirror duality between 3d sigma models in \autoref{sec:dualities:thoeries:3d} can actually be derived (in \autoref{sec:dualities:theories:3d:s-dual}).

To this end, let us first consider HW and GM theory on $M_3 \times M_1 \times \R$, where $M_3$ is a (not necessarily compact) three-manifold, and $M_1$ is the direction of time that we shall denote by $\tau$.\footnote{%
  Observables in our TQFTs are coordinate-independent; in particular,  they are independent of the coordinate directions in Euclidean spacetime. Thus, there is no preferred direction for time, so we are free to redefine the direction of time.
  \label{ft:dualities:freedom to choose direction of time}
}

\subtitle{HW Theory on $M_3 \times M_1 \times \R$ and its Effective  KW Theory on $M_3 \times \R$ at $t = -1$}

Recall how we showed, in \autoref{sec:hw:3d}, that the spectrum of states of the \emph{topological} 3d $\mathcal{N} = 4$ A-model (given by its $\mathcal{Q}$-cohomology) are effectively given by the spectrum of states of the 2d $\mathcal{N} = (4,4)$ A-model living on a time-invariant hyperslice of it.
Likewise, the spectrum of states of the \emph{topological} 5d $\mathcal{N} = 2$ HW theory (given by its $\mathcal{Q}$-cohomology) will effectively be given by the spectrum of states of the 4d $\mathcal{N} = 4$ gauge theory living on a time-invariant hyperslice of it.
We can determine what this 4d theory is through its BPS equations, which, in turn, are given by the BPS equations of HW theory (obtained by setting to zero the $\mathcal{Q}$-variations of the fermions in \eqref{eq:hw susy variations}) restricted to the hyperslice. Specifically, the 4d BPS equations of interest are obtained by setting $\partial_{\tau} \rightarrow 0$ in the BPS equations of HW theory whilst dropping all $\tau$-dependence in the fields, i.e., they are\footnote{%
  To arrive at the following expressions, we have chosen an orientation of $M_3 \times M_1$ such that it follows the orientation of $M_3$, i.e., $\epsilon_{\tau mpq} \equiv \epsilon_{mpq}$.
}
\begin{equation}
  \label{eq:dualities:hw:bps equations}
  \begin{aligned}
    F_{mn}
    - \epsilon_{mnp} D^p A^{\tau}
    - \frac{1}{2} [B_{m\tau}, B_n^{\; \tau}]
    - \frac{1}{2} [B_{mp}, B_n^{\; p}]
    - D_t B_{mn}
    &= 0
      \, ,
    & \qquad
      D_t \upsilon
    &= 0
      \, ,
    \\
    \frac{1}{2} \epsilon_{mpq} F^{pq}
    - D_m A_{\tau}
    - \frac{1}{2} [B_{\tau n}, B_m^{\; n}]
    - D_t B_{\tau m}
    &= 0
      \, ,
    & \qquad
      D_m \upsilon
    &= 0
      \, ,
    \\
    F_{tm}
    + D^n B_{nm}
    + [A^{\tau}, B_{\tau m}]
    &= 0
      \, ,
    & \qquad
      [A_{\tau}, \upsilon]
    &= 0
      \, ,
    \\
    D_t A_{\tau}
    + D^m B_{m\tau}
    &= 0
      \, ,
    & \qquad
      [B_{\mu\nu}, \upsilon]
    &= 0
      \, ,
    \\
    &
    &
    [\upsilon, \bar{\upsilon}]
    &= 0
      \, ,
  \end{aligned}
\end{equation}
where $A_{\tau}$ is now a scalar on $M_3 \times \R$.

Notice that by defining a one-form $\phi \in \Omega^1(M_3 \times \R, \text{ad}(G))$ as $\phi \coloneq - A_{\tau} dt + B_{\tau m} dx^m$, \eqref{eq:dualities:hw:bps equations} can be written as
\begin{equation}
  \label{eq:dualities:hw:bps equations:hyperslice}
  \begin{aligned}
    F_{\mu\nu}
    - [\phi_{\mu}, \phi_{\nu}]
    - \frac{1}{2} \epsilon_{\mu\nu\rho\sigma} D^{\rho} \phi^{\sigma}
    &= 0
    \, ,
    & \qquad
    D_{\mu} \upsilon
    &= 0
    \, ,
    \\
    D^{\mu} \phi_{\mu}
    &= 0
    \, ,
    & \qquad
    [\phi_{\mu}, \upsilon]
    &= 0
    \, ,
    \\
    &
    &
    [\upsilon, \bar{\upsilon}]
    &= 0
    \, ,
  \end{aligned}
\end{equation}
where $\{\mu, \nu\} \in \{m, t\}$ are indices on $M_3 \times \R$.

These are the BPS equations of a 4d $\mathcal{N} = 4$ KW theory \cite[$\S$3.2]{kapustin-2006-elect-magnet}, i.e.,
\begin{equation}
  \label{eq:dualities:kw:bps equation}
  \begin{aligned}
    (F - \phi \wedge \phi + t D\phi)^+
    &= 0
      \, ,
    & \qquad
      D \upsilon
      + t [\phi, \upsilon]
    &= 0
      \, ,
    \\
    (F - \phi \wedge \phi - t^{-1} D\phi)^-
    &= 0
      \, ,
    & \qquad
      D \upsilon
      - t^{-1} [\phi, \upsilon]
    &= 0
      \, ,
    \\
    D^{*} \phi
    - t^{-1} [\upsilon, \bar{\upsilon}]
    &= 0
    & \qquad
      D^{*} \phi
      + t [\upsilon, \bar{\upsilon}]
    &= 0
      \, ,
  \end{aligned}
\end{equation}
at $t = -1$.

This therefore means that the spectrum of HW theory on $M_3 \times M_1 \times \R$ is given by the spectrum of KW theory on $M_3 \times \R$ at $t = - 1$.
In other words, the $\mathcal{Q}$-cohomology of HW theory on $M_3 \times M_1 \times \R$ is \emph{effectively} the $\mathcal{Q}$-cohomology of KW theory on $M_3 \times \R$ at $t = - 1$.

\subtitle{GM Theory on $M_3 \times M_1 \times \R$ and its Effective  KW Theory on $M_3 \times \R$ at $t = -i$}

We repeat the analysis for 5d $\mathcal{N} = 2$ GM theory on $M_3 \times M_1 \times \R$.
The 4d $\mathcal{N} = 4$ gauge theory living on a time-invariant hyperslice of it can be determined through its BPS equations, which, in turn, are given by the BPS equations of GM theory (obtained by setting to zero the $\hat{\mathcal{Q}}$-variations of the fermions in \eqref{eq:gm:susy variations}) restricted to the hyperslice. Specifically, the 4d BPS equations of interest are obtained by setting $\partial_{\tau} \rightarrow 0$ in the BPS equations of GM theory whilst dropping all $\tau$-dependence in the fields, i.e., they are\footnote{%
  The BPS equation in the third line corresponds to the BPS equation involving the auxiliary field, i.e., $H = 0$, which via its equation of motion from \eqref{eq:gm:m4 x r:action}, implies that $D^t \phi_t + D^m \phi_m + \partial^{\tau} \phi_{\tau} + [A^{\tau}, \phi_{\tau}] = 0$.
}
\begin{equation}
  \label{eq:dualities:gm:bps equations}
  \begin{aligned}
    \overline{\mathcal{F}}_{tm}
    &= 0
      \, ,
    & \qquad
      \overline{\mathcal{D}}_{t} \overline{\mathcal{A}}_{\tau}
    &= 0
      \, ,
    \\
    \overline{\mathcal{F}}_{mn}
    &= 0
      \, ,
    & \qquad
      \overline{\mathcal{D}}_m \overline{\mathcal{A}}_{\tau}
    &= 0
      \, ,
    \\
    D^t \phi_t
    + D^m \phi_m
    + [A^{\tau}, \phi_{\tau}]
    &= 0
      \, .
  \end{aligned}
\end{equation}

Notice that by defining a complex scalar $\upsilon \in \Omega^0(M_3 \times \R, \text{ad}(G))$ as $\upsilon \coloneq \overline{\mathcal{A}}_{\tau} / \sqrt{2}$, \eqref{eq:dualities:gm:bps equations} can be written as
\begin{equation}
  \label{eq:dualities:gm:bps equations:hyperslice}
  \begin{aligned}
    \overline{\mathcal{F}}_{\mu\nu}
    &= 0
      \, ,
    &\qquad
      D_{\mu} \upsilon
      - i [\phi_{\mu}, \upsilon]
    &= 0
      \, ,
    \\
    D^{\mu} \phi_{\mu}
    + i [\upsilon, \bar{\upsilon}]
    &= 0
      \, .
  \end{aligned}
\end{equation}
These are the BPS equations \eqref{eq:dualities:kw:bps equation} of a 4d $\mathcal{N} = 4$ KW theory  at $t = - i$.

This therefore means that the spectrum of GM theory on $M_3 \times M_1 \times \R$ is given by the spectrum of KW theory on $M_3 \times \R$ at $t = -i$.
In other words, the $\hat{\mathcal{Q}}$-cohomology of GM theory on $M_3 \times M_1 \times \R$ is \emph{effectively} the $\hat{\mathcal{Q}}$-cohomology of KW theory on $M_3 \times \R$ at $t = -i$.

\subtitle{Langlands Duality Between 5d $\mathcal{N} = 2$ HW and GM Theory on $M_3 \times M_1 \times \R$ via 4d $\mathcal{N} = 4$ S-duality}

Note that as a 4d $\mathcal{N} = 4$ theory, KW theory enjoys S-duality.
In particular, KW theory of $G$ at $t = -1$ is S-dual to KW theory of $^{L}G$ at $t = -i$~\cite{kapustin-2006-elect-magnet}.
Recall that the spectrum of states of (i) HW theory and (ii) GM theory, on $M_3 \times M_1 \times \R$, are given by the spectrum of states of KW theory on $M_3 \times \R$ at (i) $t = -1$ and (ii) $t = -i$, respectively.
As such,  the 4d $\mathcal{N} = 4$ S-duality of KW theory would imply that 5d $\mathcal{N} = 2$ HW theory of $G$ is Langlands dual to 5d $\mathcal{N} = 2$ GM theory of $^{L}G$, on $M_3 \times M_1 \times \R$, just as we observed in \autoref{sec:dualities:theories:5d}.

As this will also allow us to derive, in \autoref{sec:dualities:theories:3d:s-dual}, the mirror duality between 3d sigma models in \autoref{sec:dualities:thoeries:3d}, we actually have a more fundamental derivation than in \autoref{sec:dualities:theories:5d}  of the Langlands duality between the 5d $\mathcal{N} = 2$ HW and GM theory on $M_3 \times M_1 \times \R$ of $G$ and $^{L}G$.

\subtitle{Langlands Duality Between 5d $\mathcal{N} = 2$ HW and GM Theory on $C \times \R^3$ via 4d $\mathcal{N} = 4$ S-duality}

Let us now consider our above discussion with $M_3 = C \times \R$ and $M_1 = \R$.
Then, 4d $\mathcal{N} = 4$ S-duality of KW theory would imply that 5d $\mathcal{N} = 2$ HW theory of $G$ is Langlands dual to 5d $\mathcal{N} = 2$ GM theory of $^{L}G$, on $C \times \R^3$, just as we observed in \autoref{sec:dualities:5d:r3}.

Similarly, we have a more fundamental derivation than in  \autoref{sec:dualities:5d:r3} of the Langlands duality between the 5d $\mathcal{N} = 2$ HW and GM theory on $C \times \R^3$ of $G$ and $^{L}G$.

\subsection{A Mirror Duality Between the 3d \texorpdfstring{$\mathcal{N} = 4$}{N = 4} A-model and B-model in Hitchin Moduli Spaces via 4d \texorpdfstring{$\mathcal{N} = 4$}{N = 4} S-duality}
\label{sec:dualities:theories:3d:s-dual}

Recall from (i) \autoref{sec:hw:equivalence} and (ii) \autoref{sec:gm:equivalence} that (i) HW theory of $G$ on $M_3 \times M_1 \times \R$ and (ii) GM theory of $G$ on $M_3 \times M_1 \times \R$, are equivalent to  a (i) 3d A-model with target $\mathcal{M}^G_{\text{H}}(C, \mathbf{K})$ on $I \times M_1 \times \R$ and (ii) 3d B-model with target $\mathcal{M}^G_{\text{H}}(C, \mathbf{J})$ on $I \times M_1 \times \R$, respectively.

The duality in \autoref{sec:dualities:theories:5d:s-dual} between the (i) 5d HW theory of $G$ and (ii) 5d GM theory of $^{L}G$, on $M_3 \times M_1 \times \R$, therefore implies a duality between the (i) 3d A-model with target $\mathcal{M}^G_{\text{H}}(C, \mathbf{K})$ and (ii) 3d B-model with target $\mathcal{M}^{^{L}G}_{\text{H}}(C, \mathbf{J})$, on $I \times M_1 \times \R$, just as we observed in \autoref{sec:dualities:thoeries:3d}.

Clearly, we have a more fundamental derivation than in \autoref{sec:dualities:thoeries:3d} of the mirror duality between the 3d $\mathcal{N} = 4$ A and B-model in mirror Hitchin moduli spaces.

\subsection{A Summary of the Mirror and Langlands Duality Between Theories}
\label{sec:dualities:theories:diagram}

\begin{figure}[ht]
  \savefig{fig:dualities:theories}{
    \begin{adjustbox}{
        max totalsize={\textwidth},%
      }
      \centering
      \begin{tikzpicture}[%
        auto,%
        on grid,%
        block/.style={draw, rectangle},%
        novel/.style={draw, rectangle, ultra thick},%
        every edge/.style={draw, <->},%
        relation/.style={scale=0.8, sloped, anchor=center, align=center},%
        vertRelation/.style={scale=0.8, anchor=center, align=center},%
        horRelation/.style={scale=0.8, anchor=center, align=center},%
        decoration={snake, pre length=3mm,post length=3mm},%
        shorten >=4pt,%
        shorten <=4pt,%
        ]
        \def \verRel {3} 
        \def \horRel {10} 
        \node[block] (2d-A)
        {
          \begin{tabular}{c}
            2d $\mathcal{N} = (4, 4)$ A-model
            \\
            on $\Sigma$ in $\mathcal{M}^G_{\text{H}}(C, \mathbf{K})$
          \end{tabular}
        };
        \node[block, right={\horRel} of 2d-A] (2d-B)
        {
          \begin{tabular}{c}
            2d $\mathcal{N} = (4, 4)$ B-model
            \\
            on $\Sigma$ in $\mathcal{M}^{^{L}G}_{\text{H}}(C, \mathbf{J})$
          \end{tabular}
        };
        \node[block, very thick, above={\verRel} of 2d-A] (3d-A-Sigma)
        {
          \begin{tabular}{c}
            3d $\mathcal{N} = 4$ A-model
            \\
            on $\Sigma \times \R$ in $\mathcal{M}^G_{\text{H}}(C, \mathbf{K})$
          \end{tabular}
        };
        \node[block, very thick, above={\verRel} of 2d-B] (3d-B-Sigma)
        {
          \begin{tabular}{c}
            3d $\mathcal{N} = 4$ B-model (RW model)
            \\
            on $\Sigma \times \R$ in $\mathcal{M}^{^{L}G}_{\text{H}}(C, \mathbf{J})$
          \end{tabular}
        };
        \node[block, very thick, above={\verRel} of 3d-A-Sigma] (HW-R)
        {
          \begin{tabular}{c}
            5d $\mathcal{N} = 2$ $\text{HW}^G$
            \\
            on $M_3 \times M_1 \times \R$
          \end{tabular}
        };
        \node[block, very thick, above={\verRel} of 3d-B-Sigma] (GM-R)
        {
          \begin{tabular}{c}
            5d $\mathcal{N} = 2$ $\text{GM}^{^{L}G}$
            \\
            on $M_3 \times M_1 \times \R$
          \end{tabular}
        };
        \node[block, above={\verRel} of HW-R] (KW-t1)
        {
          \begin{tabular}{c}
            4d $\mathcal{N} = 4$ $\text{KW}^G_{(t = -1)}$
            \\
            on $M_3 \times \R$
          \end{tabular}
        };
        \node[block, above={\verRel} of GM-R] (KW-ti)
        {
          \begin{tabular}{c}
            4d $\mathcal{N} = 4$ $\text{KW}^{^{L}G}_{(t = -i)}$
            \\
            on $M_3 \times \R$
          \end{tabular}
        };
        \node[block, very thick, above={\verRel} of KW-t1] (HW-R3)
        {
          \begin{tabular}{c}
            5d $\mathcal{N} = 2$ $\text{HW}^G$
            \\
            on $C \times \R^3$
          \end{tabular}
        };
        \node[block, very thick, above={\verRel} of KW-ti] (GM-R3)
        {
          \begin{tabular}{c}
            5d $\mathcal{N} = 2$ $\text{GM}^{^{L}G}$
            \\
            on $C \times \R^3$
          \end{tabular}
        };
        \draw
        (HW-R)
        edge[dashed]
        node[vertRelation, right] {
          \begin{tabular}{c}
            $M_3 = M_3' \bigcup_{\widehat{C}} M_3''$
            \\
            $\Sigma = I \times M_1$
            \\
          \end{tabular}
        }
        (3d-A-Sigma)
        (GM-R)
        edge[dashed]
        node[vertRelation, left] {
          \begin{tabular}{c}
            $M_3 = M_3' \bigcup_{\widehat{C}} M_3''$
            \\
            $\Sigma = I \times M_1$
            \\
          \end{tabular}
        }
        (3d-B-Sigma)
        (3d-A-Sigma)
        edge [double, dotted]
        node[vertRelation, left] {Hyperslice on $\R$}
        (2d-A)
        (3d-B-Sigma)
        edge [double, dotted]
        node[vertRelation, right] {Hyperslice on $\R$}
        (2d-B)
        (HW-R3.west)
        edge[dashed, bend right]
        node[vertRelation, left] {
          \begin{tabular}{c}
            $C = \widehat{C}$
            \\
            $\Sigma = \R^2$
          \end{tabular}
        }
        (3d-A-Sigma.west)
        (GM-R3.east)
        edge[dashed, bend left]
        node[vertRelation, right] {
          \begin{tabular}{c}
            $C = \widehat{C}$
            \\
            $\Sigma = \R^2$
          \end{tabular}
        }
        (3d-B-Sigma.east)
        (HW-R3.south)
        edge [double, dotted]
        node[vertRelation, left] {Hyperslice on $M_1$}
        node[vertRelation, right] {
          \begin{tabular}{c}
            $M_3 = C \times \R$
            \\
            $M_1 = \R$
          \end{tabular}
        }
        (KW-t1.north)
        (GM-R3.south)
        edge [double, dotted]
        node[vertRelation, right] {Hyperslice on $M_1$}
        node[vertRelation, left] {
          \begin{tabular}{c}
            $M_3 = C \times \R$
            \\
            $M_1 = \R$
          \end{tabular}
        }
        (KW-ti.north)
        (HW-R.north)
        edge [double, dotted]
        node[vertRelation, left] {Hyperslice on $M_1$}
        (KW-t1.south)
        (GM-R.north)
        edge [double, dotted]
        node[vertRelation, right] {Hyperslice on $M_1$}
        (KW-ti.south)
        ;
        \draw
        (2d-A)
        edge [densely dotted]
        node[relation, below] (HMS) {Enhanced HMS}
        (2d-B)
        (3d-A-Sigma)
        edge 
        node[relation, below] (3d-Sigma-Dual) {Mirror Duality}
        (3d-B-Sigma)
        (HW-R)
        edge 
        node[relation, below] (5d-R-Dual) {Langlands Duality}
        (GM-R)
        (HW-R3)
        edge 
        node[relation, below] (5d-R3-Dual) {Langlands Duality}
        (GM-R3)
        (KW-t1)
        edge 
        node[relation, below] (S-Dual) {4d $\mathcal{N} = 4$ S-duality}
        (KW-ti)
        (HMS)
        edge[decorate, double, -{Stealth[length=3mm]}]
        (3d-Sigma-Dual)
        (3d-Sigma-Dual.north east)
        edge[decorate, bend right, double, {Stealth[length=3mm]}-{Stealth[length=3mm]}]
        (5d-R-Dual.south east)
        (3d-Sigma-Dual.north west)
        edge[decorate, bend left, double, {Stealth[length=3mm]}-{Stealth[length=3mm]}]
        (5d-R3-Dual.south west)
        (S-Dual)
        edge[decorate, double, -{Stealth[length=3mm]}]
        (5d-R-Dual)
        (S-Dual)
        edge[decorate, double, -{Stealth[length=3mm]}]
        (5d-R3-Dual)
        ;
      \end{tikzpicture}
    \end{adjustbox}
    \caption{Mirror and Langlands duality between relevant theories and their physical derivations.}
  }
  \label{fig:dualities:theories}
\end{figure}
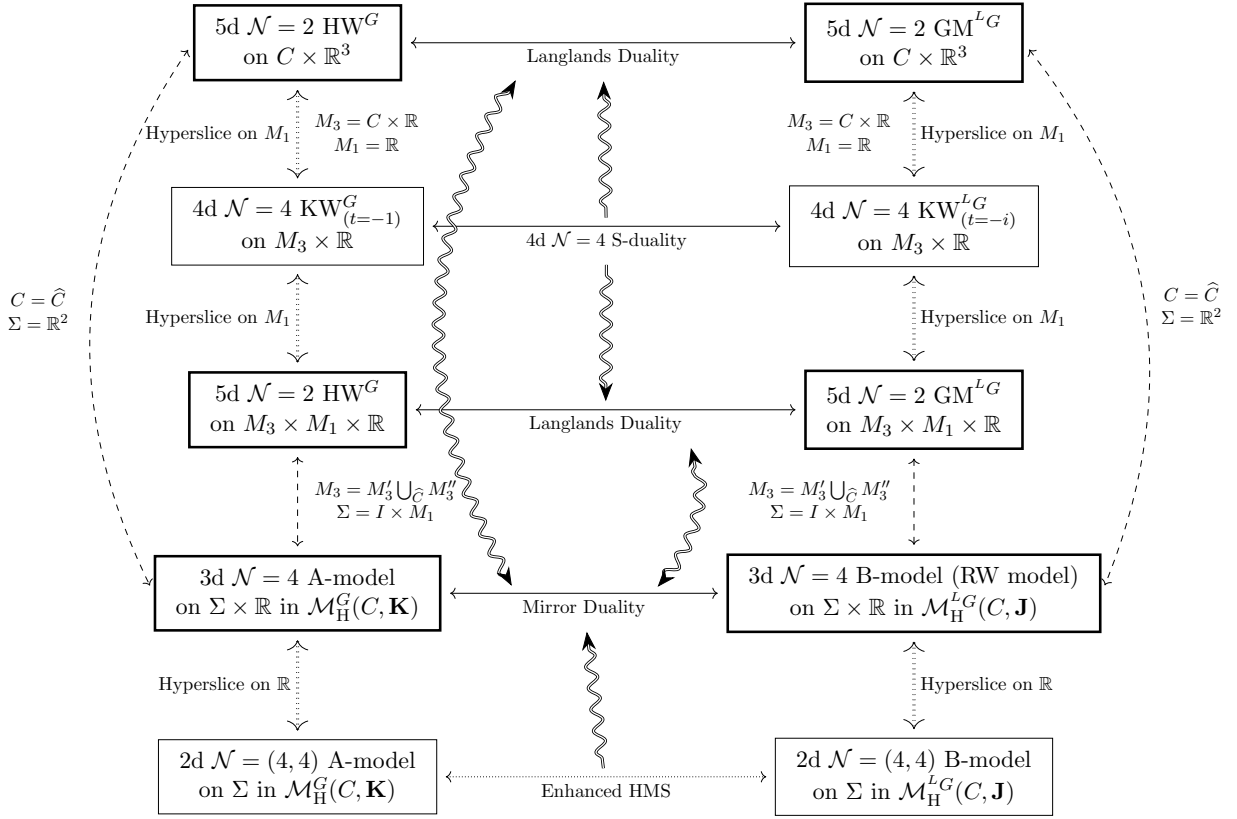

In \autoref{fig:dualities:theories}, we summarize the mirror and Langlands duality between the relevant theories and their underlying physical derivations, where
(i) dashed lines are equivalences between theories obtained via a BJSV reduction along (a Heegaard surface) $\widehat C$ of genus $g(\widehat C)\geq 2$ with variable size;
(ii) dotted lines are spectral equivalences between theories;
and (iii) wavy lines denote the implications of a relation on other relations.

\section{A Web of Langlands Dual Relations Amongst the Floer Homologies}
\label{sec:dualities:floer}

In this section, we will use the results of the previous sections and our previous work in \cite{er-2023-topol-n}, to derive a web of Langlands dual relations amongst the Floer homologies physically realized by 5d $\mathcal{N} = 2$ HW theory of $G$ and GM theory of $^{L}G$ on $M_4 \times \R$, where $M_4$ is a possibly decomposable four-manifold.

\subsection{Relations from the Topological Invariance of HW and GM Theory}
\label{sec:dualities:floer:gauge}

\subtitle{Relations Between the Gauge-theoretic Floer Homologies from GM Theory}

The topological invariance of (the $\hat{\mathcal{Q}}$-cohomology of) GM theory in all directions means that the partition functions \eqref{eq:m4 x r:partition function}, \eqref{eq:m3 x r:partition function}, and \eqref{eq:m2 x r:partition function}, are equivalent, whence we can relate them as
\begin{equation}
  \label{eq:dualities:floer:gm}
  \saveboxed{eq:dualities:floer:gm}{
    \begin{tikzcd}[
      row sep=huge,%
      column sep=huge,%
      arrows=leftrightarrow,%
      ampersand replacement=\&,%
      ]
      \sum_u \text{HHF}_{d_u}^{\text{flat}}(M_4, G_{\C})
      \arrow[r, "M_4 = M_3 \times \widehat{S}^1"] 
      \&
      \sum_v \text{HHF}_{d_v}^{\text{flat}}(M_3, G_{\HH})
      \arrow[r, "M_3 = M_2 \times \widehat{S}^1"] 
      \&
      \sum_w \text{HHF}_{d_w}^{\text{flat}}(M_2, G_{\OO})
    \end{tikzcd}
  }
\end{equation}
where $\widehat{S}^1$ denotes a circle of variable radius.

The relations in \eqref{eq:dualities:floer:gm} are consistent, in that they have a one-to-one correspondence in their summations over `$u$', `$v$', and `$w$'.
Specifically, each `$u$', `$v$', and `$w$' corresponds to a solution of (the simultaneous vanishing of the LHS and RHS of) \eqref{eq:m4 x r:gm flow}, \eqref{eq:m3 x r:gm4 flow}, and \eqref{eq:m2 x r:gm3 flow}, respectively, where \eqref{eq:m2 x r:gm3 flow} is obtained via a KK reduction of \eqref{eq:m3 x r:gm4 flow}, which in turn is obtained via a KK reduction of \eqref{eq:m4 x r:gm flow}.

\subtitle{Relations Between the Gauge-theoretic Floer Homologies from HW Theory}

In \cite[$\S$8.1]{er-2023-topol-n}, we showed how the topological invariance of (the $\mathcal{Q}$-cohomology of) HW theory in all directions meant that we could relate the partition functions of HW theory and its lower dimensional reductions in \cite[$\S$3--$\S$5]{er-2023-topol-n} as\footnote{%
  The notation of the Floer homologies to appear in the following equation differs from the notation used in \cite[$\S$3--$\S$5]{er-2023-topol-n} as follows.
  First, the superscript ``$\text{HW}_d\text{-inst}$'' refers to the BPS equation of $\text{HW}_{d}$-theory on $M_d \times \R$.
  $\text{HW}_5$ is the 5d HW theory as in \autoref{sec:hw}, and ``$\text{HW}_5\text{-inst}$'' are the HW equations \eqref{eq:hw:hw eqn} on $M_4 \times \R$; $\text{HW}_4$ is the 4d theory obtained via KK reduction of $\text{HW}_5$ along an $S^1$ circle within $M_4$, and ``$\text{HW}_4\text{-inst}$'' is the BPS equation of said theory; and likewise for $\text{HW}_3$ and ``$\text{HW}_3\text{-inst}$''.
  These BPS equations can be interpreted as instanton-type equations on $M_d \times \R$ as they involve an $F^+$ term.
  Second, the Floer homology classes of $M_2$ in \cite[$\S$5]{er-2023-topol-n} are generated by $G_{\C}$-BF configurations on $M_2$ as critical points of a (real-valued) Morse functional of $G_{\C}$-valued fields; the gauge group datum thus required to denote the classes therein was $G_{\C}$, i.e., we denoted them by $\text{HF}^{\text{HW}_3}(M_2, G_{\C})$.
  However, it is straightforward to show that whilst these Floer homology classes are generated by $G_{\C}$-valued configurations on $M_2$, the BPS equation on $M_2 \times \R$ need not be complexified. Hence, since the gauge group datum is now meant to specify the BPS equation, it can just be $G$.
  That is, the HW$_3$-instanton Floer homology used in the following equation and the rest of this paper will be specified by $G$.
  \label{ft:dualities:floer:notation}
}
\begin{equation}
  \label{eq:dualities:floer:hw}
  \begin{tikzcd}[
    row sep=huge,%
    column sep=huge,%
    arrows=leftrightarrow,%
    ampersand replacement=\&,%
    ]
    \sum_k \text{HF}^{\text{HW}_5\text{-inst}}_{d_k}(M_4, G)
    \arrow[r, "M_4 = M_3 \times {\hat S}^1"] 
    \&
    \sum_l \text{HF}_{d_l}^{\text{HW}_4\text{-inst}}(M_3, G)
    \arrow[r, "M_3 = M_2 \times {\hat S}^1"] 
    \&
    \sum_m \text{HF}_{d_m}^{\text{HW}_3\text{-inst}}(M_2, G)
    \, ,
  \end{tikzcd}
\end{equation}
where
(i) $\text{HF}^{\text{HW}_5\text{-inst}}_{d_k}(M_4, G)$,
(ii) $\text{HF}_{d_l}^{\text{HW}_4\text{-inst}}(M_3, G)$,
and (iii) $\text{HF}_{d_m}^{\text{HW}_3\text{-inst}}(M_2, G)$,
are gauge-theoretic Floer homologies generated by
(i) VW configurations on $M_4$,
(ii) 3d-Hitchin configurations on $M_3$,
and (iii) $G_{\C}$-BF configurations on $M_2$.
The BPS equations on $M_D \times \R$ defining the gradient flow equations for these Floer homologies are described in \autoref{ft:dualities:floer:notation}.

\subsection{Relations from the 5d ``S-duality'' of HW and GM Theory}
\label{sec:dualities:floer:s-dual}

\subtitle{5d ``S-duality'' of HW Theory}

In \cite[$\S$8.2]{er-2023-topol-n}, we showed how, using Tachikawa's ``S-duality'' of 5d $\mathcal{N} = 2$ SYM with an $S^1$ \cite{tachikawa-2011-s-dualit}, one can obtain Langlands duals of the (i) $\text{HW}_4$-instanton and (ii) $\text{HW}_3$-instanton Floer homologies of (i) loop group $LG$ and (ii) toroidal group $LLG$, when $G$ is nonsimply-laced.
Let us briefly describe how this is done.

First, note that HW theory of $G$-type on $M_3 \times S^1 \times \R$ can be interpreted as HW$_{4}$-theory of $LG$-type on $M_3 \times \R$, which physically realizes an $LG$-HW$_4$-instanton Floer homology of $M_3$.
5d ``S-duality'' of HW theory then implies an equivalence between the partition functions of HW$_{4}$-theory of $LG$-type and HW$_{4}$-theory of $(LG)^{\vee}$-type, where $(LG)^{\vee}$ is Langlands dual to $LG$ at the level of their loop algebras \cite[footnote 27]{er-2023-topol-n}.
Thus, we obtain the following relations amongst Floer homologies involving loop groups \cite[eqn. (8.15)]{er-2023-topol-n}
\begin{equation}
  \label{eq:dualities:s-duality:hw4}
  \begin{tikzcd}[
    row sep=normal,%
    column sep=huge,%
    arrows=leftrightarrow,%
    ampersand replacement=\&,%
    ]
    \&
    \sum_k \text{HF}_{d_k}^{\text{HW}_5\text{-inst}} \big( M_3 \times S^1, G)
    \arrow[d]
    \&
    \\
    \&
    \sum_l \text{HF}_{d_l}^{\text{HW}_4\text{-inst}} \big( M_3, LG \big)
    \arrow[r, "\text{5d ``S-duality''}"] 
    \&
    \sum_l \text{HF}_{d_l}^{\text{HW}_4\text{-inst}} \big( M_3, (LG)^{\vee} \big)
    \, .
  \end{tikzcd}
\end{equation}

Next, note that HW theory of $G$-type on $M_2 \times T^2 \times \R$ can be interpreted as HW$_3$-theory of $LLG$-type on $M_2 \times \R$, which physically realizes an $LLG$-HW$_3$-instanton Floer homology of $M_2$.
5d ``S-duality'' of the underlying 5d HW theory then implies an equivalence between the partition functions of HW$_3$ theory of $LLG$-type and HW$_3$ theory of $L(LG)^{\vee}$-type.
Thus, we obtain the following relations amongst Floer homologies involving toroidal groups \cite[eqn. (8.19)]{er-2023-topol-n}
\begin{equation}
  \label{eq:dualities:s-duality:hw3}
  \begin{tikzcd}[
    row sep=normal,%
    column sep=huge,%
    arrows=leftrightarrow,%
    ampersand replacement=\&,%
    ]
    \&
    \sum_k \text{HF}_{d_k}^{\text{HW}_5\text{-inst}} \big( M_2 \times T^2, G)
    \arrow[d]
    \&
    \\
    \&
    \sum_m \left. \text{HF}_{d_m}^{\text{HW}_3\text{-inst}} \big( M_2, LLG \big) \right|_{X, Y = 0}
    \arrow[r, "\text{5d ``S-duality''}"] 
    \arrow[d, squiggly] 
    \&
    \sum_m \left. \text{HF}_{d_m}^{\text{HW}_3\text{-inst}} \big( M_2, L(LG)^{\vee} \big) \right|_{X, Y = 0}
    \arrow[d, squiggly] 
    \\
    \&
    \sum_m \text{H}^0_{\text{dR}} \big( \mathcal{M}^{LLG_{\C}}_{\text{flat}}(M_2) \big)
    \&
    \sum_m \text{H}^0_{\text{dR}} \big( \mathcal{M}^{L(LG_{\C})^{\vee}}_{\text{flat}}(M_2) \big)
    \, .
  \end{tikzcd}
\end{equation}
Here, (i) $\text{H}^0_{\text{dR}}(\mathcal{M}^{LLG_{\C}}_{\text{flat}}(M_2))$ is a de Rham class of zero-forms in the moduli space $\mathcal{M}^{LLG_{\C}}_{\text{flat}}(M_2)$ of flat $LLG_{\C}$-connections on $M_2$,\footnote{%
  The toroidal group (that we shall denote as $\mathcal{G}$) Floer homologies in \eqref{eq:dualities:s-duality:hw3} are generated by time-invariant, isolated, and non-degenerate flat $\mathcal{G}_{\mathbb{C}}$-connections on $M_2$, i.e., time-invariant, isolated, and non-degenerate $(\mathcal{A}, \overline{\mathcal{A}})$ $\mathcal{G}_{\mathbb{C}}$-connections on $M_2$ satisfying the flatness condition.
  Such $\mathcal{G}_{\mathbb{C}}$-connections correspond to the holomorphic and antiholomorphic coordinates on $\mathcal{M}^{\mathcal{G}_{\mathbb{C}}}_{\text{flat}}(M_2)$, of zero-dimension.
  There is a spectral equivalence between the 3d gauge theory on $M_2 \times \mathbb{R}$ which realizes the aforementioned Floer homologies, and a 1d sigma model on $\mathbb{R}$ with target a zero-dimensional $\mathcal{M}^{\mathcal{G}_{\mathbb{C}}}_{\text{flat}}(M_2)$, interpreted as a Riemannian manifold.
  The spectrum of this 1d sigma model is described by the de Rham cohomology of the target, and since it is zero-dimensional, we would only have de Rham classes of zero-forms.
  \label{ft:dualities:floer:de Rham classes}
}
(ii) the squiggly lines indicate a relation due to a spectral equivalence of the underlying theory that realizes the (co)homology,
and (iii) the subscript ``$X, Y = 0$'' indicates that one is to set to zero, the scalars $X, Y$ that would have arisen from KK reduction of HW theory to HW$_3$ theory.\footnote{%
  The reason for setting to zero the scalars is as follows.
  Note that the HW$_3$-instanton Floer homology, as defined in \cite[$\S$5]{er-2023-topol-n}, is generated by $G_{\C}$-BF configurations on $M_2$, i.e., it involves an $\text{ad}(G_{\C})$-valued scalar field $Z$.
  This field is actually defined through a complexification of $\text{ad}(G)$-valued scalar fields $X$ and $Y$, i.e., $Z \coloneq X + i Y$, that were obtained via KK reduction of HW theory along two circles.
  However, as the interpretation of HW theory of $G$-type as an HW$_3$ theory of $LLG$-type does not involve any KK reduction along circles, there will be no extra scalar fields that could contribute towards the definition of $Z$.
  Thus, the HW$_3$-instanton Floer homology of $LLG$-type would actually be generated by $G_{\C}$-flat configurations on $M_2$, which is the same as $G_{\C}$-BF configurations on $M_2$ with $Z = 0$, i.e., $X, Y = 0$.
  Hence the subscripts in \eqref{eq:dualities:s-duality:hw3}.
  \label{ft:dualities:floer:x and y subscript}
}

\subtitle{5d ``S-duality'' of GM Theory}

In the same way, 5d ``S-duality'' means that from GM theory on $M_3 \times S^1 \times \R$ of nonsimply-laced $G$, we obtain the following relations amongst Floer homologies involving loop groups:
\begin{equation}
  \label{eq:dualities:s-duality:gm4}
  \saveboxed{eq:dualities:s-duality:gm4}{
    \begin{tikzcd}[
      row sep=normal,%
      column sep=huge,%
      arrows=leftrightarrow,%
      ampersand replacement=\&,%
      ]
      \sum_u \text{HHF}_{d_u}^{\text{flat}} \big( M_3 \times S^1, G_{\C})
      \arrow[d]
      \&
      \\
      \sum_v \text{HHF}_{d_v}^{\text{flat}} \big( M_3, LG_{\C} \big)
      \arrow[r, "\text{5d ``S-duality''}"] 
      \&
      \sum_v \text{HHF}_{d_v}^{\text{flat}} \big( M_3, (LG_{\C})^{\vee} \big)
    \end{tikzcd}
  }
\end{equation}

From GM theory on $M_2 \times T^2 \times \R$, we also obtain the following relations amongst Floer homologies involving toroidal groups:\footnote{%
  Notice that the group datum in the Floer homologies of (i) $M_3$ and (ii) $M_2$ are both $G_{\C}$, and not (i) $G_{\HH}$ and (ii) $G_{\OO}$, as was defined in (i) \autoref{sec:m3 x r} and (ii) \autoref{sec:m2 x r}.
  This is due to the fact that there are no KK reductions along the $S^1$'s, and thus no additional fields to complexify the BPS equations to (i) $G_{\HH}$-type on $M_3 \times \R$ or (ii) $G_{\OO}$-type on $M_2 \times \R$.
  \label{ft:dualities:s-duality:group datum of loop and toroidal}
}
\begin{equation}
  \label{eq:dualities:s-duality:gm3}
  \saveboxed{eq:dualities:s-duality:gm3}{
    \begin{tikzcd}[
      row sep=normal,%
      column sep=huge,%
      arrows=leftrightarrow,%
      ampersand replacement=\&,%
      ]
      \sum_u \text{HHF}_{d_u}^{\text{flat}} \big( M_2 \times T^2, G_{\C})
      \arrow[d]
      \&
      \\
      \sum_w \text{HHF}_{d_w}^{\text{flat}} \big( M_2, LLG_{\C} \big)
      \arrow[r, "\text{5d ``S-duality''}"] 
      \arrow[d, squiggly] 
      \&
      \sum_w \text{HHF}_{d_w}^{\text{flat}} \big( M_2, L(LG_{\C})^{\vee} \big)
      \arrow[d, squiggly] 
      \\
      \sum_w \text{H}^0_{\text{dol}} \big( \mathcal{M}^{LLG_{\C}}_{\text{BF}}(M_2) \big)
      \&
      \sum_w \text{H}^0_{\text{dol}} \big( \mathcal{M}^{L(LG_{\C})^{\vee}}_{\text{BF}}(M_2) \big)
    \end{tikzcd}
  }
\end{equation}
Here, $\text{H}^0_{\text{dol}} \big( \mathcal{M}^{LLG_{\C}}_{\text{BF}}(M_2) \big)$ is a Dolbeault (dol) class of zero-forms in the $LLG_{\C}$-BF moduli space $\mathcal{M}^{LLG_{\C}}_{\text{BF}}(M_2)$ of $M_2$.\footnote{%
  Applying the same analysis in \autoref{ft:dualities:floer:de Rham classes} to the configurations which generate the toroidal group (that we shall denote again as $\mathcal{G}$) Floer homologies in \eqref{eq:dualities:s-duality:gm3}, we find that these configurations on $M_2$ correspond to the \emph{antiholomorphic} coordinates on a zero-dimensional $\mathcal{M}^{\mathcal{G}_{\C}}_{\text{BF}}(M_2)$.
  Therefore, the target of the 1d sigma model on $\mathbb{R}$ that is spectrally equivalent to the 3d gauge theory on $M_2 \times \mathbb{R}$ which realizes these toroidal group Floer homologies, is $\mathcal{M}^{\mathcal{G}_{\mathbb{C}}}_{\text{BF}}(M_2)$, of zero-dimension, interpreted as a Kähler manifold.
  The spectrum of this 1d sigma model is described by the Dolbeault cohomology of the target, and since it is zero-dimensional, we would only have Dolbeault classes of zero-forms.
  \label{ft:dualities:floer:dolbeault classes}
}

\subsection{A Web of Langlands Dual Relations Amongst the Floer Homologies}
\label{sec:dualities:floer:diagram}

\begin{figure}[ht]
  \savefig{fig:dualities:floer}{
    \centering
    \begin{adjustbox}{
        max totalsize={\textwidth},%
      }
      \begin{tikzpicture}[%
        auto,%
        on grid,%
        block/.style={draw, rectangle},%
        novel/.style={draw, rectangle, ultra thick},%
        every edge/.style={draw, <->},%
        relation/.style={scale=0.8, sloped, anchor=center, align=center},%
        vertRelation/.style={scale=0.8, anchor=center, align=center},%
        horRelation/.style={scale=0.8, anchor=center, align=center},%
        decoration={snake, pre length=6pt,post length=6pt, amplitude=1pt, segment length=5pt},%
        shorten >=4pt,%
        shorten <=4pt,%
        ]
        \def \verRel {2} 
        \def \horRel {8} 
        \node[block, very thick] (HW) {$\sum \text{HF}_{*}^{\text{HW}_5{\text{-inst}}}(M_4, G)$};
        \node[block, below={\verRel} of HW] (4d-HW) {$\sum \text{HF}_{*}^{\text{HW}_4\text{-inst}}(M_3, G)$};
        \node[block, below={\verRel} of 4d-HW] (3d-HW) {$\sum \text{HF}_{*}^{\text{HW}_3\text{-inst}}(M_2, G)$};
        \node[block, above={\verRel} of HW] (HW-LG) {$\sum \text{HF}_{*}^{\text{HW}_4\text{-inst}}(M_3, LG)$};
        \node[block, above={\verRel} of HW-LG] (HW-LG-vee) {$\sum \text{HF}_{*}^{\text{HW}_4\text{-inst}}(M_3, (LG)^{\vee})$};
        \node[block, below={\verRel} of 3d-HW] (HW-LLG) {$\sum \text{HF}_{*}^{\text{HW}_3\text{-inst}}(M_2, LLG)_{X, Y = 0}$};
        \node[block, below={3*\verRel} of HW-LLG] (HW-LLG-vee) {$\sum \text{HF}_{*}^{\text{HW}_3{\text{-inst}}}(M_2, L(LG)^{\vee})_{X, Y = 0}$};
        \node[below={\verRel} of HW-LLG] (HW-null) {};
        \node[block, right=1 of HW-null] (HW-H-DR) {$\sum \text{H}^0_{\text{dR}}(\mathcal{M}^{LLG_{\C}}_{\text{flat}}(M_2))$};
        \node[block, below={\verRel} of HW-H-DR] (HW-H-DR-vee) {$\sum \text{H}^0_{\text{dR}}(\mathcal{M}^{L(LG_{\C})^{\vee}}_{\text{flat}}(M_2))$};
        \node[block, very thick, right={\horRel} of HW] (GC-flat) {$\sum \text{HHF}_{*}^{\text{flat}}(M_4, {^{L}G}_{\C})$};
        \node[block, below={\verRel} of GC-flat] (GH-flat) {$\sum \text{HHF}_{*}^{\text{flat}}(M_3, {^{L}G}_{\HH})$};
        \node[block, below={\verRel} of GH-flat] (GO-flat) {$\sum \text{HHF}_{*}^{\text{flat}}(M_2, {^{L}G}_{\OO})$};
        \node[block, above={\verRel} of GC-flat] (LGC-flat) {$\sum \text{HHF}_{*}^{\text{flat}}(M_3, L^{L}G_{\C})$};
        \node[block, above={\verRel} of LGC-flat] (LGC-flat-vee) {$\sum \text{HHF}_{*}^{\text{flat}}(M_3, (L^{L}G_{\C})^{\vee})$};
        \node[block, below={\verRel} of GO-flat] (LLGC-flat) {$\sum \text{HHF}_{*}^{\text{flat}}(M_2, LL^{L}G_{\C})$};
        \node[block, below={3*\verRel} of LLGC-flat] (LLGC-flat-vee) {$\sum \text{HHF}_{*}^{\text{flat}}(M_2, L(L^{L}G_{\C})^{\vee})$};
        \node[below={\verRel} of LLGC-flat] (GM-null) {};
        \node[block, left=1 of GM-null] (GM-H-Dol) {$\sum \text{H}^0_{\text{dol}}(\mathcal{M}^{LL^{L}G_{\C}}_{\text{BF}}(M_2))$};
        \node[block, below={\verRel} of GM-H-Dol] (GM-H-Dol-vee) {$\sum \text{H}^0_{\text{dol}}(\mathcal{M}^{L(L^{L}G_{\C})^{\vee}}_{\text{BF}}(M_2))$};
        \draw
        (HW)
        edge[dashed]
        node[vertRelation, left] {$M_4 = M_3 \times \widehat{S}^1$}
        (4d-HW)
        (4d-HW)
        edge[dashed]
        node[vertRelation, left] {$M_3 = M_2 \times \widehat{S}^1$}
        (3d-HW)
        (GC-flat)
        edge[dashed]
        node[vertRelation, right] {$M_4 = M_3 \times \widehat{S}^1$}
        (GH-flat)
        (GH-flat)
        edge[dashed]
        node[vertRelation, right] {$M_3 = M_2 \times \widehat{S}^1$}
        (GO-flat)
        (HW-LLG)
        edge[decorate, <->]
        (HW-H-DR)
        (HW-LLG-vee)
        edge[decorate, <->]
        (HW-H-DR-vee)
        (LLGC-flat)
        edge[decorate, <->]
        (GM-H-Dol)
        (LLGC-flat-vee)
        edge[decorate, <->]
        (GM-H-Dol-vee)
        (HW)
        edge[densely dotted]
        node[vertRelation, left] {$M_4 = M_3 \times S^1$}
        (HW-LG)
        (HW.west)
        edge[bend right,densely dotted]
        node[relation, above] {$M_4 = M_2 \times T^2$}
        (HW-LLG.west)
        (GC-flat)
        edge[densely dotted]
        node[vertRelation, right] {$M_4 = M_3 \times S^1$}
        (LGC-flat)
        (GC-flat.east)
        edge[bend left, densely dotted]
        node[relation, above] {$M_4 = M_2 \times T^2$}
        (LLGC-flat.east)
        (HW-LG)
        edge [loosely dotted]
        node[vertRelation, left] (HW-LG-S-duality) {
          \begin{tabular}{c}
            5d ``S-duality''
            \\
            $G$ nonsimply-laced
          \end{tabular}
        }
        (HW-LG-vee)
        (HW-LLG.south west)
        edge [loosely dotted]
        node[vertRelation, right] {
          \begin{tabular}{c}
            5d ``S-duality''
            \\
            $G$ nonsimply-laced
          \end{tabular}
        }
        (HW-LLG-vee.north west)
        ($(HW-LLG.north west)!0.5!(HW-LLG-vee.north west)$.west)
        edge [bend left, densely dotted]
        node [relation, above, rotate=180] {$M_3 = M_2 \times S^1$}
        (HW-LG-S-duality.west)
        (LGC-flat)
        edge [loosely dotted]
        node[vertRelation, right] (LGC-flat-S-duality) {
          \begin{tabular}{c}
            5d ``S-duality''
            \\
            $G$ nonsimply-laced
          \end{tabular}
        }
        (LGC-flat-vee)
        (LLGC-flat.south east)
        edge [loosely dotted]
        node[vertRelation, left] {
          \begin{tabular}{c}
            5d ``S-duality''
            \\
            $G$ nonsimply-laced
          \end{tabular}
        }
        (LLGC-flat-vee.north east)
        ($(LLGC-flat.south east)!0.5!(LLGC-flat-vee.north east)$.east)
        edge [bend right, densely dotted]
        node [relation, above, rotate=180] {$M_3 = M_2 \times S^1$}
        (LGC-flat-S-duality.east)
        (HW-H-DR)
        edge [loosely dotted]
        (HW-H-DR-vee)
        (GM-H-Dol)
        edge [loosely dotted]
        (GM-H-Dol-vee)
        (HW)
        edge 
        node[relation,above] {$M_4 = M_3 \times M_1$}
        (GC-flat)
        (4d-HW)
        edge 
        (GH-flat)
        (3d-HW)
        edge 
        (GO-flat)
        (HW-LG)
        edge 
        (LGC-flat)
        (HW-LG-vee)
        edge 
        (LGC-flat-vee)
        (HW-LLG)
        edge 
        (LLGC-flat)
        (HW-LLG-vee)
        edge 
        (LLGC-flat-vee)
        (HW-H-DR)
        edge 
        (GM-H-Dol)
        (HW-H-DR-vee)
        edge 
        (GM-H-Dol-vee)
        ;
      \end{tikzpicture}
    \end{adjustbox}
    \caption{Web of Langlands dual relations amongst Floer homologies from HW and GM theory.}
  }
  \label{fig:dualities:floer}
\end{figure}
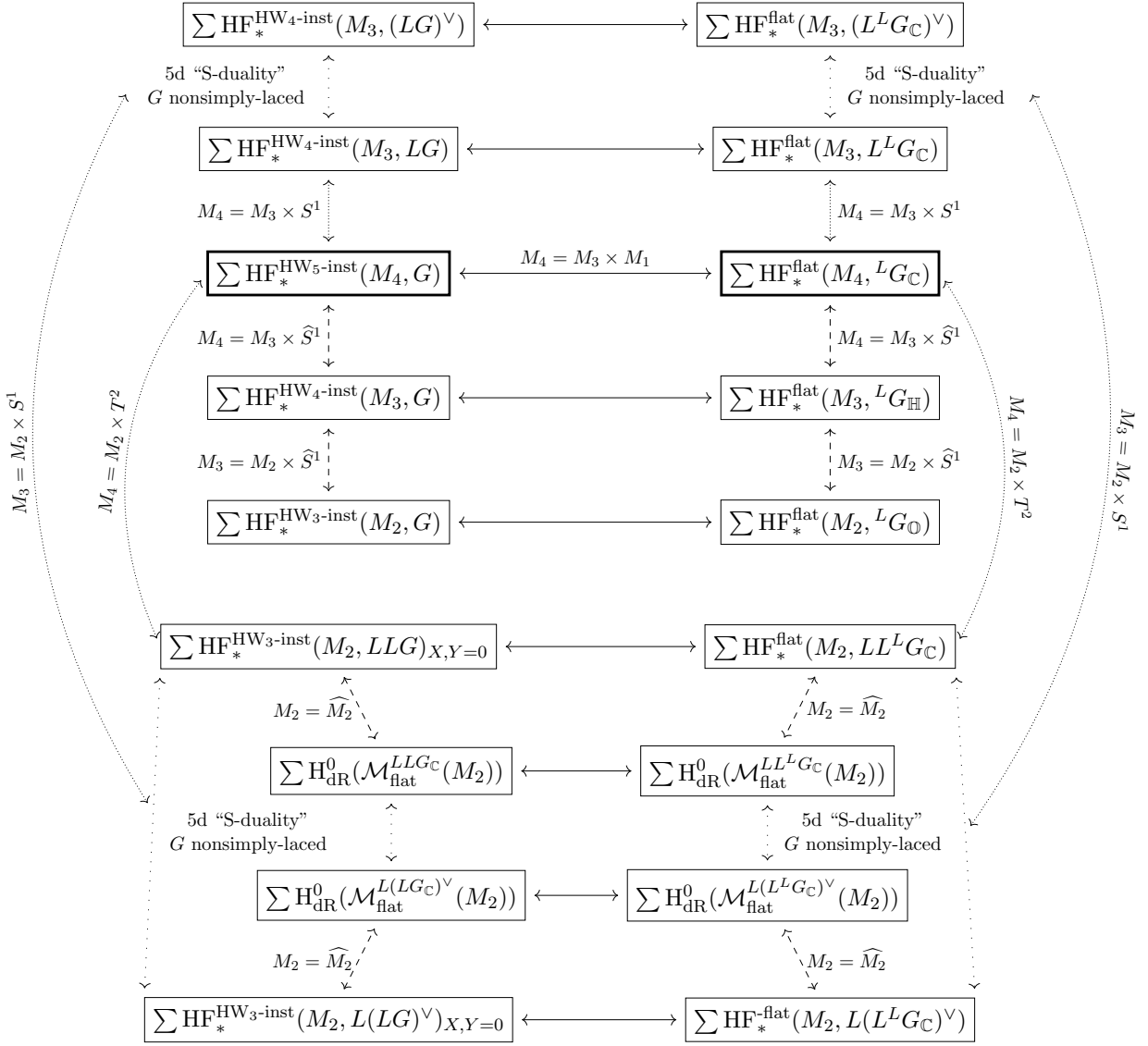

Applying the Langlands duality between HW theory of $G$-type and GM theory of $^{L}G$-type on $M_3 \times M_1 \times \R$ in \autoref{sec:dualities:theories:5d:s-dual} to the results in \autoref{sec:dualities:floer:gauge} and \autoref{sec:dualities:floer:s-dual}, we obtain a web of Langlands dual relations amongst the various gauge-theoretic Floer homologies as shown in \autoref{fig:dualities:floer}, where
(i) the radii of the $\widehat{S}^1$ circles can be varied;
(ii) horizontal undashed lines indicate a relation that is due to the Langlands duality between HW and GM theory;
(iii) dashed lines indicate a relation that is due to an equivalence under dimensional reduction;
(iv) squiggly lines indicate a relation that is due to a spectral equivalence of the underlying theory that realizes the (co)homology;
(v) dotted lines indicate a relation due to an ``S-duality'' of the underlying 5d theory that realizes the Floer homology;
and (vi) finely dotted lines are used to specify the manifolds.

\section{Mirror Symmetry and Langlands Duality of \texorpdfstring{$A_\infty$}{A-infinity}-categories of Floer Homologies}
\label{sec:dualities:cats}

In this section, we will use the results of the previous sections and our previous works in \cite{er-2023-topol-n, er-2024-topol-gauge} to derive a mirror symmetry and Langlands duality of the $A_{\infty}$-categories physically realized by 5d $\mathcal{N} = 2$ HW theory of $G$ and GM theory of $^{L}G$ on $M_3 \times \R^2$, where $M_3$ is a possibly decomposable three-manifold.

\subsection{Mirror Symmetry and Langlands Duality Between Fukaya-Seidel-type and Orlov-type \texorpdfstring{$A_{\infty}$}{A-infinity}-1-categories}
\label{sec:dualities:cats:1-cat:gauge}

\subtitle{HW Theory and a Gauge-theoretic Fukaya-Seidel-type $A_{\infty}$-1-category}

In \cite[$\S$9]{er-2023-topol-n}, we showed how HW theory of $G$-type on $M_3 \times \R^2$ physically realizes a gauge-theoretic Fukaya-Seidel-type (FS-type) $A_{\infty}$-1-category of ($\theta$-deformed) 3d-Hitchin configurations on $M_3$, thereby 1-categorifying the ($\theta$-deformed) $G$-HW$_4$-instanton Floer homology $\text{HF}^{\text{HW}_4\text{-inst}}(M_3, G)$ of $M_3$.

\subtitle{GM Theory and a Gauge-theoretic Orlov-type $A_{\infty}$-1-category}

In \autoref{sec:m3 x r2}, we showed how GM theory of $G$-type on $M_3 \times \R^2$ physically realizes a gauge-theoretic Orlov-type $A_{\infty}$-1-category of ($\theta$-deformed) $G_{\HH}$-BF configurations on $M_3$, thereby 1-categorifying the ($\theta$-deformed) holomorphic $G_{\HH}$-flat Floer homology $\text{HHF}^{\text{flat}}(M_3, G_{\HH})$ of $M_3$.

\subtitle{A Langlands Duality Between the Gauge-theoretic 1-Categories}

Applying the Langlands duality between HW theory of $G$-type and GM theory of $^{L}G$-type on $M_3 \times \R^2$ (see~\autoref{sec:dualities:theories:5d:s-dual} with $M_1 = \R$), we obtain a Langlands dual relation between the gauge-theoretic FS-type $A_{\infty}$-1-category that 1-categorifies $\text{HF}^{\text{HW}_4\text{-inst}}(M_3, G)$ and the gauge-theoretic Orlov-type $A_{\infty}$-1-category that 1-categorifies $\text{HHF}^{\text{flat}}(M_3, {^{L}G_{\HH}})$.

Recall from \autoref{sec:dualities:floer}, in particular~\autoref{fig:dualities:floer}, that the objects of these  1-categories can be identified with each other.
Thus, this relation between 1-categories is indeed a consistent one.

\subtitle{A Langlands Dual Gauge-theoretic Generalization of HMS for LG Models and a Mirror Symmetry of Gauge-theoretic 1-Categories}

At this point, note that a relation between an FS 1-category of \emph{vanishing cycles} and an Orlov 1-category of \emph{singularities} is the statement of HMS for LG models \cite{orlov-2012-landau-ginzb}.
As such, since the objects of our related 1-categories are \emph{gauge-theoretic Floer homologies}, what we actually have is a Langlands dual \emph{gauge-theoretic generalization} of HMS for LG models!

In turn, this means that the Langlands dual relation between the gauge-theoretic FS-type $A_{\infty}$-1-category that 1-categorifies $\text{HF}^{\text{HW}_4\text{-inst}}(M_3, G)$ and the gauge-theoretic  Orlov-type $A_{\infty}$-1-category that 1-categorifies $\text{HHF}^{\text{flat}}(M_3, {^{L}G_{\HH}})$, is \emph{also} a mirror symmetric one!

\subsection{Mirror Symmetry and Langlands Duality Between Fueter and RW \texorpdfstring{$A_{\infty}$}{A-infinity}-2-categories}
\label{sec:dualities:cats:2-cat}

\subtitle{HW Theory and a Fueter $A_{\infty}$-2-category}

In \cite[$\S$7]{er-2024-topol-gauge}, we showed how, via a Heegaard split of $M_3$ along a Heegaard surface $C$, HW theory of $G$-type on $M_3 \times \R^2$ physically realizes a Fueter $A_{\infty}$-2-category of complex-Lagrangian branes $L$ of $\mathcal{M}^G_{\text{H}, \theta}(C, \mathbf{K})$.

\subtitle{GM Theory and an RW $A_{\infty}$-2-category}

In \autoref{sec:m3 x r2:rw}, we showed how, via a Heegaard split of $M_3$ along a Heegaard surface $C$, GM theory of $G$-type  on $M_3 \times \R^2$ physically realizes an RW $A_{\infty}$-2-category of complex-Lagrangian branes $\widehat{L}$ of $\mathcal{M}^G_{\text{H}, \theta}(C, \mathbf{J})$.

\subtitle{A Langlands Duality between the 2-Categories}

Applying the Langlands duality between HW theory of $G$-type and GM theory of $^{L}G$-type on $M_3 \times \R^2$ (see \autoref{sec:dualities:theories:5d:s-dual} with $M_1 = \R$), we obtain a Langlands dual relation between the Fueter $A_{\infty}$-2-category of complex-Lagrangian branes $L$ of $\mathcal{M}^G_{\text{H}, \theta}(C, \mathbf{K})$ and the RW $A_{\infty}$-2-category of complex-Lagrangian branes $\widehat{L}$ of $\mathcal{M}^{^{L}G}_{\text{H}, \theta}(C, \mathbf{J})$.

Since $L$ and $\widehat{L}$ can be identified under the enhanced HMS of the mirror manifolds  $\mathcal{M}^G_{\text{H}, \theta}(C, \mathbf{K})$ and  $\mathcal{M}^{^{L}G}_{\text{H}, \theta}(C, \mathbf{J})$~\cite{hausel-2021-enhan-mirror}, this relation between 2-categories is indeed a consistent one.

\subtitle{A Mirror Symmetry of 2-Categories}

Note that the above Fueter and RW 2-category of branes, defined by a Fueter and constant map, are just higher-categorical generalizations of a Fukaya and Derived 1-category of branes, defined by a holomorphic and constant map, respectively. Indeed, if we were to dimensionally reduce the 3d A/B topological sigma models which physically realize the former, we would end up with the 2d A/B topological sigma models which physically realize the latter, where the 3d Fueter and constant map BPS equations would reduce to the 2d holomorphic and constant map BPS equations.

Thus, as (enhanced) HMS~\cite{hausel-2021-enhan-mirror} relates the Fukaya and Derived 1-category of branes, it would mean that the Langlands dual relation between the Fueter $A_{\infty}$-2-category of complex-Lagrangian branes $L$ of $\mathcal{M}^G_{\text{H}, \theta}(C, \mathbf{K})$ and the RW $A_{\infty}$-2-category of complex-Lagrangian branes $\widehat{L}$ of $\mathcal{M}^{^{L}G}_{\text{H}, \theta}(C, \mathbf{J})$, can \emph{also} be regarded a mirror symmetric one!

\subtitle{A Physical Proof of Bousseau-Doan-Rezchikov's Mirror Conjecture}

At any rate, a relation between a Fueter $A_{\infty}$-2-category and a KRS 2-category of branes in mirror hyperkähler spaces was conjectured by Bousseau-Doan-Rezchikov (B-DR) in \cite{bousseau-2024-holom-floer, doan-2022-holom-floer}.
Via the correspondence between RW $A_{\infty}$-2-categories and KRS 2-categories shown in~\autoref{sec:m3 x r2:rw}, it would also mean that we have a relation between a Fueter $A_{\infty}$-2-category and a KRS 2-category of complex-Lagrangian branes in mirror hyperkähler Hitchin moduli spaces.

In other words, we have a purely physical proof of the B-DR mirror conjecture!

\subsection{Mirror Symmetry and Langlands Duality Between Fueter-type and RW-type \texorpdfstring{$A_{\infty}$}{A-infinity}-2-categories}
\label{sec:dualities:cats:2-cat:gauge}

\subtitle{HW Theory and a Gauge-theoretic Fueter-type $A_{\infty}$-2-category}

In \cite[$\S$3]{er-2024-topol-gauge}, we showed how HW theory of $G$-type on $M_2 \times \R^3$ physically realizes a gauge-theoretic Fueter-type $A_{\infty}$-2-category of ($\theta$-deformed) $G_{\C}$-BF configurations on $M_2$, thereby 2-categorifying the ($\theta$-deformed) $G$-HW$_3$-instanton Floer homology $\text{HF}^{\text{HW}_3{\text{-inst}}}(M_2, G)$ of $M_2$.

\subtitle{GM Theory and a Gauge-theoretic RW-type $A_{\infty}$-2-category}

In \autoref{sec:m2 x r3}, we showed how GM theory of $G$-type on $M_2 \times \R^3$ physically realizes a gauge-theoretic RW-type $A_{\infty}$-2-category of ($\theta$-deformed) $G_{\OO}$-BF configurations on $M_2$, thereby 2-categorifying the ($\theta$-deformed) holomorphic $G_{\OO}$-flat Floer homology $\text{HHF}^{\text{flat}}(M_2, G_{\OO})$ of $M_2$.

\subtitle{A Langlands Duality Between the Gauge-theoretic 2-Categories}

Applying the Langlands duality between HW theory of $G$-type and GM theory of $^{L}G$-type on $M_2 \times \R^3$ (see \autoref{sec:dualities:theories:5d:s-dual}), we obtain a Langlands dual relation between a Fueter-type $A_{\infty}$-2-category that 2-categorifies $\text{HF}^{\text{HW}_3{\text{-inst}}}(M_2, G)$ and an RW-type $A_{\infty}$-2-category that 2-categorifies $\text{HHF}^{\text{flat}}(M_2, {^LG}_{\OO})$.

Recall from \autoref{sec:dualities:floer}, in particular~\autoref{fig:dualities:floer}, that the objects of these 2-categories can be identified with each other.
Thus, this relation between 2-categories is indeed a consistent one.

\subtitle{A Mirror Symmetry of Gauge-theoretic 2-Categories}

Note that the above Fueter and RW-type 2-category of Floer homologies, defined by a Fueter and constant map, are just higher-categorical generalizations of an FS and Orlov-type 1-category of Floer homologies, defined by a holomorphic and constant map, respectively. Indeed, if we were to take a hyperslice of the 3d A/B-twisted LG models which physically realize the former, we would end up with the 2d A/B-twisted LG models which physically realize the latter, where the 3d Fueter and constant map BPS equations would reduce to the 2d holomorphic and constant map BPS equations.

Hence, as a gauge-theoretic generalization of HMS for LG models relates the FS and Orlov-type 1-category of Floer homologies (see \autoref{sec:dualities:cats:1-cat:gauge}), it would mean that the Langlands dual relation between a Fueter-type $A_{\infty}$-2-category that 2-categorifies $\text{HF}^{\text{HW}_3{\text{-inst}}}(M_2, G)$ and an RW-type $A_{\infty}$-2-category that 2-categorifies $\text{HHF}^{\text{flat}}(M_2, {^LG}_{\OO})$, can \emph{also} be regarded as a mirror symmetric one!

\subtitle{A Gauge-theoretic Generalization of the B-DR Mirror Conjecture}

Since the objects of our related categories are \emph{gauge-theoretic Floer homologies}, what we actually have is a gauge-theoretic generalization of our result in \autoref{sec:dualities:cats:2-cat}. In other words, we have a \emph{gauge-theoretic generalization} of the B-DR mirror conjecture!

\subsection{Mirror Symmetry and Langlands Duality of \texorpdfstring{$A_\infty$}{A-infinity}-categories of Floer Homologies}
\label{sec:dualities:cats:diagram}

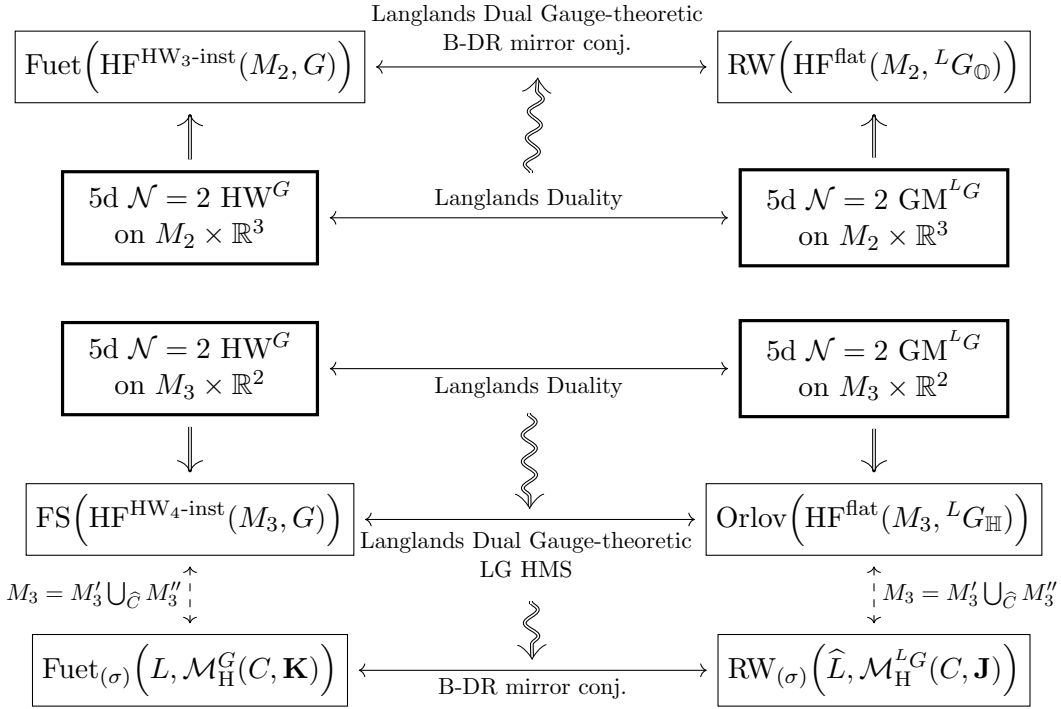
\begin{figure}[ht]
  \savefig{fig:dualities:cats}{
    \centering
    \begin{adjustbox}{
        max totalsize={\textwidth},%
      }
      \begin{tikzpicture}[%
        auto,%
        on grid,%
        block/.style={draw, rectangle},%
        novel/.style={draw, rectangle, ultra thick},%
        every edge/.style={draw, <->},%
        relation/.style={scale=0.8, sloped, anchor=center, align=center},%
        vertRelation/.style={scale=0.8, anchor=center, align=center},%
        horRelation/.style={scale=0.8, anchor=center, align=center},%
        decoration={snake, pre length=3pt,post length=7pt},%
        shorten >=4pt,%
        shorten <=4pt,%
        ]
        \def \verRel {2} 
        \def \horRel {9} 
        \node[block, very thick] (HW-R2)
        {
          \begin{tabular}{c}
            5d $\mathcal{N} = 2$ $\text{HW}^G$
            \\
            on $M_3 \times \R^2$
          \end{tabular}
        };
        \node[block, very thick, right={\horRel} of HW-R2] (GM-R2)
        {
          \begin{tabular}{c}
            5d $\mathcal{N} = 2$ $\text{GM}^{^{L}G}$
            \\
            on $M_3 \times \R^2$
          \end{tabular}
        };
        \node[block, very thick, above={\verRel} of HW-R2] (HW-R3)
        {
          \begin{tabular}{c}
            5d $\mathcal{N} = 2$ $\text{HW}^G$
            \\
            on $M_2 \times \R^3$
          \end{tabular}
        };
        \node[block, very thick, right={\horRel} of HW-R3] (GM-R3)
        {
          \begin{tabular}{c}
            5d $\mathcal{N} = 2$ $\text{GM}^{^{L}G}$
            \\
            on $M_2 \times \R^3$
          \end{tabular}
        };
        \node[block, below={\verRel} of HW-R2] (FS)
        {$\text{FS} \Big( \text{HF}^{\text{HW}_4\text{-inst}}(M_3, G) \Big)$};
        \node[block, below={\verRel} of GM-R2] (Orlov)
        {
          $\text{Orlov} \Big( \text{HHF}^{\text{flat}}(M_3, {^{L}G_{\HH}}) \Big)$
        };
        \node[block, below={\verRel} of FS] (Fueter)
        {$\text{Fuet}_{(\sigma)} \Big( L, \mathcal{M}^G_{\text{H}}(C, \mathbf{K}) \Big)$};
        \node[block, below={\verRel} of Orlov] (RW)
        {
          $\text{RW}_{(\sigma)} \Big( \widehat{L}, \mathcal{M}^{^{L}G}_{\text{H}}(C, \mathbf{J}) \Big)$
        };
        \node[block, above={\verRel} of HW-R3] (Fueter-type)
        {$\text{Fuet} \Big( \text{HF}^{\text{HW}_3\text{-inst}}(M_2, G) \Big)$};
        \node[block, above={\verRel} of GM-R3] (RW-type)
        {$\text{RW} \Big( \text{HHF}^{\text{flat}}(M_2, {^{L}G_{\OO}}) \Big)$};
        \draw
        (HW-R2) edge [->, double]
        (FS)
        (GM-R2) edge [->, double]
        (Orlov)
        (FS) edge[dashed]
        node[vertRelation, left] {$M_3 = M_3' \bigcup_{\widehat{C}} M_3''$}
        (Fueter)
        (Orlov) edge[dashed]
        node[vertRelation, right] {$M_3 = M_3' \bigcup_{\widehat{C}} M_3''$}
        (RW)
        (HW-R2) edge 
        node[relation, below] (Duality) {Langlands Duality}
        (GM-R2)
        (FS) edge 
        node[relation, below] (HMS) {
          \begin{tabular}{c}
            Langlands Dual Gauge-theoretic
            \\
            LG HMS
          \end{tabular}
        }
        (Orlov)
        (Fueter) edge 
        node[relation, below] (B-DR) {B-DR mirror conj.}
        (RW)
        (Duality) edge[decorate, double, -{Classical TikZ Rightarrow[length=2mm]}]
        (HMS)
        (HMS) edge[decorate, double, -{Classical TikZ Rightarrow[length=2mm]}]
        (B-DR)
        (HW-R3) edge 
        node[relation, above] (R3-Duality) {Langlands Duality}
        (GM-R3)
        (HW-R3) edge [->, double]
        (Fueter-type)
        (GM-R3) edge [->, double]
        (RW-type)
        (Fueter-type) edge 
        node[relation, above] (Gauge-B-DR) {
          \begin{tabular}{c}
            Langlands Dual Gauge-theoretic
            \\
            B-DR mirror conj.
          \end{tabular}
        }
        (RW-type)
        (R3-Duality) edge[decorate, double, -{Classical TikZ Rightarrow[length=2mm]}]
        (Gauge-B-DR)
        ;
      \end{tikzpicture}
    \end{adjustbox}
    \caption{Mirror symmetry and Langlands duality of the $A_{\infty}$-categories from HW and GM theory.}
  }
  \label{fig:dualities:cats}
\end{figure}

Finally, we summarize the various mirror symmetric and Langlands dual relations between the $A_{\infty}$-categories obtained hitherto in \autoref{fig:dualities:cats}, where
(i) the size of $\widehat{C}$ can be varied,
(ii) dashed lines indicate a relation that is due to an equivalence under dimensional/topological reduction,
and (iii) wavy lines denote the implications of a relation on other relations.

\appendix

\section{Verifying the Jacobi Identity}
\label{app:jacobi}

In this appendix, we will verify the Jacobi identity of the commutator brackets for the $\mathfrak{g}_{\mathbb{H}}$ quaternionified and $\mathfrak{g}_{\mathbb{O}}$ octonionified Lie algebras introduced in \autoref{sec:m3 x r} and \autoref{sec:m2 x r}, respectively.

\subsection{Verifying the Jacobi Identity for Quaternionified Lie Algebras}
\label{app:jacobi:g-h}

Consider $\mathcal{X}, \mathcal{Y} \in \mathfrak{g}_{\mathbb{C}}$, where $\mathfrak{g}_{\mathbb{C}}$ is the complexification of a real and compact Lie algebra $\mathfrak{g}$; for example, the complexification of $\mathfrak{g} = \mathfrak{su}(n)$ is $\mathfrak{g}_{\mathbb{C}} = \mathfrak{sl}(n, \mathbb{C})$.
Let the complexification be done with the imaginary number $i$, i.e., $\mathcal{X} = \Re(\mathcal{X}) + i \Im(\mathcal{X})$ and $\mathcal{Y} = \Re(\mathcal{Y}) + i \Im(\mathcal{Y})$.

Consider $\mathcal{X} + j \mathcal{Y} \eqcolon \mathscr{Z} \in \mathfrak{g}_{\mathbb{H}}$, where $j$ satisfies the quaternionic relations with $i$, i.e., $ij = - ji = k$ and $i^2 = j^2 = k^2 = -1$.
The $\mathfrak{g}_{\mathbb{H}}$ Lie algebra is equipped with an $\mathbb{R}$-bilinear bracket $[\cdot, \cdot]_{\mathbb{H}}: \mathfrak{g}_{\mathbb{H}} \times \mathfrak{g}_{\mathbb{H}} \rightarrow \mathfrak{g}_{\mathbb{H}}$ that is defined as \cite[eqn.~(2.11)]{tosiaki-2020-quart-compl}
\begin{equation}
  \label{app:eq:g-h:bracket}
  \begin{aligned}
    [\mathscr{Z}_1, \mathscr{Z}_2]_{\mathbb{H}}
    = [\mathcal{X}_1 + j \mathcal{Y}_1, \mathcal{X}_2 + j \mathcal{Y}_2]_{\mathbb{H}}
    &= \mathcal{X}_1 \mathcal{X}_2 - \mathcal{X}_2 \mathcal{X}_1 - \overline{\mathcal{Y}}_1 \mathcal{Y}_2 + \overline{\mathcal{Y}}_2 \mathcal{Y}_1
    \\
    &\quad
    + j \left( \mathcal{Y}_1 \mathcal{X}_2 - \mathcal{Y}_2 \mathcal{X}_1 + \overline{\mathcal{X}}_1 \mathcal{Y}_2 - \overline{\mathcal{X}}_2 \mathcal{Y}_1 \right)
    \, ,
  \end{aligned}
\end{equation}
where an overhead bar indicates a complex conjugation, e.g., $\overline{\mathcal{X}} = \Re(\mathcal{X}) - i \Im(\mathcal{X})$.
It is easy to see from \eqref{app:eq:g-h:bracket} that the bracket is antisymmetric, i.e., $[\mathscr{Z}_1, \mathscr{Z}_2]_{\mathbb{H}} = - [\mathscr{Z}_2, \mathscr{Z}_1]_{\mathbb{H}}$.

We shall now verify the Jacobi identity for this bracket.
To do so, let us introduce the shorthand $\mathscr{J}_{123} \coloneq [\mathscr{Z}_1 , [\mathscr{Z}_2, \mathscr{Z}_3]_{\mathbb{H}}]_{\mathbb{H}}$.
Expanding this in terms of the $\mathcal{X}$'s and $\mathcal{Y}$'s, we get
\begin{equation}
  \label{app:eq:g-h:jacobi}
  \begin{aligned}
    \mathscr{J}_{123}
    &= \mathcal{X}_1 \mathcal{X}_2 \mathcal{X}_3
    + \mathcal{X}_3 \mathcal{X}_2 \mathcal{X}_1
    + \mathcal{X}_2 \overline{\mathcal{Y}}_3 \mathcal{Y}_1
    + \mathcal{X}_1 \overline{\mathcal{Y}}_3 \mathcal{Y}_2
    + \overline{\mathcal{Y}}_2 \mathcal{Y}_3 \mathcal{X}_1
    + \overline{\mathcal{Y}}_1 \mathcal{Y}_3 \mathcal{X}_2
    + \overline{\mathcal{Y}}_2 \overline{\mathcal{X}}_3 \mathcal{Y}_1
    + \overline{\mathcal{Y}}_1 \overline{\mathcal{X}}_3 \mathcal{Y}_2
    \\
    &\quad
    - \mathcal{X}_2 \mathcal{X}_3 \mathcal{X}_1
    - \mathcal{X}_1 \mathcal{X}_3 \mathcal{X}_2
    - \mathcal{X}_1 \overline{\mathcal{Y}}_2 \mathcal{Y}_3
    - \mathcal{X}_3 \overline{\mathcal{Y}}_2 \mathcal{Y}_1
    - \overline{\mathcal{Y}}_1 \mathcal{Y}_2 \mathcal{X}_3
    - \overline{\mathcal{Y}}_3 \mathcal{Y}_2 \mathcal{X}_1
    - \overline{\mathcal{Y}}_1 \overline{\mathcal{X}}_2 \mathcal{Y}_3
    - \overline{\mathcal{Y}}_3 \overline{\mathcal{X}}_2 \mathcal{Y}_1
    \\
    &\quad
    + j \Bigl(
      \mathcal{Y}_1 \mathcal{X}_2 \mathcal{X}_3
      + \mathcal{Y}_3 \mathcal{X}_2 \mathcal{X}_1
      + \mathcal{Y}_2 \overline{\mathcal{Y}}_3 \mathcal{Y}_1
      + \mathcal{Y}_1 \overline{\mathcal{Y}}_3 \mathcal{Y}_2
      + \overline{\mathcal{X}}_1 \mathcal{Y}_2 \mathcal{X}_3
      + \overline{\mathcal{X}}_3 \mathcal{Y}_2 \mathcal{X}_1
      + \overline{\mathcal{X}}_1 \overline{\mathcal{X}}_2 \mathcal{Y}_3
      + \overline{\mathcal{X}}_3 \overline{\mathcal{X}}_2 \mathcal{Y}_1
    \Bigr)
    \\
    &\quad
    - j \Bigl(
      \mathcal{Y}_2 \mathcal{X}_3 \mathcal{X}_1
      + \mathcal{Y}_1 \mathcal{X}_3 \mathcal{X}_2
      + \mathcal{Y}_1 \overline{\mathcal{Y}}_2 \mathcal{Y}_3
      + \mathcal{Y}_3 \overline{\mathcal{Y}}_2 \mathcal{Y}_1
      + \overline{\mathcal{X}}_2 \mathcal{Y}_3 \mathcal{X}_1
      + \overline{\mathcal{X}}_1 \mathcal{Y}_3 \mathcal{X}_2
      + \overline{\mathcal{X}}_2 \overline{\mathcal{X}}_3 \mathcal{Y}_1
      + \overline{\mathcal{X}}_1 \overline{\mathcal{X}}_3 \mathcal{Y}_2
    \Bigr)
    \, .
  \end{aligned}
\end{equation}

From this, we find that
\begin{equation}
  \label{app:eq:g-h:jacobi:identity}
  [\mathscr{Z}_1, [\mathscr{Z}_2, \mathscr{Z}_3]_{\mathbb{H}}]_{\mathbb{H}}
  + [\mathscr{Z}_2, [\mathscr{Z}_3, \mathscr{Z}_1]_{\mathbb{H}}]_{\mathbb{H}}
  + [\mathscr{Z}_3, [\mathscr{Z}_1, \mathscr{Z}_2]_{\mathbb{H}}]_{\mathbb{H}}
  = \mathscr{J}_{123} + \mathscr{J}_{231} + \mathscr{J}_{312}
  = 0
  \, .
\end{equation}
As such, the $\mathfrak{g}_{\mathbb{H}}$ bracket does indeed satisfy the Jacobi identity.

\subsection{Verifying the Jacobi Identity for Octonionified Lie Algebras}
\label{app:jacobi:g-o}

Consider $\mathscr{X}, \mathscr{Y} \in \mathfrak{g}_{\mathbb{H}}$, where the quaternionified Lie algebra $\mathfrak{g}_{\mathbb{H}}$ is the complexification of $\mathfrak{g}_{\mathbb{C}}$; for example, the complexification of $\mathfrak{g}_{\mathbb{C}} = \mathfrak{sl}(n, \mathbb{C})$ is $\mathfrak{g}_{\mathbb{H}} = \mathfrak{sl}(n, \mathbb{H})$ \cite{tosiaki-2020-quart-compl}.
Let the quaternionification be done with the imaginary numbers $\mathbf{e}_{01}, \mathbf{e}_{02}, \mathbf{e}_{03}$ satisfying the quaternionic relations $\mathbf{e}_{0i} \mathbf{e}_{0j} = - \delta_{ij} + \epsilon_{ijk} \mathbf{e}_{0k}$.

Consider $\mathscr{X} + \mathbf{e}_0 \mathscr{Y} \eqcolon \mathsf{Z} \in \mathfrak{g}_{\mathbb{O}}$, where $\mathbf{e}_0$ satisfies the octonionic relations with $\mathsf{e}_{0i}$ as defined in \eqref{eq:m2 x r:octonionic relations}.
Just like the $\mathfrak{g}_{\mathbb{H}}$ Lie algebra, the $\mathfrak{g}_{\mathbb{O}}$ Lie algebra is equipped with an $\mathbb{R}$-bilinear bracket $[\cdot, \cdot]_{\mathbb{O}}: \mathfrak{g}_{\mathbb{O}} \times \mathfrak{g}_{\mathbb{O}} \rightarrow \mathfrak{g}_{\mathbb{O}}$ that we can define as
\begin{equation}
  \label{app:eq:g-o:bracket}
  \begin{aligned}
    [\mathsf{Z}_1, \mathsf{Z}_2]_{\mathbb{O}}
    = [\mathscr{X}_1 + \mathbf{e}_0 \mathscr{Y}_1, \mathscr{X}_2 + \mathbf{e}_0 \mathscr{Y}_2]_{\mathbb{O}}
    &= \mathscr{X}_1 \mathscr{X}_2 - \mathscr{X}_2 \mathscr{X}_1 - \overline{\mathscr{Y}}_1 \mathscr{Y}_2 + \overline{\mathscr{Y}}_2 \mathscr{Y}_1
    \\
    &\quad
    + \mathbf{e}_0 \left( \mathscr{Y}_1 \mathscr{X}_2 - \mathscr{Y}_2 \mathscr{X}_1 + \overline{\mathscr{X}}_1 \mathscr{Y}_2 - \overline{\mathscr{X}}_2 \mathscr{Y}_1 \right)
    \, ,
  \end{aligned}
\end{equation}
where an overhead bar indicates a quaternion conjugation, e.g., $\overline{a + \mathbf{e}_{01} b + \mathbf{e}_{02} c + \mathbf{e}_{03} d} = a - \mathbf{e}_{01} b - \mathbf{e}_{02} c - \mathbf{e}_{03} d$.
It is easy to see from \eqref{app:eq:g-o:bracket} that the bracket is antisymmetric, i.e., $[\mathsf{Z}_1, \mathsf{Z}_2]_{\mathbb{O}} = - [\mathsf{Z}_2, \mathsf{Z}_1]_{\mathbb{O}}$.

We shall now verify the Jacobi identity for this bracket.
To do so, let us introduce the shorthand $\mathsf{J}_{123} \coloneq [\mathsf{Z}_1 , [\mathsf{Z}_2, \mathsf{Z}_3]_{\mathbb{O}}]_{\mathbb{O}}$.
Expanding this in terms of the $\mathscr{X}$'s and $\mathscr{Y}$'s, we get
\begin{equation}
  \label{app:eq:g-o:jacobi}
  \begin{aligned}
    &\mathsf{J}_{123}
    \\
    &= \mathscr{X}_1 \mathscr{X}_2 \mathscr{X}_3
    + \mathscr{X}_3 \mathscr{X}_2 \mathscr{X}_1
    + \mathscr{X}_2 \overline{\mathscr{Y}}_3 \mathscr{Y}_1
    + \mathscr{X}_1 \overline{\mathscr{Y}}_3 \mathscr{Y}_2
    + \overline{\mathscr{Y}}_2 \mathscr{Y}_3 \mathscr{X}_1
    + \overline{\mathscr{Y}}_1 \mathscr{Y}_3 \mathscr{X}_2
    + \overline{\mathscr{Y}}_2 \overline{\mathscr{X}}_3 \mathscr{Y}_1
    + \overline{\mathscr{Y}}_1 \overline{\mathscr{X}}_3 \mathscr{Y}_2
    \\
    &\quad
    - \mathscr{X}_2 \mathscr{X}_3 \mathscr{X}_1
    - \mathscr{X}_1 \mathscr{X}_3 \mathscr{X}_2
    - \mathscr{X}_1 \overline{\mathscr{Y}}_2 \mathscr{Y}_3
    - \mathscr{X}_3 \overline{\mathscr{Y}}_2 \mathscr{Y}_1
    - \overline{\mathscr{Y}}_1 \mathscr{Y}_2 \mathscr{X}_3
    - \overline{\mathscr{Y}}_3 \mathscr{Y}_2 \mathscr{X}_1
    - \overline{\mathscr{Y}}_1 \overline{\mathscr{X}}_2 \mathscr{Y}_3
    - \overline{\mathscr{Y}}_3 \overline{\mathscr{X}}_2 \mathscr{Y}_1
    \\
    &\quad
    + \mathbf{e}_0 \Bigl(
      \mathscr{Y}_1 \mathscr{X}_2 \mathscr{X}_3
      + \mathscr{Y}_3 \mathscr{X}_2 \mathscr{X}_1
      + \mathscr{Y}_2 \overline{\mathscr{Y}}_3 \mathscr{Y}_1
      + \mathscr{Y}_1 \overline{\mathscr{Y}}_3 \mathscr{Y}_2
      + \overline{\mathscr{X}}_1 \mathscr{Y}_2 \mathscr{X}_3
      + \overline{\mathscr{X}}_3 \mathscr{Y}_2 \mathscr{X}_1
      + \overline{\mathscr{X}}_1 \overline{\mathscr{X}}_2 \mathscr{Y}_3
      + \overline{\mathscr{X}}_3 \overline{\mathscr{X}}_2 \mathscr{Y}_1
    \Bigr)
    \\
    &\quad
    - \mathbf{e}_0 \Bigl(
      \mathscr{Y}_2 \mathscr{X}_3 \mathscr{X}_1
      + \mathscr{Y}_1 \mathscr{X}_3 \mathscr{X}_2
      + \mathscr{Y}_1 \overline{\mathscr{Y}}_2 \mathscr{Y}_3
      + \mathscr{Y}_3 \overline{\mathscr{Y}}_2 \mathscr{Y}_1
      + \overline{\mathscr{X}}_2 \mathscr{Y}_3 \mathscr{X}_1
      + \overline{\mathscr{X}}_1 \mathscr{Y}_3 \mathscr{X}_2
      + \overline{\mathscr{X}}_2 \overline{\mathscr{X}}_3 \mathscr{Y}_1
      + \overline{\mathscr{X}}_1 \overline{\mathscr{X}}_3 \mathscr{Y}_2
    \Bigr)
    \, .
  \end{aligned}
\end{equation}

From this, we find that
\begin{equation}
  \label{app:eq:g-o:jacobi:identity}
  [\mathsf{Z}_1, [\mathsf{Z}_2, \mathsf{Z}_3]_{\mathbb{O}}]_{\mathbb{O}}
  + [\mathsf{Z}_2, [\mathsf{Z}_3, \mathsf{Z}_1]_{\mathbb{O}}]_{\mathbb{O}}
  + [\mathsf{Z}_3, [\mathsf{Z}_1, \mathsf{Z}_2]_{\mathbb{O}}]_{\mathbb{O}}
  = \mathsf{J}_{123} + \mathsf{J}_{231} + \mathsf{J}_{312}
  = 0
  \, .
\end{equation}
As such, the $\mathfrak{g}_{\mathbb{O}}$ bracket does indeed satisfy the Jacobi identity.

\printbibliography%

\end{document}